\documentclass[usenatbib,usegraphicx]{mn2e}
\input epsf
\usepackage{graphics}
\usepackage{epsfig}
\usepackage{url}

\newcommand{\Min}{${}^{\prime}$}
\newcommand{\Ha} {H$\alpha$}
\newcommand{\ha} {H$\alpha$}

\newcommand{\Deg}{${}^{\circ}$}

\newcommand{\Sun}{${}_{\odot}$}

\newcommand{\kms}{~km~s$^{-1}$}
\newcommand{\kmsMpc}{km~s$^{-1}~$Mpc$^{-1}$}
\def\degr{\hbox{$^\circ$}}
\def\deg{\hbox{$^\circ$}}
\newcommand{\PVM}{position-velocity diagram}
\newcommand{\pvm}{position-velocity diagram}
\newcommand{\PVMs}{position-velocity diagrams}

\newcommand{\PA}{position angle}

\newcommand{\PAs}{position angles}
\newcommand{\vfs}{velocity fields}
\newcommand{\vf}{velocity field}
\newcommand{\VF}{velocity field}
\newcommand{\VFs}{velocity fields}
\newcommand{\rc}{rotation curve}
\newcommand{\SNR}{signal-to-noise ratio}
\newcommand{\snr}{signal-to-noise ratio}
\newcommand{\RC}{rotation curve}
\newcommand{\rcs}{rotation curves}
\newcommand{\RCs}{rotation curves}
\newcommand{\FOV}{field-of-view}
\newcommand{\TF}{Tully-Fisher}

\title[GHASP : An \Ha~kinematic survey of spiral and irregular galaxies - VI.]
     {GHASP : An \Ha~kinematic survey of spiral and irregular galaxies
- VI. New \ha~data cubes for 108 galaxies.}

\author[B. Epinat et al.]
{Epinat, B.$^{1}$, Amram, P.$^{1}$, Marcelin M.$^{1}$, Balkowski
C.$^{2}$, Daigle, O.$^{1,3}$,
\newauthor Hernandez, O.$^{3,1}$, Chemin, L.$^{2}$, Carignan, C.$^{3}$, Gach, J.-L.$^{1}$, Balard, P.$^{1}$\\
$^{1}$Laboratoire d'Astrophysique de Marseille, OAMP, Universit\'e
de Provence \&
CNRS, 2 Place Le Verrier, 13248 Marseille Cedex 04 France\\
$^{2}$GEPI, Observatoire de Paris-Meudon, Universit\'e Paris VII,
5 Place Jules Janssen, 92195 Meudon, France.\\
$^{3}$LAE et Observatoire du mont M\'egantic, Universit\'e de
Montr\'eal, C. P. 6128 succ. centre ville, Montr\'eal, Qu\'ebec,
Canada H3C 3J7\\}

\date{Accepted. Received; in original form }

\pubyear{2008}
\begin{document}
\maketitle

\label{firstpage}

\begin{abstract}
We present the Fabry-Perot observations obtained for a new set of 108 galaxies in the frame of the GHASP survey (Gassendi HAlpha survey of SPirals). The GHASP survey consists of 3D \ha~data cubes for 203 spiral and irregular galaxies, covering a large range in morphological types and absolute magnitudes, for kinematics analysis. The new set of data presented here completes the survey. The GHASP sample is by now the largest sample of Fabry-Perot data ever published. The analysis of the whole GHASP sample will be done in forthcoming papers. Using adaptive binning techniques based on Vorono\"i tessellations, we have derived \ha~data cubes from which are computed \ha~maps, radial \VFs~as well as residual \VFs, \PVMs, \RCs~and the kinematical parameters for almost all galaxies.
Original improvements in the determination of the kinematical parameters, \RCs~and their uncertainties have been implemented in the reduction procedure. This new method is based on the whole 2D \VF~and on the power spectrum of the residual \VF~rather than the classical method using successive crowns in the \VF.
Among the results, we point out that morphological \PAs~have systematically higher uncertainties than kinematical ones, especially for galaxies with low inclination. Morphological inclination of galaxies having no robust determination of their morphological \PA~cannot be constrained correctly. Galaxies with high inclination show a better agreement between their kinematical inclination and their morphological inclination computed assuming a thin disk. The consistency of the velocity amplitude of our \RCs~have been checked using the \TF~relationship. Our data are in good agreement with previous determinations found in the literature. Nevertheless, galaxies with low inclination have statistically higher velocities than expected and fast rotators are less luminous than expected.





\end{abstract}


\section{Introduction}
\label{intro}


As it is nowadays largely admitted, 2D \VFs~allow the computation of 1D \RCs~in a
more robust way than long slit spectrography. Indeed, first of
all, the spatial coverage is larger and moreover, the kinematical
parameters are determined \textit{a posteriori} instead of \textit{a priori}
in long slit spectrography.
In that context, we have undertaken the kinematical 3D GHASP survey (acronym for Gassendi \ha~survey of SPirals).
The GHASP survey consists of a sample of 203 spiral and irregular galaxies, mostly located in nearby low density environments, observed with a scanning Fabry-Perot for studying their kinematical and dynamical properties through the ionized hydrogen
component.

Studying the links between parameters reflecting the dynamical
state of a galaxy will help us to have a better understanding of
the evolution of galaxies. This sample has been constituted in
order:
\begin{enumerate}
  \item to compute the local \TF~relation;
  \item to compare the kinematics of galaxies in
different environments (field, pairs, compact groups, galaxies in
cluster) for discriminating secular evolution from an external
origin (e.g. \citealp{Garrido:2005});
  \item to study the distribution of
luminous and dark halo components along the Hubble sequence for high and low surface brightness galaxies, for a
wide range of luminosities in combining the optical data with the
radio ones (e.g. \citealp{Spano:2008,Barnes:2004});
  \item to model the effect of non axisymmetric
structures like bars, spiral arms, oval distortions, lopsidedness
in the mass distribution using both N-body + hydrodynamic
numerical simulations (e.g. Hernandez et al. in preparation), kinemetric
analysis (e.g. \citealp{Krajnovic:2006}) and Tremaine-Weinberg method to measure the bar, spiral and inner structure pattern speeds (\citeauthor{Hernandez:2005a} 2005a);
  \item to analyze the gaseous velocity dispersion
and to link it with the stellar one;
  \item to create templates \RCs~(e.g. \citealp{Catinella:2006,Persic:1991,Persic:1996}) and templates \vfs;
  \item to map the 2D mass distribution using 2D \vf, broad band imagery
and spectrophotometric evolutionary models;
  \item to search for links between the kinematics (shape of
\rcs, angular momentum, ...) and the other physical properties of galaxies like star
formation rate (e.g. comparison with star forming galaxies like
blue compact galaxies, ...);
  \item to produce a reference sample of nearby galaxies to compare to the kinematics of high redshift galaxies \citep{Puech:2006,Epinat:2007}. Indeed, it is necessary to disentangle the effects of galaxy evolution from spatial (beam smearing) and spectral resolution effects.
\end{enumerate}

This paper is the sixth of a series called hereafter Paper I to V
\citep{Garrido:2002,Garrido:2003,Garrido:2004,Garrido:2005,Spano:2008} presenting the data
obtained in the frame of the GHASP survey. The data gathered with
the seven first observing runs have been published from Paper I to
IV. Dark matter distribution in a sub-sample of 36 spiral galaxies
have been presented in Paper V. This paper presents the last
unpublished 101 \ha~data cubes of the GHASP survey. It includes
108 galaxies (seven data cubes contain two galaxies), providing 106 \VFs~and 93 \RCs~resulting from
observational runs eight to fourteen. This represents the largest set of galaxies observed with Fabry-Perot techniques ever published in the same paper (\citealp{Schommer:1993} sample consists of 75 cluster galaxies in the southern hemisphere observed with Fabry-Perot techniques and was the largest sample published to date). Including the previous papers (Paper I to IV), the GHASP survey
totalizes Fabry-Perot data for 203 galaxies observed from 1998 to
2004. The GHASP sample is by now the largest sample of Fabry-Perot data ever published.

In section \ref{data}, the selection criteria of the GHASP sample, the
instrumental set-up of the instrument and the data reduction
procedure are described. In section \ref{analysis}, different
momenta of the data cubes are presented as well as the new method
to build the \RCs~and to determine the uncertainties. An analysis
of the residual \VFs~and of the kinematical parameters is thus
given. In section \ref{tullyfisher}, the \TF~relation is
plotted for the GHASP galaxies presented in this paper. In section
\ref{conclusion}, we give the summary and conclusions. In Appendix
\ref{method}, we present some details on the method used to
compute the \RCs. In Appendix \ref{notes}, the comments for each
individual galaxy are given. In Appendix \ref{tables}, the
different tables are given while in Appendix \ref{maps} the
individual maps and \PVM s are shown. The \RCs~are finally
displayed in Appendix \ref{rc} while the numerical tables
corresponding to the \RCs~are given in Appendix \ref{rc_tables}.

When the distances of the galaxies are not known, a Hubble
constant H$_{0}$=75\kms~Mpc$^{-1}$ is used throughout this paper.

\section{Observations and Data Reduction}
\label{data}
\subsection{The GHASP sample}

The GHASP survey was originally selected to be a sub-sample
complementing the radio survey WHISP (Westerbork survey of HI in
SPirals galaxies) providing HI distribution and velocity maps for
about 400 galaxies (\url{http://www.astro.rug.nl/~whisp}). The first
set of GHASP galaxies was selected from the first WHISP website
list but some of them have never been observed by WHISP. Thus only
130 galaxies have finally been observed in both surveys. The
comparison between the kinematics of neutral and ionized gas
coming from the GHASP and the WHISP datasets is possible for a
sub-sample of 31 dwarf galaxies studied by \citet{Swaters:2002}
and for another sub-sample of 19 early-type galaxies analyzed by
\citet{Noordermeer:2005}, the remaining part of the WHISP sample
being yet unpublished but most of the HI maps are nevertheless
available on the WHISP website.

Most of the GHASP targets were chosen in nearby low-density
environments. Nevertheless, some galaxies in pairs or in low
density groups have been observed mainly when they were selected
by WHISP (see individual comments in Appendix \ref{notes}). Seven galaxies of the GHASP sample are located in
nearby clusters (UGC 1437 in the cluster Abell 262, UGC 4422 and 4456 in the
Cancer cluster, UGC 6702 in the cluster Abell 1367, UGC 7021, 7901 and 7985 in the
Virgo cluster). More Virgo cluster galaxies observed with the same instrument have been published elsewhere \citep{Chemin:2006}.

\begin{figure}
\begin{center}
\includegraphics[width=7.5cm]{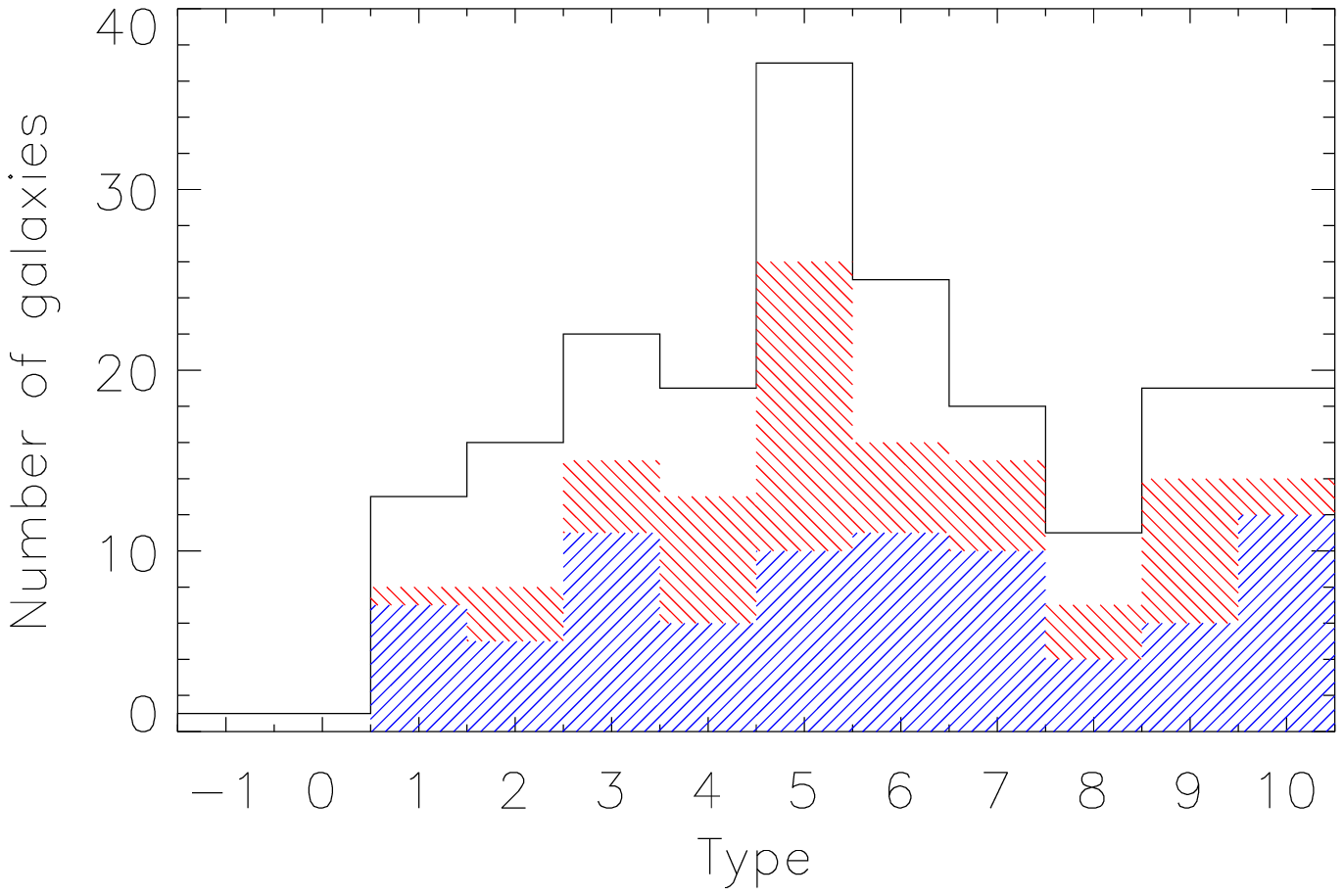}
\includegraphics[width=7.5cm]{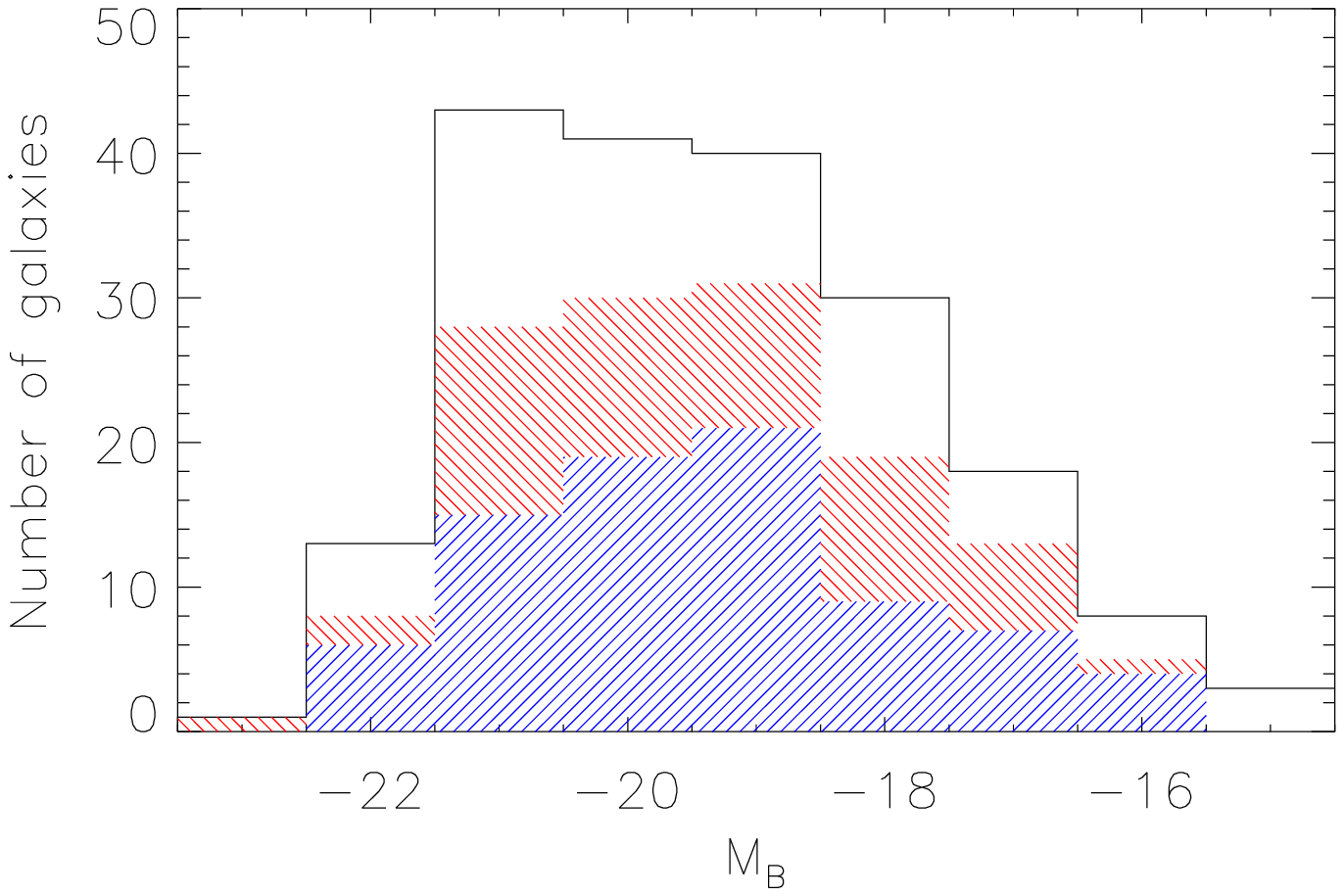}
\includegraphics[width=7.5cm]{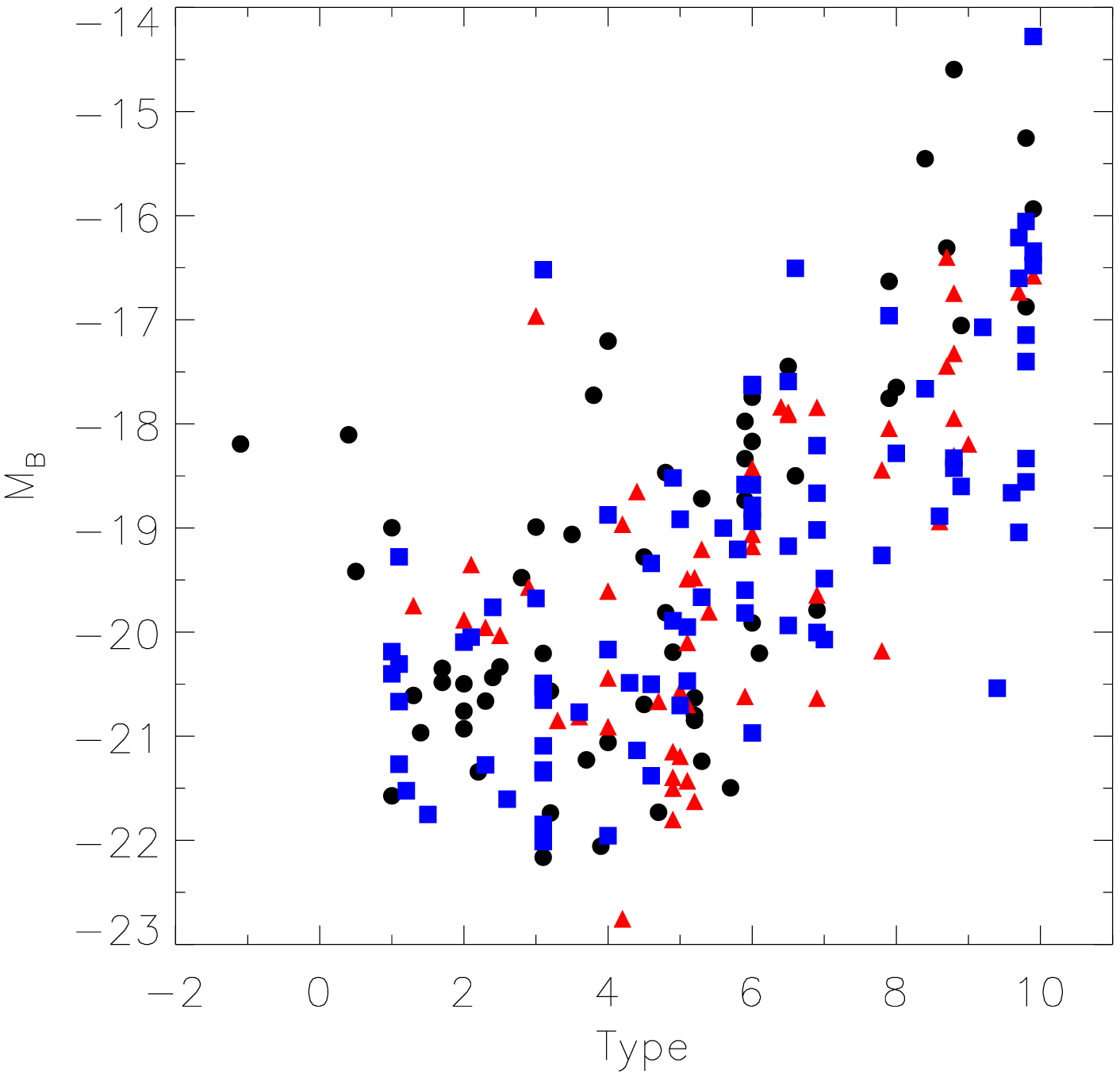}
\end{center}
\caption{\textbf{Top:} distribution of morphological type for almost all
of the GHASP sample (201/203 galaxies). \textbf{Middle:} distribution of
the absolute B-band magnitude for almost all of the GHASP sample
(198/203 galaxies). For both top and middle, the blue hash, red
hash and residual white represent respectively the strongly
barred, the moderately barred and the non-barred galaxies. \textbf{Bottom:}
distribution for almost all of the GHASP sample (198/203 galaxies)
in the "magnitude-morphological type" plane distinguishing
strongly barred (blue squares), moderately barred (red triangles)
and unbarred galaxies (black circles).} \label{sample}
\end{figure}

Figure \ref{sample} displays the distribution of the whole GHASP
survey in the magnitude-morphological type plane. We have found in
the literature measurements for both M$_{B}$ and Hubble type for
198 galaxies (over the 203 galaxies). Among the sample of 203
galaxies, 83 are strongly barred galaxies (SB or IB) and 53 are
moderately barred galaxies (SAB or IAB). The GHASP sample provides a
wide coverage in galaxy luminosities (-16$\leq$M$_{b}$$\leq$-22),
thus in galaxy masses (10$^{9}$M$\odot$ -- 5 10$^{11}$M$\odot$)
and in morphological types (from Sa to Irregular). The well-known
relation for spiral and irregular galaxies between the
morphological type and the absolute magnitude is observed through
the GHASP sample. Within the dispersion of this relation, the 203
GHASP galaxies are reasonably distributed through the plane down
to low magnitudes ($\geq$-16) and to early-types spiral ($\leq$0,
Sa). The whole GHASP dataset of galaxies is reasonably
representative of the LEDA sample (see Paper IV).

The journal of the observations for the 107 new galaxies is given in
Table \ref{tablelog}. The 108th galaxy, UGC 11300 has been
observed for the third time in order to check the consistency of
the new data reduction method (the second observation published in
Paper IV was already done in order to compare the new GaAs camera
with the "S20" photocathode), leading to 107 new galaxies. Note
that the right ascension and the declination given in table
\ref{tablelog} are the coordinates of the kinematical center (and
not the morphological ones given in HyperLeda or in LEDA, except if stated otherwise).

\subsection{The instrumental setup}

In order to map the flux distribution and the \VFs~of the sample of galaxies, high spectral resolution 3D data cubes in the \Ha~line have been obtained. This has been achieved using a focal reducer containing the scanning Fabry-Perot
interferometer attached at the Cassegrain focus of the 1.93 m OHP
telescope (Observatoire de Haute Provence).
The instrument principles and characteristics are the same as for
papers I, II, III, IV and V. The detector, a new generation image
photon counting system (IPCS) is the same as the one used for
Paper IV (with a GaAs photocathode). The pixel size is 0.68\arcsec
(however the angular resolution of our data is limited by the
seeing, about $\sim$3\arcsec, e.g. Table \ref{tablelog}), the field of view is 5.8 square
arcmin and the velocity sampling is $\sim$5\kms~(for a resolution of $\sim$10\kms).

\subsection{The data reduction}

The Fabry-Perot technique provides an \ha~profile inside each
pixel, so that a typical \VF~of a GHASP galaxy contains thousands
of velocity points. For most of the galaxies observed with GHASP,
the velocity field is not limited to the HII regions but covers
most of the diffuse emission of the disk, as can be seen on the
figures. The detection limit of our device is about 10$^{-18}$ erg
cm$^{-2}$ s$^{-1}$ arcsec$^{-2}$, with a S/N ratio between 1 and 2
for a typical 2 hours exposure time according to figure 2 of \citet{Gach:2002}. This insures a good detection of the \ha~diffuse
emission of the disk for most of the galaxies since most of the
\ha~emission found below 1.6 10$^{-16}$ erg cm$^{-2}$ s$^{-1}$
arcsec$^{-2}$ may be considered as filamentary and/or diffuse
according to \citet{Ferguson:1996}. The way to derive the
different moment maps of the 3D data cube (\ha~line maps and \ha~\VFs) are explained in D2006. The \ha~image is a pure
monochromatic image of the galaxy (continuum and [NII] free).

In a few cases, when the velocity amplitude is comparable to or
higher than the width of the interference filter, its transmission
is not necessarily centered on the systemic velocity and one side
of the galaxy may be better transmitted than the other side,
leading to an artificial asymmetry in the intensity of the
\ha~emission (see Paper IV for additional explanations as well as in individual comments on each galaxy in Appendix \ref{notes}).

The data processing and the measurements of the kinematical
parameters are different from those used in Paper I to V. The data
processing used in this paper is basically the same as the one
described by \citeauthor{Daigle:2006b} (2006b) (hereafter D2006). One of the
main improvements implemented in this data processing is the use
of adaptive spatial binning, based on the 2D-Vorono\"i tessellations
method applied to the 3D data cubes, allowing to optimize the
spatial resolution to the \SNR. Let us also mention that when
enough stars or bright HII regions were available in the field of
view, we corrected the observation from telescope drift (or
instrumental flexures) when necessary. Hereafter, we just point
out the main difference between the method described in D2006 and
the method used in this paper.

The main difference is the criterion used to fix the size of the
bins. Indeed, with the spatial adaptive binning technique, a bin
is accreting new pixels until it has reached a given criterion
given \emph{a priori}. In D2006, the criterion is the \SNR~of the
emission line within the bin. For each bin, the noise is
determined from the r.m.s of the continuum (the line free region
of the whole spectrum). The \SNR~is thus the ratio between the
flux in the line and the r.m.s. in the continuum. While for the
data described in D2006, the number of channels scanned is large
enough (36 channels) to determine properly the noise in the
continuum, this is not anymore the case here where the number of
channels scanned is smaller (24 channels). For the same given
spectral resolution (fixed by the interference order and the
Finesse of the Fabry-Perot interferometer used), the ratio between
the number of channels containing the continuum and the number of
channels containing the line is thus larger (by a factor 3/2) in
D2006 than here. Furthermore, the criterion in D2006 is not
relevant anymore for the GHASP data. Instead of the \SNR, the
criterion used here is simply the square root of the flux in the
line, that is an estimate of the Poisson noise in the line flux.

A second major improvement is the suppression of the ghosts due to
reflection at the interfaces air/glass of the interferometer
(\citealp{Georgelin:1970}). The reduction routine takes into account the
front reflection (between the interference filter and the
interferometer) and the back reflection (between the
interferometer and the camera detector window). The ghosts are
calibrated and then subtracted thanks to bright stars.

Another major improvement is an automatic cleaning of the \VFs. The outskirts of a galaxy, where there is no more diffuse
\ha~emission, has to be delimited. Due to residual night sky lines
and background emissions (after subtraction), adaptive binning
produces large sky bins with a given \SNR~or given flux. These
bins containing only sky emission are separated from the bins of
the galaxy thanks to a velocity continuity process. The \VF~is divided in several regions where the velocity difference between contiguous bins is lower than a given cutoff value. The regions with too low monochromatic flux and too large bins are erased. The given cutoff fixed \emph{a priori} is about one tenth of the total amplitude of the \VF~(let's say 30\kms~for a \VF~with an overall amplitude of $\sim$300\kms).




\textsc{Karma} (\citealp{Gooch:1996}) and its routine \textsc{Koords} have
been used to compute the astrometry. XDSS Blue band images or XDSS
Red band images when blue image was not available are displayed
(see individual captions in Figures D1 to
D106). Systematic comparison between these broad-band
images and the field stars in high resolution continuum images
(with no adaptive binning) were made in order to find the correct
World Coordinate System for each image.

\section{Data Analysis}
\label{analysis}

\subsection{Different maps from the 3D data cube}

For each galaxy, in Appendix \ref{maps}, from Figure D1
to D106, we present up to five frames per figure: the
XDSS blue (or red) image (top/left), the \ha~\VF~(top/right), the \ha~monochromatic image (middle/left, eventually
the \ha~residual \VF~(middle/right) and finally the
\PVM~along the major axis (bottom) when it can be computed.
The white and black cross indicates the center used for the
kinematic analysis (given in Table \ref{tablelog}, e.g. Appendix \ref{method} for determination) while the black
line traces the kinematical major axis deduced from the
\VF~analysis (e.g. section \ref{vfanalysis}) or the morphological
one (taken from HyperLeda) when no position angle of the
kinematical major axis could be derived using the kinematic (e.g. Table \ref{tablemod}).
This line ends at the radius D$_{25}$/2 corresponding to the
isophotal level 25 mag arcsec$^{-2}$ in the B-band (given in Table
\ref{tabletf}) in order to compare the \VF~extent with the optical
disk of the galaxies. Position-velocity diagrams are computed
along the axis defined by this black line, using a virtual slit
width of seven pixels, and the red line on the \PVM~is the \RC~deduced from the model \VF~(see next section) along this virtual slit. When no fit is satisfying (generally because of poor \snr), we used the real \VF~instead of the model (see individual captions in Figures D1 to D106).
The \RCs~are found in Appendix \ref{rc} (figures) and \ref{rc_tables} (tables). Colour version of the \RCs~in Appendix \ref{rc} are only available online. Rotation curves are computed and
displayed following the method described in section
\ref{vfanalysis}. These figures are also available on the
Web site of GHASP:\\
\url{http://FabryPerot.oamp.fr}.\\
In order to illustrate the printed version of the paper we have chosen to display the diversity through four galaxies having different morphological types, see Appendix \ref{ugc3740} (UGC 3740, SAB(r)c pec), \ref{ugc4820} (UGC 4820, S(r)ab), \ref{ugc5786} (UGC 5786, SAB(r)b), \ref{ugc7154} (UGC 7154, SBcd). The Appendix \ref{maps} maps of the other galaxies are only available on line. 

Only the first page of Appendix \ref{rc_tables} that contains the tables corresponding to the \RCs~of the two first galaxies is displayed on the printed version of the paper, the remaining part of Appendix \ref{rc_tables} being available on line.

\subsection{Construction of the Rotation Curves and Determination of the Uncertainties}
\label{vfanalysis}

A new automatic fitting method has been developed to derive
automatically a rotation curve from the 2D velocity field. This
method makes the synthesis between (i) the method used in Paper I
to IV, (ii) the method based on tilted-ring models found for
instance in the \textsc{ROTCUR} routine of \textsc{Gipsy} \citep{Begeman:1987} and (iii) the method used by \citet{Barnes:2003}.
\par
Warps are mainly seen in galactic disks at radii R $>$ R$_{opt}$.
Tilted-ring models have been developed to model the distribution
of neutral hydrogen for which warps of the HI disk may be more or less
severe. In case of a warp, a monotonic change of the major axis
position angle ($PA$) and of the inclination ($i$) is observed.
\par
On the other hand, within the optical disk, the kinematic
parameters $PA$ and $i$ do not vary significantly and change
monotonically with the radius (Paper I to IV and \citeauthor{Hernandez:2005} 2005b). The variation of $PA$ and $i$ with the radius is more
likely due to non circular motions in the plane of the disk (e.g.
bars) than to motions out of the plane (like warps) and looks like
oscillations around a median value. Thus, we do not allow $PA$ nor
$i$ to vary with the radius.
\par
Using tilted-ring models, the errors on the parameters are
the dispersion of the kinematical parameters over the rings. The
method developed here uses the whole residual \VF~to estimate the
dispersion induced by non circular motions and not only the
segmented information within each ring as it is the case in
tilted-ring models.
\par
Our fitting method is similar to \citet{Barnes:2003} method.
Two differences may nevertheless be pointed out. A minor
difference is that they use a non parametric profile while we fit
an analytic function (more details on the building of the \RC~are
given in Appendix \ref{method}). The major improvement is the
computation of the kinematic uncertainties. Indeed, the
statistical uncertainty on the fit is unrealistically small
\citep{Barnes:2003}, because the noise on the data is
considered as a blank random noise. That is not the case because
the noise in the residual \VF~is mainly due to non circular
motions (bar, oval distortions, spiral arms, local inflows and
outflows, ...) and to the intrinsic turbulence of the gas that
have characteristic correlation lengths. In order to take it into
account, we compute the errors with the power spectrum of the
residual velocity field, applying a Monte-Carlo method (see
Appendix \ref{method}).
\par
Rotation curves for the barred galaxies of our sample have been
plotted without correction for non-circular motions along the
bar.
\par
The \RCs~are sampled with rings. Within the transition
radius (defined in Appendix \ref{method}), the width of the rings is set to
match half the seeing. Beyond that radius, each ring contains from
16 to 25 velocity bins.
\par
The curves are plotted with both sides superimposed in the same
quadrant, using different symbols for the receding (crosses) and
approaching (dots) side (with respect to the center). The black
vertical arrow on the x-axis represents the radius $D_{25}/2$
while the smaller grey arrow on the x-axis represents the
transition radius, always smaller than $D_{25}/2$ by definition.
\par
For galaxies seen almost edge-on (inclination higher than
75$\degr$) our model does not describe accurately the rotation of
a galaxy since the thickness and the opacity of the disk cannot be
neglected anymore. Indeed, on the one hand it is well known that,
due to inner galactic absorption, edge-on galaxies tend to display
smoother inner \VF~and \RC~gradients than galaxies with low or
intermediate inclinations and, on the other hand, due to the
actual thickness of the disk, using a simple rotation model in the
plane of the galaxy disk, motion out of the disk are wrongly
interpreted as circular motions in the disk. As a consequence,
for most of highly inclined galaxies, the fit converges towards
unrealistic low inclination values, leading to modelled \VFs~and
\RCs~having too high velocity amplitudes. Thus, for NGC 542, UGC
5279, UGC 5351, UGC 7699, UGC 9219, UGC 10713 and UGC 11332, no
\RC~has been plotted. For them, the \pvm~gives a more suitable information than the \RC~and allows to follow the peak-to-peak or peak-to-valley velocity distribution along the major axis.


\subsection{Residual velocity fields}
\label{section_residuals}

As detailed in the previous paragraph and in Appendix
\ref{method}, the main assumption necessary to derive a \RC~from
the observed \VF~is that rotation is dominant and that all non circular
motions are not part of a large-scale pattern. The five
kinematical parameters computed from the \VF~to draw the \RC~are
determined from different symmetry properties of the radial \VF.
The influence of small errors in these parameters is to produce
patterns with characteristic symmetries in the residual \VF. This
was first illustrated by \citet{Warner:1973} and by \citet{van-der-Kruit:1978}. In their schematic representation of the residual
motions in disk galaxies (the modelled \VF~computed from the
\RC~has been subtracted to the observed \VF), a bad determination
of one or several kinematical parameters leads to typical
signatures in the residual \VF~(e.g. velocity asymmetry around the
major axis in case of a bad position angle determination, ...).
The residual \VFs~plotted for each galaxy in Appendix \ref{maps}
clearly show that these typical signatures are not seen, this
means that the best determination of the kinematical parameters has
been achieved.
\par
The deviation from purely circular velocity can be
large. In a forthcoming paper these residual \VFs~will be
analyzed in terms of bars and oval distortions, warps, spiral arms
(streaming motions), outflows and inflows, ... (e.g. \citealp{Fathi:2007}). 

\begin{figure}
\begin{center}
\includegraphics[width=7.5cm]{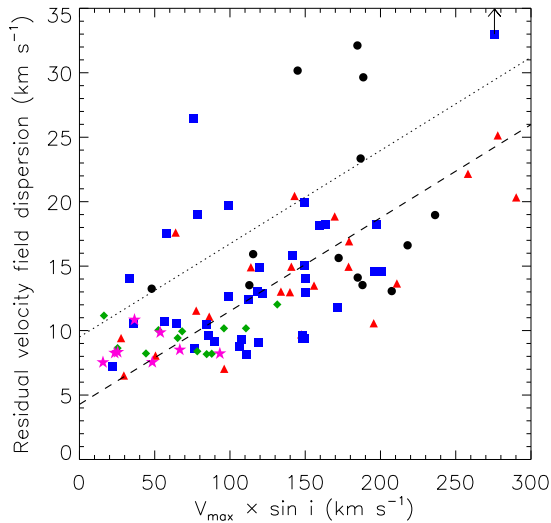}
\end{center}
\caption{Dispersion in residual \VF~versus maximum velocity,
sorted by Hubble morphological type: black circles 0$\leq$t$<$2,
red triangles 2$\leq$t$<$4, blue squares 4$\leq$t$<$6, green rhombuses
6$\leq$t$<$8 and pink stars 8$\leq$t$<$10. The dashed line represents the linear regression on the data. The points above the dotted line are discussed in section \ref{section_residuals}. UGC 3334 labelled with an arrow has actually a residual velocity dispersion of 54\kms~(see Table \ref{tablemod}).} \label{resi}
\end{figure}

The mean velocity dispersion on each residual \VF~has been
computed for each galaxy and tabulated in Table \ref{tablemod},
they range from 4 to 54\kms~with a mean value around 13\kms.
Figure \ref{resi} shows that the residual velocity dispersion is correlated with the maximum amplitude of the \VF~(shown by the dashed linear regression), this trend remains if we display the residual velocity dispersion versus the maximum circular velocity (not plotted). Surprisingly, barred galaxies do not have, in average, a higher mean residual velocity dispersion than unbarred galaxies (not plotted). This may be explained by the fact that the number of bins contaminated by the bar is usually rather low with respect to the total bins of the disk. Indeed, this is not the case for disks dominated by a bar. Compared to the general trend, we observe a set of  about a dozen of galaxies with a high residual velocity dispersion (points above the dotted line in Figure \ref{resi}). These points correspond to galaxies having strong bar or spiral structure and to data of lower quality:
 (i) galaxies dominated by strong bars (UGC 89 and UGC 11407), or strong spiral structures (UGC 5786 and UGC 3334) are not correctly described by our model which does not take into account non axisymmetric motions; (ii) the \VF~of the lower quality data (UGC 1655, UGC 3528, IC 476, UGC 4256 , UGC 4456,  IC 2542, UGC 6277, UGC 9406 and UGC 11269) present a mean size of the bins greater than 25 pixels and an integrated total \Ha~flux lower than 4.5 W m$^{-2}$ (a rough calibration of the total \Ha~flux of GHASP galaxies using the 26 galaxies we have in common with \citealp{James:2004} has been made, assuming a spectral ratio \Ha~over [NII] of 3:1). Figure \ref{resi} also shows that, for a given velocity amplitude, this correlation does not clearly depend on the morphological type. We note the well known fact that late-type galaxies have in average a lower velocity amplitude than early-type ones.

For most of the galaxies seen almost edge-on ($i$ higher than 75$\degr$), due to
the thickness of the disk, no model has been fitted (see previous subsection) thus
no residual velocity fields can be plotted.

\subsection{Kinematical parameters}

Table \ref{tablemod} gives the input (morphological) parameters of
the fits and the results of the fits (output parameters, $\chi^2$,
and parameters of the residual maps). Table \ref{tabletf} gives
some fundamental parameters of the galaxies compiled in the
literature (morphological and Hubble type, distance, M$_B$,
D$_{25}/2$, axis ratio, HI maps available in literature), together with maximum velocity parameters computed from the \RCs~(V$_{max}$, quality flag on V$_{max}$).
The four galaxies larger than
our field of view are flagged in the Table \ref{tabletf}. For some
galaxies for which the \SNR~or the spatial coverage is too low, the
fit could not converge correctly and one or two parameters ($i$
and $PA$) were usually fixed to the morphological values to
achieve the fit. These galaxies are flagged with an asterisk
($^{*}$) in the Table \ref{tablemod}. When it is not the case,
parameter determinations are discussed in Appendix \ref{notes}.
For some extreme cases, even when $i$ and $PA$ were fixed, the fit
does not converge. In particular for galaxies having high
inclinations, then no model was computed (see section \ref{vfanalysis}).

As underlined in Paper I, \citet{Garrido:thesis} and Paper IV,
due to the difference in wavelength between the calibration and the
redshifted \ha~lines, the coating of the Fabry-Perot interferometer induces
a small systematic bias (phase shift effect) to the absolute systemic velocities.
We tabulate the systemic velocities without
correcting them from this phase shift because the dispersion by
the phase effect is typically of the same order of magnitude than
the dispersion of the systemic velocities found in HyperLeda
(Paper IV) and also because the forthcoming analysis and in
particular the \RCs~do not depend on this effect.

\begin{figure}
\begin{center}
\includegraphics[width=7.5cm]{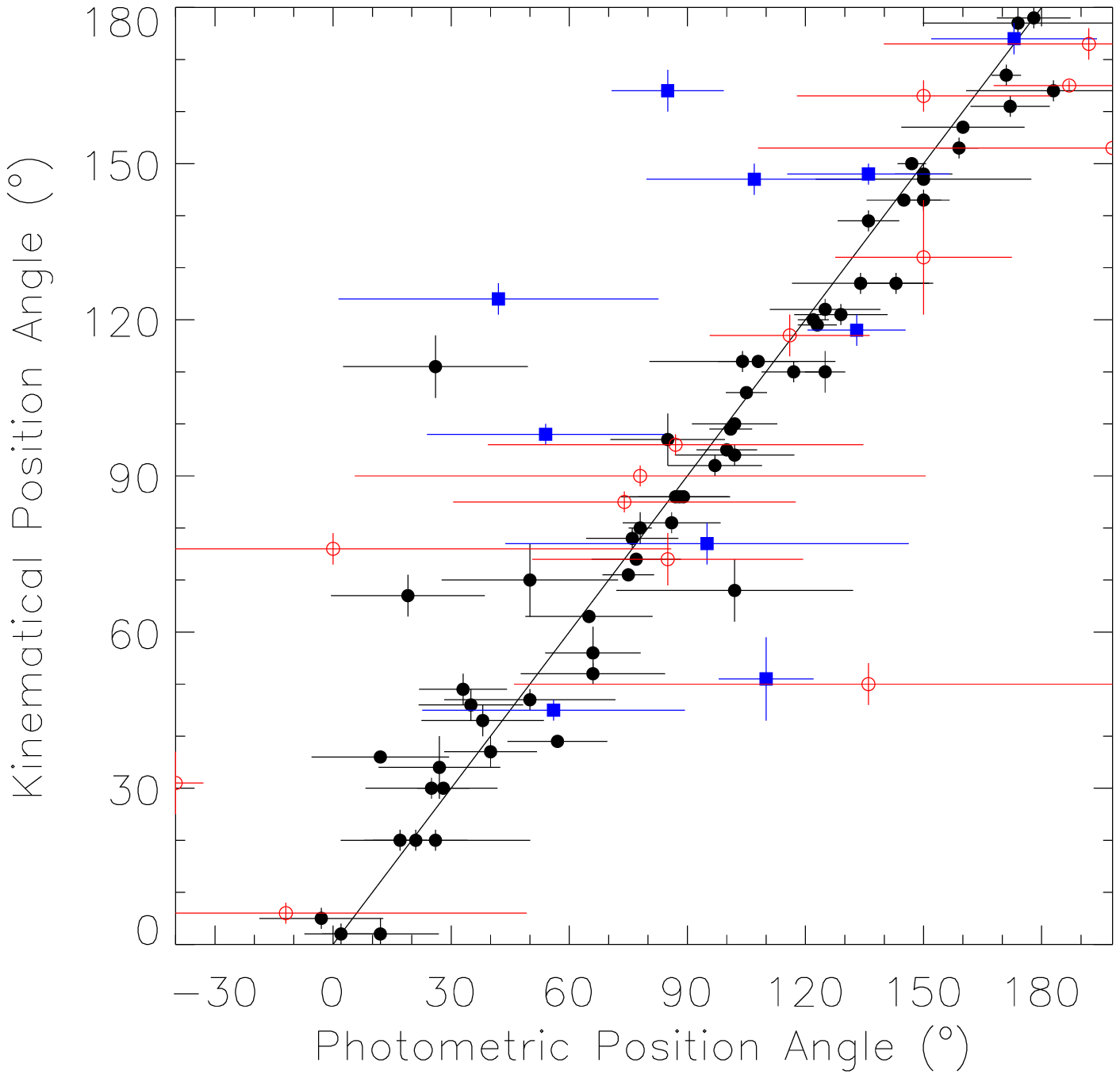}
\includegraphics[width=7.5cm]{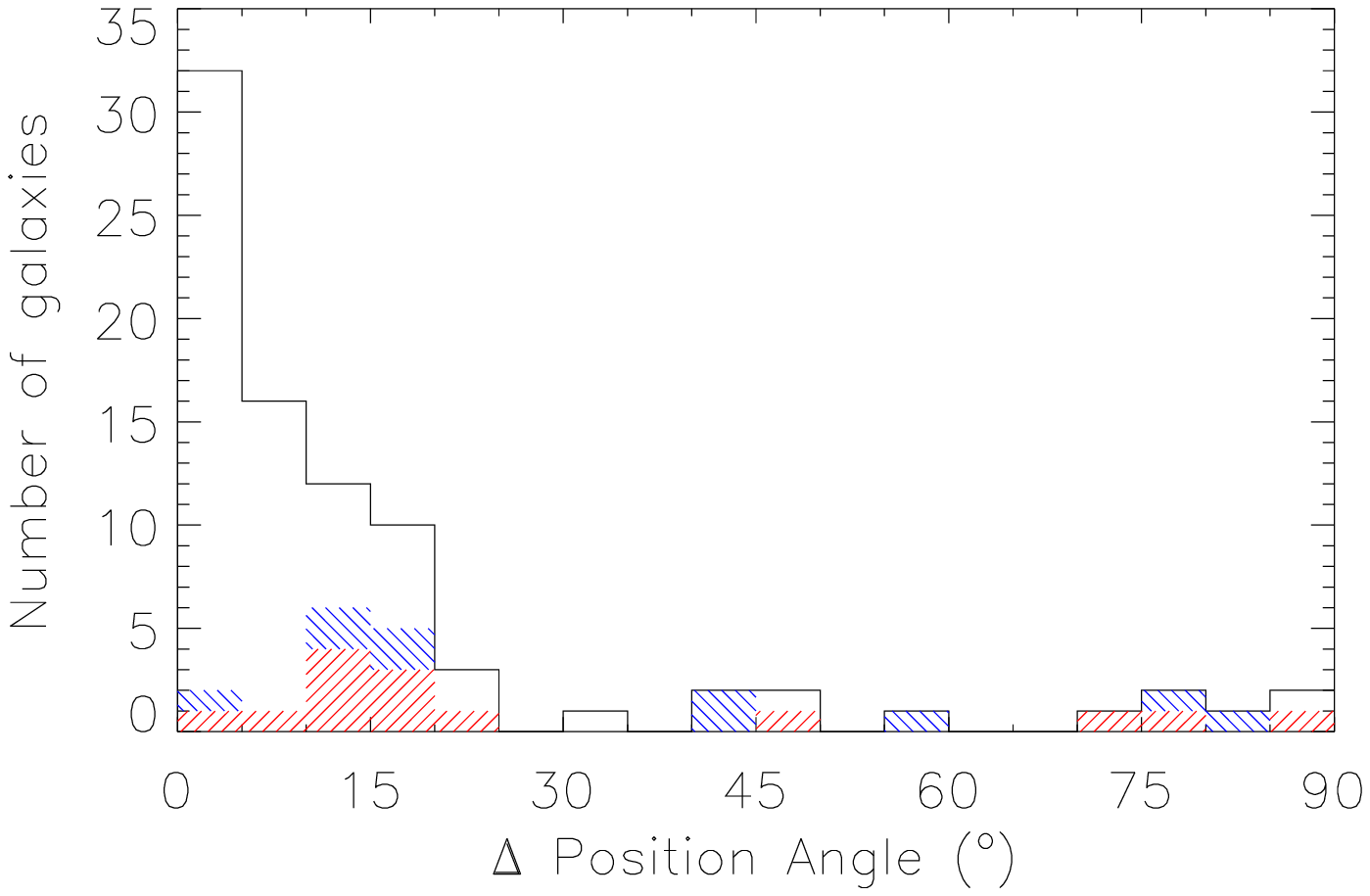}
\end{center}
\caption{\textbf{Top:} kinematical versus morphological (HyperLeda)
\PAs~of the major axis. Galaxies for which no accurate
morphological \PA~has been computed are shown by red open circles;
galaxies having an inclination lower than 25\degr~are displayed by
blue squares; the other galaxies are represented by black circles.
\textbf{Bottom:} histogram of the variation between kinematical and
morphological \PAs. The red hash, blue hash and residual white
represent respectively the galaxies for which no accurate \PA~has
been measured, for which inclination is lower than 25\degr~and the
other galaxies of the sample.} \label{pa}
\end{figure}

In Figure \ref{pa}, the kinematical \PAs~obtained by GHASP are
compared with the photometric \PAs~ (found in HyperLeda). The
error bar on the morphological \PA, which is generally not
homogenously given in the literature (or not given at all), has
been estimated using the axis ratio and optical radius
uncertainties. The galaxy disk in the sky plane is modelled by an
ellipse of axis ratio $b/a$ where $a$ is equal to D$_{25}$. Given
the uncertainty on D$_{25}$, $\Delta$D$_{25}$, a circle of
diameter D$_{25}$-$\Delta$D$_{25}$/2 having the same center than
the ellipse is considered. A line passing through the intersection
between the ellipse and the circle and their common center is thus
defined. The angle formed between the major axis of the ellipse
and the previously defined line represents the 1-$\sigma$
uncertainty on the \PA.

For all galaxies, HyperLeda references a list of \PAs~from which
they often computed one \PA~value. HyperLeda does not compute a
value when the dispersion or the uncertainty is too large. Indeed,
the \PA~may be quite different from a study to another, depending
on (i) the method, (ii) the size of the disk and (iii) the broad
band colors considered by the different authors (non homogeneity
in radius and colors measurements). When no value is computed in
HyperLeda, we put the whole list in Table \ref{tablemod}.
Moreover, to make it readable and to minimize the dispersion on
Figure \ref{pa}, we only plot the morphological value found closest
from the kinematical \PA. In Figure \ref{pa}, we have distinguished
the bulk of galaxies (black circles) for which the agreement is
rather good (lower than 20\degr~see \ref{pa}, bottom) from (i) the
galaxies for which no accurate morphological \PA~has been computed
(red open circles) and (ii) the galaxies having an inclination
lower than 25\degr~(blue squares). Indeed, some galaxies present a
disagreement between kinematical and photometric \PA~larger than
20\degr. Most of these galaxies have (i) a bad morphological
determination of the \PA~or (ii) have kinematical inclinations
lower than 25\deg or (iii) are specific cases (namely UGC 3740, IC
476, UGC 4256, UGC 4422) and are discussed in Appendix
\ref{notes}. On the other hand, Figure \ref{pa} shows that
morphological \PAs~have systematically higher uncertainties than
kinematical ones, this is specially true for galaxies with low
inclination. Quantitatively, for kinematical inclinations greater than 25\degr, the mean error on morphological \PAs~is $\sim$13\degr~while the mean error on kinematical \PAs~is $\sim$2\degr. For inclinations lower than 25\degr~the difference in the methods is even larger: the mean error on morphological \PAs~is $\sim$27\degr~while the mean error on kinematical \PAs~is $\sim$3\degr. For comparison, \citet{Barnes:2003}, using the difference between morphological and kinematical parameters, estimated that nonaxisymmetric features introduce inclination and \PA~uncertainties of 5\degr~on average.

The histogram of the variation between kinematical and
morphological \PAs~given in Figure \ref{pa} (bottom) indicates that
(i) for more than 60\% of these galaxies, the agreement is better
than 10\degr; (ii) for more than 83\%, the agreement is better
than 20\degr; (iii) the disagreement is larger than 30\degr~for
15\% of these galaxies.

In other words, the position of the slit in long slit
spectroscopy (which is usually based on the major axis determined
from broad-band imagery) with respect to the actual \PA~may be not
negligible, highlighting the strength of the integral field
spectroscopy methods to determine the \PAs~(see also illustrations in Paper IV, \citealp{Chemin:2006} and \citeauthor{Daigle:2006a} 2006a).

\begin{figure}
\begin{center}
\includegraphics[width=7.5cm]{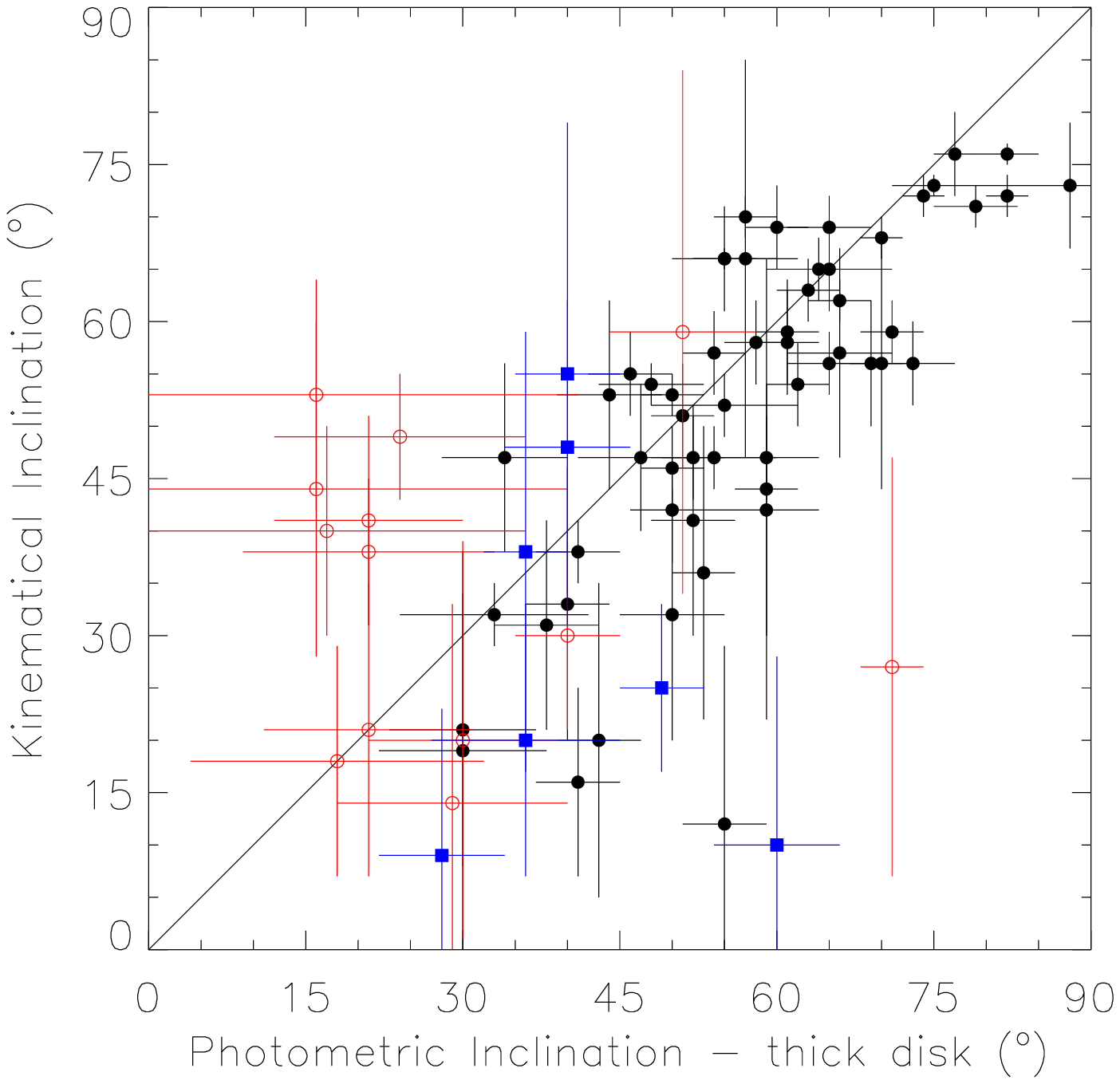}
\includegraphics[width=7.5cm]{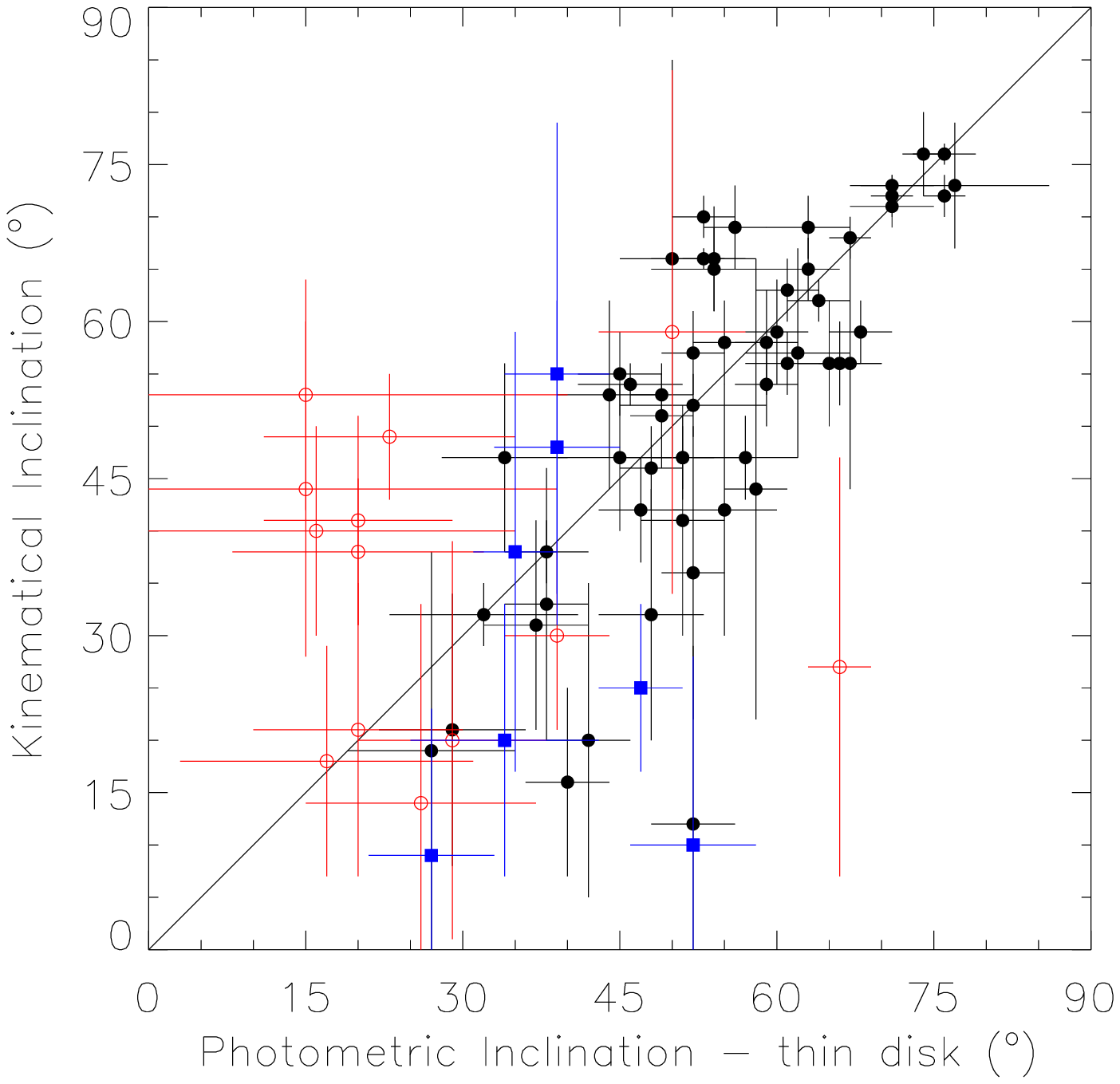}
\includegraphics[width=7.5cm]{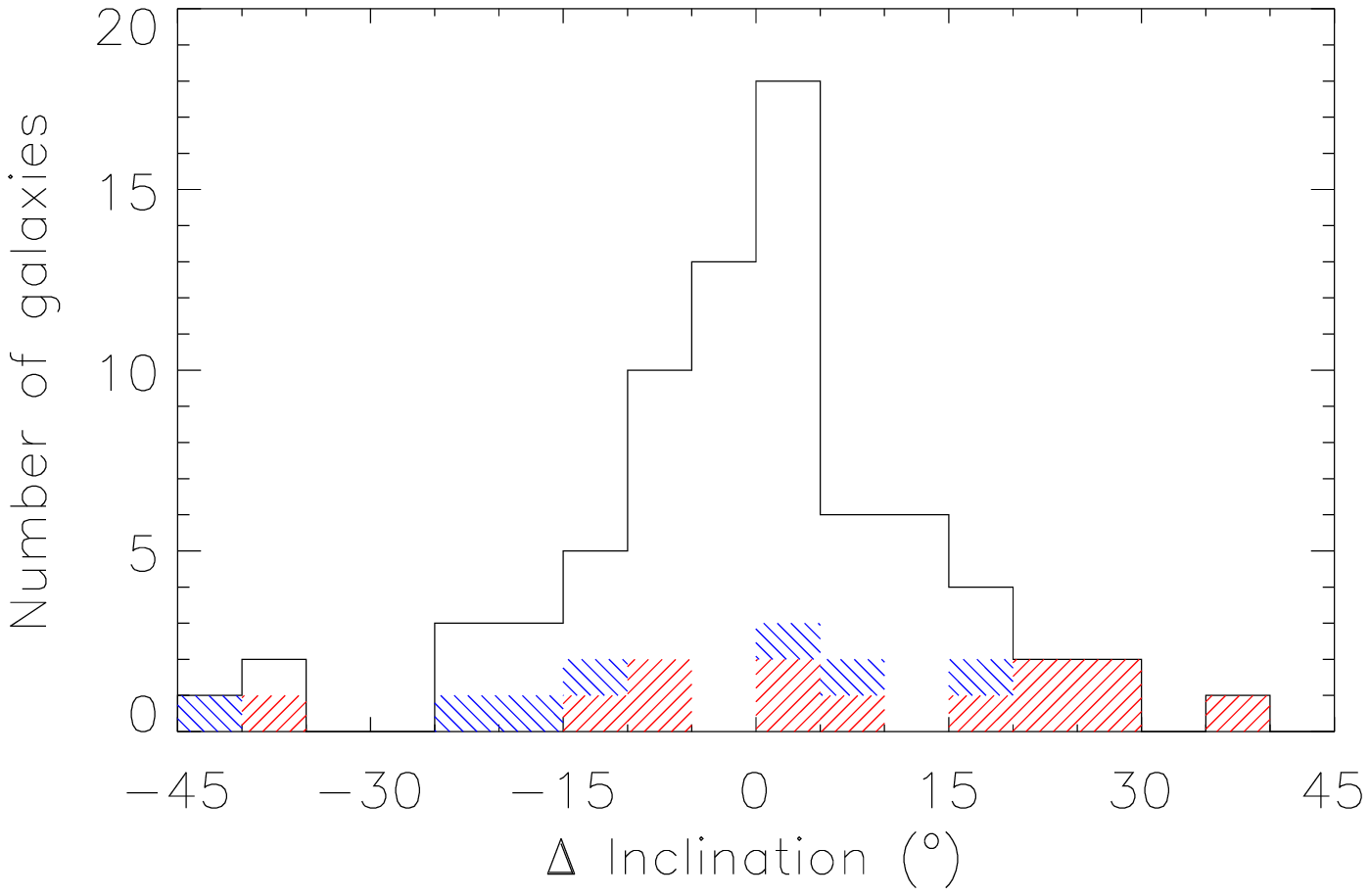}
\end{center}
\caption{\textbf{Top:} kinematical versus thick disk morphological inclinations.
\textbf{Middle:} kinematical versus thin disk morphological inclinations. \textbf{Top and Middle:} Galaxies for which no accurate morphological
\PA~has been computed are shown by red open circles; galaxies with
a difference between the kinematical and morphological \PAs~larger
than 20\degr~are displayed with blue squares; the other galaxies are
represented by black circles. \textbf{Bottom:} histogram of the variation
between kinematical and morphological inclinations. The red hash,
blue hash and residual white represent respectively the galaxies
for which no accurate \PA~has been measured, for which the
difference between the kinematical and morphological \PAs~is
larger than 20\degr~and the other galaxies of the sample.}
\label{inclination}
\end{figure}

In Figure \ref{inclination}, the inclinations obtained by GHASP are
compared with the photometric inclinations. On the top panel the
photometric inclination is the one computed using a correction
factor depending on the morphological type \citep{Hubble:1926}:
$$\sin^2{i}=\frac{1-10^{-2\log{r_{25}}}}{1-10^{-2\log{r_0}}}$$
where $r_{25}$ is the apparent flattening, and
$\log{r_0}=0.43+0.053~t$ for the de Vaucouleurs type $t$ ranging
from -5 to 7, and $\log{r_0}=0.38$ for $t$ higher than 7 \citep{Paturel:1997}.

On the middle panel the photometric inclination $i$ is derived
from the axis ratio $b/a$ without any correction ($\cos{i}=b/a$).
As for the \PAs, red open circles are the galaxies for which the
morphological \PA~could not be determined accurately. Blue squares
are the galaxies for which the difference between morphological
and kinematical \PA~exceeds 20\degr.

The dispersion around the $y=x$ line (equality between the
morphological and kinematical inclinations) decreases with the
inclination. The main discrepancy is found for low inclinations.
The corresponding galaxies are discussed in Appendix \ref{notes}
(notes on individual galaxies).
On the one hand, the morphological inclination of
the galaxies having no robust determination of their morphological
\PA~cannot be constrained correctly. On the other hand, galaxies for which the \PA~disagreement is
relatively high have a high dispersion and their morphological inclination is statistically overestimated.


Excluding these galaxies which have a bad \PA~estimation, for low
inclination systems, kinematical methods may underestimate the
inclination or alternatively, morphological estimations may be
overestimated. In average, the errors on morphological inclinations ($\sim$6\degr) and on kinematical inclinations ($\sim$8\degr) are comparable. Whatever the method used, the determination of the inclination of galaxies having a low inclination remains less
accurate than for more inclined galaxies.


The comparison of the two plots in Figure \ref{inclination} (top and
middle) shows that galaxies with high inclination have a better agreement between
their kinematical inclination and their morphological inclination computed considering a thin disk.
The actual thickness of the disk may not be reproduced by our simple
thin disk \VF~modelling. If it is the case, the kinematical
inclination may be systematically underestimated. Alternatively,
the good agreement between thin disk morphological inclination and
kinematical inclination may mean that the morphological thickness
corrections are overestimated.


The histogram of the difference between morphological and
kinematical inclinations (Figure \ref{inclination}, bottom) shows
that a difference of inclination larger than 10\degr~is found for
40\% of the sample.

\section{The Tully-Fisher relation}
\label{tullyfisher}

\begin{figure}
\begin{center}
\includegraphics[width=7.5cm]{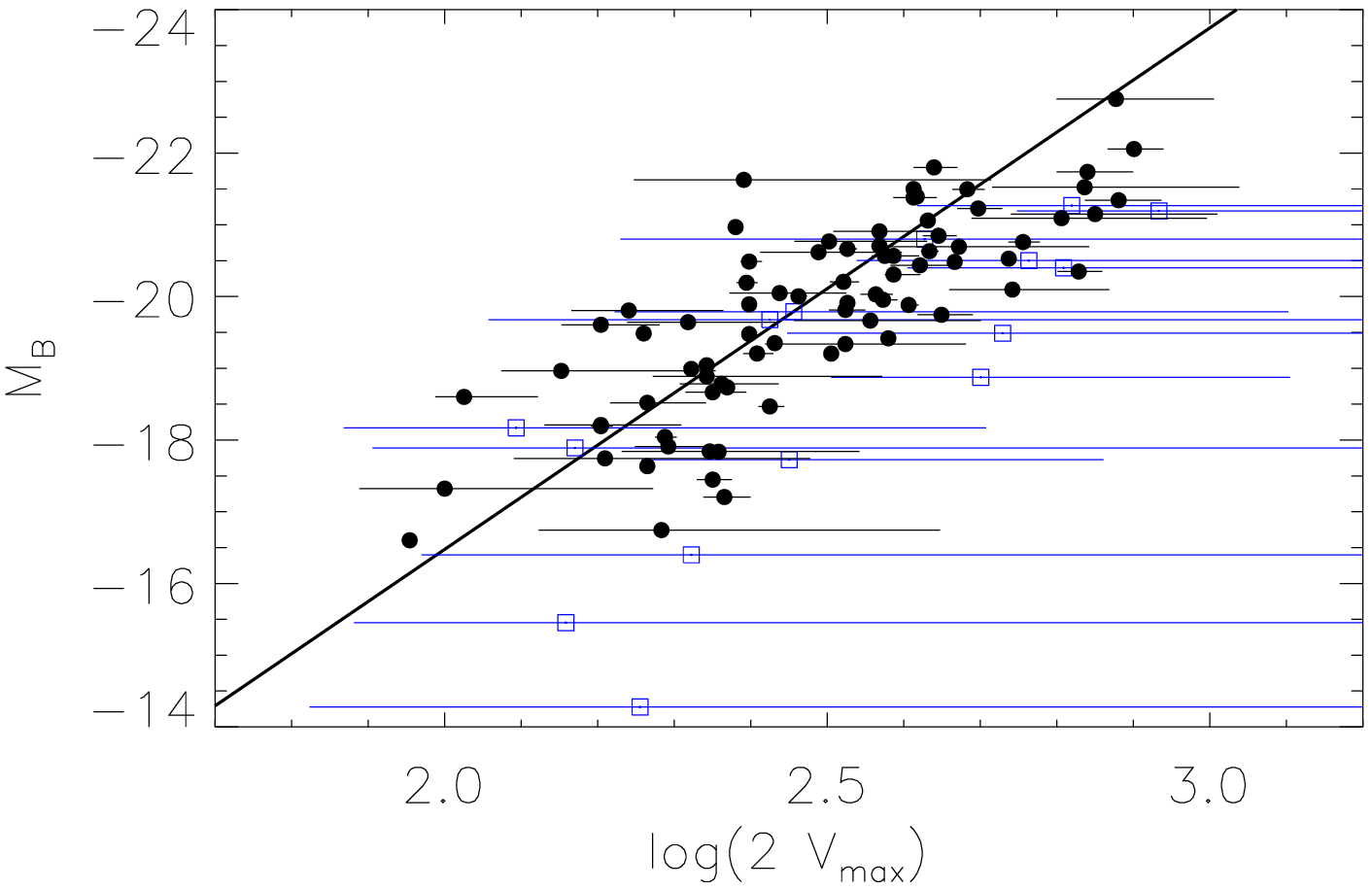}
\includegraphics[width=7.5cm]{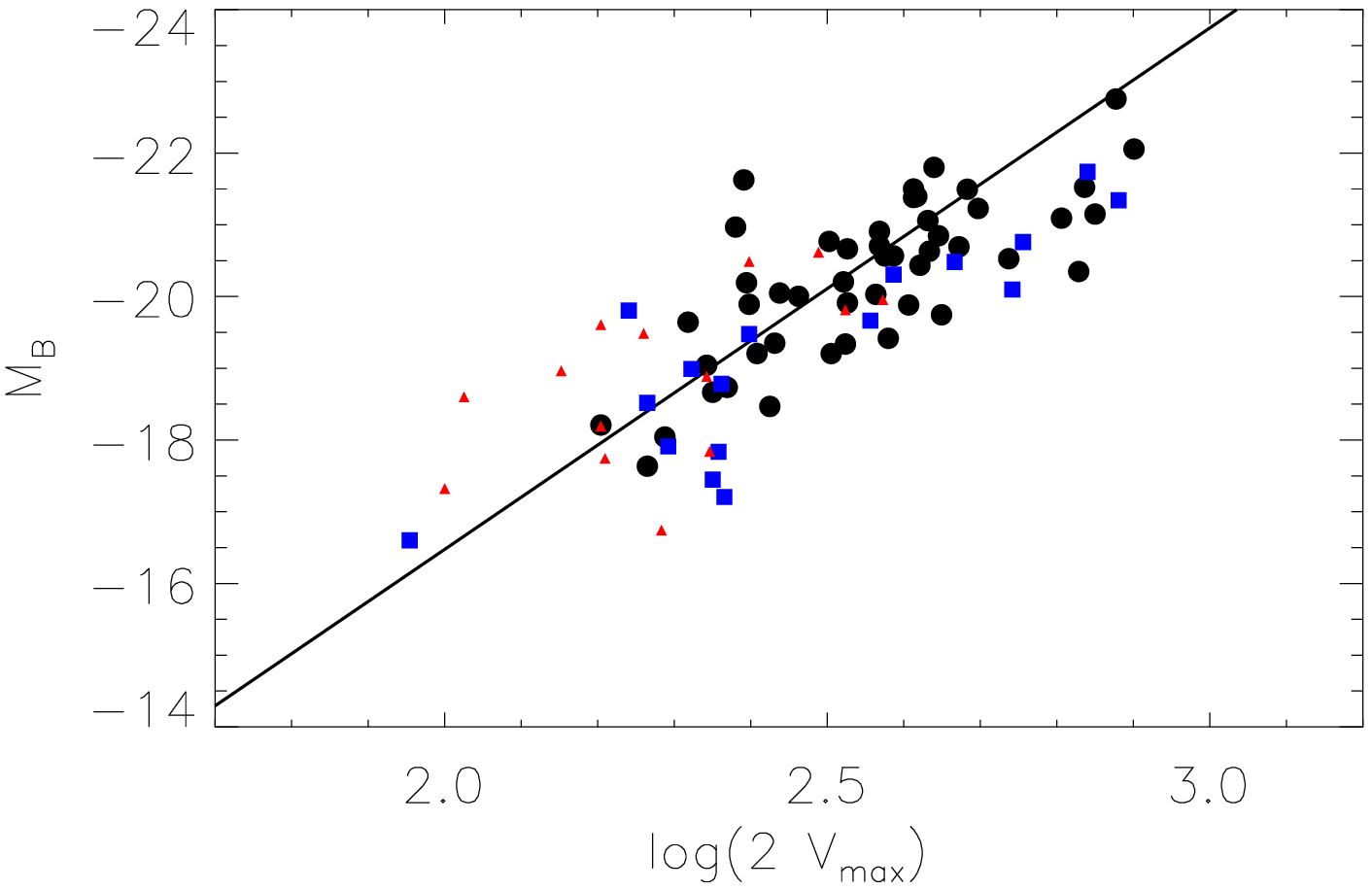}
\includegraphics[width=7.5cm]{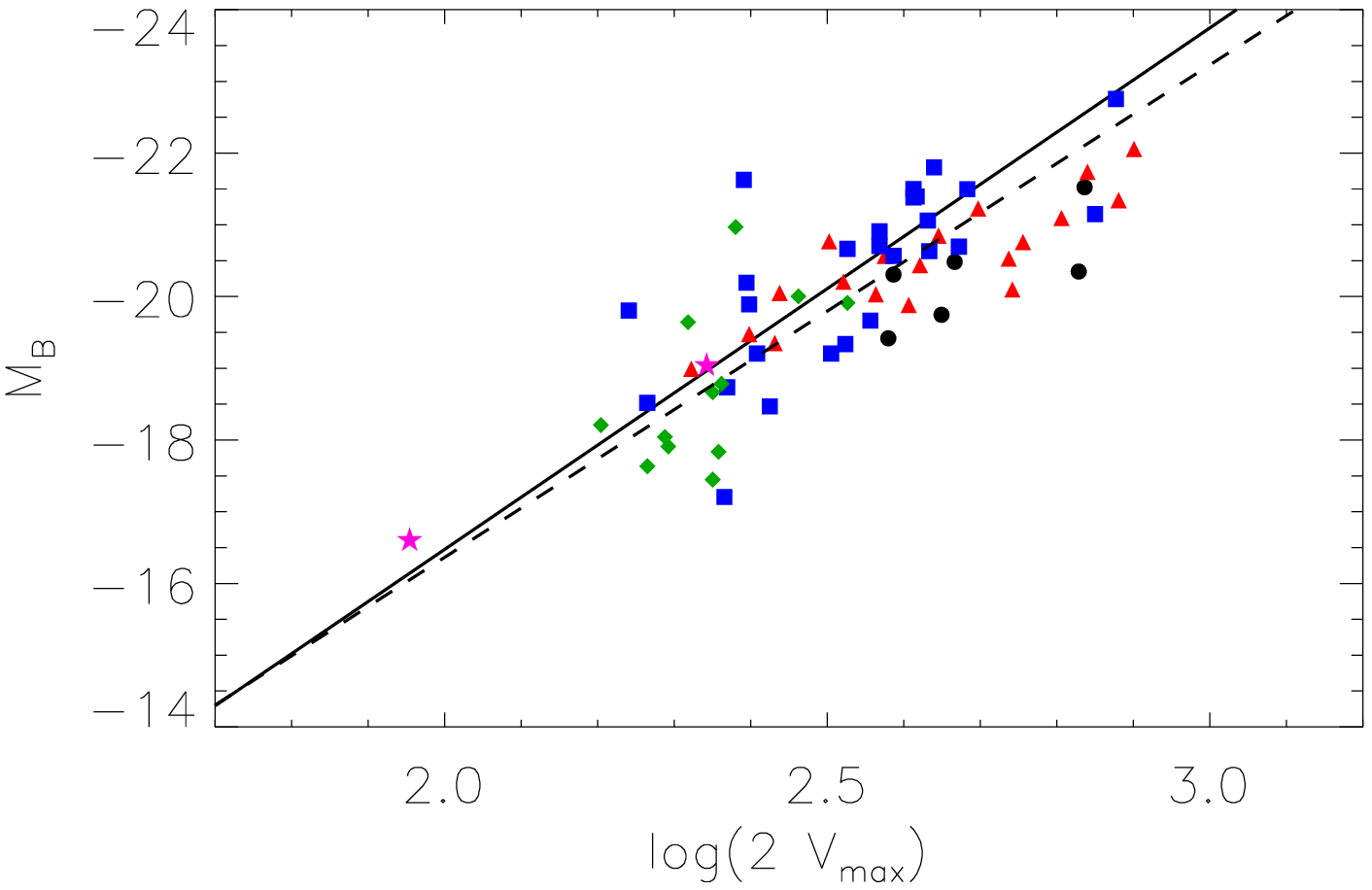}
\end{center}
\caption{TF relation for our sample of galaxies. The
solid line represents the B magnitude \TF~relation determined by \citet{Tully:2000} from nearby galaxies in clusters (Ursa Major, Pisces filament, Coma).
\textbf{Top:} sorted by inclination - low inclination galaxies
($i<$25\degr): blue squares; other galaxies ($i$$\geq$25\degr):
black circles.
\textbf{Middle:} sorted by V$_{max}$ flags - V$_{max}$ reached:
black dots, large size; V$_{max}$ probably reached: blue squares, medium size; V$_{max}$ probably not
reached: red triangles, small size.
\textbf{Bottom:} sorted by morphological type - black circles from 0 to 2;
red triangles from 2 to 4; blue squares from 4 to 6; green
rhombuses from 6 to 8; pink stars from 8 to 10; the dotted line represents the best linear fit on the data.}
\label{tullyfisher_plot}      
\end{figure}

Among the present sample of 108 galaxies, we have plotted the \TF~relation (\citealp{Tully:1977}, M$_{B}$ as a function of $\log{2 V_{max}}$) for a sub-sample of 94 galaxies in Figure \ref{tullyfisher_plot}. The 14 other galaxies are not considered in the present discussion because (i) for five galaxies the \RC~does not reach the maximum rotation velocity (UGC 1655, UGC 4393, UGC 6523, UGC 8898 and UGC 9406); (ii) no B magnitude is available for one galaxy (UGC 3685) and (iii) no velocity measurement neither on the \RC~nor on the \PVM~is possible for eight other galaxies (see Table \ref{tabletf}).

The maximum velocity V$_{max}$ has been obtained from the fit to the \VF. The error on V$_{max}$
is the quadratic combination of the error due to the uncertainty
on the inclination (the product V$_{max} \times \sin{i}$ is
constant) and the median dispersion in the rings of the \RC~beyond
D$_{25}$/10. In the cases where the \RC~has no point beyond that
radius, we replace this term by the intrinsic uncertainty on the
velocity determination due to the spectral resolution (8\kms).
For the highly inclined galaxies for
which no correct fit was possible with our method (because it does
not take into account the thickness of the disk, see section \ref{vfanalysis}), we computed
V$_{max}$ from the \Ha~\PVM~corrected from the photometric
inclination. For them, the error on V$_{max}$ is simply the
intrinsic uncertainty on the velocity determination. For the
particular case of UGC 5786, the fit is not good enough to use it
to compute V$_{max}$ because of the long blue northern tail and
because of the strong bar. We estimated V$_{max}$ to 80\kms~by eye
inspection of the \RC. These galaxies are flagged in Table \ref{tabletf}.

The solid line in Figure \ref{tullyfisher_plot} is the relation found by \citet{Tully:2000}:
\begin{displaymath}
M_{B}=-7.3[\log{2 V_{max}} - 2.5] -20.1
\end{displaymath}

In Figure \ref{tullyfisher_plot} (Top), the error bars on the velocity are displayed and galaxies with inclination lower than 25\degr~are distinguished (blue open squares). We clearly notice that these galaxies have statistically higher velocities than expected from the \citet{Tully:2000} relation. This effect is due to the link between inclination and velocity determination. Indeed, on the \VFs, we observe the projected velocity on the line of sight: $V_{rot}\times \sin{i}$. A given underestimate of the inclination thus leads to a higher overestimate on maximum velocity for low inclination galaxies than for high inclination galaxies. This also explains the strong trend for low inclination galaxies to exhibit large error bars. Considering this effect, we choose to exclude the 15 galaxies with inclinations lower than 25\degr~from the \TF~analysis.

Among the 79 remaining galaxies, the maximum velocity V$_{max}$ is reached for 48 of them (black dots, large size), probably reached for 17 of them (blue squares, medium size) and probably not reached for 14 of them (red triangles, small size). They are distinguished in Figure \ref{tullyfisher_plot} (Middle) and flagged in Table \ref{tabletf}. The quality flag on the maximum velocity is deduced from (i) the inspection of the shape of the \ha~\RCs~and \PVM s; (ii) from the comparison with HI \VFs~and \RCs~when available (see Table \ref{tabletf}); (iii) from the comparison of the \ha~\VFs~amplitudes with HI line widths (see individual comments in Appendix \ref{notes}). It appears from this last point that the HI line width at 20\%~has most often the best agreement with the \ha~\VF~amplitude (better than the line width at 50\%). Figure \ref{tullyfisher_plot} (Middle) confirms the two classifications "V$_{max}$ probably reached" and "V$_{max}$ probably not reached" since for the majority of each class the points are respectively in agreement and above the \citet{Tully:2000} relation.
From the two classes "V$_{max}$ reached" and "V$_{max}$ probably reached", we find the following relation:
\begin{equation}
M_{B}=(-6.9\pm 1.6)[\log{2 V_{max}} - 2.5] -(19.8\pm0.1)
\label{tfghasp}
\end{equation}
This relation is displayed as a dotted line in Figure \ref{tullyfisher_plot}, in which morphological types are distinguished for the two best classes (black circles from 0 to 2, red triangles from 2 to 4, blue squares from 4 to 6, green rhombuses from 6 to 8 and pink stars from 8 to 10).
Coefficients have been computed using the mean of the coefficients obtained (i) using a fit on the absolute magnitudes (as dependant variables) and (ii) using a fit on the velocities (as dependant variables). The difference in the slope determination by these two methods is quite large due to a strong scatter in our data (the error on the parameter in equation \ref{tfghasp} is half that difference). Indeed, usually one uses local calibrators for which distance measurements are accurate (based on Cepheids, red giants branch, members of a same cluster, ...) leading to a small scatter in the data. From our data, the main difficulty is that the distance determination is mostly based on the systemic velocity corrected from Virgo infall (see Table \ref{tabletf}), and that no error bar on the magnitude can be easily estimated. Thus we have no reason to be more confident on the velocity measurements (mainly affected by inclination determination) than on absolute magnitude measurements.
However, despite the dispersion in our data, the resulting parameters using the mean of the two fits are in good agreement with \citet{Tully:2000}, even if our slope is a bit lower. A lower slope had already been observed in HI (\citealp{Yasuda:1997}; \citealp{Federspiel:1998}) and more systematically in optical studies (e.g. \citealp{Courteau:1997}; \citealp{Rubin:1999}; \citealp{Marquez:2002}; papers III and IV).

For the \TF~relation we derived, on the one hand we observe that fast rotators (V$_{max}>300$\kms: UGC 89, UGC 4422, UGC 4820, UGC 5532, UGC 8900, UGC 8937 and UGC 11470) are less luminous than expected, except maybe for UGC 3334 which is one of the fastest disk rotators \citep{Rubin:1979} (see discussion in Appendix \ref{notes}). This trend can also be observed in several optical studies (\citealp{Marquez:2002}; Papers III and IV). Interestingly, these fast rotators are not observed in HI samples (\citealp{Tully:2000}; \citealp{Federspiel:1998}). This may be explained by the shape of the \RCs~of fast rotators: except for UGC 8900, the \RC~always reaches the maximum velocity within the first five arcseconds, (i.e. within our seeing). This inner maximum may be missed in HI because of beam smearing, averaging the maximum velocity reached in the center. Note that the \RC~of UGC 5532 is clearly decreasing while the other ones are flat.
On the other hand slow rotators have a small velocity gradient and within the optical regions the maximum could not be reached whereas HI observations would be able to measure it without any doubt.
These two effects could explain the trend observed in optical \TF~relations.

The result obtained for the \TF~relation is in agreement with the one obtained with the previous samples (Papers III and IV). The analysis of the whole GHASP sample will be done in a forthcoming paper (Epinat et al. in preparation).

\section{Summary and perspectives}
\label{conclusion}

The knowledge of the links between the kinematical and dynamical
state of galaxies will help us to have a better understanding of
the physics and evolution of galaxies. The GHASP sample, which
consists of 203 spiral and irregular galaxies covering in a wide
range of morphological types and absolute magnitudes, has been
constituted in order to provide a kinematical reference sample of
nearby galaxies. The GHASP galaxies have been observed in the
\ha~line with a scanning Fabry-Perot, providing data cubes.

We present in this paper the last set of 108 galaxies leading to
106 \VFs~and 93 \RCs.  By now, this work consists of the largest
sample of galaxies observed with Fabry-Perot techniques ever
presented in the same publication.  Added to the four previous
sets already obtained in the frame of this survey (Paper I to IV),
GHASP represents the largest sample of 2D \VFs~of galaxies
observed at \ha~wavelength. For each galaxy, we have presented the
\ha~\VF, the \ha~monochromatic image and eventually the
\ha~residual \VF, the \PVM~along the major axis and the \RC~when
available.

Major improvements in the reduction and in the analysis have been
developed and implemented:
\begin{itemize}
    \item in order to optimize the spatial resolution for a given \SNR,
adaptative binning method, based on the 2D-Vorono\"i
tessellations, was used to derive the 3D \Ha~data cubes and to extract from it the line maps and the radial velocity fields;
    \item the ghosts due to reflections at the interfaces air/glass of the interferometer, have been removed in the data cubes;
    \item the analysis of the faint outskirts or diffuse regions is automatic;
    \item the kinematical parameters and their error bars are directly derived from the \VF;
    \item the uncertainties are estimated from the analysis of the residual \VF~power spectrum;
    \item the whole 2D \VF~has been used rather than successive crowns in tilted-ring
models to compute the \RC~and the error bars.
\end{itemize}

The main results of this paper are summarized by the following
items.
\begin{itemize}
    \item The absence of typical and well known bias in the residual \VFs~
means that the best determination of the kinematical parameters has
been achieved.
    \item The mean velocity dispersion on each residual \VF~ranges from 6 to
23\kms~with a mean value around 13\kms~and is strongly
correlated with the maximum amplitude of the \VF. For a given
velocity amplitude, this correlation does not clearly depend on
the morphological type. Only strongly barred galaxies have a
higher residual velocity dispersion than mild-barred or non barred
galaxies. Peculiar galaxies also show a high residual velocity
dispersion.
    \item The kinematical \PAs~obtained by GHASP are compared with the
photometric \PAs. Morphological \PAs~have systematically higher
uncertainties than kinematical ones, this is specially true for
galaxies with low inclination. 
When using long slit spectroscopy, the \PA~should be known \emph{a priori}. This is usually done using morphological determinations based on broad band imagery. We have shown that in some cases the difference between the \PA~determined using 2D kinematics and morphologies may be as large as 90\degr~and that in any case the \PAs~are better determined by 2D kinematics. Thus, large differences between morphological and kinematical \PAs~may lead to incorrect \RC~and maximum velocity determination when using long slit spectroscopy. This may strongly bias mass distribution models and Tully-Fisher studies, highlighting the strength of integral field spectroscopy with a Fabry-Perot.
    \item The morphological inclination of the galaxies having no robust
determination of their morphological \PA~cannot be constrained
correctly. Galaxies for which the \PA~disagreement is relatively
high have a high dispersion and their morphological inclination is
statistically overestimated. Galaxies with high inclination have a
better agreement between their kinematical inclination and their
morphological inclination computed assuming a thin disk. For galaxies with intermediate disk inclinations (higher than 25\degr~and lower than 75\degr), to reduce the degrees of freedom in kinematical models, the inclination could be fixed to the morphological value. This is specially true when only low quality kinematical data are available as it is the case for high redshift galaxies.
    \item The \TF~relation found with this new set of data is
in good agreement with \citet{Tully:2000}, even if our slope is a
bit lower. This trend for a lower slope has already been observed in HI
by \citet{Yasuda:1997} and \citet{Federspiel:1998}. Galaxies with inclination lower than
25\degr, have statistically higher velocities than expected from the
TF relation derived by \citet{Tully:2000}. Fast rotators (V$_{max}>$300\kms) are less
luminous than expected. This may be explained by the shape of the
\RCs~of fast rotators.
\end{itemize}


\section*{acknowledgements} {The authors warmly thank Dr O.
Garrido for leading or participating to most of the observations.
They also thank the Programme National Galaxies for supporting the
GHASP project in allocating continuously observing time during
several years, the Observatoire de Haute-Provence team for its
technical assistance during the observations, O. Boissin for
his technical help during the observing runs, and J. Boulesteix for permanent support. They thank I. J\'egouzo and C. Surace for building the Fabry-Perot Database. This research has
made use of the GOLD Mine Database and
of the NASA/IPAC Extragalactic Database (NED) which is
operated by the Jet Propulsion Laboratory, California Institute of
Technology, under contract with the National Aeronautics and Space
Administration. The authors have also made an extensive use of the
HyperLeda Database (\url{http://leda.univ-lyon1.fr}). The Digitized Sky Surveys were produced at the Space Telescope Science Institute under U.S. Government grant NAG W-2166. The images of these surveys are based on photographic data obtained using the Oschin Schmidt Telescope on Palomar Mountain and the UK Schmidt Telescope. The plates were processed into the present compressed digital form with the permission of these institutions.}

\bibliographystyle{mn2e}
\bibliography{biblio}

\appendix

\clearpage
\section{Building a rotation curve}
\label{method}
\subsection{The Model.}

For each of the $N$ independent bins covering the field of view of
the galaxy, the vector velocity in the frame of the galactic plane
is described by two components lying in the plane of the galaxy:
\begin{enumerate}
    \item $V_{\mathrm{rot}}(R)$: the rotation velocity;
    \item $V_{\mathrm{exp}}(R)$: the expansion velocity;
\end{enumerate}
plus one component perpendicular to this plane:
\begin{enumerate}
\setcounter{enumi}{2}
    \item $V_{\mathrm{z}}(R)$: the vertical motions velocity.
\end{enumerate}

The observed radial velocities $V_{\mathrm{obs}}(R)$ is linked to
$V_{\mathrm{rot}}(R)$, $V_{\mathrm{exp}}(R)$ and
$V_{\mathrm{z}}(R)$ through 5 additional parameters:
\begin{enumerate}
\setcounter{enumi}{3}
    \item $PA$: the position angle of the major axis of the galaxy
    (measured counterclockwise from the North to the direction of receding side of the
    galaxy);
    \item $i$: the inclination of the galactic disk with respect to
    the sky plane;
    \item $V_{\mathrm{sys}}$: the systemic velocity of the galaxy
    \item $\alpha$: the right ascension of the rotation center;
    \item $\delta$: the declination of the rotation center;
\end{enumerate}
through the following equation:

 \begin{eqnarray}
 \label{vlos_disk}
V_{\mathrm{obs}} & = & V_{\mathrm{sys}} + V_{\mathrm{rot}}(R)\cos\theta\sin i \nonumber\\
& & + V_{\mathrm{exp}}(R)\sin\theta\sin i\\
& & + V_{\mathrm{z}}(R)\cos i \nonumber
 \end{eqnarray}

$R$ and $\theta$ being the polar coordinates in the plane of the
galaxy. The angle in the plane of the galaxy, $\theta$,  is linked
to the position angle $PA$, the inclination $i$, the position
$x,y$ and center $x_c,y_c$ in the sky by the set of equations
\ref{links_s} to \ref{links_e}:
 \begin{eqnarray}
 \label{links_s}
 \cos{\theta}= R \cos{\psi}\\
 \sin{\theta}= R \frac{\sin{\psi}}{\cos{i}}\\
 \cos{\psi}=\frac{(y-y_c) \cos{PA} - (x-x_c) \sin{PA}}{r}\\
 \sin{\psi}=-\frac{(x-x_c) \cos{PA} + (y-y_c) \sin{PA}}{r}\\
 r=\sqrt{(x-x_c)^2+(y-y_c)^2}\\
 R=r\sqrt{\cos^2{\psi}+\frac{\sin^2{\psi}}{\cos^2{i}}}
 \label{links_e}
 \end{eqnarray}
$\psi$ being the counterclockwise angle in the plane of the sky
from the North.

Formally, one has to solve a system of $N$ equations (as many
equations as the number $N$ of pixels taken into account) with
$8N$ unknowns.

If one makes the assumption that, at the first order, for spiral
galaxies, the expansion and vertical motions are negligible with respect
to the rotation velocity, the equation \ref{vlos_disk} becomes:
$$
V_{\mathrm{obs}}(R) = V_{\mathrm{sys}}(R) +
V_{\mathrm{rot}}(R)\cos\theta\sin i
$$

This leads to solve a system of $N$ equations with $6N$ unknowns.

A usual solution to solve this degenerate system is to fix
$V_{\mathrm{sys}}$ to a unique value for a given galaxy, and to
consider that $i$, $PA$, $x_c$ and $y_c$ only depend on the
galactic radius (to take warps into account), as it is the case
for $V_{\mathrm{rot}}(R)$. The field is decomposed in a certain
number of elliptical rings (at given radii, with a given width)
for which a set of parameters is computed for the corresponding
radii (Begeman, 1987). The number of rings being at least one
order of magnitude less than $N$, the system of equation is no
more degenerate. The physical width of the rings is typically
ranging from 3 to 6 pixels ($\sim$ 2\arcsec to 4\arcsec).

We decided to use a new method. To solve this degenerate system of
equations, the number of unknowns is reduced by introducing
physical constraints:
$V_{\mathrm{sys}}$, $i$, $PA$, $x_c$ and $y_c$  are fixed to a unique value for a given galaxy as warps are hardly observed within optical radius.


Moreover, the rotation velocity is approximated by a function with
only four parameters:
\begin{equation}
V_{\mathrm{rot}}(R)=V_t \frac{\left( R/r_t \right) ^g}{1+\left(
R/r_t\right) ^a} \label{zhao}
\end{equation}




\subsection{Method.}


The method implemented is a $\chi^2$ minimization method on the
velocity field, using the IDL routine \emph{lmfit}.  This routine,
based on the Levenberg-Marquardt method (as described in section
15.5 of \citealp{numericalrecipes}) quickly converges towards the best
model.

The starting set of parameters is chosen as follows: the rotation
center ($x_c$, $y_c$) is supposed to be the nucleus identified on
our continuum image (when no nucleus can be seen, neither on our
images nor in other bands, it is chosen as the center of symmetry
of our velocity field), $i$ is derived from the axis ratio found
in the literature, the position angle is computed from the
photometry (or eye-defined by the outer parts of our velocity
field when not available), the systemic velocity is taken from the
literature.


In order to have a pretty good estimate of the analytical function
parameters (\ref{zhao}), we fit them on this preliminary rotation
curve.


We then let free all the parameters. We compute an iterative
3.5-sigma clipping on the velocity field to reject points that
have not been cleaned.


For well behaved galaxies, the iterative process was quite easy
and rapidly converged. However, for some irregular galaxies with
asymmetric rotation curves it was hard to converge and we had a
strong uncertainty on the kinematical parameters, more especially
the inclination which is the less constrained. In some specific
case, marked in table \ref{tablemod} with an asterisk ($^{*}$), we
then fixed i and/or PA during the fit process.


The rotation curves are computed from the velocity fields, using
the previously found projection parameters $i, PA, x_c$ and $y_c$.
A sector of 22.5$^{\circ}$ in the plane of the galaxy around the
minor axis is excluded, and the points are weighted according to
their corresponding $|\cos{\theta}|$. The rotation curves are
sampled with rings.

In the inner parts, the width of the rings is set to match half the seeing, in order to respect Shannon sampling criteria. The transition radius $r_t$ is defined by the first ring that contains more than 25 uncorrelated bins. If $r_t$ is not reached before D$_{25}$/10, $r_t$ is set to D$_{25}$/10. In the outer parts, the \RC~is computed in successive rings containing the same number of uncorrelated bins, except eventually for the last ring of each side.
The number of bins in the rings of the outer parts is set to the number between 16 and 25 that maximises the number of uncorrelated bins of the last rings of each side. This range (16 to 25) is found to be the best compromise between the \snr~and the spatial coverage in each ring. Thus the width in each ring is variable. The velocity computed for each ring is then always the average of the same number of velocity points.


On the plots of \RC, for each individual ring, the vertical error
bars are given by the dispersion of the rotation velocities inside
the ring, normalized to the number of uncorrelated points inside
the ring; the horizontal error bars represent the $\pm$1$\sigma$
radius dispersion weighted by cos($\theta$).

\subsection{Error estimation of the parameters.}

This method provides a determination of parameters errors, by
associating a random noise to the data for which the amplitude can
be fixed. However, this error seems unrealistically small. Indeed,
the residual velocity field (difference between model and real
velocity field) does not appear to be uniformly randomized. It
contains the effects of non axisymmetric motions such as
expansion, spiral arms, bars or gas bubble expansion (local
expansions),..., that is to say real physical effects that cannot
be described by our simple model. To be more realistic in
parameters errors determination, we simulate residual velocity
fields from the real residual velocity field. We compute its power
spectrum and put a random phase. As a result, the new residual
field contains the same kind of structure, but placed differently.
Then we use a Monte Carlo method to estimate the errors: we
compute the standard deviation in the parameters found over 250
simulated velocity fields. The advantage of this method is that it
is completely automatic, and that it enables to use the whole
information to compute global parameters and their errors. We also
add 0.5 degrees to the position angle uncertainty because of the
uncertainty on the astrometry.

The typical accuracy we reach is about 1 arcsec for the position
of the rotation center, 2 to 3\kms~for the systemic velocity, 2
degrees for the position angle of the major axis, but only 5 to 10
degrees for the inclination.

\clearpage
\section{Notes on individual galaxies}
\label{notes}
\noindent \textbf{UGC 12893}. The \ha~emission is very weak, as
confirmed by \citet{James:2004} who find a total surface
brightness of 0.4 $10^{-16}$ W.m$^{-2}$. The quality of our \RC~is
thus rather poor, anyway a maximum rotation velocity
($\sim$72\kms) seems to be reached within the optical limit,
adopting the inclination of 19\degr~deduced from our
\VF~(HyperLeda gives 30\degr~from the photometry). The width of
the HI profile at 20\%~(75\kms~from \citealp{Giovanelli:1993} and
$\sim$90\kms~from \citealp{Schneider:1990}) is in agreement with the
amplitude of our \ha~\VF~and suggests that we actually reach the
maximum rotation velocity with our \ha~\RC.
\\%
\noindent \textbf{UGC 89 (NGC 23)}. Because of the presence of a
strong bar in this galaxy, the kinematics method used to determine
the inclination is biased. Nevertheless our kinematical
inclination (33$\pm$13\degr) is compatible with the photometric
inclination (40$\pm$4\degr) and with the value of 45\degr~from
\citet{Fridman:2005} (\ha~Fabry Perot observations) as well as with
the value of 50\degr~from \citet{Noordermeer:2005} (HI data from
WHISP). Our \ha~\RC~reaches a plateau at $\sim$350\kms,
compatible with its morphological type (SBa) and with the
\PVM~from \citet{Noordermeer:2005}. The steep rise of the \ha~\RC~is also
in agreement with the HI observation of \citet{Noordermeer:2005} and in
very good agreement with the \ha~\VF~of \citet{Fridman:2005}.
Our \ha~\VF~does not show any obvious evidence for interaction
with its companion, UGC 94.
\\
\noindent \textbf{UGC 94 (NGC 26)}. The steep rise of the
\ha~\RC~is in agreement with the HI observation \citep{Noordermeer:2005}. The \ha~\RC~reaches a plateau at $\sim$210\kms, compatible
with its morphological type (Sab). No \ha~emission can be seen as
a counter part of the HI extension to the South-East \citep{Noordermeer:2005}. There is no obvious evidence for interaction with
its companion, UGC 89, on our \ha~\VF.
%
\\
\noindent \textbf{UGC 1013 (NGC 536) \& NGC 542}. An optical
\RC~has been obtained by \citet{Vogt:2004} for UGC
1013. Because of bad weather, we could not get a sufficient SNR
when observing that galaxy with GHASP. On the other hand, we have
detected, in the same field of view, \ha~emission from its
companion NGC 542 (systemic velocity around 4660\kms~from
HyperLeda) with enough signal to get a reliable \VF.
%
\\
\noindent \textbf{UGC 1317 (NGC 697)}. It is the brightest galaxy
of the group NGC 677-697. HI data have been obtained by \citet{Giovanardi:1985}, by \citet{Rhee:1996} and by WHISP
(website). A good agreement with HI observations within the first
2\arcmin~is observed. Outside 2\arcmin, the HI rotation
velocities of the receding side increase but we have no
\ha~emission there to check that. There is no obvious evidence for
interaction with its companion, NGC 677, on our \ha~\VF.
%
\\
\noindent \textbf{UGC 1437 (NGC 753)}. Relatively close to the
center of the A262 cluster, it is not HI deficient however. It has
been already observed in the optical by \citet{Rubin:1980}, \citet{Amram:1994} (Fabry-Perot data), \citet{Courteau:1997} and by \citet{Vogt:2004} who found roughly the same kinematical parameters and
\RCs. From their HI \RC, \citet{Bravo-Alfaro:1997} confirm the
flatness of the \ha~\RC~even beyond the optical radius
($D_{25}/2$). Although less extended, our \ha~\VF~is in good
agreement with the HI observations by \citet{Broeils:1994} and by WHISP (website).
%
%
%
\\
%
%
\noindent \textbf{UGC 1655 (NGC 828)}. \ha~emission is detected
only in its very center, so that only the rising part of the
\ha~\RC~can be plotted. The \ha~image by \citet{Hattori:2004}
shows two bright patches on each side of the nucleus. Because of
the limited extension of the \RC, we could not determine the
inclination from the kinematics and adopted the value found in
HyperLeda. Anyway, a plateau seems to be reached at
$\sim$20\arcsec~from the center with a velocity $\sim$205\kms.
The \RC~derived by \citet{Marquez:2002} is in agreement with ours
and extends a bit further but with a strong dispersion beyond 20".
\citet{Wang:1991} provide a CO \RC~limited to a 10" radius, also
in agreement with our \ha~\RC. No HI map is available in the
literature but the width of the HI profile at 20\%~(427\kms~from
\citealp{Bottinelli:1990}; 556\kms~from \citealp{Springob:2005}) is
about twice our \ha~\VF~amplitude, showing that our \ha~\RC~is far
from reaching the maximum of the rotation velocity.
\\
%
%
%
\noindent \textbf{UGC 1810 \& UGC 1813}. Faint \ha~detection in
UGC 1810 despite two hours of integration in good conditions while
a strong \ha~emitting blob is found in the central region of UGC
1813. Interestingly, two compact \ha~emitters are detected away
from the two galaxies on the eastern and western edges of our
field of view (02h21m38s, 39\degr21\arcmin35\arcsec and 02h21m20s,
39\degr21\arcmin35\arcsec respectively). These two objects are
related with the two galaxies for which we estimate a systemic
velocity around 7550$\pm$100\kms, in agreement with WHISP
(website) whereas HyperLeda gives a systemic velocity of
7356\kms~$\pm$~57. They may be intergalactic HII regions. A
faint velocity gradient ($\sim$30\kms) is observed in the \VF~of
the western object, suggesting that it may by a tidal dwarf galaxy
candidate with a systemic velocity around 7680\kms~(modulo
378\kms~which is the free spectral range). At the opposite, no
velocity gradient is seen in the eastern object having a systemic
velocity of 7400\kms~(modulo 378\kms).
%
\\
%
%
%
\noindent \textbf{UGC 3056 (NGC 1569, Arp 210)}. Magellanic
irregular starburst galaxy (Seyfert I type). We detect a strong
\ha~emission in the center, causing ghosts in the
southern outskirts of the galaxy on our data. This prevented us
from computing an accurate \vf~in that region, even though there
is some real emission there, as confirmed by \citet{Hunter:2004} line map. Also, we miss some extended filaments, being too
faint or out of our field of view. Our \ha~\vf~is almost uniform and
does not show any evidence for rotation. Thus we could not
fit any rotation model to this \vf~and do not show any \RC~here.
However, the \vf~and the \pvm~shows markedly higher velocities in the center and
lower velocities on the northwestern side, a feature not clearly
seen in previous studies of the ionized gas with long slit
spectroscopy (\citealp{Tomita:1994}, \citealp{Martin:1998}) mostly focused on the
filamentary structures. Our \pvm~is compatible with the HI \rc~of
\citet{Stil:2002} that does not show any clearly rising part
until 50\arcsec~radius.

%
%
%
\noindent \textbf{UGC 3334 (NGC 1961, Arp 184)}. NGC 1961 is a very bright and massive distorted LINER 2 SAB(rs)b galaxy showing highly irregular outer spiral arms and a pathological disk \citep{Arp:1966}.  It does not show any nearby companion and no clear double nucleus indicates a merger in progress.  Nevertheless, NGC 1961 is the central member of the small group of nine galaxies located in the same velocity interval with a projected separation of 1 Mpc. Two long straight arms tangent to north-following side of galaxy point toward an extended HI counterpart \citep{Shostak:1982}. The amplitude of the WHISP \VF~(website) as well as the width of the HI profile at 20\%~($\sim$700\kms~from the WHISP website and 690\kms~from \citealp{Bottinelli:1990}) are fully compatible with the \ha~\VF~amplitude.  The overall resemblance between \ha~\vf~and HI \vf~is pretty good within the optical disk taking into account the low spatial resolution in the HI data.  However, \ha~kinematics present perturbations all over the disk leading to an asymmetric and wavy \rc.
The southern spiral arm and the knotty regions in the northern arm present unexpected velocities leading to a model of \rc~with strong residuals for which it was necessary to constrain the inclination.  Our \rc~is in reasonable agreement with the \rc~along the major axis from \citet{Rubin:1979} but our \rc~is almost twice extended on the receding side. \citet{Rubin:1979} claimed that NGC 1961, with its total mass greater than $10^{12}$ M\Sun, was the most massive spiral known. Due to the uncertainties on the inclination and to waves in the \rc, the maximum rotation velocity (377$\pm$85\kms) could even be higher. From their optically derived spectra, they concluded to unexplained motions within the system.  The \ha~lines continuously display a double profile (not resolved by \citealp{Rubin:1979}) from the centre to the outermost points of the approaching side (as it can be seen on the \ha~\pvm, the most external velocities reach respectively $\sim$3500\kms~and 3720\kms).  The maximum velocity has been chosen as the mean external velocity.  These double profiles in the disk are an additional evidence for the complex history of this galaxy (merging, interaction, stripping) which still needs to be modeled taking into account its disturbed and asymmetric HI distribution and X-ray emission.
\\
\noindent \textbf{UGC 3382}. This early type SBa galaxy presents a
lack of \ha~emission in the center, we thus miss the inner part of
the \VF~within the first 3 kpc. Due to faint \ha~SNR in the rest
of the disk, our \RC~is based on a limited number of velocity
bins. Despite of this, both sides of the \RC~agree fairly well. We
excluded from the analysis the outermost part of the approaching
side because it has no counterpart on the receding side. An inner velocity gradient is well
seen in the HI \pvm~\citep{Noordermeer:2005}, not in the \ha one.
Their diagram shows that the \RC~rises steeply in
the center and suggests that the maximum velocity is rapidly
reached, at about 1'. The amplitude of the HI velocities is
in good agreement with that of our \ha~\VF, suggesting that we
actually reach the maximum rotation velocity within the optical
radius, at the end of our \ha~\RC. The morphological and
\ha~kinematical inclinations are the same (21\degr) while
16\degr~is found from HI data.
%
\\
\noindent \textbf{UGC 3463 (KIG 168)}. Our \ha~map is in agreement
with \citet{James:2004} but suffers from bad seeing conditions
leading to the confusion of several HII regions. Our \RC~shows a
bump at $\sim$10\arcsec~likely due to a bar. No HI \RC~is
available. The width of the HI profile at 20\%~(341\kms~from
\citealp{Springob:2005} and 334\kms~from \citealp{Bottinelli:1990}) is
in agreement with the amplitude of our \ha~\VF, confirming that
the maximum velocity is reached before the optical radius
$D_{25}/2$.
\\
\noindent \textbf{UGC 3521}. The \ha~\RC~is poorly defined in the
center because of the faint \ha~emission. It rises slowly up to
$\sim$165\kms~and extends up to about two third of the optical
radius ($D_{25}/2$) without being certain to reach the maximum
rotation velocity. No HI \RC~is available but the width of the HI
profile at 20\%~(381\kms~from \citealp{Springob:2005}) is higher (by
$\sim$50\kms) than that of our \ha~\VF, suggesting that the
plateau is almost reached but not yet.
\\
\noindent \textbf{UGC 3528}. The \ha~emission is faint. It is
sufficient however to derive a \RC~apparently reaching a maximum
although it barely reaches half the optical radius. The width of
the HI profile at 20\%~(344\kms~from \citealp{Springob:2005}) is in
good agreement with the amplitude of our \ha~\VF, confirming that
we do reach the maximum rotation velocity ($D_{25}/2$).
\\
\noindent \textbf{UGC 3618 (NGC2308)}. No \ha~emission is
detected thus no image is displayed.
%
\\
\noindent \textbf{UGC 3685}. Flocculent SBb galaxy whose bar
terminates at a well-defined circular ring. The bar is aligned
with the kinematic major axis and there is no clear signature of
it on the \ha~\VF~although a bump can be seen on the \ha~\rc~at
about 1 kpc. \ha~emission is mainly seen in a wide ring and in
short spiral arms. No \ha~emission can be seen in the center
except along the bar, in agreement with \citet{James:2004}
\ha~image, but we miss some faint emitting region to the
southeast. As seen in the \pvm, a large velocity dispersion is observed in the nucleus of the galaxy.
The photometric inclination is 55\Deg~from HyperLeda,
32\Deg~from NED and 33\Deg~from \citet{James:2004}. We agree
with \citet{James:2004} measurement, but we may estimate the uncertainty
at $\sim$10\Deg. Furthermore, if we take into account the very
faint outer arms, the disk is rounder. On the other hand, the HI
disk, which is about five times more extended than the optical
one, is rather circular. It is possible that the inclination
should vary with the radius, this should be done on the HI data
(not yet published). Moreover, the \ha~ring is also almost
circular. This probably means that the inclination of the galaxy
may still be lower. We fit a kinematical inclination of
12$\pm$16\Deg. Within the error bars, the kinematical inclination
is compatible with the morphological one ($\sim$33$\pm$10\Deg).
The \RC~rises rapidly within the first 2 kpc as expected for early
type galaxies despite the fact that the bar, aligned with the
kinematical major axis should lower the inner gradient
(Hernandez et al., in preparation). If we exclude this inner structure, the
\RC~seems to grow with a solid body behaviour. The outermost
\ha~regions disconnected from the main body of the disk extend
beyond the optical radius $D_{25}/2$. The amplitude of the WHISP
\VF~(website) as well as the width of the HI profile at
20\%~($\sim$100\kms~from the WHISP website, 118\kms~from \citealp{Springob:2005} and 103\kms~from \citealp{Bottinelli:1990}) are almost twice
the \ha~\VF~amplitude. Moreover, regarding
the HI velocity gradient, slowly growing up to the HI outskirts,
our \ha~\RC~probably does not reach the maximum velocity. 
\\
\noindent \textbf{UGC 3708 (NGC 2341)}. It forms a pair with UGC
3709. Its \ha~distribution is asymmetric, brighter on the eastern
side. Our \ha~\VF~suggests an inclination of $44\degr\pm16\degr$,
higher than the photometric inclination ($16\degr\pm24\degr$) but
compatible with the error bars. No HI \VF~is available. The width
of the HI profile at 20\%~(324\kms, \citealp{Bottinelli:1990}) is in
agreement with our \ha~\VF~amplitude, confirming that we reach the
maximum rotation velocity.
%
%
\\
\noindent \textbf{UGC 3709 (NGC 2342)}. It forms a pair with UGC
3708. The \ha~\RC~reaches a plateau with a lower maximum rotation
velocity than the one given by \citet{Karachentsev:1984y},
around 250\kms~instead of 292\kms. No HI \VF~is available. The
width of the HI profile at 20\%~(396\kms, \citealp{Bottinelli:1990})
is in agreement with our \ha~\VF~amplitude, confirming that the
maximum velocity is reached.
%
%
%
\\
\noindent \textbf{UGC 3826 (KIG 188)}. It has a faint
\ha~emission, in agreement with the map presented by \citet{James:2004}. The rising part of the \ha~\RC~is ill defined, but a
plateau seems to be reached within the optical radius. The HI
\VF~(WHISP, website) shows a velocity amplitude in agreement with
our \ha~\VF~and gives the same overall orientation for the major
axis position angle. However, the pattern of the HI isovelocity
lines in the central part points at a quite different orientation
compared with that suggested by our \ha~\VF. The width of the HI
profile at 20\%~given by \citet{Springob:2005} (101\kms) is
significantly larger than the amplitude of our \VF~(almost double)
and seems abnormally large when compared with the HI profile
obtained by WHISP (website).
%
%
\\
\noindent \textbf{UGC 3740 (NGC 2276, Arp 25)}. Its spiral pattern
is unusual, perhaps because of a tidal encounter with the probable
companion NGC 2300 \citep{Karachentsev:1972} or more likely, due to
tidal stripping. Indeed, it is a member of a group where stripping
has been evidenced \citep{Rasmussen:2006}. The western side of
our \ha~image and the \VF~show compression due to stripping by the
intragroup medium. Our \ha~data also reveal low surface brightness
filaments extending towards the east, in agreement with \citet{James:2004}. Our \ha~\RC~is very peculiar and asymmetric.
A steep velocity rise is observed in the galaxy core. The width
of the HI profile at 20\%~(167\kms~from \citealp{Springob:2005}) is
slightly larger than our \ha~\VF~amplitude. Nevertheless, the
shape of our \ha~\RC~suggests that the maximum is reached just
after the optical radius ($D_{25}/2$). The CO emission \citep{Elfhag:1996} is distributed in a lopsided fashion, with more emission
towards the northwestern region.
%
\\
\noindent \textbf{UGC 3876 (KIG 193)}. Diffuse \ha~emission can be
seen all over the disc, in agreement with \citet{James:2004}. The
\ha~\RC~rises slowly beyond the optical radius ($D_{25}/2$) so
that we are not sure to reach the maximum rotation velocity. No HI
\VF~is available. However, the width of the HI profile
at 20\%~(208\kms, \citealp{Bottinelli:1990}) is in good agreement
with our \ha~\VF~amplitude, suggesting that the maximum rotation
velocity is effectively reached with our \ha~\RC.
%
\\
\noindent \textbf{UGC 3915}. Strong \ha~emission can be seen all
over the disc. Our \ha~\RC~rises steeply and reaches a plateau
around 200\kms~at about 0.5~$D_{25}/2$. No strong signature of the
bar (aligned with the major axis) can be seen on our \VF. A small
bump, seen on our \RC~in the 5 inner arcsec could be due to the
bar. No HI \VF~is available in the literature. The
width of the HI profile at 20\%~(323\kms, \citealp{Bottinelli:1990})
is in agreement with our \ha~\VF~amplitude.
\\
\noindent \textbf{IC 476}. 
We detected some \ha~emission in IC 476, the small companion of UGC 402
(both observed in the same \FOV). We derive its \RC~up
to about 0.5 $D_{25}/2$, thus it is not sure that the
maximum rotation velocity is reached. No HI data are available
in the literature.
\\
\noindent \textbf{UGC 4026 (NGC 2449)}. We detected faint
\ha~emission in the low surface brightness galaxy UGC 4026.
However, its \ha~emission is sufficient to derive with confidence a \RC~up
to about 0.5 $D_{25}/2$. As it seems that we observe the plateau of the \RC, the maximum velocity may be reached. A kinematical inclination of
56$\pm$4\Deg~has been computed, lower than the morphological one
of 73$\pm$4\degr. The kinematical inclination is very uncertain
due to the very low SNR of our \ha~data. No HI data are available
in the literature.
%
%
\\
\noindent \textbf{UGC 4165 (NGC 2500, KIG 224)}. This galaxy
belongs to a quartet of galaxies \citep{Sandage:1994}. Diffuse
\ha~emission is observed in our \ha~map, in agreement with \citet{James:2004}. Its short bar is almost aligned with its minor
kinematical axis. Within the error bar, a good agreement is
observed between the kinematical and morphological inclination.
Our \ha~\VF~is in very good agreement with the WHISP data
(website). The width of the HI profile at 20\%~(101\kms~from
\citealp{Springob:2005}, 114\kms~from \citealp{Bottinelli:1990} and
100.9\kms~from \citealp{Haynes:1998}) is in agreement with our
\ha~\VF~amplitude.
%
%
%
%
\\
\noindent \textbf{UGC 4256 (NGC 2532, KIG 232)}. It presents
patchy \ha~emission along its spiral arms. The general pattern of
the \ha~\VF~is in good agreement with the HI \VF~(WHISP, website).
The position angle of the kinematical major axis of the
\ha~\VF~(116\degr) is in agreement with the HI one, but is very
different from the value given in HyperLeda (26\degr) and in the
RC3 (10\degr) as already noticed by \citet{Marquez:1996}.
Indeed it is clear that the outermost contours of the galaxy
measured from broadband imaging are elongated along the minor
kinematical axis. Furthermore, morphological and kinematical
inclinations are determined using position angles separated by
90\Deg. Nevertheless the HI disk is elongated along its
kinematical major axis, leading to an inclination $\sim$30\Deg.
Our \ha~\RC~rapidly reaches a plateau climbing up to a maximum
velocity $\sim$100\kms~around the optical radius ($D_{25}/2$) in
agreement with \citet{Marquez:1996} from \ha~slit spectroscopy.
%
\\
\noindent \textbf{UGC 4393 (KIG 250)}. A strong \ha~emission can
be seen along the bar (aligned with the kinematical major axis)
and in the southwestern spiral arm, in agreement with \citet{James:2004} \ha~map. Due to the bar, the \ha~\VF~is strongly perturbed
in the center. As a consequence, the \ha~\RC~is strongly
perturbed, with counter-rotation motions in the center. No HI
\VF~is available in the literature. The HI width at
20\%~(160\kms~from \citealp{Springob:2005}) is more than twice our
\ha~\VF~amplitude (70\kms). This means that the maximum velocity
is not reached in \ha. Moreover, due to the presence of the strong
bar, the inclination is probably overestimated. Indeed, the
external axis ratio of the outermost isophotes leads to an
inclination around 35\Deg. However, even with this lower value of
inclination, the HI maximum rotation velocity remains slightly
lower than expected for such a galaxy, according to the Tully
Fisher relationship.
%
%
%
\\
\noindent \textbf{UGC 4422 (NGC 2595)}. Located in the Cancer
cluster, this barred spiral exhibits a prominent nucleus and
distorted outer regions extending up to 70\arcsec~($\sim$18 kpc).
\ha~emission is observed in the center, in the ring and in the
beginning of the bar, in agreement with \citet{Gavazzi:1998}. Our \RC~rapidly rises and reaches a plateau at almost 350\kms~in the innermost 5\arcsec. The
shape of the \RC~is consistent with the Fabry-Perot observations
from \citet{Amram:1992} and we derive the same set of parameters
within the uncertainties. There is a strong discrepancy between
the photometric (49$\pm$4\degr) and kinematic inclination
(25$\pm$8\degr). No HI velocity map is available. The corrected HI
profile width of 321\kms~from \citet{Springob:2005} is in
agreement with our \ha~\VF~amplitude.
\\
\noindent \textbf{UGC 4456 (KIG 260)}. A difference of 82\Deg~is
observed between the kinematical and morphological major axis
position angle. This difference is due to its low inclination
leading to an uncertain morphological determination. Within the
error bar, a good agreement is observed between the kinematical
and morphological inclination. No HI velocity map is available.
The width of the HI profile at 20\%~(110\kms~from \citealp{Bottinelli:1990} and 105\kms~from \citealp{Lewis:1987}) is larger than our
\ha~\VF~amplitude although our \ha~\RC~rapidly reaches a plateau
extending well beyond the optical limit.
%
\\
\noindent \textbf{UGC 4555 (NGC 2649, KIG 281)}. No HI velocity map is
available. The width of the HI profile at 20\%~(251\kms~from
\citealp{Bottinelli:1990} and 256\kms~from \citealp{Lewis:1987}) is in
agreement with our \ha~\VF~amplitude.
%
\\
\noindent \textbf{UGC 4770 (NGC 2746)}. In agreement with its
morphological type (SBa), it shows a weak and asymmetric
\ha~emission. In particular, no \ha~is detected within the bar.
The \ha~\PVM~and \RC~are badly defined due to faint SNR in the
data, specially on the receding side. The maximum velocity
$V_{CO}$=207\kms~deduced from the molecular component \citep{Sauty:2003} is in agreement with our \ha~\VF~amplitude. No HI velocity
map is available. The width of the HI profile at
20\%~(264\kms~from \citealp{Bottinelli:1990} and 270\kms~from \citealp{Lewis:1987}) suggests that \ha~and the CO \RCs~do not reach the maximum velocity.
%
%
\\
\noindent \textbf{UGC 4820 (NGC 2775, KIG 309)}. \ha~imaging from
\citet{Hameed:2005} is comparable with our data, showing a
flocculent ring of \ha~emission. Stellar dynamics from \citet{Kregel:2005} suggest a maximum velocity of 283\kms~lower than that
suggested by the \ha~kinematics. No HI velocity map is available
in the literature. The width of the HI profile at
20\%~(435\kms~from \citealp{Bottinelli:1990}) is in agreement with
our \ha~\VF~amplitude.
%
\\
\noindent \textbf{UGC 5045}. It is a triple-arm barred galaxy that
suffers from global distortion and shows UV excess (KISO survey,
\citealp{Takase:1991}). Its arms are knotty with many HII regions
distributed asymmetrically in the disk and no \ha~emission is
detected in the very center. The rising part of our \ha~\RC~is
thus missing and the curve looks like a plateau around 400\kms,
which seems quite high for such an Sc type galaxy. Indeed, our
\ha~\VF~suggests a faint inclination (17\Deg~only, with an
uncertainty of 10\degr) whereas the photometry suggests
41$\pm$4\Deg~(HyperLeda). Adopting this last value would lower the
plateau of our \RC~around a more normal value of 200\kms, thus
casting a doubt on the inclination deduced from our kinematics. No
HI velocity map is available but the width of the HI profile at
20\%~(200\kms, \citealp{Springob:2005}) is in agreement with our
\ha~\VF~amplitude. Taking into account the methodology used, and
the fact that the \VF~which presents a good SNR is not strongly
disturbed (no bar, symmetric \RC), we adopt the kinematic
inclination.
%
\\
\noindent \textbf{UGC 5175 (NGC 2977, KIG 363)}. This galaxy shows
a strong \ha~emission. Our \ha~\RC~is flat, with a plateau at
$\sim$190\kms, suggesting a maximum velocity rotation lower than
that found by \citet{Karachentsev:1984i} in the optical. It has
been Observed in CO by \citet{Sauty:2003} who find a line width
at 50\%~of 312\kms. No HI velocity map is available but the
line width at 20\%~of 356\kms~\citep{Theureau:1998} is comparable
with our \ha~\VF~amplitude.
\\
%
%
\noindent \textbf{UGC 5228}. The \ha~\RC~of this strong
\ha~emitter, rises up to a plateau at $\sim$125\kms. The HI line
width of $\sim$270\kms~measured by \citet{Doyle:2005} is in good
agreement with our \ha~\VF~amplitude.
%
\\
\noindent \textbf{UGC 5251 (NGC 3003)}. As already noted by \citet{Rossa:2003}, except in the nucleus and in several bright HII
regions, the \ha~emission is rather faint and its distribution
asymmetric, like the spiral arms of the galaxy. Such an asymmetry
suggests that this galaxy may be disturbed by a dwarf companion.
The \VF~is also rather asymmetric. No HI \VF~is available. The
width of the HI profile at 20\%~(305\kms~from \citealp{Springob:2005}
and 289\kms~from \citealp{Bottinelli:1990}) is slightly higher (by
$\sim$30\kms) than the velocity amplitude of our \ha~\VF. Thus,
the maximum rotation velocity is probably not reached with our
\ha~\RC.
%
%
\\
\noindent \textbf{UGC 5279 (NGC 3026, KIG 377)}.
No HI \VF~is available. The
width of the HI profile at 20\%~($\sim$220\kms, \citealp{Bottinelli:1990}) is lower than our \ha~\VF~amplitude, suggesting that the
maximum velocity is reached on our \ha~\PVM.
%
%
\\
\noindent \textbf{UGC 5319 (NGC 3061, KIG 382)}. Its bar, almost
aligned with the major axis, shows no significant signature in our
\ha~\VF. Our \ha~\RC~shows an inclined plateau, continuously
rising within the optical limits. The steep rising in the first
kpc of the \RC~for this relatively low mass galaxy may be the
signature of a bar. No HI \VF~is available but the width of the
HI profile at 20\%~(233\kms~from \citealp{Springob:2005} and
272\kms~from \citealp{Lang:2003}, HIJASS survey) is significantly
higher than the amplitude of our \ha~\VF~($\sim$180\kms). However
the corrected velocity from \citet{Springob:2005} is about
198\kms~and the width of the HI profile at 50\%~from \citet{Lang:2003} is 189\kms~which is in better agreement. This
disagreement between HI and \ha~data could indicate that the
galaxy is embedded in an HI complex extending much further out
than the optical limit.
\\
\noindent \textbf{UGC 5351 (NGC 3067)}. This galaxy has been
studied in the optical by \citet{Rubin:1982}, the position angle
of the major axis and the outer velocity gradients agree with both
sets of data. Taking into account the difference in distance
adopted for that galaxy (28.3 vs 19.3 Mpc), the extensions of the
\RCs~agree. The \RC~in the inner parts from stellar kinematics
\citep{Heraudeau:1999} is in very good agreement with the inner
part of the \ha~\pvm. The full resolution HI \VF~(WHISP,
\citealp{Noordermeer:2005}) is still too low and does not allow a
straightforward comparison with the \ha. In particular the bar is
not seen in HI while the signature of the bar is clearly seen in
the inner region of the \ha~\VF. The HI and \ha~velocity amplitude
and gas extension are similar but their behaviour along the major axis
seems different. The \ha~\pvm~suggests that a constant velocity is reached after radius $\sim$5\arcsec\ 
whereas  the HI \pvm~does not show such a plateau. The outermost isophotes of the HST
image \citep{Carollo:1998} suggest an almost edge-on galaxy. Due
to its high inclination, to avoid contamination due to the
thickness of the disk (outer regions along the minor axis), we
have fixed the inclination to 82$\degr$ from the morphology. The
maximum rotation velocity has been computed taking into account
this inclination and no \RC~has been plotted.
%
\\
\noindent \textbf{UGC 5373 (Sextans B, KIG 388)}. This dwarf
galaxy is part of the local group. Our \ha~map is in good
agreement with the \ha~maps from \citet{Hunter:2004} and
\citet{James:2004}. The \ha~\VF~presents a low amplitude velocity
gradient ($\sim$30\kms) barely visible on the \pvm. Both major axis orientation and inclination are
difficult to determine, nevertheless the kinematical major axis
seems to be quite different ($\sim$30\degr) from the photometric
major axis probably due to non circular motions. The \ha~\RC~is
ill defined in the central part, with a possible counter rotation
within 15\arcsec~from the center, but rises rapidly beyond 30
arcsec and seems to reach a plateau at about 50 arcsec, in
agreement with the HI data from \citet{Hoffman:1996} who miss the
rising part because of their poor resolution (and possibly lack of
HI in the center). We also agree that this galaxy is almost face
on (from the kinematics we find 10\degr, with an uncertainty of
18\Deg~and \citep{Hoffman:1996} find 18\degr) whereas the photometry
suggests 60\Deg~(HyperLeda). However, our very low inclination
value may lead to overestimate the rotation velocities.
%
\\
\noindent \textbf{UGC 5398 (NGC 3077)}. As a member of the M81
group of galaxies it is strongly disrupted by the interaction with
M81 and M82. As shown by \citet{Walter:2002}, its optical image
is offset with respect to a prominent HI tidal arm lying at about
4\arcmin~East from the galaxy. Our \ha~map is in good agreement
with the one from \citet{James:2004}. The \ha~\VF~shows no
evidence for rotation, although some velocity gradient can be seen
on the edges of the disk, with the lowest velocities observed on
the western side (in agreement with \citealp{Walter:2002}
observations, in HI, \ha~and CO). No \ha~\RC~could be derived from
our data as one can see it on the \VF~and on the \pvm.
%
%
\\
\noindent \textbf{IC 2542 (KIG 399)}. Kinematic and photometric
data lead to a major axis position angle in good agreement but a
difference of 20\Deg~is observed on the inclination, nevertheless
the difference on the inclination is compatible with the error
bars. Our \ha~\RC~seems to reach a plateau at almost
300\kms~within the optical radius ($D_{25}/2$). No HI data are
available in the literature.
%
\\
\noindent \textbf{UGC 5510 (NGC 3162, NGC 3575) }. Our
\ha~\RC~seems to reach a plateau just before the optical radius
($D_{25}/2$) although the velocities for the receding side are
still increasing beyond. No HI \VF~is available. The width of the
HI profile at 20\%~(204\kms~from \citealp{van-Driel:2001} and
187\kms~from \citealp{Bottinelli:1990}) is in agreement with our
\ha~\VF.
\\
\noindent \textbf{UGC 5532 (NGC 3147)}. The \ha~emission is very
weak in the nuclear region and does not allow us to plot the
rising part of the \ha~\RC~which rapidly reaches (within 1 kpc) a
slightly decreasing plateau starting at almost 400\kms. No HI
velocity map is available. The width of the HI profile at
20\%~(455\kms~from \citealp{Richter:1991}, and 403\kms~from
\citealp{Lang:2003}) is in agreement with our \ha~\VF.
%
\\
\noindent \textbf{UGC 5556 (NGC 3187, Arp 316, HCG 44D)}.
It is a member of
the famous compact group HCG 44 \citep{Hickson:1989} strongly
interacting with NGC 3190 \citep{Sandage:1994}.
 \ha~emission is only observed along the lenticular central region (the bar) and in the inner arms.
The arms are obviously driven by streaming motions due to the
interaction with its companion. The position angle of the
kinematical major axis of the galaxy is almost perpendicular to
the bar . Thus, the \VF~traces the kinematics of the
bar and of the streaming motions in the arms but not the
kinematics of the disk. Furthermore we cannot compute a \RC. The velocity amplitude perpendicular to the bar accross the
\VF~is $\sim$150\kms, lower than the width of the HI
profile at 20\%~(296\kms~from \citealp{van-Driel:2001} and
257\kms~from \citealp{Bottinelli:1990}).
%
%
\\
\noindent \textbf{UGC 5786 (NGC 3310, Arp 217)}. The unusual
smooth outer plume on the western side of the galaxy is probably
the result of a recent merger \citep{Balick:1981}. The plume
has a smaller radial velocity than the galaxy.\citet{Wehner:2005} evidenced a closed loop in the V and R-band that may be
tidal debris. Diffuse \ha~emission, not visible by \citet{James:2004}, is detected on our \ha~image all around the galaxy. A
large portion of the star formation is located in a central ring
surrounding an off-centered nucleus. The central region of the
\ha~\VF~displays a S-shape pattern encircling two velocity peaks
leading to two severe bumps in the \RC. The bright nucleus exhibits 
a steep velocity rise. Outside the nuclear region, the velocity decrases 
and then remains flat along the galaxy major axis. Despite the evidence for
perturbations, the \RC~is fairly symmetric although showing some
oscillations. Outflows are observed in the central region of the
galaxy ($\sim$1 kpc) by \citet{Schwartz:2006} where we measure
large \ha~linewidths. The width of the HI profile at 20\%~(330\kms
from \citealp{Springob:2005}) is significantly larger than the
amplitude of our \ha~\VF, suggesting that we cannot reach the
maximum rotation velocity with our \ha~data. Indeed, our
\ha~\RC~clearly reaches a maximum at about 120\kms~with its
central bump at 0.5 kpc but the behavior of the curve in the outer
parts is too chaotic (with a total divergence between receding and
approaching side beyond the optical limit) for concluding anything
about the true maximum. The kinematical inclination has been
determined in excluding the central spiral structure within the
first kpc, leading to a rather high inclination of 53$\pm$11\Deg.
This is significantly higher than the morphological inclination of
16$\pm$25\Deg~(HyperLeda) but the difference remains compatible
with the error bars. We finally choose the kinematical
inclination.
%
\\
\noindent \textbf{UGC 5840 (NGC 3344, KIG 435)}. The \ha~map is in
agreement with \citet{Knapen:2004}, however, we miss the
outermost parts of the galaxy because of our \FOV~(limited to a
4\arcmin~diameter in that case because of the use of a 2 inches
circular filter). Also, the warp observed in HI (WHISP, website)
cannot be seen within our \FOV. Despite the fact that this ringed
spiral galaxy is fairly regular, the \VF~is not at all
symmetric with respect to the minor axis. The nuclear region shows
a large velocity dispersion, as seen in the \PVM~diagram. The \ha~\RC~exhibits a
valley where the main spiral structure vanishes in the optical.
The width of the HI profile at 20\%~(166\kms~from \citealp{Bottinelli:1990}) is in agreement with the amplitude of our \ha~\VF. The
maximum velocity found on our \ha~\RC~is rather high for such a
small galaxy ($D_{25}/2\sim$~6 kpc) of Sbc type.
%
\\
\noindent \textbf{UGC 5842 (NGC 3346, KIG 436}. The \ha~emission
seems relatively poor on the receding side because the
interference filter we have used was not perfectly centered on the
galaxy. Nevertheless we were able to derive a quite acceptable
\RC. The high dispersion of our rotation velocities in the center
is probably due to the bar, then (from 1 to 6 kpc) our \RC~is slowly rising like a solid body. No HI \VF~is available.
The width of the HI profile at 20\%~(166\kms~from \citealp{Bottinelli:1990} and $\sim$180\kms~from \citealp{Tift:1988}) is in
agreement with our \ha~\VF~suggesting that the maximum rotation
velocity is reached. We conclude that a plateau (if any) must
begin just beyond the optical radius ($D_{25}/2$) where our
\ha~\RC~ends after an almost continuously rising tendency.
%
\\
\noindent \textbf{UGC 6118 (NGC 3504)}. This early type galaxy
presents an almost circular outer ring, and a thin bar (more
visible in J-H-K band from NED) embedded in a rather oval
structure (axis ratio $\sim$0.5). A bulge is visible on the near
infrared images. The agreement between our \ha~map and the one
from \citet{Hameed:2005} is very good. The main bar has a
size $\sim$1.2 arcmin. A secondary bar may be suspected in the
first $\sim$20\arcsec around the very bright \ha~nuclei. An
\ha~spiral structure is observed within the oval structure
starting at the end of the inner bar. We find the major
kinematical axis to be almost parallel to the bar and the oval
structure. Our \ha~\VF~is almost limited to the central oval
structure and the bar, with only a few points in the outer ring.
A steep inner rise of the velocities is observed in the galaxy core, 
otherwise the \ha~velocities remain roughly constant before 45\arcsec, 
radius beyond which they begin to increase slightly. 
This is not observed in the HI \pvm~from WHISP \citep{Noordermeer:2005}, 
and the HI \RC~slightly increases all along the first arcminute. 
Their HI \VF~perfectly covers the outer ring and leads to an
inclination of 39\Deg. The morphological inclination deduced from
the outer ring is $\sim$27\Deg, while the inclination of the oval
structure is $\sim$45\Deg. The kinematical inclination of 52\Deg
deduced from our \ha~\VF~is consistent with the inclination of the
oval distortion. Nevertheless we computed the \RC~using the HI
inclination because our kinematical inclination mainly relies on
velocities measured within the central oval structure. The central
region of the \VF~within the bar displays an S-shape pattern
encircling two symmetric velocity peaks leading to two strong
bumps on the \RC~(at $\sim$0.5-1 kpc). Our \ha~\RC~is in good
agreement with the long-slit observations of \citet{Usui:2001}.
%
%
\\
\noindent \textbf{UGC 6277 (NGC 3596, KG 472)}. Except for the
first central 15\arcsec, it is a low surface brightness galaxy
displaying everywhere weak \ha~emission.
As a result, our \ha~\RC~seems to be limited to its rising part.
Nevertheless, the velocity amplitude is in agreement with the HI
profile of \citet{Kornreich:2000}, suggesting that our
\ha~\RC~probably reaches the maximal rotation velocity despite its
limited extent (about half the optical radius).
%
%
\\
\noindent \textbf{UGC 6419 (NGC 3664, Arp 05)}. It is one of the
prototypical strongly barred magellanic spirals and has a nearby
companion, UGC 6418, at 6.2\arcmin. HI VLA observations \citep{Wilcots:2004} show that the current interactions affect the
morphology and the kinematics of the main galaxy. The HI \VF~of
both galaxies are connected, with a large extension
($\sim$10\arcmin) compared with the optical one
($\sim$1\arcmin). The position angle of the major axis of our
\ha~\VF~is almost perpendicular to the HI one. Indeed, the
gradient of our \ha~\VF~is almost aligned along the bar, which
cannot be seen in the HI data due to the low spatial resolution.
It is therefore possible that our \ha~\RC~does not reflect the
rotation of the galaxy but more probably the kinematics of the
bar, likely to be affected by non circular motions. Anyway, the
shape of our \RC~(solid body type) is not surprising for a galaxy
of this type, but we find a rather low rotation velocity
considering the luminosity of that galaxy. This suggests that the
kinematical inclination (66\Deg) as well as the inclination
determined from the disk shape when excluding the tidal tail
(57\Deg) are too high. Indeed, the axial ratio of the disk
including the tidal tail is close to one, suggesting that the
galaxy is seen almost face on. We conclude that the velocities of
our \ha~\RC~are probably underestimated. Furthermore, the HI
\VF~amplitude observed by \citet{Wilcots:2004} ($\sim$150\kms) is
about twice ours ($\sim$70\kms), suggesting that we do not reach
the maximum of the rotation velocity.
%
%
%
%
%
\\
\noindent \textbf{UGC 6521 (NGC 3719)}. No evidence for interaction
with its companion, UGC 6523, can be seen in our \ha~\VF~which is
fairly symmetric. The resulting \RC~is slightly decreasing beyond
6 kpc. No HI \VF~is available. The width of the HI profile at
20\%~(397\kms, \citealp{Theureau:1998}) is somewhat higher than the
amplitude of our \ha~\VF~but the shape of our \RC~leaves no doubt
that the maximum is effectively reached within the optical limits.
%
\\
\noindent \textbf{UGC 6523 (NGC 3720)}. Nearby companion of UGC
6521. The \ha~emission is limited to the central regions (about
one third the optical radius), thus our \ha~\RC~only shows the
central rising part of the curve and no clear sign of interaction
can be seen. No HI \VF~is available. The width of the HI profile
at 20\%~(393\kms~from \citealp{Bottinelli:1990}) is much larger
(more than three times) than our \ha~\VF~amplitude, indicating
that we are far from reaching the maximum of the rotation
velocity.
%
%
\\
\noindent \textbf{UGC 6787 (NGC 3898)}. Because of the
interference filter used for this observation, which was not
perfectly centered on the systemic velocity of the galaxy, the
\ha~emission of the eastern side was not transmitted through the
filter (a good \ha~image could be found in \citealp{Pignatelli:2001}). 
As a consequence, our \ha~\RC~is only traced with the
approaching side but seems acceptable anyway. It reaches a plateau
within the optical radius ($D_{25}/2$) and seems to decrease
beyond, in agreement with the HI data from WHISP \citep{Noordermeer:2005}. 
However, the HI data suggest that the \RC~climbs
again beyond 2 arcmin to recover the same velocity level.
Also, the steep inner rise of the velocities seen in the HI \pvm~from WHISP 
is not detected in our \ha~\pvm, which is probably due to a
too low signal-to-noise ratio of the optical observations. 
The \ha~velocities remain constant in the approaching half 
of the diagram whereas the HI velocities increase in the same region.
%
%
\\
\noindent \textbf{UGC 7021 (NGC 4045, NGC 4046)}. The \ha~emission
of this barely barred galaxy is found along the inner ring and in
the center. No diffuse \ha~is detected in otherwise in the disk,
which is not surprising for an SAB(r)a galaxy. Our \ha~map is very
similar to the one found in GOLD Mine \citep{Gavazzi:2003}.
Because \ha~emission is only detected in the very
central area of the optical disk, we use the photometric
inclination (56\degr) rather than the kinematical one (34\degr).
Indeed the kinematical inclination is based only on the inner ring
which is supposed to be circular, thus biasing the inclination
determination. Our \ha~\RC~shows a slowly decreasing plateau
starting at $\sim$220\kms~and $\sim$2 kpc, corresponding to the
tip of the bar. The core of the galaxy exhibits two different velocity 
components, one at $\sim$1920\kms and another one at $\sim$2000\kms.
No HI \VF~is available. The width of the HI profile at 20\%~(337\kms, 
\citealp{Springob:2005}) is in agreement
with our \ha~\VF~amplitude, confirming that we actually reach the
maximum of the velocity rotation although our \ha~\RC~is far from
reaching the optical radius ($D_{25}/2$).
\\
\noindent \textbf{UGC 7045 (NGC 4062)}. This galaxy is a strong
\ha~emitter, as already reported by \citet{James:2004}. A perfect
agreement between photometric and kinematical parameters is
observed. Slit spectroscopy observations using \ha~and [NII]
\citep{Sofue:1998} are compatible with ours, showing an
\ha~\RC~that reaches a slowly rising plateau after about
20\arcsec. The \PVM~derived in HI by \citet{Broeils:1994} is
affected by beam smearing effects, since the inner part shows a
solid body rotation up to almost 1.5\arcmin. At larger radii
however the optical and radio data are compatible. The velocity
dispersion in the central region of our \ha~\RC~is rather high,
probably due to the bar. The linewidth of the HI profile at
20\%~(309\kms, \citealp{Springob:2005}) is in good agreement with our
\ha~\VF~amplitude.
%
%
\\
\noindent \textbf{UGC 7154 (NGC 4145)}. This galaxy, having a bar
embedded in a large elliptical bulge, is one of the principal
galaxies in the Ursa Major Cluster paired with A1208+40 (Holm
342b) located at 13\arcmin. The \ha~\RC~beyond the optical radius
($D_{25}/2$) is mainly traced from emission regions of the spiral
arms, leading to large wiggles in the \RC. The grand design of the
WHISP HI \VF~(website) is compatible with the \ha~one. The HI
maximum rotation velocity of 171\kms~\citep{Warmels:1988}, assuming a
42\Deg~inclination, is compatible with our \ha~value (about 150
\kms~assuming a 65\Deg~inclination). Due to beam smearing effects,
the inner velocity gradient in the HI is nevertheless much lower
than in the \ha~and the HI \PVM~suggests a
solid body rotation curve up to 2 arcmin from the center.
%
%
\\
\noindent \textbf{UGC 7429 (NGC 4319)}. A very faint \ha~emission
has been detected in two spots for this spiral galaxy, companion
of NGC 4291. This detection needs to be confirmed. No HI emission
has been detected by \citet{Sengupta:2006}.
\\
\noindent \textbf{UGC 7699}.
No HI \VF~is available in the literature. The linewidth of the HI
profile at 20\%~(205\kms, \citealp{Broeils:1994}) is in agreement
with our \ha~\VF~amplitude, their \PVM~shows a solid body rotation
having the same velocity amplitude as ours but the higher spatial
resolution of the \ha~\VF~enables us to observe deviations from a
pure solid body rotation.
%
\\
\noindent \textbf{UGC 7766 (NGC 4559)}. The bar of this galaxy is
almost aligned with the major axis, its signature can be seen on
the \VF~as well as on the \RC~within the first 30\arcsec.
\citet{Meyssonnier:1984} obtained a \RC~from slit spectroscopy which is
in agreement, although it has a much higher dispersion than our
\ha~\RC. Our \RC~is more extended since our \VF~reaches more outer
regions although our \FOV~is limited by the size of the
interference filter.\citet{Krumm:1979} found a flat HI
\RC~from 2\arcmin~to 7\arcmin. It has been observed more recently
by WHISP (website), their \PVM~confirms the flat behavior of the
\RC~up to 9\arcmin~together with the amplitude determined by the
previous authors. The width of the HI profile at 20\%~(254\kms,
\citealp{Springob:2005}) is in agreement with our \ha~\VF~amplitude.
\\
%
%
%
%
\noindent \textbf{UGC 7831 (NGC 4605)}. Diffuse \ha~emission is
observed in the outer disk of this galaxy. Our \ha~\VF~and
\RC~exhibit a strong asymmetry likely to be explained by the bar.
Such an asymmetry is confirmed by the \ha~+[NII] \RC~from \citet{Rubin:1980} and by \citet{Sofue:1998}. I-band image (XDSS) clearly
provides the morphological center of the galaxy and leads to the
\RC~presented here. On the other hand, the kinematical center for
this galaxy appears to be shifted by about 10\arcsec~eastward from
the morphological one but none of them leads to a symmetric inner
\RC. The receding side of the \RC~displays a plateau after 40
arcsec, following a solid body shape in the center, while the
approaching side is continuously climbing (note also that the
receding side is less luminous than the approaching one). The
method described in the paper leads to the \RC~presented here
because it minimizes the dispersion. Nevertheless this solution is
unphysical since it leads to negative rotation velocities for the
receding side in the innermost region. To avoid this, the
systemic velocity should be somewhat lowered by $\sim20$\kms,
leading to a worst disagreement between both sides of the \RC,
another solution would be to consider another rotation center but
it does not solve the asymmetry problem as explained above. This
galaxy has been observed in HI by WHISP (website), their \PVM~is
asymmetric, in agreement with our \ha~\PVM~and \RC, and the HI \VF~suggests
a warp in the outer parts of the disk.
\\
%
%
%
\noindent \textbf{UGC 7853 (NGC 4618, Arp 23)}. It forms a
physical pair with UGC 7861 (NGC 4625). Assuming a distance for
both galaxies of 8.9 Mpc \citep{Moustakas:2006}, their
physical separation is only $\sim$22 kpc (8.5\arcmin). Indeed,
these galaxies show obvious signs of interaction (e.g. strong
tidal unique tail for both galaxies). Its bar is not centered on
the nucleus \citep{Eskridge:2002}. Due to the strong southern
arm, the morphological center is offset from the kinematical one.
The \ha~\RC~displays a solid body shape up to its end
($\sim$150\arcsec) suggesting that the maximum rotation velocity
is probably not reached. The HI \RC~\citep{van-Moorsel:1983} decreases
beyond 3\arcmin~from the center (note however that their beam is
larger than 1\arcmin). HI has been also observed by WHISP
(website), their \PVM~displays a solid body \RC~up to
$\sim$1.5\arcmin~for the approaching side and a plateau beyond
1\arcmin~for the receding side. Their low resolution \VF~shows a
severely warped disk (the position angle of the major axis rotates
by 90$\degr$ between the inner regions and the outermost parts of
the HI disk). Taking into account the beam smearing effect, the
agreement between the HI and our \ha~\PVM~and \RC~is quite acceptable.
\\
\noindent \textbf{UGC 7861 (NGC 4625)}. This galaxy is the
companion of UGC 7853 (see notes above). Here again the center of
the galaxy is offset from the center of the \VF~due to the strong
southern tidal arm. Our \ha~\RC~shows a strong dispersion in the
central part (up to 10\arcsec~from the center) then rises slowly
as a solid body up to the optical limit (about 40\arcsec)
suggesting that the maximum rotation velocity is probably not
reached. Fabry-Perot observations of this galaxy, with the 3.6m
CFHT, have been published by \citeauthor{Daigle:2006a} (2006a), their \VF~and
\RC~are in good agreement with ours. HI has been observed by WHISP
(website), their \PVM~suggests a \RC~with a plateau extending
beyond the optical limit, from 1\arcmin~up to 4\arcmin.
%
\\
\noindent \textbf{UGC 7876 (NGC 4635)}. This isolated galaxy is a
member of the Coma1 Cloud (according to GOLD Mine, \citealp{Gavazzi:2003}). The
\ha~\RC~is asymmetric and the fitting method probably led to an
underestimation of the systemic velocity $\sim$10\kms. This may
be caused by the bar.
Its CO emission is very faint \citep{Sauty:2003} and no HI \VF~is
available but the linewidth of the HI profile at
20\%~(173\kms~Springob:2005) is in agreement with our
\ha~\VF~amplitude.
\\
\noindent \textbf{UGC 7901 (NGC 4651, Arp 189)}. Member of the
Virgo cluster, it is a strong \ha~emitter, particularly in an
inner ring from where start the arms. A very good agreement is
observed between our \ha~map and the ones of \citet{James:2004},
\citet{Koopmann:2001}, and GOLD Mine \citep{Gavazzi:2003}. A faint signature
of a bar can be seen in the inner regions of our \ha~\VF. Our
\ha~\RC~is in good agreement with that observed by \citet{Rubin:1999} from long slit spectroscopy except that we do not observe
the rising trend they observe for the outermost part of the
receding side. Such a behavior seems suspect, all the more since
our \ha~\RC~is perfectly symmetric and extends further out, owing
to velocities collected far from the major axis. No HI velocity
map is available. The width of the HI profile at 20\%~(395\kms,
\citealp{Springob:2005}) is in agreement with our \ha~\VF~amplitude.
%
%
\\
\noindent \textbf{UGC 7985 (NGC 4713)}. Member of the Virgo cluster
showing a strong \ha~emission as shown by the same authors as for
UGC 7901. The signature of a bar is seen in the inner part of our
\ha~\VF~within a radius of $\sim$30\arcsec. Our \ha~\RC~extends
further out than the \RC~observed by \citet{Rubin:1999}, beyond
the optical radius ($D_{25}/2$). Our \RC~is fairly symmetric and
does not show the strong decrease observed by \citet{Rubin:1999} for the
receding part around 40\arcsec. No HI velocity map is available
but the HI linewidth at 20\%~(217\kms, \citealp{Springob:2005}) is in agreement with our \ha~\VF~amplitude.
%
\\
\noindent \textbf{UGC 8334 (NGC 5055, M 63)}. M 63 is a very well
studied flocculent spiral galaxy. Our observations are fully
compatible with the \ha~Fabry-Perot data observed with the 1.6m
Mont Megantic Telescope (\citeauthor{Daigle:2006a} 2006a). Their \RC~is more
extended, due to a larger of view, but does not reach the optical
radius ($\sim$6\arcmin). It possesses a huge HI disk
($\sim$36\arcmin~diameter) strongly warped beyond the optical
disk \citep{Battaglia:2006}. The agreement between our
\ha~\RC~and their HI \RC~is very good but we find a better
symmetry for receding and approaching sides from 2 to 10 kpc
radius. The \pvm~shows that the velocities steeply rise
in the inner 10\arcsec.
%
\\
\noindent \textbf{UGC 8403 (NGC 5112)}. Paired with NGC 5107 at
13.5\Min, its \Ha~map shows a very good agreement with that of
\citet{James:2004}. The signature of a weak bar is visible in
the inner region of our \ha~\VF~but it is difficult to see the
signature of an interaction with its companion in its \VF. An
attempt of CO detection proofed dubious \citep{Braine:1993}. It
has been observed in HI by \citet{Springob:2005}, and by WHISP
(website). There is a good agreement between the WHISP \VF~and
ours and our \ha~emission is almost as extended as the HI disk.
The shape of the HI \PVM~found by WHISP is in very good agreement
with our \ha~\PVM~and \RC. Also, the HI linewidth at 20\%~(226\kms,
\citealp{Springob:2005}) is in good agreement with our
\ha~\VF~amplitude.
%
\\
\noindent \textbf{NGC 5296}. This galaxy has no UGC number. It is
the small companion of UGC 8709 (discussed hereafter). Our \ha~map
is deeper than the one by \citet{Rossa:2003} but nevertheless
does not show any \ha~emission in the tidal arms seen on their
unsharped-mask R-band image, the emission being restricted to the
central regions ($\sim$1/3 $D_{25}/2$). As a consequence, our
\ha~\RC~only shows the inner rising part. It is difficult to make
a direct comparison with the \RC~obtained by \citet{Rampazzo:1995} because they choose to align their slit in the direction of
UGC 8709, which is about 30\Deg~different from the true major
axis. However, simulating a slit with their orientation along our
\ha~\VF~shows that our results are consistent. On the WHISP HI
\VF~(website) one can hardly distinguish NGC 5296 close to UGC
8709, although it seems to appear as a point source on the average
resolution map.
\\
\noindent \textbf{UGC 8709 (NGC 5297)}. It is the large companion
of NGC 5296. \ha~emission is detected in the northern arm but only
in the beginning of the southern tidal arm, in agreement with
\citet{Rossa:2003}. The clear signature of a bar can be seen
in the center of the \ha~\VF. From the kinematics, we find an
inclination (76$\pm$1\degr) slightly smaller than the photometric
one (82$\pm$3\degr). It seems that the photometric inclination has
been computed from the axis ratio of the disk including the tidal
arms. The width of the HI profile at 20\%~(418\kms~from \citealp{van-Driel:2001}, 413\kms~from \citealp{Bottinelli:1990}) is in
agreement with our \ha~\VF~amplitude. The HI \VF~has been observed
by WHISP (website) and is consistent with ours. The HI \PVM~shows
a slightly decreasing plateau beyond 1', in agreement with our
\ha~\RC. The long slit \RC~observed by \citet{Rampazzo:1995} also
shows a decreasing trend in the outer parts. No CO has been
detected in this galaxy \citep{Elfhag:1996}.
%
%
%
%
\\
\noindent \textbf{UGC 8852 (NGC 5376)}. This galaxy is included in
a group together with UGC 8860 (NGC 5379) and UGC 8866 (NGC 5389).
Strong \ha~emission is seen, in particular in a ring located
15\arcsec~from the center. No HI \VF~is available. The width of
the HI profile at 20\%~(402\kms, \citealp{Theureau:1998}) is
significantly higher than our \ha~\VF~amplitude ($\sim$320\kms).
Since our \ha~\RC~is still rising in the outer parts, before we
reach the optical radius, this suggests that the maximum velocity
is not reached.
\\
\noindent \textbf{UGC 8863 (NGC 5377)}. The detection of \ha~emission in this
early type SBa galaxy is limited to two faint lobes at about
1\arcmin~on each side of the disk. As a consequence our \ha~\RC~is
reduced to two points (which nevertheless represent 34 independent bins, i.e. about a thousand of pixels), one for the receding side and one for the
approaching side and we only get a lower limit for the \ha~maximum
rotation velocity of $\sim$190\kms. Moreover, even if the \SNR~does not allow to compute another velocity bin in the center of the galaxy, the \PVM~suggests a faint \ha~emission (yellow spots) allowing us to measure a central gradient of $\sim$40\kms$/$arcsec (i.e. $\sim$350\kms~kpc$^{-1}$). Due to beam smearing limitation, this strong gradient cannot be seen on the WHISP \PVM~\citep{Noordermeer:2005}. Indeed, the maximum velocity is reached at about 1\arcmin~from the center in HI data leading to a lower velocity gradient of $\sim$3\kms$/$arcsec (i.e. $\sim$25\kms~kpc$^{-1}$). 
Our maximum velocity is probably close
to the actual maximum velocity rotation since the width of the HI
profile at 20\%~(391\kms) found by \citet{Noordermeer:2005} is close to our \ha~\VF~amplitude.
%
\\
\noindent \textbf{UGC 8898 \& UGC 8900 (NGC 5394 \& NGC 5395, Arp
84)}. The nuclei of this interacting pair of galaxies are
separated by only 1.9 \arcmin~($\sim$27 kpc). No \ha~emission is
detected in the tidal arms of UGC 8898 while \ha~emission is more
extended in UGC 8900, in agreement with \citet{Kaufman:1999} from
Fabry-Perot imaging. The photometric inclination for UGC 8898 is
70\degr$\pm$3\degr, quite different from our kinematical
inclination (27\degr$\pm$20\degr) which is in fact based on the
very central part only (about 15\arcsec~diameter).
Despite its limited extent, our \ha~\VF~of UGC 8898 clearly
suggests a \PA~of the major axis of 31\degr, different by
66\degr~from the one used by \citet{Marquez:2002} for long slit
spectroscopy, which they confess is not adequate to elaborate the
\RC. Indeed they did not observe any clear rotation within the
first 10\arcsec. Our \ha~\RC~extends only up to
$\sim$10\arcsec~and we do not reach neither a plateau nor,
probably, the maximum rotation velocity. Because our observations
of UGC 8898 have been done through the edge of the transmission
function of the interference filter, the spatial coverage of our
data is smaller than that of \citet{Kaufman:1999}. They have also
observed UGC 8898 in the CO line \citep{Kaufman:2002} and find a
very good agreement between \ha~and CO data.
In agreement with the Fabry-Perot \ha~\VF~of \citet{Kaufman:1999}, our \ha~\VF~of UGC 8900 is not well enough defined to show
straightforward evidence for interaction, except that the \VF~is
uncompleted and the kinematical major axis is shifted by 11\degr
from the morphological one, which is a significant difference
considering the high inclination of this galaxy.\citet{Marquez:2002} have aligned the slit of their spectrograph almost along
the morphological major axis, explaining part of the differences
between our \RCs. Their \RC~is slightly more extended but much
more chaotic and asymmetric, with a markedly smaller velocity
amplitude than ours. Both sides of our \ha~\RC~are fairly
symmetric and show a slowly rising trend (almost solid body like
between 20\arcsec~and 60\arcsec) without ever reaching a plateau.
Such a behavior is unexpected for an Sa type galaxy and could be
due to the interaction with its companion. The kinematical and
morphological inclinations of UGC 8900 are compatible within the
error bars (they differ by only 9\degr). HI single dish
observations cannot disentangle UGC 8898 from its companion UGC
8900 (\citealp{van-Driel:2001}; \citealp{Theureau:1998}). The pair has
also been observed in the HI by WHISP (website) and by \citet{Kaufman:1999}. Taking into account the lower HI spatial resolution,
the HI and \ha~kinematics are compatible, and the HI \PVM~suggests
that we actually reach the maximum of the \RC~at the
end of our \ha~\RC.
\\
%
%
%
\noindent \textbf{UGC 8937 (NGC 5430)}. Our \ha~map shows low
emission \ha~regions that were not detected by \citet{Garcia-Barreto:1996}. 
The signature of the strong central bar can be seen on
our \ha~\VF~(S shape signature). Our \RC~reaches a plateau within
a few arcsec. A strong velocity gradient of $\sim$300\kms is observed 
between -5\arcsec and 5\arcsec, as can be seen in the \pvm.
The HI line width at 20\%~(371\kms~from \citealp{Springob:2005}, 
344\kms~from \citealp{Theureau:1998}) is slightly smaller
than the amplitude of our \ha~\VF~($\sim$400\kms).
\\
\noindent \textbf{UGC 9013 (NGC 5474)}. This late-type peculiar
dwarf galaxy is the nearest companion of M101 (44\arcmin) and is
tidally deformed into a very asymmetric and disturbed object. Our
\ha~\VF~cannot help us finding the rotation center as it shows a
solid body shape.\citet{Knapen:2004} show an \ha~map in good
agreement with ours. The \ha~\RC~derived from long slit
spectroscopy data by \citet{Catinella:2005} cannot be compared
directly with ours, because their center has been chosen to be the
"pseudonucleus". Several HI studies are available in the
literature (\citealp{van-der-Hulst:1979}; \citealp{Huchtmeier:1979}; \citealp{Rownd:1994}; \citealp{Kornreich:2000}).\citet{Kornreich:2000} and \citet{Rownd:1994} used the same dataset from the VLA
array (35\arcsec~beam). The optical part of the galaxy is not
affected by the severe warp seen in HI and the solid body rotation
seen in \ha~is compatible with HI data. These authors place the
kinematical center as the symmetry center of the warp. It is
closer to the center of the outermost optical isophotes than the
"pseudonucleus".\citet{Rownd:1994} assumed an inclination of
21\degr~from a tilted ring model. Their \PA~of the major axis in
the center ($\sim$158\deg) is in agreement with our but almost
perpendicular to the photometric \PA. Due to the presence of the
strong tidal arm and the relative low spatial \ha~coverage, the
inclination and center are in fact difficult to recover. That is
why we preferred using the HI kinematical center and inclination
(21\deg) consistent with the external axis ratio of the outermost
isophotes from the XDSS image. The strange behavior of the
resulting \ha~\RC~in the first 40\arcsec~could be the signature of
strong non circular motions. We find a maximum velocity rotation
$\sim$120\kms~at about 70\arcsec, significantly higher than the
velocities found by \citet{Kornreich:2000} even when assuming a
very low inclination for the HI disk. Also the amplitude of our
\ha~\VF~is higher than the HI \VF~amplitude.
\\
\noindent \textbf{UGC 9179 (NGC 5585)}. It is a satellite of M101
\citep{Sandage:1994}. A good agreement is observed between our \ha~map
and that of \citet{James:2004}, however we miss an \ha~region in
the North East because of our smaller \FOV. The \ha~\RC~is
perturbed, due to the presence of the bar and an inner arm like
structure. This galaxy has been observed in HI by \citet{Cote:1991} who obtained a \VF~and a \RC~in agreement with ours
although their HI extent is larger and our \ha~resolution higher.
Our \ha~rotation velocities are higher because we assume an
inclination of 36\degr~(derived from our \ha~\VF) whereas the HI
data led to a higher inclination of 51.5\degr. Both values are
nevertheless compatible within the \ha~error bars. Although the
shape of our \ha~\RC~suggests that we reach the maximum rotation
velocity at the optical limit, the HI \RC~from \citet{Cote:1991} shows that the true maximum is reached a bit further.
\\
\noindent \textbf{UGC 9219 (NGC 5608)}. It has been observed in
\ha~by \citet{van-Zee:2000} and \citet{James:2004}, their maps are in
agreement with ours.
Our \ha~\VF~amplitude is nevertheless compatible with the HI line
width at 20\%~of 130\kms from \citet{Bottinelli:1990}. No HI
velocity map is available in the literature for this galaxy.
%
%
\\
\noindent \textbf{UGC 9248 (NGC 5622)}. The \ha~emission is
asymmetric and much brighter on the western side. Nevertheless
the resulting \RC~is fairly symmetric and reaches a plateau within
the optical limit. No HI velocity map is available in the
literature but HI line width measurements at 20\%~have been done
and are compatible with our \ha~\VF~amplitude although somewhat
larger (357\kms~by \citealp{Theureau:1998}, 349\kms~by \citealp{Springob:2005}).
\\
\noindent \textbf{UGC 9358 (NGC5678)}.\citet{Marquez:2002} derived a \RC~from long slit spectroscopy
which is in good agreement with our \ha~\RC, for both sides.
However, since we adopted a slightly lower inclination, our
velocities are slightly higher. Their \RC~is more extended, but
the dispersion of their points in the outer parts is quite high,
specially on the receding side. Our \ha~\VF~is perturbed and
clearly shows the signature of a bar in the center (S shape of the
isovelocity lines). No HI velocity map is available in the
literature but the width of the HI profile at 20\%~(424\kms,
\citealp{Springob:2005}) is in good agreement with our
\ha~\VF~amplitude.
\\
%
\noindent \textbf{UGC 9363 (NGC 5668)}. Our observations are in agreement with previous \ha~Fabry-Perot observations by \citet{Jimenez-Vicente:2000}. They adopted an inclination of 18\degr~derived by \citet{Schulman:1996} from a tilted ring model applied to VLA HI data. This inclination is lower than expected from morphology (33\deg). Our data confirm this tendency, explaining why morphological and
kinematical major axis are found to be quite different (40\deg). Our kinematical estimate of the inclination is close to 0\deg and leads to unrealistically high rotational velocities. Thus, as \citet{Jimenez-Vicente:2000}, we fixed the inclination to the HI value of 18\degr.
\citet{Schulman:1996} HI data show a warp starting at 120\arcsec, which is
the outermost limit of our \ha~\RC. Indeed, it is possible that
the outermost points of our \ha~\RC~are affected by this warp,
since it shows a clear trend to increase from 80\arcsec~to
120\arcsec~whereas the HI observations by \citet{Schulman:1996}
suggest that the plateau of the curve is already reached at
100\arcsec~when correcting for the warp. The width of the HI
profile at 20\%~(122\kms, \citealp{Springob:2005}) is in agreement
with our \ha~\VF~amplitude.
%
\\
\noindent \textbf{UGC 9406 (NGC 5693)}. A short bright bar and an
asymmetric disk (with one knotty arm) are observed. This galaxy is
paired with UGC 9399 (NGC 5689) at 11.8\arcmin, which may explain
the presence of a single arm. Poor \ha~emission is detected and
our \ha~\VF~shows a strong dispersion, so that it is difficult to
draw any reliable \RC~from our data. The photometric inclination
found in the literature varies from 33\deg~\citep{Vorontsov-Velyaminov:1994} to 51\deg~(HyperLeda). Even when adopting the lowest
value of inclination, we find a maximum rotation velocity which
remains abnormally low considering the absolute magnitude of this
galaxy, suggesting that our \ha~\VF~is far from reaching the
maximum velocity amplitude. Indeed the width of the HI profile at
20\%~is 74\kms~\citep{Bottinelli:1990} whereas our
\ha~\VF~amplitude is smaller than 50\kms. No HI velocity map is
available for this galaxy in the literature.
%
\\
\noindent \textbf{UGC 9465 (NGC 5727)}. A good agreement is
observed between our \ha~map and that of \citet{James:2004}. No
signature of the central bar can be seen on our \ha~\VF. It has
been observed in HI by \citet{Pisano:2002} and their \VF~is in
good agreement with ours. These authors adopted an inclination of
61\Deg~(close to the inclination of 65\Deg~we deduced from our
\ha~\VF) and found a maximum rotation velocity of 93\kms~in good
agreement with our \ha~value (98\kms). The inclination of
90\deg~given in HyperLeda is certainly wrong because this galaxy
seems far from being seen edge-on. The inclination deduced from
our \ha~\VF~exactly matches the value suggested by the axis ratio
given in NED.
%
%
%
%
\\
\noindent \textbf{UGC 9576 (NGC 5774)}. This galaxy is the companion
of NGC 5775. There is a good agreement between our \ha~map and
that of \citet{James:2004}.\citet{Marquez:2002} derived a
\RC~from long slit spectroscopy. However their slit was
20\Deg~away from the true kinematical major axis, maybe explaining
why their \RC~is more chaotic than ours. Our \ha~\RC~is much more
symmetric and extends further out. The width of the HI profile at
20\%~(280\kms) from \citet{Springob:2005} is almost twice our
\ha~\VF~amplitude, which is quite surprising since the shape of
our \ha~\RC~suggests that we reach the maximum rotation velocity.
The value given by \citet{Springob:2005} seems nevertheless
suspicious as \citet{Bottinelli:1990} and \citet{Irwin:1994}
respectively found 152\kms~and 179\kms~(uncorrected for galaxy
inclination) for the width of the HI profile at 20\%, which is
quite compatible with our \ha~\VF~amplitude. The detailed HI
\VF~of the pair NGC 5774-5775 has been observed at the VLA by
\citet{Irwin:1994}. Her \VF~and \RC~for NGC 5774 are in good agreement
with our \ha~data and have about the same spatial extension. She
used the same method as us to determine the kinematical parameters
and they are in very good agreement with ours.
%
%
\\
\noindent \textbf{UGC 9736 (NGC 5874)}. We detect poor and
asymmetric \ha~emission in that galaxy. Our \ha~\RC~is
nevertheless fairly symmetric and almost reaches the optical
limit, with a trend to flatten in its outermost parts. The width
of the HI profile at 20\%~(324\kms~from \citealp{Springob:2005},
315\kms~from \citealp{Bottinelli:1990}) is in agreement with our
\ha~\VF~amplitude and suggests that we actually reach the maximum
of the \RC. No HI velocity map is available in the
literature.
%
%
\\
\noindent \textbf{UGC 9866 (NGC 5949)}. Comparing our \ha~map with
that of \citet{James:2004} shows that our data suffer from bad
seeing conditions leading to non resolved HII regions. However the
\ha~emission is strong enough so that we have a complete \VF~all
over the disk. The \ha~emission is asymmetric, stronger on the
receding side (northwest). Our \RC~is in very good agreement with
the radial velocities measured by \citet{Karachentsev:1990}
from slit spectroscopy when correcting from the inclination.
\citet{Courteau:1997}, also from slit spectroscopy, finds a slightly
smaller extension for the \RC~but the velocity width he measures
from the flux-weighted rotation profile (213\kms) is in agreement
with our maximum \VF~amplitude. No HI velocity map is available
but the width of the HI profile at 20\%~(216\kms~from \citealp{Springob:2005}, 197\kms~from \citealp{Bottinelli:1990}) is also in good
agreement with our \ha~\VF~amplitude.
%
%
\\
\noindent \textbf{UGC 9943 (NGC 5970)}. It forms a pair with IC
1131 at 8\arcmin. \ha~emission is strong in particular in an inner
ring. Our \ha~\RC~is in good agreement with the slit spectroscopy
one of \citet{Marquez:2002}. A plateau is clearly reached around 3
kpc. No HI velocity map is available in the literature but the
width of the HI profile at 20\%~(338\kms~from \citealp{Springob:2005}, 326\kms~from \citealp{Bottinelli:1990}) is in good agreement
with our \ha~\VF~amplitude.
%
\\
\noindent \textbf{UGC 10075 (NGC 6015)}. There is a good agreement
between our \ha~map and that of \citet{James:2004}. We detect
however some faint emission in an outer spiral arm to the
southeast that they do not detect. Our \ha~\RC~is in good
agreement with the slit spectroscopy observations of \citet{Carozzi:1976} when correcting from the different inclination adopted.
However, her detection was not as good, and our \RC~is almost
twice more extended, clearly showing a slowly rising plateau
beyond 50 \arcsec~radius. No HI velocity map is available in the
literature but the width of the HI profile at 20\%~(315\kms~from
\citealp{Springob:2005}, 310\kms~from \citealp{Bottinelli:1990}) is in
good agreement with our \ha~\VF~amplitude. It has been observed in
CO by \citet{Braine:1993} who find a strong CO emission at
40\arcsec~(2.7 kpc) from the center.
\\
\noindent \textbf{UGC 10521 (NGC 6207)}. There is a good agreement
between our \ha~map and that of \citet{James:2004}. Our
\ha~\RC~is in good agreement with the slit spectroscopy
observations by \citet{Carozzi:1976} ($PA$=15\degr) and \citet{Marquez:2002} (with $PA$=22\degr) when correcting from the different
inclinations adopted. No HI velocity map available in the
literature but the width of the HI profile at 20\%~(255\kms~from
\citealp{Springob:2005}, 240\kms~from \citealp{Bottinelli:1990}) is in
good agreement with our \ha~\VF~amplitude.
%
\\
\noindent \textbf{UGC 10652 (NGC 6283)}. This galaxy is
asymmetric, with some bright \ha~spots and diffuse \ha~emission
all over the disk. An inner \ha~ring, some 7\arcsec~radius, can be
seen in the center. The morphological inclination of
30$\pm$7\deg~is compatible, within the error bars, with the
kinematical inclination (16$\pm$12\deg) deduced from our \ha~\VF.
The position of the kinematical major axis differs from the
photometric one by 12\degr. Our \ha~\RC~seems to reach a plateau
at about 2 kpc (23\arcsec) within the optical limit. No useful HI
data are available in the literature for that galaxy.
\\
\noindent \textbf{UGC 10713}. We adopted an inclination of
90\degr~for that galaxy, otherwise our method led to a clearly
wrong value below 70\degr, probably because of the odd pattern of
our \ha~\VF~in the center.
It has been observed in HI by WHISP (website) and their \VF~is in
agreement with ours, but much more extended. Their HI
\PVM~suggests that the \RC~reaches a plateau at about
1\arcmin~radius, just beyond the limits of our \ha~\RC~which is
limited to the solid body rising part. The HI line width at
20\%~(268\kms~by \citealp{Theureau:1998} and 260\kms~by \citealp{Springob:2005}) is in agreement with our \ha~\VF~amplitude.
%
\\
\noindent \textbf{UGC 10757}. This galaxy has a velocity of
1168\kms, it is located in a triple subgroup with UGC 10762 (NGC
6340) an S0/a galaxy with a peculiar morphology with a series of
tightly wound, almost circular, multiple-fragment, relatively thin
outer arms surrounding a bulge and a lens at 6.4\Min and UGC 10769
which has an Sb pec morphological type located at 6.1\Min with a
velocity of 1283 km/s. This system is clearly in interaction as
shown by the HI cloud in which the 3 galaxies are embedded (e.g.
WHISP website). This galaxy is asymmetric, with a brighter
\ha~emission on the beginning of the northern spiral arm, also
seen on the optical (XDSS) and UV (GALEX) images. The WHISP HI
\VF~corresponding to the optical extent of the galaxy (also
coinciding with the brightest part of the HI complex) is in
agreement with our \ha~\VF. However, the pattern of the whole
extent of the HI \VF~is quite odd, apparently because of several
galaxies interacting there. As a result, the width of the HI
profile at 20\%~found in the literature for that galaxy
(283\kms~from \citealp{Springob:2005}, 276\kms~from \citealp{Lang:2003},
222\kms~from \citealp{Theureau:1998}) is larger than the amplitude of
the \ha~\VF~because a more extended region is embedded in the HI.
The shape of the \ha~\RC~suggests that we are not far from
reaching the maximum rotation velocity within the optical limit.
%
\\
\noindent \textbf{UGC 10769}. This galaxy has a diffuse disk and
no spiral pattern visible. It has a higher velocity than the two
other galaxies in the triple system (see UGC 10757 for a detailed
discussion). We detect \ha~emission only in the north-eastern edge
of the disk so that no rotation curve can be derived. The
secondary peak in the HI distribution corresponds to the same
region where the \ha~is detected.
%
\\
\noindent \textbf{UGC 10791}. This Low Surface Brightness galaxy
has two companions according to the WHISP website. We observe a
faint diffuse \ha~emission throughout the disk insufficient to
determine unambiguously its inclination. This galaxy is classified
face-on in HyperLeda, nevertheless it does not look face-on on the
XDSS image (see Figure D93), in addition the \VF~displays
a clear rotation compatible with the HI \VF~(e.g. same position
angle and velocity amplitude in the central region). Thus we fixed
the inclination to the WHISP value of 34\degr. The \ha~distribution
extends only until half of the optical radius and thus
does not reach the maximum rotation velocity observable in the HI
(FWHM: 199\kms~from \citealp{Springob:2005}, 156\kms~from \citealp{Theureau:1998}; amplitude of the WHISP \VF~of $\sim$150\kms).
\\
\noindent \textbf{UGC 11012 (NGC 6503)}. This galaxy has a strong
\ha~emission all over its disk, with a faint extended feature on
the western edge ($\sim$2\arcmin~from the center). This faint
extension cannot be seen on the \ha~map from \citet{Strickland:2004}. Our \RC~clearly reaches a plateau around 80\arcsec, well
before the optical limit. Several optical \RCs are
found in the literature but none of them takes this extension into
account and their extension is limited to $\sim$80\arcsec~whereas
ours extends up to 150\arcsec~(for the receding side, owing to the
mentioned \ha~extension). Nevertheless these \RCs
(\citealp{Karachentsev:1990}; \citealp{de-Vaucouleurs:1982}) are
in very good agreement with our \ha~\RC. H$\beta$ observations
made by \citet{Bottema:1989} also show a \RC~in good agreement with
ours. The HI map obtained by \citet{Begeman:1987} is more extended than
our \ha~map otherwise both \VFs~are in good agreement, as well as
the derived \RCs. The parameters computed in HI with a tilted ring
model ( $PA$=-59.4\degr, $i$=73.8\degr) are in very good agreement with
our own parameters.
%
%
\\
\noindent \textbf{UGC 11269 (NGC 6667)}. Despite a short exposure
time ($\sim$1 hour), our \ha~observations perfectly match
\citet{James:2004} \ha~data who find faint patchy emission. It
has been observed in HI by WHISP \citep{Noordermeer:2005} and their
\VF~and \RC~extend more than four times the optical radius. 
Our \ha~\pvm~does not show the steep velocity gradient 
observed on their HI \pvm~in the innermost 40\arcsec due to a 
very noisy and patchy distribution of the ionized gas.
The maximum rotation velocity is reached in \ha, as
confirmed by the width of the HI profile at 20\%~found by
different authors (415\kms~from \citealp{Noordermeer:2005},
394\kms~from \citealp{Springob:2005}, 406\kms~from \citealp{Bottinelli:1990}). 
Our \ha~data suggest that the maximum rotation velocity is reached within
the first kpc instead of the first $\sim$6 kpc as suggested by the
HI data \citep{Noordermeer:2005}.
%
%
%
\\
\noindent \textbf{UGC 11300 (NGC 6689, NGC 6690)}. 
This galaxy has been published in paper IV. See specific comment
at the end of this section.
%
%
%
%
\\
\noindent \textbf{UGC 11332 (NGC 6654A)}. There is a good
agreement between our \ha~map and that of \citet{James:2004}.
However, our image is affected by very bad seeing conditions (see
Table \ref{tablemod}) resulting in a diffuse emission around the
galaxy that is most probably an artifact. No HI velocity map is
available in the literature. The width of the HI profile at
20\%~(331\kms~from \citealp{Springob:2005}, 315\kms~from \citealp{Bottinelli:1990}) is significantly larger than our \ha~\VF~amplitude
(about 200\kms), suggesting that our \ha~\VF~does not reach the
maximum rotation velocity although it extends up to the optical
radius.
%
%
\\
\noindent \textbf{UGC 11407 (NGC 6764)}. The \ha~emission is
conspicuous along the bar and in the arms (particularly the
northern arm). Our \ha~\RC~has been drawn using the photometric
position angle of the major axis and the photometric inclination.
It is much chaotic in the center, up to 20\arcsec~(probably
because of the strong bar), then looks in average like that of a
solid body rotating disk, although it is markedly asymmetric. The
central bar is almost aligned along the photometric major axis and
the HI \VF~derived from VLA observations \citep{Wilcots:2001}
leads to the same position angle. In the central parts, the
isovelocity lines pattern show the signature of the strong central
bar on both \ha~and HI \VFs. The \pvm~shows a steep velocity rise
in the galaxy core. The width of the HI profile at 20
\%~(291\kms~from \citealp{Springob:2005}, 293\kms~from \citealp{Bottinelli:1990}, 
$\sim$300\kms~from \citealp{Wilcots:2001}) is in agreement
with our \ha~\VF~amplitude.
%
%
%
\\
%
%
\noindent \textbf{UGC 11466}. We detect a strong \ha~emission
along a bar like feature. This galaxy has been observed in HI by
WHISP (website) and their high resolution \VF~is in agreement with
our \ha~\VF. Their lower resolution maps show a much larger HI
disk but no greater velocity amplitude, suggesting that our
ha~\RC~reaches the maximum velocity in the outer parts of the
optical disk (also, the WHISP \PVM~is in agreement with our \Ha~\PVM~and \RC). This is confirmed by the width of the HI profile
at 20\%~measured by different authors (247\kms~from \citealp{Springob:2005}, 239\kms~from \citealp{Theureau:1998}, 251\kms~from \citealp{Kamphuis:1996}) which is in perfect agreement with our
\ha~\VF~amplitude.
%
\\
\noindent \textbf{UGC 11470 (NGC 6824)}. The \ha~emission is
rather faint, and hardly detected on the eastern side of the
galaxy because of the interference filter transmission mismatching
the systemic velocity. As a result, the \RC~could be drawn almost
only from the approaching side. No HI velocity map is available in
the literature. The width of the HI profile at 20\%~(574\kms~from
\citealp{Springob:2005}) is in agreement with the \ha~velocity
amplitude showing that the maximum velocity is reached at a small
radius ($\sim$3 kpc) compared to the optical radius ($\sim$18
kpc), in agreement with what is expected for such a bright early
type galaxy ($M_b$=-21.3; Sab). There has been no detection of CO
in that galaxy by \citet{Elfhag:1996}.
%
\\
\noindent \textbf{UGC 11496}. The \ha~emission is very faint in
that galaxy, sufficient however for drawing a \VF~and deriving an
acceptable \RC. The HI \VF~(WHISP, website) is in agreement with
our \ha~\VF~but much more extended. The width of the HI profile at
20\%~(121\kms~from \citealp{Bottinelli:1990}) is in agreement with
our \ha~\VF~amplitude and confirms that we reach the maximum
rotation velocity as already suggested by the shape of our
\ha~\RC, showing a trend to reach a plateau in the outer parts.
%
\\
\noindent \textbf{UGC 11498}. The \ha~emission is faint and
asymmetric in this early type galaxy (SBb). However, although
there is only a few \ha~emission on the receding side, the shape
of our \RC~suggests that a plateau is reached on both sides. No HI
velocity map is available in the literature. The width of the HI
profile at 20\%~(509\kms~from \citealp{Springob:2005}, 500\kms~from
\citealp{Bottinelli:1990}) perfectly matches our \ha~\VF~amplitude.
We find a maximum rotation velocity of 274\kms~which seems a bit
high when taking into account its luminosity.
%
\\
\noindent \textbf{UGC 11597 (NGC 6946)}. 
NGC 6946 is a very well studied nearby spiral galaxy. Our
\ha~Fabry-Perot observations are in perfect agreement with the
\ha~Fabry-Perot data recently published by \citeauthor{Daigle:2006a} (2006a)
who had a larger field of view (we miss the outer parts of the
optical disk). We find a kinematical inclination of
40$\pm$10\degr close to their value (8.4$\pm$3\degr)
and compatible with the photometric value (17$\pm$19\degr)
within the error bars.
%
\\
\noindent \textbf{UGC 11670 (NGC 7013)}. The \ha~emission is
mainly concentrated in the central part of the disk (with two
bright blobs, on each side of the central bulge, along the major
axis) and some emission can be seen on the northwestern edge of
the optical disk. Our \ha~\RC~rapidly reaches a maximum, at about
20\arcsec~from the center ($\sim$1 kpc), followed by a slight
decrease and a plateau. The velocity gradient observed in the 
inner 20\arcsec of our \pvm~is consistent with that observed 
in the HI \pvm~from WHISP \citep{Noordermeer:2005} and
their \VF~is in agreement with our \ha~\VF~but much more extended.
The width of the HI profile at 20\%~(342\kms~from \citealp{Noordermeer:2005}, 363\kms~from \citealp{Springob:2005}, 355\kms~from
\citealp{Bottinelli:1990}) is in good agreement with our
\ha~\VF~amplitude. The morphological inclination of 90\deg~found
in HyperLeda assumes a thick disk. However, assuming a thin disk,
the inclination is found to be 68\degr~with a tilted ring fit to
the HI \VF~\citep{Noordermeer:2005} in agreement with the value
deduced from our \ha~\VF~(65$\pm$~2\degr) as well as with the
inclination deduced from the axis ratio (69\degr) or other
morphological determinations \citep{Vorontsov-Velyaminov:1994}.
%
\\
\noindent \textbf{UGC 11872 (NGC 7177)}. Our \ha~map has a mottled
appearance, in agreement with that of \citet{James:2004}, however
our observations suffer from bad seeing conditions. Our
\ha~\RC~rapidly reaches a well defined and symmetric plateau,
extending up to the optical radius. The \RC~obtained by \citet{Marquez:2002} reaches the plateau more rapidly (10\arcsec~instead of
20\arcsec) but they find it at a lower value since they assume a
surprisingly high inclination (79\degr~while we find 47\degr~from
our \ha~\VF, which is closer to the morphological value of
54\degr). Furthermore, slit spectroscopy of the inner part of the
galaxy by \citet{Heraudeau:1999} confirms that the maximum
velocity is not reached before 20\arcsec. No HI velocity map is
available in the literature, however HI line widths have been
measured (314\kms~from \citealp{Springob:2005}, 317\kms~from
\citealp{Bottinelli:1990}) and are in agreement with our
\ha~\VF~amplitude. Moreover, a radio synthesis observation has
been done \citep{Rhee:1996} to derive a \PVM. It shows a
solid body rotation as far out as 1\arcmin~which is likely to be
explained by beam smearing effects.
%
\\
%
%
\noindent \textbf{UGC 12082}. The \ha~emission in this low
luminosity nearby galaxy is rather patchy. Our \ha~map is in
agreement with that of \citet{James:2004}. The HI \VF~(WHISP,
website) confirms our position angle determination for the major
axis in the central part. It is however much more extended than
the \ha~one and shows a strong warp of the HI disk. The WHISP
\PVM~suggests that we do not reach the maximum velocity within the
optical radius although the shape of our \ha~\RC~shows a trend to
reach a plateau in the outer parts. The HI line width at
20\%~confirms that point, since all the values found in the
literature (95\kms~from \citealp{Braun:2003}, 103\kms~from \citealp{Springob:2005}, 79\kms~from \citealp{Bottinelli:1990}) are markedly
larger than the amplitude of our \ha~\VF.
%
\\
\noindent \textbf{NB.}The following targets, already published in
the previous GHASP papers, have been observed again in different conditions
(filters, seeing, transparency, exposure time, ...) in order to
check if the quality of the data may be improved: UGC 2023, UGC
2034, UGC 3734, UGC 11300 and UGC 12060. The results are fully
consistent with the previous set of observation without any
significant improvement so that we do not present the new data in
this paper, except for UGC 11300.

We present the new observation for UGC 11300 mainly to compare the
new method of reduction presented in this paper with the previous
GHASP papers. We
observe a general good agreement in velocities as well as in
radial extension. The adaptive spatial binning enables us now: 1)
to plot velocity measurement in the outskirts of the approaching
side; 2) to increase the spatial resolution within the inner
30\arcsec ($\sim$1 kpc) 3) to underline that the error bars are
correlated with the spatial resolution and the quality of the data
(the error bars are generally smaller and a wide range of
amplitude is observed).

\clearpage
\section{Tables}
\label{tables}
\begin{table*}
\caption{Log of the observations.}
\begin{tabular}{ccccccccc}
\noalign{\medskip} \hline
N\Deg & N\Deg &  $\alpha$    & $\delta$   &   $\lambda_{c}$&  FWHM & date & exposure time& seeing\\
 UGC        &  NGC       &  (2000) & (2000) & \AA & \AA & & s &"\\
 (1)&(2)&(3)&(4)&(5)&(6)&(7)&(8)&(9)\\
\hline
12893 &  & 00$\rm^{h}$00$\rm^{m}$28.0$\rm^{s}$ & 17$\degr$13'09" & 6582.8 & 10.6 & Oct, 25 2003 & 8640 &  2.6 \\
89 & 23 & 00$\rm^{h}$09$\rm^{m}$53.4$\rm^{s}$ & 25$\degr$55'24" & 6660.3 & 20.0 & Sep, 05 2002 & 720 &  2.6 \\
 &  &  &  & 6660.3 & 20.0 & Sep, 06 2002 & 1920 &  4.3 \\
 &  &  &  & 6660.2 & 20.0 & Sep, 07 2002 & 6000 &  2.6 \\
94 & 26 & 00$\rm^{h}$10$\rm^{m}$25.9$\rm^{s}$ & 25$\degr$49'54" & 6660.3 & 20.0 & Sep, 03 2002 & 6240 &  2.9 \\
1013 & 536 & \textit{01$\rm^{h}$26$\rm^{m}$21.8$\rm^{s}$} & \textit{34\degr42'11"} & 6675.0 & 20.0 & Sep, 10 2002 & 1920 &  4.8 \\
 &  &  &  & 6675.2 & 20.0 & Sep, 08 2002 & 1920 &  2.4 \\
NGC 542 & 542 & \textit{01$\rm^{h}$26$\rm^{m}$21.8$\rm^{s}$} & \textit{34\degr42'11"} & 6675.0 & 20.0 & Sep, 10 2002 & 1920 &  4.8 \\
 &  &  &  & 6675.2 & 20.0 & Sep, 08 2002 & 1920 &  2.4 \\
1317 & 697 & 01$\rm^{h}$51$\rm^{m}$17.6$\rm^{s}$ & 22$\degr$21'28" & 6628.4 & 20.0 & Sep, 10 2002 & 4800 &  3.4 \\
1437 & 753 & 01$\rm^{h}$57$\rm^{m}$42.2$\rm^{s}$ & 35$\degr$54'58" & 6660.2 & 20.0 & Sep, 03 2002 & 6720 &  3.8 \\
1655 & 828 & 02$\rm^{h}$10$\rm^{m}$09.7$\rm^{s}$ & 39$\degr$11'25" & 6693.1 & 22.6 & Oct, 24 2000 & 12720 &  3.9 \\
1810 &  & \textit{02$\rm^{h}$21$\rm^{m}$28.7$\rm^{s}$} & \textit{39\degr22'32"} & 6735.0 & 20.0 & Sep, 07 2002 & 7200 &  3.6 \\
3056 & 1569 & \textit{04$\rm^{h}$30$\rm^{m}$49.2$\rm^{s}$} & \textit{64\degr50'53"} & 6560.3 & 12.0 & Nov, 21 2001 & 6000 &  2.7 \\
3334 & 1961 & 05$\rm^{h}$42$\rm^{m}$04.6$\rm^{s}$ & 69$\degr$22'43" & 6654.9 & 20.6 & Nov, 20 2001 & 5040 &  3.1 \\
3382 &  & 05$\rm^{h}$59$\rm^{m}$47.7$\rm^{s}$ & 62$\degr$09'28" & 6658.0 & 20.0 & Oct, 25 2003 & 4080 &  3.6 \\
 &  &  &  & 6658.0 & 20.0 & Oct, 30 2003 & 7920 &  4.6 \\
3463 &  & 06$\rm^{h}$26$\rm^{m}$55.8$\rm^{s}$ & 59$\degr$04'47" & 6627.6 & 20.0 & Mar, 06 2003 & 6960 &  7.2 \\
3521 &  & 06$\rm^{h}$55$\rm^{m}$00.1$\rm^{s}$ & 84$\degr$02'30" & 6659.3 & 20.0 & Mar, 08 2003 & 8400 &  3.4 \\
3528 &  & 06$\rm^{h}$56$\rm^{m}$10.6$\rm^{s}$ & 84$\degr$04'44" & 6659.3 & 20.0 & Mar, 08 2003 & 8400 &  3.4 \\
3618 & 2308 & \textit{06$\rm^{h}$58$\rm^{m}$37.6$\rm^{s}$} & \textit{45\degr12'38"} & 6689.5 & 20.0 & Mar, 04 2003 & 6000 &  2.4 \\
3685 &  & 07$\rm^{h}$09$\rm^{m}$05.9$\rm^{s}$ & 61$\degr$35'44" & 6603.2 & 12.0 & Mar, 17 2002 & 5520 &  3.0 \\
3708 & 2341 & 07$\rm^{h}$09$\rm^{m}$12.0$\rm^{s}$ & 20$\degr$36'11" & 6674.2 & 20.0 & Mar, 02 2003 & 6480 &  3.3 \\
3709 & 2342 & 07$\rm^{h}$09$\rm^{m}$18.1$\rm^{s}$ & 20$\degr$38'10" & 6674.2 & 20.0 & Mar, 02 2003 & 6480 &  3.3 \\
3826 &  & 07$\rm^{h}$24$\rm^{m}$28.0$\rm^{s}$ & 61$\degr$41'38" & 6601.6 & 12.0 & Mar, 20 2002 & 4560 &  3.2 \\
3740 & 2276 & 07$\rm^{h}$27$\rm^{m}$13.1$\rm^{s}$ & 85$\degr$45'16" & 6613.7 & 11.1 & Mar, 15 2002 & 4800 &  3.4 \\
3876 &  & 07$\rm^{h}$29$\rm^{m}$17.5$\rm^{s}$ & 27$\degr$54'00" & 6581.7 & 10.6 & Mar, 18 2004 & 6000 &  3.4 \\
 &  &  &  & 6581.7 & 10.6 & Mar, 19 2004 & 9600 &  3.4 \\
3915 &  & 07$\rm^{h}$34$\rm^{m}$55.8$\rm^{s}$ & 31$\degr$16'34" & 6659.2 & 20.0 & Mar, 03 2003 & 6000 &  3.4 \\
IC 476 &  & 07$\rm^{h}$47$\rm^{m}$16.5$\rm^{s}$ & 26$\degr$57'03" & 6659.2 & 20.0 & Mar, 07 2003 & 6240 &  3.0 \\
4026 & 2449 & 07$\rm^{h}$47$\rm^{m}$20.4$\rm^{s}$ & 26$\degr$55'48" & 6659.2 & 20.0 & Mar, 07 2003 & 6240 &  3.0 \\
4165 & 2500 & 08$\rm^{h}$01$\rm^{m}$53.2$\rm^{s}$ & 50$\degr$44'15" & 6573.7 & 11.4 & Mar, 16 2002 & 4800 &  3.0 \\
 &  &  &  & 6573.6 & 11.4 & Mar, 17 2002 & 4800 &  2.5 \\
4256 & 2532 & 08$\rm^{h}$10$\rm^{m}$15.2$\rm^{s}$ & 33$\degr$57'24" & 6674.3 & 20.0 & Mar, 17 2004 & 6960 &  2.4 \\
4393 &  & 08$\rm^{h}$26$\rm^{m}$04.4$\rm^{s}$ & 45$\degr$58'02" & 6611.1 & 11.1 & Mar, 20 2004 & 6960 &  3.5 \\
4422 & 2595 & 08$\rm^{h}$27$\rm^{m}$42.0$\rm^{s}$ & 21$\degr$28'45" & 6658.4 & 20.0 & Mar, 09 2003 & 7200 &  2.7 \\
4456 &  & 08$\rm^{h}$32$\rm^{m}$03.5$\rm^{s}$ & 24$\degr$00'39" & 6688.7 & 20.0 & Mar, 22 2004 & 6720 &  4.8 \\
4555 & 2649 & 08$\rm^{h}$44$\rm^{m}$08.4$\rm^{s}$ & 34$\degr$43'02" & 6658.7 & 20.0 & Mar, 23 2004 & 6960 &  5.9 \\
4770 & 2746 & 09$\rm^{h}$05$\rm^{m}$59.4$\rm^{s}$ & 35$\degr$22'38" & 6718.7 & 20.0 & Mar, 24 2004 & 7920 &  4.3 \\
4820 & 2775 & 09$\rm^{h}$10$\rm^{m}$20.1$\rm^{s}$ & 07$\degr$02'17" & 6593.0 & 10.1 & Mar, 06 2003 & 8400 &  6.3 \\
5045 &  & 09$\rm^{h}$28$\rm^{m}$10.2$\rm^{s}$ & 44$\degr$39'52" & 6734.3 & 20.0 & Mar, 09 2003 & 6960 &  2.5 \\
5175 & 2977 & 09$\rm^{h}$43$\rm^{m}$46.8$\rm^{s}$ & 74$\degr$51'35" & 6628.5 & 20.0 & Mar, 23 2004 & 5520 &  6.9 \\
5228 &  & 09$\rm^{h}$46$\rm^{m}$03.8$\rm^{s}$ & 01$\degr$40'06" & 6602.9 & 12.0 & Mar, 04 2003 & 9120 &  2.8 \\
5251 & 3003 & 09$\rm^{h}$48$\rm^{m}$36.4$\rm^{s}$ & 33$\degr$25'17" & 6593.9 & 10.1 & Mar, 15 2002 & 1680 &  3.1 \\
5279 & 3026 & \textit{09$\rm^{h}$50$\rm^{m}$55.1$\rm^{s}$} & \textit{28\degr33'05"} & 6594.2 & 10.1 & Mar, 16 2004 & 6000 &  2.7 \\
5319 & 3061 & 09$\rm^{h}$56$\rm^{m}$12.0$\rm^{s}$ & 75$\degr$51'59" & 6613.9 & 11.1 & Mar, 17 2004 & 7200 &  2.4 \\
\hline
\label{tablelog}
\end{tabular}
\end{table*}
\begin{table*}
\contcaption{}
\begin{tabular}{ccccccccc}
\noalign{\medskip} \hline
N\Deg & N\Deg &  $\alpha$    & $\delta$   &   $\lambda_{c}$&  FWHM & date & exposure time& seeing\\
 UGC        &  NGC       &  (2000) & (2000) & \AA & \AA & & s &"\\
 (1)&(2)&(3)&(4)&(5)&(6)&(7)&(8)&(9)\\
\hline
5351 & 3067 & \textit{09$\rm^{h}$58$\rm^{m}$21.3$\rm^{s}$} & \textit{32\degr22'12"} & 6593.7 & 10.1 & Mar, 07 2003 & 7200 &  4.1 \\
5373 &  & 10$\rm^{h}$00$\rm^{m}$00.5$\rm^{s}$ & 05$\degr$19'58" & 6571.0 & 11.4 & Mar, 22 2004 & 7680 &  4.6 \\
5398 & 3077 & \textit{10$\rm^{h}$03$\rm^{m}$20.0$\rm^{s}$} & \textit{68\degr44'01"} & 6563.7 & 12.0 & Mar, 08 2003 & 7920 &  3.2 \\
IC 2542 &  & 10$\rm^{h}$07$\rm^{m}$50.5$\rm^{s}$ & 34$\degr$18'55" & 6689.7 & 20.0 & Mar, 21 2004 & 7440 &  4.1 \\
5510 & 3162 & 10$\rm^{h}$13$\rm^{m}$31.7$\rm^{s}$ & 22$\degr$44'14" & 6592.3 & 10.1 & Mar, 02 2003 & 6240 &  3.5 \\
5532 & 3147 & 10$\rm^{h}$16$\rm^{m}$53.5$\rm^{s}$ & 73$\degr$24'03" & 6630.3 &  9.6 & Mar, 19 2002 & 4560 &  3.3 \\
 &  &  &  & 6623.1 &  8.6 & Mar, 20 2002 & 1920 &  3.3 \\
5556 & 3187 & \textit{10$\rm^{h}$17$\rm^{m}$47.9$\rm^{s}$} & \textit{21\degr52'24"} & 6593.9 & 10.1 & Mar, 17 2002 & 5760 &  3.1 \\
5786 & 3310 & 10$\rm^{h}$38$\rm^{m}$45.9$\rm^{s}$ & 53$\degr$30'12" & 6584.2 & 10.6 & Mar, 20 2002 & 4080 &  4.1 \\
5840 & 3344 & 10$\rm^{h}$43$\rm^{m}$31.1$\rm^{s}$ & 24$\degr$55'21" & 6573.7 & 11.4 & Mar, 20 2002 & 5760 &  3.3 \\
5842 & 3346 & 10$\rm^{h}$43$\rm^{m}$39.0$\rm^{s}$ & 14$\degr$52'18" & 6584.2 & 10.6 & Mar, 20 2004 & 6000 &  3.5 \\
6118 & 3504 & 11$\rm^{h}$03$\rm^{m}$11.3$\rm^{s}$ & 27$\degr$58'20" & 6593.9 & 10.1 & Mar, 18 2002 & 5760 &  2.4 \\
6277 & 3596 & 11$\rm^{h}$15$\rm^{m}$06.2$\rm^{s}$ & 14$\degr$47'12" & 6580.8 & 10.6 & Mar, 19 2004 & 10320 &  2.9 \\
6419 & 3664 & 11$\rm^{h}$24$\rm^{m}$24.6$\rm^{s}$ & 03$\degr$19'36" & 6593.9 & 10.1 & Apr, 27 2003 & 6960 &  3.5 \\
6521 & 3719 & 11$\rm^{h}$32$\rm^{m}$13.4$\rm^{s}$ & 00$\degr$49'09" & 6689.2 & 20.0 & Mar, 05 2003 & 6960 &  3.7 \\
6523 & 3720 & 11$\rm^{h}$32$\rm^{m}$21.6$\rm^{s}$ & 00$\degr$48'14" & 6689.2 & 20.0 & Mar, 05 2003 & 6960 &  3.7 \\
6787 & 3898 & 11$\rm^{h}$49$\rm^{m}$15.6$\rm^{s}$ & 56$\degr$05'04" & 6584.3 & 10.6 & Mar, 19 2002 & 3600 &  2.9 \\
 &  &  &  & 6584.2 & 10.6 & Mar, 20 2002 & 5760 &  3.3 \\
7021 & 4045 & 12$\rm^{h}$02$\rm^{m}$42.3$\rm^{s}$ & 01$\degr$58'36" & 6602.6 & 12.0 & Mar, 04 2003 & 7200 &  4.3 \\
7045 & 4062 & 12$\rm^{h}$04$\rm^{m}$03.8$\rm^{s}$ & 31$\degr$53'42" & 6581.1 & 10.6 & Mar, 18 2004 & 6000 &  2.1 \\
7154 & 4145 & 12$\rm^{h}$10$\rm^{m}$01.4$\rm^{s}$ & 39$\degr$53'02" & 6584.2 & 10.6 & Mar, 21 2002 & 5520 &  4.2 \\
7429 & 4319 & \textit{12$\rm^{h}$21$\rm^{m}$43.1$\rm^{s}$} & \textit{75\degr19'22"} & 6592.1 & 10.1 & Mar, 03 2003 & 6960 &  4.3 \\
7699 &  & \textit{12$\rm^{h}$32$\rm^{m}$48.0$\rm^{s}$} & \textit{37\degr37'18"} & 6573.9 & 14.0 & Mar, 22 2004 & 6720 &  4.6 \\
7766 & 4559 & 12$\rm^{h}$35$\rm^{m}$57.3$\rm^{s}$ & 27$\degr$57'38" & 6583.3 & 10.6 & Jun, 17 2002 & 2400 &  1.9 \\
7831 & 4605 & 12$\rm^{h}$39$\rm^{m}$59.7$\rm^{s}$ & 61$\degr$36'29" & 6565.7 & 12.0 & Jun, 15 2002 & 3840 &  2.2 \\
7853 & 4618 & 12$\rm^{h}$41$\rm^{m}$33.1$\rm^{s}$ & 41$\degr$09'05" & 6573.6 & 11.4 & Mar, 18 2002 & 5040 &  2.6 \\
7861 & 4625 & 12$\rm^{h}$41$\rm^{m}$52.9$\rm^{s}$ & 41$\degr$16'25" & 6575.7 & 11.4 & Jun, 16 2002 & 3840 &  3.3 \\
7876 & 4635 & 12$\rm^{h}$42$\rm^{m}$39.3$\rm^{s}$ & 19$\degr$56'44" & 6584.3 & 10.6 & Mar, 17 2004 & 6720 &  3.0 \\
7901 & 4651 & 12$\rm^{h}$43$\rm^{m}$42.7$\rm^{s}$ & 16$\degr$23'35" & 6581.7 & 10.6 & Mar, 19 2004 & 8160 &  3.8 \\
7985 & 4713 & 12$\rm^{h}$49$\rm^{m}$57.9$\rm^{s}$ & 05$\degr$18'42" & 6574.1 & 11.4 & Apr, 26 2003 & 6240 &  5.4 \\
8334 & 5055 & 13$\rm^{h}$15$\rm^{m}$49.4$\rm^{s}$ & 42$\degr$01'46" & 6572.9 & 13.7 & Jun, 14 2002 & 4320 &  2.3 \\
8403 & 5112 & 13$\rm^{h}$21$\rm^{m}$56.6$\rm^{s}$ & 38$\degr$44'05" & 6584.2 & 10.6 & Mar, 21 2002 & 4080 &  4.4 \\
NGC 5296 & 5296 & 13$\rm^{h}$46$\rm^{m}$18.7$\rm^{s}$ & 43$\degr$51'04" & 6615.4 & 11.1 & Jun, 13 2002 & 6000 &  3.2 \\
8709 & 5297 & 13$\rm^{h}$46$\rm^{m}$23.7$\rm^{s}$ & 43$\degr$52'20" & 6615.4 & 11.1 & Jun, 13 2002 & 6000 &  3.2 \\
8852 & 5376 & 13$\rm^{h}$55$\rm^{m}$16.1$\rm^{s}$ & 59$\degr$30'25" & 6610.2 & 11.1 & Mar, 05 2003 & 2880 &  2.8 \\
 &  &  &  & 6609.9 & 11.1 & Mar, 07 2003 & 4800 &  3.7 \\
8863 & 5377 & 13$\rm^{h}$56$\rm^{m}$16.7$\rm^{s}$ & 47$\degr$14'08" & 6601.1 & 12.0 & Mar, 26 2004 & 5040 &  3.3 \\
8898 & 5394 & 13$\rm^{h}$58$\rm^{m}$33.7$\rm^{s}$ & 37$\degr$27'12" & 6651.1 & 20.6 & Mar, 18 2002 & 4320 &  2.7 \\
8900 & 5395 & 13$\rm^{h}$58$\rm^{m}$38.0$\rm^{s}$ & 37$\degr$25'28" & 6651.1 & 20.6 & Mar, 18 2002 & 4320 &  2.7 \\
8937 & 5430 & 14$\rm^{h}$00$\rm^{m}$45.8$\rm^{s}$ & 59$\degr$19'43" & 6628.2 & 20.0 & Mar, 03 2003 & 8880 &  4.4 \\
9013 & 5474 & 14$\rm^{h}$05$\rm^{m}$02.0$\rm^{s}$ & 53$\degr$39'08" & 6565.2 & 12.0 & Jun, 14 2002 & 5040 &  2.3 \\
9179 & 5585 & 14$\rm^{h}$19$\rm^{m}$48.1$\rm^{s}$ & 56$\degr$43'45" & 6570.7 & 11.4 & Mar, 20 2004 & 8160 &  3.1 \\
9219 & 5608 & \textit{14$\rm^{h}$23$\rm^{m}$17.5$\rm^{s}$} & \textit{41\degr46'34"} & 6573.7 & 11.4 & Mar, 22 2004 & 6480 &  4.7 \\
9248 & 5622 & 14$\rm^{h}$26$\rm^{m}$12.2$\rm^{s}$ & 48$\degr$33'51" & 6655.2 & 20.6 & Mar, 21 2004 & 10800 &  4.8 \\
9358 & 5678 & 14$\rm^{h}$32$\rm^{m}$05.6$\rm^{s}$ & 57$\degr$55'16" & 6603.0 & 12.0 & Mar, 19 2004 & 7680 &  4.3 \\
9363 & 5668 & 14$\rm^{h}$33$\rm^{m}$24.4$\rm^{s}$ & 04$\degr$27'02" & 6594.1 & 10.1 & Apr, 26 2003 & 8160 &  2.8 \\
9406 & 5693 & 14$\rm^{h}$36$\rm^{m}$11.1$\rm^{s}$ & 48$\degr$35'06" & 6612.6 & 11.1 & Apr, 28 2003 & 6000 &  3.3 \\
9465 & 5727 & 14$\rm^{h}$40$\rm^{m}$26.1$\rm^{s}$ & 33$\degr$59'23" & 6594.1 & 10.1 & Mar, 17 2004 & 6480 &  3.4 \\
9576 & 5774 & 14$\rm^{h}$53$\rm^{m}$42.5$\rm^{s}$ & 03$\degr$34'57" & 6594.2 & 10.1 & Apr, 27 2003 & 6840 &  4.1 \\
9736 & 5874 & 15$\rm^{h}$07$\rm^{m}$51.9$\rm^{s}$ & 54$\degr$45'08" & 6629.1 & 20.0 & Mar, 02 2003 & 3840 &  4.8 \\
\hline
\end{tabular}
\end{table*}
\begin{table*}
\contcaption{}
\begin{tabular}{ccccccccc}
\noalign{\medskip} \hline
N\Deg & N\Deg &  $\alpha$    & $\delta$   &   $\lambda_{c}$&  FWHM & date & exposure time& seeing\\
 UGC        &  NGC       &  (2000) & (2000) & \AA & \AA & & s &"\\
 (1)&(2)&(3)&(4)&(5)&(6)&(7)&(8)&(9)\\
\hline
9866 & 5949 & 15$\rm^{h}$28$\rm^{m}$00.6$\rm^{s}$ & 64$\degr$45'46" & 6572.2 & 11.4 & Mar, 24 2004 & 5280 &  6.9 \\
9943 & 5970 & 15$\rm^{h}$38$\rm^{m}$30.0$\rm^{s}$ & 12$\degr$11'11" & 6602.5 & 12.0 & Mar, 04 2003 & 8400 &  4.3 \\
10075 & 6015 & 15$\rm^{h}$51$\rm^{m}$25.3$\rm^{s}$ & 62$\degr$18'36" & 6580.9 & 10.6 & Mar, 18 2004 & 5760 &  3.2 \\
10521 & 6207 & 16$\rm^{h}$43$\rm^{m}$03.7$\rm^{s}$ & 36$\degr$49'55" & 6582.1 & 10.6 & Apr, 28 2003 & 2400 &  3.0 \\
 &  &  &  & 6582.1 & 10.6 & Apr, 29 2003 & 5280 &  3.9 \\
10652 & 6283 & 16$\rm^{h}$59$\rm^{m}$26.5$\rm^{s}$ & 49$\degr$55'20" & 6584.2 & 10.6 & Apr, 26 2003 & 2880 &  3.2 \\
 &  &  &  & 6584.2 & 10.6 & Apr, 27 2003 & 3120 &  4.1 \\
10713 &  & \textit{17$\rm^{h}$04$\rm^{m}$33.7$\rm^{s}$} & \textit{72\degr26'45"} & 6585.5 & 10.6 & Sep, 04 2002 & 5040 &  2.4 \\
10757 &  & 17$\rm^{h}$10$\rm^{m}$13.4$\rm^{s}$ & 72$\degr$24'38" & 6586.0 & 10.6 & Jun, 17 2002 & 5280 &  2.4 \\
10769 &  & \textit{17$\rm^{h}$11$\rm^{m}$33.5$\rm^{s}$} & \textit{72\degr24'07"} & 6592.9 & 10.1 & Jun, 17 2002 & 5280 &  2.1 \\
10791 &  & 17$\rm^{h}$14$\rm^{m}$38.5$\rm^{s}$ & 72$\degr$23'56" & 6593.1 & 10.1 & Jun, 15 2002 & 6000 &  2.6 \\
11012 & 6503 & 17$\rm^{h}$49$\rm^{m}$26.3$\rm^{s}$ & 70$\degr$08'40" & 6562.2 &  8.8 & Sep, 06 2002 & 5040 &  2.6 \\
11269 & 6667 & 18$\rm^{h}$30$\rm^{m}$39.7$\rm^{s}$ & 67$\degr$59'14" & 6615.5 & 11.1 & Jun, 17 2002 & 3840 &  2.3 \\
11300 & 6689 & 18$\rm^{h}$34$\rm^{m}$49.9$\rm^{s}$ & 70$\degr$31'28" & 6584.4 & 10.6 & Sep, 05 2002 & 5040 &  3.7 \\
11332 & 6654A & \textit{18$\rm^{h}$39$\rm^{m}$25.2$\rm^{s}$} & \textit{73\degr34'48"} & 6594.7 & 10.1 & Sep, 10 2002 & 7200 &  5.5 \\
11407 & 6764 & 19$\rm^{h}$08$\rm^{m}$16.4$\rm^{s}$ & 50$\degr$55'59" & 6615.0 & 11.1 & Sep, 11 2002 & 6480 &  2.6 \\
11466 &  & 19$\rm^{h}$42$\rm^{m}$59.1$\rm^{s}$ & 45$\degr$17'58" & 6583.0 & 10.6 & Jun, 16 2002 & 5040 &  2.3 \\
11470 & 6824 & 19$\rm^{h}$43$\rm^{m}$40.8$\rm^{s}$ & 56$\degr$06'34" & 6628.6 & 20.0 & Sep, 06 2002 & 3600 &  2.2 \\
11496 &  & 19$\rm^{h}$53$\rm^{m}$01.8$\rm^{s}$ & 67$\degr$39'54" & 6611.3 & 11.1 & Jun, 18 2002 & 5520 &  1.9 \\
11498 &  & 19$\rm^{h}$57$\rm^{m}$15.1$\rm^{s}$ & 05$\degr$53'24" & 6628.0 & 20.0 & Oct, 25 2003 & 8400 &  4.3 \\
11597 & 6946 & 20$\rm^{h}$34$\rm^{m}$52.5$\rm^{s}$ & 60$\degr$09'12" & 6562.3 &  8.8 & Jun, 14 2002 & 5280 &  2.4 \\
11670 & 7013 & 21$\rm^{h}$03$\rm^{m}$33.7$\rm^{s}$ & 29$\degr$53'50" & 6581.9 & 10.6 & Sep, 11 2002 & 7440 &  2.7 \\
11872 & 7177 & 22$\rm^{h}$00$\rm^{m}$41.2$\rm^{s}$ & 17$\degr$44'18" & 6584.8 & 10.6 & Sep, 10 2002 & 7200 &  4.7 \\
12082 &  & 22$\rm^{h}$34$\rm^{m}$11.3$\rm^{s}$ & 32$\degr$51'42" & 6582.1 & 10.6 & Sep, 02 2002 & 7440 &  2.8 \\
\hline
\end{tabular}
\\(1) Name of the galaxy in the UGC catalog except for NGC 542, IC 476, IC 2542 and NGC 5296 that do not have UGC name. (2) Name in the NGC catalog when available. (3\&4) Coordinates (in 2000) of the center of the galaxy used for the kinematic study except those in italic (taken from HyperLeda). (5) Central wavelength of the interference filter used. (6) FWHM of the interference filter. (7) Date of the observations. (8) Total exposure time in second. (9) Seeing in arcsec.
\end{table*}

\begin{table*}
\caption{Model parameters.}
\begin{tabular}{cccccccccc}
\noalign{\medskip} \hline 
N\Deg & V$_{sys\_Leda}$ & V$_{sys\_FP}$ & i$_{Morph}$ & i$_{Kin}$ & P.A.$_{Morph}$ & P.A.$_{Kin}$ & $\overline{Res}$ & $\sigma_{res}$ & $\chi^{2}_{red}$\\
 UGC & \kms & \kms & \Deg & \Deg & \Deg & \Deg & 10$^{-3}$\kms & \kms &\\
(1)&(2)&(3)&(4)&(5)&(6)&(7)&(8)&(9)&(10)\\
\hline
12893 & 1102$\pm$7 & 1097$\pm$2 & 30$\pm$8 & 19$\pm$19 & 95$\pm$51 & 77$^\#$$\pm$5 & -28.1 & 8 & 1.2 \\
89 & 4564$\pm$3 & 4510$\pm$5 & 40$\pm$4 & 33$\pm$13 & 174$\pm$24 & 177$\pm$4 & 42.5 & 23 & 8.6 \\
94 & 4592$\pm$4 & 4548$\pm$2 & 50$\pm$4 & 42$\pm$5 & 102$\pm$15 & 94$\pm$2 & -3.9 & 13 & 2.7 \\
1013 & 5190$\pm$4 &  & 80$\pm$3 &  & 68$\pm$5 &  &  &  &  \\
NGC 542 & 4660$\pm$8 &  & 90 &  & 143$\pm$6 &  &  &  &  \\
1317 & 3111$\pm$5 & 3090$\pm$2 & 75$\pm$4 & 73$\pm$1 & 105$\pm$5 & 106$\pm$1 & -1.3 & 15 & 3.3 \\
1437 & 4893$\pm$4 & 4858$\pm$2 & 52$\pm$3 & 47$\pm$4 & 134$\pm$17 & 127$^\#$$\pm$2 & -0.8 & 18 & 5.2 \\
1655 & 5340$\pm$16 & 5427$\pm$7 & 45$\pm$9 & 45$\pm$18$^*$ &  & 138$^\#$$\pm$6 & -2744.9 & 30 & 15.4 \\
1810 & 7556$\pm$21 &  & 75$\pm$3 &  & 42$\pm$6 &  &  &  &  \\
3056 & -100$\pm$7 &  & 65$\pm$7 &  & 120$\pm$12 &  &  &  &  \\
3334 & 3934$\pm$4 & 3952$\pm$13 & 47$\pm$7 & 47$\pm$14$^*$ & 85$\pm$15 & 97$^\#$$\pm$6 & 9.9 & 54 & 46.0 \\
3382 & 4497$\pm$6 & 4489$\pm$2 & 21$\pm$10 & 21$\pm$14 & 30$^{2M}$/168$^{Pa}$$\pm$61 & 6$^\#$$\pm$2 & -11.5 & 16 & 4.1 \\
3463 & 2692$\pm$4 & 2679$\pm$3 & 63$\pm$3 & 63$\pm$3 & 117$\pm$8 & 110$\pm$2 & -0.1 & 15 & 3.6 \\
3521 & 4426$\pm$7 & 4415$\pm$2 & 61$\pm$3 & 58$\pm$5 & 76$\pm$12 & 78$^\#$$\pm$3 & 3.2 & 16 & 4.1 \\
3528 & 4421$\pm$18 & 4340$\pm$5 & 59$\pm$5 & 42$\pm$12 & 38$\pm$16 & 43$^\#$$\pm$4 & -17.8 & 32 & 17.6 \\
3618 & 5851$\pm$6 &  & 49$\pm$5 &  & 171$\pm$16 &  &  &  &  \\
3685 & 1796$\pm$4 & 1795$\pm$1 & 55$\pm$4 & 12$\pm$17 & 133$\pm$12 & 118$^\#$$\pm$4 & -1.3 & 9 & 1.4 \\
3708 & 5201$\pm$26 & 5161$\pm$4 & 16$\pm$24 & 44$\pm$16 & 136$^{Ni}$$\pm$90 & 50$^\#$$\pm$4 & -11.6 & 18 & 5.4 \\
3709 & 5223$\pm$50 & 5292$\pm$4 & 46$\pm$4 & 55$\pm$4 & 66$\pm$18 & 52$^\#$$\pm$2 & -4.7 & 18 & 5.3 \\
3826 & 1733$\pm$3 & 1724$\pm$2 & 30$\pm$9 & 20$\pm$19 & 160$^{2M}$/85$^{Ni}$$\pm$34 & 74$^\#$$\pm$5 & 135.4 & 9 & 1.2 \\
3740 & 2417$\pm$6 & 2416$\pm$2 & 40$\pm$6 & 48$\pm$14 & 19$\pm$20 & 67$^\#$$\pm$4 & $<$0.1 & 11 & 1.7 \\
3876 & 860$\pm$7 & 854$\pm$2 & 61$\pm$3 & 59$\pm$5 & 178$\pm$9 & 178$^\#$$\pm$3 & -0.2 & 10 & 1.6 \\
3915 & 4679$\pm$7 & 4659$\pm$3 & 59$\pm$5 & 47$\pm$4 & 25$\pm$17 & 30$\pm$2 & 6.8 & 13 & 2.7 \\
IC 476 & 4734$\pm$32 & 4767$\pm$3 & 40$\pm$5 & 55$\pm$24 & 102$\pm$30 & 68$\pm$6 & -5.9 & 18 & 5.5 \\
4026 & 4782$\pm$11 & 4892$\pm$3 & 73$\pm$4 & 56$\pm$4 & 136$\pm$8 & 139$\pm$2 & 4.0 & 19 & 5.7 \\
4165 & 515$\pm$4 & 504$\pm$1 & 21$\pm$9 & 41$\pm$10 & 74$^{Pa}$$\pm$43 & 85$^\#$$\pm$2 & -0.4 & 10 & 1.6 \\
4256 & 5252$\pm$5 & 5252$\pm$3 & 36$\pm$4 & 38$\pm$21 & 26$\pm$23 & 111$^\#$$\pm$6 & 15.9 & 26 & 11.1 \\
4393 & 2126$\pm$5 & 2119$\pm$4 & 50$\pm$5 & 50$\pm$9$^*$ & 50$\pm$22 & 70$^\#$$\pm$7 & 0.6 & 11 & 1.7 \\
4422 & 4333$\pm$4 & 4321$\pm$2 & 49$\pm$4 & 25$\pm$8 & 12$\pm$17 & 36$\pm$2 & -18.3 & 20 & 6.2 \\
4456 & 5497$\pm$23 & 5470$\pm$1 & 28$\pm$6 & 9$\pm$14 & 42$\pm$41 & 124$\pm$3 & 7.3 & 14 & 3.1 \\
4555 & 4235$\pm$6 & 4235$\pm$2 & 21$\pm$12 & 38$\pm$7 & 140$^{2M}$/78$^{SDSS}$$\pm$72 & 90$\pm$2 & 0.5 & 15 & 3.5 \\
4770 & 7063$\pm$9 & 7026$\pm$3 & 36$\pm$9 & 20$\pm$13 & 54$\pm$30 & 98$^\#$$\pm$3 & -36.8 & 14 & 3.0 \\
4820 & 1355$\pm$4 & 1350$\pm$2 & 41$\pm$4 & 38$\pm$3 & 160$\pm$16 & 157$\pm$2 & $<$0.1 & 13 & 2.7 \\
5045 & 7716$\pm$23 & 7667$\pm$2 & 41$\pm$4 & 16$\pm$9 & 136$\pm$21 & 148$\pm$2 & -2.8 & 13 & 2.7 \\
5175 & 3052$\pm$11 & 3049$\pm$2 & 65$\pm$4 & 56$\pm$3 & 145$\pm$9 & 143$\pm$2 & -2.3 & 13 & 2.9 \\
5228 & 1873$\pm$7 & 1869$\pm$2 & 82$\pm$2 & 72$\pm$2 & 122$\pm$4 & 120$\pm$2 & -0.7 & 9 & 1.3 \\
5251 & 1481$\pm$3 & 1465$\pm$3 & 88$\pm$9 & 73$\pm$6 & 78$\pm$3 & 80$^\#$$\pm$3 & $<$0.1 & 15 & 3.5 \\
5279 & 1488$\pm$4 &  & 90 &  & 83$\pm$4 &  &  &  &  \\
5319 & 2448$\pm$9 & 2439$\pm$1 & 40$\pm$5 & 30$\pm$9 & 125$^{2M}$/7$^{Pa}$$\pm$19 & 165$^\#$$\pm$2 & -21.0 & 9 & 1.3 \\
5351 & 1473$\pm$4 &  & 82$\pm$6 &  & 105$\pm$4 &  &  &  &  \\
5373 & 302$\pm$3 & 291$\pm$2 & 60$\pm$6 & 10$\pm$18 & 110$\pm$12 & 51$\pm$8 & -0.5 & 8 & 0.9 \\
5398 & 14$\pm$5 &  & 40$\pm$10 &  & 45$\pm$23 &  &  &  &  \\
IC 2542 & 6113$\pm$20 & 6111$\pm$2 & 43$\pm$4 & 20$\pm$15 & 173$\pm$21 & 174$\pm$3 & -0.9 & 20 & 6.3 \\
5510 & 1301$\pm$3 & 1298$\pm$2 & 38$\pm$5 & 31$\pm$10 & 26$\pm$24 & 20$^\#$$\pm$3 & -0.3 & 10 & 1.4 \\
5532 & 2812$\pm$8 & 2802$\pm$1 & 33$\pm$9 & 32$\pm$3 & 150$\pm$27 & 147$\pm$1 & $<$0.1 & 14 & 2.9 \\
5556 & 1581$\pm$3 &  & 75$\pm$2 &  & 105$\pm$5 &  &  &  &  \\
5786 & 990$\pm$3 & 992$\pm$4 & 16$\pm$25 & 53$\pm$11 & 18$^{SDSS}$$\pm$90 & 153$\pm$5 & -574.4 & 18 & 4.9 \\
5840 & 582$\pm$4 & 580$\pm$1 & 18$\pm$14 & 18$\pm$11 &  & 153$^\#$$\pm$3 & $<$0.1 & 12 & 2.1 \\
5842 & 1261$\pm$10 & 1245$\pm$1 & 34$\pm$6 & 47$\pm$9 & 104$\pm$24 & 112$^\#$$\pm$2 & -1.5 & 10 & 1.7 \\
6118 & 1539$\pm$5 & 1525$\pm$2 & 27$\pm$7 & 39$\pm$8$^*$ & 150$^{2M}$/57$^{Pa}$$\pm$32 & 163$^\#$$\pm$3 & 16.3 & 11 & 1.9 \\
6277 & 1192$\pm$3 & 1191$\pm$3 & 17$\pm$16 & 17$\pm$17$^*$ & 0$^{2M}$$\pm$86 & 76$\pm$3 & -5.1 & 19 & 5.7 \\
6419 & 1365$\pm$24 & 1381$\pm$2 & 57$\pm$5 & 66$\pm$19 & 27$\pm$15 & 34$\pm$7 & -1.8 & 8 & 0.9 \\
6521 & 5879$\pm$5 & 5842$\pm$2 & 50$\pm$3 & 46$\pm$4 & 21$\pm$13 & 20$\pm$2 & -0.9 & 17 & 4.5 \\
6523 & 5913$\pm$13 & 5947$\pm$2 & 24$\pm$7 & 24$\pm$14$^*$ & 36$^{Va}$/12$^{Pa}$/51$^{Pa}$$\pm$52 & 173$^\#$$\pm$3 & -6.4 & 13 & 2.8 \\
6787 & 1173$\pm$3 & 1157$\pm$3 & 57$\pm$3 & 70$\pm$2 & 108$\pm$10 & 112$\pm$2 & -0.7 & 17 & 4.4 \\
7021 & 1979$\pm$8 & 1976$\pm$3 & 56$\pm$4 & 56$\pm$7$^*$ & 89$\pm$11 & 86$^\#$$\pm$2 & -0.6 & 14 & 3.1 \\
7045 & 770$\pm$6 & 758$\pm$1 & 70$\pm$2 & 68$\pm$2 & 101$\pm$5 & 99$\pm$2 & 0.2 & 10 & 1.5 \\
7154 & 1011$\pm$28 & 1009$\pm$1 & 64$\pm$3 & 65$\pm$3 & 100$\pm$8 & 95$^\#$$\pm$2 & $<$0.1 & 12 & 2.3 \\
7429 & 1476$\pm$24 &  & 73$\pm$3 &  & 162$\pm$7 &  &  &  &  \\
\hline
\label{tablemod}
\end{tabular}
\end{table*}
\begin{table*}
\contcaption{}
\begin{tabular}{cccccccccc}
\noalign{\medskip} \hline 
N\Deg & V$_{sys\_Leda}$ & V$_{sys\_FP}$ & i$_{Morph}$ & i$_{Kin}$ & P.A.$_{Morph}$ & P.A.$_{Kin}$ & $\overline{Res}$ & $\sigma_{res}$ & $\chi^{2}_{red}$\\
 UGC & \kms & \kms & \Deg & \Deg & \Deg & \Deg & 10$^{-3}$\kms & \kms &\\
(1)&(2)&(3)&(4)&(5)&(6)&(7)&(8)&(9)&(10)\\
\hline
7699 & 496$\pm$1 &  & 78$\pm$2 &  & 32$\pm$4 &  &  &  &  \\
7766 & 814$\pm$3 & 807$\pm$1 & 65$\pm$4 & 69$\pm$3 & 150$\pm$7 & 143$^\#$$\pm$2 & $<$0.1 & 12 & 2.4 \\
7831 & 147$\pm$4 & 136$\pm$3 & 70$\pm$3 & 56$\pm$12 & 125$\pm$5 & 110$^\#$$\pm$5 & $<$0.1 & 9 & 1.2 \\
7853 & 537$\pm$4 & 530$\pm$2 & 58$\pm$5 & 58$\pm$28$^*$ & 40$\pm$12 & 37$^\#$$\pm$4 & 12.8 & 8 & 1.1 \\
7861 & 611$\pm$4 & 598$\pm$1 & 47$\pm$6 & 47$\pm$24$^*$ & 116$^{Pa}$/30$^{SDSS}$$\pm$20 & 117$^\#$$\pm$4 & $<$0.1 & 11 & 1.9 \\
7876 & 955$\pm$8 & 944$\pm$1 & 44$\pm$5 & 53$\pm$9 & 3$\pm$22 & 164$^\#$$\pm$3 & 0.3 & 8 & 1.1 \\
7901 & 799$\pm$2 & 788$\pm$2 & 50$\pm$3 & 53$\pm$2 & 77$\pm$11 & 74$^\#$$\pm$2 & $<$0.1 & 12 & 2.2 \\
7985 & 653$\pm$3 & 642$\pm$2 & 24$\pm$12 & 49$\pm$6 & 110$^{2M}$/87$^{Ha}$/153$^{Pa}$/100$^{Ni}$/88$^{SDSS}$$\pm$48 & 96$^\#$$\pm$3 & -0.1 & 8 & 1.0 \\
8334 & 508$\pm$3 & 484$\pm$1 & 55$\pm$5 & 66$\pm$1 & 102$\pm$11 & 100$\pm$1 & -0.1 & 11 & 1.7 \\
8403 & 969$\pm$4 & 975$\pm$2 & 54$\pm$3 & 57$\pm$4 & 129$\pm$12 & 121$\pm$2 & -0.1 & 9 & 1.3 \\
NGC 5296 & 2243$\pm$3 & 2254$\pm$2 & 65$\pm$6 & 65$\pm$4 & 12$\pm$15 & 2$\pm$3 & -6.2 & 4 & 0.3 \\
8709 & 2407$\pm$13 & 2405$\pm$3 & 82$\pm$3 & 76$\pm$1 & 147$\pm$4 & 150$^\#$$\pm$2 & 0.1 & 15 & 3.3 \\
8852 & 2023$\pm$17 & 2075$\pm$1 & 55$\pm$7 & 52$\pm$3 & 65$\pm$16 & 63$\pm$2 & 0.9 & 10 & 1.5 \\
8863 & 1796$\pm$7 & 1789$\pm$4 & 77$\pm$4 & 77$\pm$13$^*$ & 38$\pm$5 & 38$^\#$$\pm$7$^*$ & -41.4 & 14 & 3.3 \\
8898 & 3464$\pm$10 & 3448$\pm$2 & 71$\pm$3 & 27$\pm$20 & 30$^{2M}$/140$^{Pa}$/117$^{SDSS}$$\pm$7 & 31$\pm$6 & 41.5 & 7 & 0.7 \\
8900 & 3466$\pm$11 & 3511$\pm$3 & 66$\pm$5 & 57$\pm$10 & 172$\pm$10 & 161$\pm$2 & 5.8 & 20 & 6.6 \\
8937 & 2968$\pm$9 & 2961$\pm$5 & 50$\pm$5 & 32$\pm$12 & 177$\pm$16 & 5$^\#$$\pm$3 & 2361.4 & 19 & 5.7 \\
9013 & 255$\pm$23 & 262$\pm$1 & 50$\pm$4 & 21$\pm$16$^*$ & 85$\pm$14 & 164$\pm$4 & 1.8 & 7 & 0.8 \\
9179 & 302$\pm$2 & 293$\pm$2 & 53$\pm$3 & 36$\pm$14 & 33$\pm$11 & 49$\pm$4 & $<$0.1 & 9 & 1.4 \\
9219 & 666$\pm$11 &  & 81$\pm$6 &  & 99$\pm$9 &  &  &  &  \\
9248 & 3867$\pm$6 & 3865$\pm$2 & 58$\pm$3 & 58$\pm$4 & 86$\pm$12 & 81$^\#$$\pm$2 & 3.5 & 15 & 3.6 \\
9358 & 1907$\pm$4 & 1912$\pm$3 & 62$\pm$3 & 54$\pm$4 & 2$\pm$9 & 2$^\#$$\pm$2 & 2.3 & 15 & 3.5 \\
9363 & 1584$\pm$3 & 1577$\pm$1 & 33$\pm$6 & 18$\pm$14$^*$ & 107$\pm$27 & 147$\pm$3 & 3.2 & 8 & 1.1 \\
9406 & 2279$\pm$2 & 2281$\pm$2 & 51$\pm$7 & 59$\pm$25 & 60$^{2M}$/150$^{SDSS}$$\pm$22 & 132$\pm$12 & 78.0 & 11 & 2.0 \\
9465 & 1495$\pm$3 & 1485$\pm$2 & 90 & 65$\pm$4 & 143$\pm$9 & 127$\pm$3 & 0.3 & 8 & 1.1 \\
9576 & 1565$\pm$5 & 1555$\pm$2 & 52$\pm$4 & 41$\pm$11 & 125$\pm$14 & 122$\pm$3 & -0.4 & 10 & 1.6 \\
9736 & 3128$\pm$4 & 3135$\pm$2 & 51$\pm$3 & 51$\pm$5 & 57$\pm$13 & 39$^\#$$\pm$2 & 2.9 & 14 & 3.2 \\
9866 & 427$\pm$4 & 430$\pm$1 & 69$\pm$3 & 56$\pm$6 & 150$\pm$7 & 148$\pm$2 & -5.1 & 7 & 0.8 \\
9943 & 1958$\pm$4 & 1946$\pm$1 & 48$\pm$5 & 54$\pm$2 & 87$\pm$14 & 86$^\#$$\pm$2 & -0.1 & 9 & 1.4 \\
10075 & 831$\pm$3 & 827$\pm$1 & 66$\pm$3 & 62$\pm$2 & 28$\pm$7 & 30$^\#$$\pm$1 & $<$0.1 & 9 & 1.4 \\
10521 & 852$\pm$2 & 832$\pm$2 & 71$\pm$3 & 59$\pm$3 & 17$\pm$7 & 20$\pm$2 & 0.7 & 9 & 1.2 \\
10652 & 1092$\pm$26 & 1089$\pm$1 & 30$\pm$7 & 21$\pm$13 & 56$\pm$33 & 45$^\#$$\pm$3 & 0.6 & 8 & 1.0 \\
10713 & 1073$\pm$4 &  & 90 &  & 8$\pm$4 &  &  &  &  \\
10757 & 1168$\pm$8 & 1210$\pm$2 & 59$\pm$3 & 44$\pm$22 & 66$\pm$12 & 56$\pm$6 & 0.7 & 11 & 1.8 \\
10769 & 1230$\pm$13 &  & 57$\pm$4 &  & 41$\pm$16 &  &  &  &  \\
10791 & 1328$\pm$6 & 1318$\pm$3 & 0 & 34$\pm$20$^*$ &  & 92$\pm$4 & 726.5 & 10 & 1.7 \\
11012 & 36$\pm$12 & 25$\pm$1 & 74$\pm$2 & 72$\pm$2 & 123$\pm$5 & 119$^\#$$\pm$2 & 0.4 & 8 & 1.0 \\
11269 & 2590$\pm$6 & 2563$\pm$6 & 60$\pm$3 & 69$\pm$4 & 97$\pm$12 & 92$^\#$$\pm$3 & 3.8 & 30 & 14.4 \\
11300 & 488$\pm$3 & 480$\pm$2 & 77$\pm$2 & 76$\pm$4 & 171$\pm$4 & 167$\pm$3 & $<$0.1 & 10 & 1.6 \\
11332 & 1569$\pm$25 &  & 82$\pm$2 &  & 65$\pm$3 &  &  &  &  \\
11407 & 2412$\pm$4 & 2402$\pm$8 & 64$\pm$3 & 64$\pm$22$^*$ & 65$\pm$9 & 65$\pm$10$^*$ & 1.0 & 20 & 6.6 \\
11466 & 820$\pm$9 & 826$\pm$3 & 55$\pm$3 & 66$\pm$5 & 35$\pm$13 & 46$^\#$$\pm$3 & -0.3 & 13 & 2.6 \\
11470 & 3530$\pm$40 & 3546$\pm$5 & 47$\pm$6 & 47$\pm$7 & 50$\pm$22 & 47$\pm$3 & 258.0 & 25 & 10.7 \\
11496 & 2105$\pm$6 & 2115$\pm$2 & 0 & 44$\pm$16 &  & 167$\pm$4 & 135.8 & 9 & 1.2 \\
11498 & 3266$\pm$8 & 3284$\pm$4 & 79$\pm$4 & 71$\pm$2 & 75$\pm$7 & 71$^\#$$\pm$2 & -23.4 & 22 & 7.9 \\
11597 & 46$\pm$3 & 40$\pm$2 & 17$\pm$19 & 40$\pm$10 &  & 61$^\#$$\pm$3 & 0.4 & 13 & 2.5 \\
11670 & 778$\pm$3 & 776$\pm$3 & 90 & 65$\pm$2 & 159$\pm$5 & 153$^\#$$\pm$2 & 0.8 & 16 & 3.9 \\
11872 & 1147$\pm$5 & 1140$\pm$1 & 54$\pm$6 & 47$\pm$3 & 88$\pm$13 & 86$\pm$2 & -0.1 & 13 & 2.7 \\
12082 & 803$\pm$2 & 792$\pm$2 & 29$\pm$11 & 14$\pm$19 &  & 143$\pm$5 & 0.6 & 8 & 1.1 \\
\hline
\end{tabular}
\\(1) Name in the UGC catalog (see table \ref{tablelog}). (2) Systemic velocity found in HyperLeda data base. (3) Systemic velocity deduced from our velocity field analysis. (4) Morphological inclination from HyperLeda \citep{Paturel:1997}. (5) Inclination deduced from the analysis of our \VF; those marked with an asterisk ($^*$) have been fixed equal to morphological value, except UGC 6118, UGC 9013, UGC 9363 and UGC 10791 for which we used inclinations determined from HI data (see table \ref{tabletf}). (6) Morphological position angle from HyperLeda, except for those marked ($Ha$: \citealp{Haynes:1999}; $Ni$: \citealp{Nilson:1973}; $Pa$: \citealp{Paturel:2000}; $SDSS$: 2006 Sloan Digital Sky Survey, DR5; $2M$: Two Micron All Sky Survey team 2003, 2MASS extended objects; $Va$: \citealp{Vauglin:1999}). (7) Position angle deduced from our \VF; those marked with an asterisk ($^*$) have been fixed equal to morphological value. The symbol $^\#$ indicates that the position angle refers to the approaching side. (8) Mean residual velocity on the whole \VF. (9) Residual velocity dispersion on the whole \VF. (10) Reduced $\chi^{2}$ of the model.
\end{table*}

\begin{table*}
\caption{Galaxy parameters.}
\begin{tabular}{ccccccccccc}
\noalign{\medskip} \hline 
N\Deg & t & Type & D & M$_{b}$ & b/a & i$_{b/a}$ & D$_{25}$/2 & V$_{max}$ & V$_{max}$& HI data\\
UGC & & & Mpc & mag & & \Deg & "/kpc & \kms & flag & \\
(1)&(2)&(3)&(4)&(5)&(6)&(7)&(8)&(9)&(10)&(11)\\
\hline
12893 & 8.4$\pm$0.8 & Sd & 12.5$^{Ja}$ & -15.5 & 0.89$\pm$0.06 & 27$\pm$7 & 34$\pm$5/2.1$\pm$0.3 & 72$\pm$67 & 2 &  \\
89 & 1.2$\pm$0.6 & SBa & 64.2$^{Mo}$ & -21.5 & 0.79$\pm$0.04 & 38$\pm$3 & 46$\pm$4/14.5$\pm$1.4 & 343$\pm$117 & 1 & W$^{N05}$ \\
94 & 2.4$\pm$0.6 & S(r)ab & 64.2$^{Mo}$ & -20.4 & 0.68$\pm$0.04 & 47$\pm$3 & 34$\pm$3/10.5$\pm$0.8 & 209$\pm$21 & 1 & W$^{N05}$ \\
1013 & 3.1$\pm$0.2 & SB(r)b pec & 70.8 & -22.0 & 0.31$\pm$0.03 & 72$\pm$2 & 88$\pm$6/30.1$\pm$2.0 &  &  & W$^{Web}$ \\
NGC 542 & 2.8$\pm$3.9 & Sb pec & 63.7 & -19.5 & 0.22$\pm$0.03 & 77$\pm$2 & 34$\pm$6/10.4$\pm$1.7 & 125$\pm$8$^{PV}$ & 2 &  \\
1317 & 4.9$\pm$0.7 & SAB(r)c & 42.2 & -21.5 & 0.33$\pm$0.04 & 71$\pm$2 & 114$\pm$7/23.4$\pm$1.5 & 205$\pm$9 & 1 & W$^{Web}$ \\
1437 & 4.9$\pm$1.0 & SABc & 66.8 & -21.8 & 0.63$\pm$0.03 & 51$\pm$2 & 43$\pm$5/13.8$\pm$1.7 & 218$\pm$15 & 1 & W$^{Web}$ \\
1655 & 1.0$\pm$0.5 & Sa & 73.0 & -21.6 & 0.75$\pm$0.09 & 42$\pm$8 & 86$\pm$10/30.3$\pm$3.5 & 205$\pm$64 & 4 &  \\
1810 & 3.1$\pm$0.6 & Sb pec & 102.4 & -22.2 & 0.36$\pm$0.02 & 69$\pm$1 & 52$\pm$4/26.0$\pm$2.0 &  &  & W$^{Web}$ \\
3056 & 9.6$\pm$1.2 & IB & 2.5$^{Oc}$ & -18.7 & 0.57$\pm$0.06 & 55$\pm$4 & 119$\pm$9/1.4$\pm$0.1 &  &  &  \\
3334 & 4.2$\pm$1.0 & SABb & 55.6 & -22.8 & 0.70$\pm$0.07 & 45$\pm$6 & 132$\pm$8/35.7$\pm$2.2 & 377$\pm$85 & 1 & W$^{Web}$ \\
3382 & 1.0$\pm$0.4 & SB(r)a & 62.8 & -20.4 & 0.94$\pm$0.05 & 20$\pm$9 & 38$\pm$4/11.5$\pm$1.1 & 322$\pm$207 & 2 & W$^{N05}$ \\
3463 & 4.7$\pm$0.9 & SABc & 38.6 & -20.7 & 0.49$\pm$0.03 & 61$\pm$2 & 66$\pm$4/12.4$\pm$0.8 & 168$\pm$9 & 1 &  \\
3521 & 4.8$\pm$1.8 & Sc & 62.6 & -19.8 & 0.52$\pm$0.03 & 59$\pm$2 & 35$\pm$4/10.7$\pm$1.1 & 167$\pm$12 & 3 &  \\
3528 & 2.0$\pm$0.3 & SBab & 61.8 & -20.1 & 0.58$\pm$0.06 & 55$\pm$4 & 41$\pm$5/12.2$\pm$1.6 & 276$\pm$66 & 2 &  \\
3618 & 2.0$\pm$0.3 & Sab & 80.0 & -20.9 & 0.70$\pm$0.05 & 46$\pm$4 & 44$\pm$4/16.9$\pm$1.4 &  &  &  \\
3685 & 3.0$\pm$0.4 & SB(r)b & 26.3$^{Ja}$ & -19.7 & 0.61$\pm$0.04 & 52$\pm$3 & 57$\pm$4/7.3$\pm$0.5 & 133$\pm$177 & 3 & W$^{Web}$ \\
3708 & 4.5$\pm$1.7 & Sbc pec & 70.0 & -20.7 & 0.96$\pm$0.11 & 15$\pm$23 & 25$\pm$6/8.3$\pm$2.0 & 235$\pm$69 & 1 &  \\
3709 & 5.7$\pm$1.6 & Sc & 70.7 & -21.5 & 0.71$\pm$0.05 & 45$\pm$4 & 35$\pm$3/12.1$\pm$1.1 & 241$\pm$14 & 1 &  \\
3826 & 6.5$\pm$0.8 & SABc & 25.7$^{Ja}$ & -17.9 & 0.87$\pm$0.07 & 29$\pm$8 & 98$\pm$9/12.2$\pm$1.2 & 74$\pm$66 & 1 & W$^{Web}$ \\
3740 & 5.4$\pm$0.6 & SAB(r)c pec & 17.1$^{Sh}$ & -19.8 & 0.78$\pm$0.06 & 39$\pm$5 & 67$\pm$5/5.5$\pm$0.4 & 87$\pm$20 & 2 & W$^{Web}$ \\
3876 & 6.5$\pm$0.8 & Scd & 14.5$^{Ja}$ & -17.4 & 0.50$\pm$0.03 & 60$\pm$2 & 57$\pm$4/4.0$\pm$0.3 & 112$\pm$10 & 2 &  \\
3915 & 4.6$\pm$1.6 & SBc & 63.6 & -21.4 & 0.55$\pm$0.06 & 57$\pm$4 & 34$\pm$6/10.3$\pm$1.7 & 205$\pm$16 & 1 &  \\
IC 476 & 4.2$\pm$2.6 & SABb & 63.9 & -19.0 & 0.78$\pm$0.05 & 39$\pm$5 & 18$\pm$3/5.7$\pm$0.8 & 71$\pm$22 & 3 &  \\
4026 & 2.0$\pm$0.4 & Sab & 64.7 & -20.8 & 0.41$\pm$0.03 & 66$\pm$2 & 43$\pm$4/13.5$\pm$1.2 & 285$\pm$14 & 2 &  \\
4165 & 6.9$\pm$0.4 & SBcd & 11.0$^{Mo}$ & -18.2 & 0.94$\pm$0.05 & 20$\pm$8 & 74$\pm$5/3.9$\pm$0.2 & 80$\pm$18 & 1 & W$^{Web}$ \\
4256 & 5.2$\pm$0.6 & SABc & 71.7 & -21.6 & 0.82$\pm$0.04 & 35$\pm$4 & 50$\pm$4/17.3$\pm$1.3 & 123$\pm$59 & 1 & W$^{Web}$ \\
4393 & 4.6$\pm$1.3 & SBc & 31.5$^{Ja}$ & -19.3 & 0.66$\pm$0.05 & 49$\pm$4 & 44$\pm$7/6.7$\pm$1.1 & 47$\pm$10 & 4 &  \\
4422 & 4.9$\pm$0.6 & SAB(r)c & 58.1 & -21.1 & 0.68$\pm$0.04 & 47$\pm$3 & 51$\pm$5/14.4$\pm$1.4 & 354$\pm$95 & 1 &  \\
4456 & 5.2$\pm$0.6 & S(r)c & 74.0 & -20.8 & 0.89$\pm$0.04 & 27$\pm$5 & 31$\pm$3/11.0$\pm$1.1 & 212$\pm$321 & 1 &  \\
4555 & 4.0$\pm$0.6 & SABb & 58.0 & -20.9 & 0.94$\pm$0.07 & 20$\pm$11 & 45$\pm$5/12.7$\pm$1.4 & 185$\pm$30 & 1 &  \\
4770 & 1.1$\pm$0.6 & SBa & 95.9 & -21.3 & 0.83$\pm$0.07 & 34$\pm$8 & 48$\pm$5/22.3$\pm$2.3 & 330$\pm$195 & 3 &  \\
4820 & 1.7$\pm$0.8 & S(r)ab & 17.1$^{Sh}$ & -20.3 & 0.79$\pm$0.04 & 38$\pm$4 & 127$\pm$6/10.6$\pm$0.5 & 337$\pm$20 & 1 &  \\
5045 & 5.0$\pm$0.5 & SAB(r)c & 105.1 & -21.2 & 0.76$\pm$0.04 & 40$\pm$4 & 35$\pm$3/18.0$\pm$1.5 & 429$\pm$228 & 1 &  \\
5175 & 3.2$\pm$0.7 & Sb & 44.1 & -20.6 & 0.48$\pm$0.05 & 61$\pm$3 & 62$\pm$5/13.2$\pm$1.1 & 188$\pm$10 & 1 &  \\
5228 & 4.9$\pm$0.5 & SBc & 24.7 & -19.9 & 0.25$\pm$0.02 & 76$\pm$1 & 68$\pm$5/8.2$\pm$0.6 & 125$\pm$9 & 1 &  \\
5251 & 4.3$\pm$0.8 & SBbc pec & 21.5 & -20.5 & 0.22$\pm$0.01 & 77$\pm$1 & 142$\pm$7/14.8$\pm$0.7 & 125$\pm$9 & 3 & W$^{Web}$ \\
5279 & 9.7$\pm$1.1 & IB & 21.3 & -19.0 & 0.27$\pm$0.02 & 74$\pm$1 & 67$\pm$5/6.9$\pm$0.5 & 110$\pm$8$^{PV}$ & 1 &  \\
5319 & 5.3$\pm$0.6 & SB(r)c & 35.8 & -19.7 & 0.77$\pm$0.05 & 39$\pm$4 & 47$\pm$3/8.2$\pm$0.6 & 180$\pm$47 & 2 &  \\
5351 & 2.1$\pm$0.6 & SABa & 19.3$^{Sh}$ & -19.4 & 0.32$\pm$0.04 & 71$\pm$2 & 62$\pm$3/5.8$\pm$0.3 & 135$\pm$8$^{PV}$ & 1 & W$^{N05}$ \\
5373 & 9.9$\pm$0.3 & IB & 1.4$^{Ka}$ & -14.3 & 0.62$\pm$0.05 & 52$\pm$4 & 148$\pm$10/1.0$\pm$0.1 & 90$\pm$162 & 2 &  \\
5398 & 7.9$\pm$3.8 & Sd & 3.8$^{Ka}$ & -17.8 & 0.81$\pm$0.08 & 36$\pm$8 & 162$\pm$13/3.0$\pm$0.2 &  &  &  \\
IC 2542 & 4.6$\pm$1.3 & SBc & 83.4 & -20.5 & 0.75$\pm$0.04 & 42$\pm$4 & 31$\pm$3/12.4$\pm$1.2 & 290$\pm$192 & 2 &  \\
5510 & 4.6$\pm$1.0 & SAB(r)c & 18.6 & -19.3 & 0.80$\pm$0.05 & 37$\pm$5 & 64$\pm$6/5.8$\pm$0.5 & 167$\pm$44 & 1 &  \\
5532 & 3.9$\pm$0.6 & Sbc & 41.1 & -22.1 & 0.85$\pm$0.08 & 32$\pm$8 & 122$\pm$10/24.2$\pm$1.9 & 398$\pm$24 & 1 & W$^{Web}$ \\
5556 & 5.0$\pm$0.8 & SBc pec & 22.2 & -18.9 & 0.32$\pm$0.02 & 71$\pm$1 & 67$\pm$4/7.2$\pm$0.4 &  &  & W$^{Web}$ \\
5786 & 4.0$\pm$0.1 & SAB(r)b & 14.2$^{Sh}$ & -19.6 & 0.96$\pm$0.11 & 15$\pm$23 & 54$\pm$6/3.7$\pm$0.4 & 80$\pm$15 & 3 & W$^{Web}$ \\
5840 & 4.0$\pm$0.3 & SB(r)bc & 6.9$^{Ka}$ & -18.9 & 0.95$\pm$0.07 & 17$\pm$13 & 200$\pm$10/6.7$\pm$0.3$^*$ & 251$\pm$138 & 1 & W$^{Web}$ \\
5842 & 6.0$\pm$0.4 & SBc & 15.2$^{Sh}$ & -18.8 & 0.83$\pm$0.05 & 34$\pm$5 & 79$\pm$5/5.8$\pm$0.4 & 115$\pm$18 & 2 &  \\
6118 & 2.1$\pm$0.6 & SB(r)ab & 19.8$^{Sh}$ & -20.0 & 0.90$\pm$0.05 & 26$\pm$7 & 74$\pm$5/7.1$\pm$0.4 & 137$\pm$24 & 1 & W$^{N05}$ \\
6277 & 5.1$\pm$0.5 & SABc & 16.9 & -19.5 & 0.96$\pm$0.07 & 17$\pm$15 & 106$\pm$9/8.7$\pm$0.7 & 268$\pm$257 & 2 & V$^{K00}$ \\
6419 & 8.9$\pm$0.9 & SBm & 18.8 & -18.6 & 0.65$\pm$0.04 & 50$\pm$3 & 44$\pm$4/4.0$\pm$0.4 & 53$\pm$11 & 3 & V$^{W04}$ \\
6521 & 3.7$\pm$0.9 & S(r)bc & 78.6 & -21.2 & 0.67$\pm$0.03 & 48$\pm$2 & 50$\pm$3/19.1$\pm$1.2 & 249$\pm$18 & 1 &  \\
6523 & 1.4$\pm$1.1 & Sa & 80.0 & -21.0 & 0.92$\pm$0.04 & 23$\pm$7 & 32$\pm$3/12.5$\pm$1.3 & 118$\pm$63 & 4 &  \\
6787 & 1.7$\pm$0.8 & Sab & 18.9 & -20.5 & 0.60$\pm$0.03 & 53$\pm$2 & 104$\pm$6/9.5$\pm$0.5 & 232$\pm$11 & 2 & W$^{N05}$ \\
7021 & 1.3$\pm$0.8 & SAB(r)a & 26.8 & -19.7 & 0.62$\pm$0.04 & 52$\pm$3 & 77$\pm$5/10.0$\pm$0.6 & 223$\pm$18 & 1 &  \\
7045 & 5.3$\pm$0.6 & SABc & 11.4$^{Mo}$ & -19.2 & 0.39$\pm$0.03 & 67$\pm$2 & 124$\pm$6/6.9$\pm$0.3 & 160$\pm$9 & 1 &  \\
7154 & 6.9$\pm$0.4 & SBcd & 16.2 & -20.0 & 0.46$\pm$0.03 & 63$\pm$2 & 139$\pm$9/10.9$\pm$0.7 & 145$\pm$9 & 1 & W$^{Web}$ \\
7429 & 2.4$\pm$0.7 & SB(r)ab & 23.7 & -19.8 & 0.40$\pm$0.03 & 67$\pm$2 & 73$\pm$5/8.4$\pm$0.6 &  &  &  \\
\hline
\label{tabletf}
\end{tabular}
\end{table*}
\begin{table*}
\contcaption{}
\begin{tabular}{ccccccccccc}
\noalign{\medskip} \hline 
N\Deg & t & Type & D & M$_{b}$ & b/a & i$_{b/a}$ & D$_{25}$/2 & V$_{max}$ & V$_{max}$& HI data\\
UGC & & & Mpc & mag & & \Deg & "/kpc & \kms & flag & \\
(1)&(2)&(3)&(4)&(5)&(6)&(7)&(8)&(9)&(10)&(11)\\
\hline
7699 & 6.0$\pm$0.6 & SBc & 9.3 & -17.6 & 0.28$\pm$0.01 & 74$\pm$1 & 108$\pm$6/4.9$\pm$0.3 & 92$\pm$8$^{PV}$ & 1 &  \\
7766 & 6.0$\pm$0.4 & SBc & 13.0 & -21.0 & 0.46$\pm$0.05 & 63$\pm$3 & 317$\pm$15/20.0$\pm$1.0$^*$ & 120$\pm$9 & 1 & W$^{Web}$ \\
7831 & 4.9$\pm$0.4 & SBc & 5.2$^{Ka}$ & -18.5 & 0.39$\pm$0.03 & 67$\pm$2 & 177$\pm$8/4.5$\pm$0.2 & 92$\pm$15 & 2 & W$^{Web}$ \\
7853 & 8.6$\pm$1.1 & SBm & 8.9$^{Mo}$ & -18.9 & 0.64$\pm$0.05 & 50$\pm$3 & 106$\pm$6/4.6$\pm$0.3 & 110$\pm$35 & 3 & W$^{Web}$ \\
7861 & 8.8$\pm$0.7 & SAB(r)m pec & 10.2$^{Mo}$ & -17.3 & 0.75$\pm$0.05 & 41$\pm$4 & 41$\pm$4/2.0$\pm$0.2 & 50$\pm$21 & 3 & W$^{Web}$ \\
7876 & 6.5$\pm$0.9 & SABc & 14.5 & -17.9 & 0.72$\pm$0.05 & 44$\pm$4 & 58$\pm$7/4.1$\pm$0.5 & 98$\pm$14 & 2 &  \\
7901 & 5.2$\pm$0.6 & Sc pec & 20.7$^{Sh}$ & -20.6 & 0.66$\pm$0.03 & 49$\pm$3 & 115$\pm$6/11.6$\pm$0.6 & 215$\pm$10 & 1 &  \\
7985 & 6.9$\pm$0.5 & SBcd & 13.7$^{Mo}$ & -18.7 & 0.92$\pm$0.08 & 23$\pm$11 & 51$\pm$5/3.4$\pm$0.3 & 112$\pm$13 & 1 &  \\
8334 & 4.0$\pm$0.2 & Sbc & 9.8 & -21.1 & 0.61$\pm$0.06 & 53$\pm$4 & 356$\pm$20/16.9$\pm$1.0 & 214$\pm$9 & 1 &  \\
8403 & 5.8$\pm$0.6 & SBc & 19.1$^{Ja}$ & -19.2 & 0.61$\pm$0.04 & 52$\pm$3 & 90$\pm$6/8.3$\pm$0.6 & 128$\pm$10 & 1 & W$^{Web}$ \\
NGC 5296 & -1.1$\pm$0.8 & S0-a & 32.8 & -18.2 & 0.58$\pm$0.04 & 54$\pm$3 & 28$\pm$3/4.5$\pm$0.5 & 80$\pm$9 & 3 & W$^{Web}$ \\
8709 & 4.9$\pm$0.8 & SABc pec & 35.0 & -21.4 & 0.24$\pm$0.02 & 76$\pm$1 & 112$\pm$7/19.0$\pm$1.1 & 207$\pm$9 & 1 & W$^{Web}$ \\
8852 & 2.3$\pm$0.6 & SAB(r)a & 30.6 & -20.0 & 0.62$\pm$0.08 & 52$\pm$6 & 77$\pm$9/11.4$\pm$1.3 & 187$\pm$10 & 3 &  \\
8863 & 1.1$\pm$0.4 & SBa & 25.5$^{Ko}$ & -20.3 & 0.39$\pm$0.03 & 67$\pm$2 & 108$\pm$5/13.4$\pm$0.6 & 193$\pm$13 & 2 & W$^{N05}$ \\
8898 & 3.1$\pm$0.6 & SBb pec & 49.0 & -20.5 & 0.41$\pm$0.03 & 66$\pm$2 & 79$\pm$6/18.7$\pm$1.3 & 65$\pm$45 & 4 & W$^{Web}$ \\
8900 & 3.2$\pm$0.6 & Sb pec & 49.2 & -21.7 & 0.47$\pm$0.06 & 62$\pm$4 & 75$\pm$8/17.8$\pm$1.8 & 346$\pm$37 & 2 & W$^{Web}$ \\
8937 & 3.1$\pm$0.4 & SBb & 49.0$^{Mo}$ & -21.1 & 0.67$\pm$0.06 & 48$\pm$4 & 69$\pm$6/16.4$\pm$1.4 & 320$\pm$105 & 1 &  \\
9013 & 6.0$\pm$0.3 & Sc pec & 7.2$^{Ka}$ & -18.2 & 0.66$\pm$0.04 & 49$\pm$3 & 72$\pm$5/2.5$\pm$0.2 & 62$\pm$45 & 2 & V$^{R94}$ \\
9179 & 6.9$\pm$0.4 & SABc & 5.7$^{Ka}$ & -17.8 & 0.61$\pm$0.03 & 52$\pm$2 & 128$\pm$8/3.5$\pm$0.2 & 111$\pm$36 & 3 &  \\
9219 & 9.7$\pm$1.4 & IB & 10.2$^{Ja}$ & -16.6 & 0.44$\pm$0.03 & 64$\pm$2 & 49$\pm$4/2.4$\pm$0.2 & 45$\pm$8$^{PV}$ & 2 &  \\
9248 & 3.1$\pm$0.5 & Sb & 54.9 & -20.2 & 0.57$\pm$0.03 & 55$\pm$2 & 40$\pm$4/10.6$\pm$1.0 & 166$\pm$11 & 1 &  \\
9358 & 3.3$\pm$0.8 & SABb & 29.1 & -20.8 & 0.52$\pm$0.03 & 59$\pm$2 & 94$\pm$6/13.3$\pm$0.9 & 221$\pm$14 & 1 &  \\
9363 & 6.9$\pm$0.4 & S(r)cd & 22.3 & -19.8 & 0.84$\pm$0.05 & 33$\pm$6 & 57$\pm$5/6.2$\pm$0.5 & 143$\pm$105 & 1 & V$^{S96}$ \\
9406 & 6.9$\pm$0.4 & SB(r)cd & 33.8 & -19.0 & 0.64$\pm$0.08 & 50$\pm$6 & 44$\pm$8/7.3$\pm$1.3 & 19$\pm$10 & 4 &  \\
9465 & 7.9$\pm$0.9 & SABd & 26.4$^{Ja}$ & -18.0 & 0.40$\pm$0.03 & 67$\pm$2 & 23$\pm$3/3.0$\pm$0.4 & 97$\pm$9 & 1 &  \\
9576 & 6.9$\pm$0.4 & SABc pec & 27.4$^{Ja}$ & -19.6 & 0.63$\pm$0.05 & 51$\pm$4 & 51$\pm$4/6.8$\pm$0.6 & 104$\pm$25 & 1 & V$^{I94}$ \\
9736 & 5.0$\pm$0.7 & SABc & 45.4 & -20.6 & 0.65$\pm$0.03 & 49$\pm$2 & 71$\pm$4/15.7$\pm$1.0 & 193$\pm$16 & 1 &  \\
9866 & 4.0$\pm$0.3 & S(r)bc & 7.4$^{Ja}$ & -17.2 & 0.41$\pm$0.03 & 65$\pm$2 & 55$\pm$4/2.0$\pm$0.1 & 116$\pm$11 & 2 &  \\
9943 & 5.0$\pm$0.6 & SB(r)c & 28.0 & -20.7 & 0.69$\pm$0.05 & 46$\pm$4 & 82$\pm$5/11.1$\pm$0.7 & 185$\pm$10 & 1 &  \\
10075 & 6.0$\pm$0.4 & Sc & 14.7$^{Ja}$ & -19.9 & 0.44$\pm$0.04 & 64$\pm$3 & 174$\pm$9/12.4$\pm$0.7 & 168$\pm$9 & 1 &  \\
10521 & 4.9$\pm$0.7 & Sc & 18.0$^{Mo}$ & -20.2 & 0.38$\pm$0.03 & 68$\pm$2 & 106$\pm$9/9.3$\pm$0.7 & 124$\pm$9 & 1 &  \\
10652 & 3.8$\pm$2.6 & S(r)bc & 18.2 & -17.7 & 0.87$\pm$0.05 & 29$\pm$6 & 33$\pm$3/2.9$\pm$0.3 & 141$\pm$82 & 2 &  \\
10713 & 3.0$\pm$0.4 & Sb & 18.3 & -19.0 & 0.19$\pm$0.02 & 79$\pm$1 & 54$\pm$7/4.8$\pm$0.7 & 105$\pm$8$^{PV}$ & 2 & W$^{Web}$ \\
10757 & 6.0$\pm$0.4 & Sc & 19.5 & -17.7 & 0.53$\pm$0.04 & 58$\pm$2 & 36$\pm$4/3.4$\pm$0.4 & 81$\pm$33 & 3 & W$^{Web}$ \\
10769 & 3.0$\pm$0.5 & SABb & 20.0 & -17.0 & 0.59$\pm$0.04 & 54$\pm$3 & 28$\pm$3/2.7$\pm$0.3 &  &  & W$^{Web}$ \\
10791 & 8.8$\pm$0.5 & SABm & 21.7 & -16.7 & 1.00$\pm$0.13 & 0$\pm$0 & 56$\pm$9/5.9$\pm$0.9 & 96$\pm$49 & 3 & W$^{Web}$ \\
11012 & 5.9$\pm$0.7 & Sc & 5.3$^{Ka}$ & -18.7 & 0.33$\pm$0.03 & 71$\pm$2 & 185$\pm$11/4.7$\pm$0.3 & 117$\pm$9 & 1 &  \\
11269 & 2.0$\pm$0.5 & SABa & 35.0$^{Ja}$ & -19.9 & 0.56$\pm$0.04 & 56$\pm$3 & 56$\pm$5/9.5$\pm$0.8 & 202$\pm$13 & 1 & W$^{N05}$ \\
11300 & 6.4$\pm$0.9 & SABc & 8.4$^{Ja}$ & -17.8 & 0.28$\pm$0.01 & 74$\pm$1 & 99$\pm$5/4.0$\pm$0.2 & 114$\pm$9 & 2 & W$^{Web}$ \\
11332 & 7.0$\pm$0.5 & SBcd & 23.0$^{Ja}$ & -19.5 & 0.21$\pm$0.01 & 78$\pm$1 & 63$\pm$5/7.1$\pm$0.5 & 91$\pm$8$^{PV}$ & 3 &  \\
11407 & 3.6$\pm$0.6 & SBbc & 35.8 & -20.8 & 0.49$\pm$0.03 & 61$\pm$2 & 75$\pm$6/13.0$\pm$1.0 & 159$\pm$31 & 1 & V$^{W01}$ \\
11466 & 4.8$\pm$1.9 & Sc & 14.2 & -18.5 & 0.59$\pm$0.03 & 54$\pm$2 & 45$\pm$4/3.1$\pm$0.3 & 133$\pm$10 & 1 & W$^{Web}$ \\
11470 & 2.2$\pm$0.6 & Sab & 50.8 & -21.3 & 0.71$\pm$0.07 & 45$\pm$5 & 72$\pm$9/17.8$\pm$2.2 & 380$\pm$40 & 2 &  \\
11496 & 8.8$\pm$0.5 & Sm & 31.9 &  & 1.00$\pm$0.13 & 0$\pm$0 & 57$\pm$9/8.9$\pm$1.3 & 96$\pm$29 & 2 & W$^{Web}$ \\
11498 & 3.1$\pm$0.7 & SBb & 44.9 & -20.5 & 0.32$\pm$0.04 & 71$\pm$2 & 84$\pm$9/18.2$\pm$1.9 & 273$\pm$9 & 1 &  \\
11597 & 5.9$\pm$0.3 & SABc & 5.9$^{Ka}$ & -20.6 & 0.96$\pm$0.09 & 16$\pm$18 & 342$\pm$14/9.8$\pm$0.4$^*$ & 154$\pm$32 & 3 &  \\
11670 & 0.5$\pm$1.0 & S(r)a & 12.8 & -19.4 & 0.33$\pm$0.03 & 71$\pm$2 & 125$\pm$7/7.7$\pm$0.4 & 190$\pm$9 & 1 & W$^{N05}$ \\
11872 & 2.5$\pm$0.5 & SAB(r)b & 18.1$^{Ko}$ & -20.0 & 0.63$\pm$0.07 & 51$\pm$5 & 85$\pm$6/7.4$\pm$0.5 & 183$\pm$12 & 1 &  \\
12082 & 8.7$\pm$0.8 & SABm & 10.1$^{Ja}$ & -16.4 & 0.90$\pm$0.07 & 26$\pm$10 & 81$\pm$9/3.9$\pm$0.4 & 105$\pm$137 & 3 & W$^{Web}$ \\
\hline
\end{tabular}
\\(1) Name of the galaxy in the UGC catalog (see table \ref{tablelog}). (2) Morphological type from the de Vaucouleurs classification \citep{de-Vaucouleurs:1979} in HyperLeda data base. (3) Morphological type from HyperLeda data base. (4) Distance D, deduced from the systemic velocity taken in NED corrected from Virgo infall, assuming H$_{o}$ = 75 \kmsMpc, except for those marked ($Ja$: \citealp{James:2004}; $Ka$: \citealp{Karachentsev:2004}; $Ko$: \citealp{Koopmann:2006}; $Mo$: \citealp{Moustakas:2006}; $Oc$: \citealp{OConnell:1994}; $Sh$: \citealp{Shapley:2001}). (5) Absolute B magnitude from D and apparent corrected B magnitude (HyperLeda). (6) Axis ratio from HyperLeda. (7) Inclination derived from the axis ratio ($\arccos{b/a}$). (8) Isophotal radius at the limiting surface brightness of 25 B mag/sq arcsec, from HyperLeda \citep{Paturel:1991} in arcsecond and kpc adopting the distance given in column 4; an asterisk ($^*$) indicates that the galaxy is larger than GHASP field of view. (9) Maximum velocity, V$_{max}$, derived from the fit of the velocity field discussed in section \ref{vfanalysis}, or from the \PVM~(marked with $^{PV}$). (10) Quality flag on V$_{max}$ (1: reached; 2: probably reached; 3 probably not reached; 4: not reached). (11) Aperture synthesis HI data references: W for WHISP data ($S02$: \citealp{Swaters:2002}; $N05$: \citealp{Noordermeer:2005}; $web$: \url{http://www.astro.rug.nl/~whisp}); V for VLA data ($I94$: \citealp{Irwin:1994}; $R94$: \citealp{Rownd:1994}; $S96$: \citealp{Schulman:1996}; $K00$: \citealp{Kornreich:2000}; $W01$: \citealp{Wilcots:2001}; $W04$: \citealp{Wilcots:2004}).
\end{table*}

\clearpage
\section{Individual maps and position-velocity diagrams}
\label{maps}

\setcounter{figure}{18}
\begin{figure*}
\begin{minipage}{180mm}
\begin{center}
   \includegraphics[width=17cm]{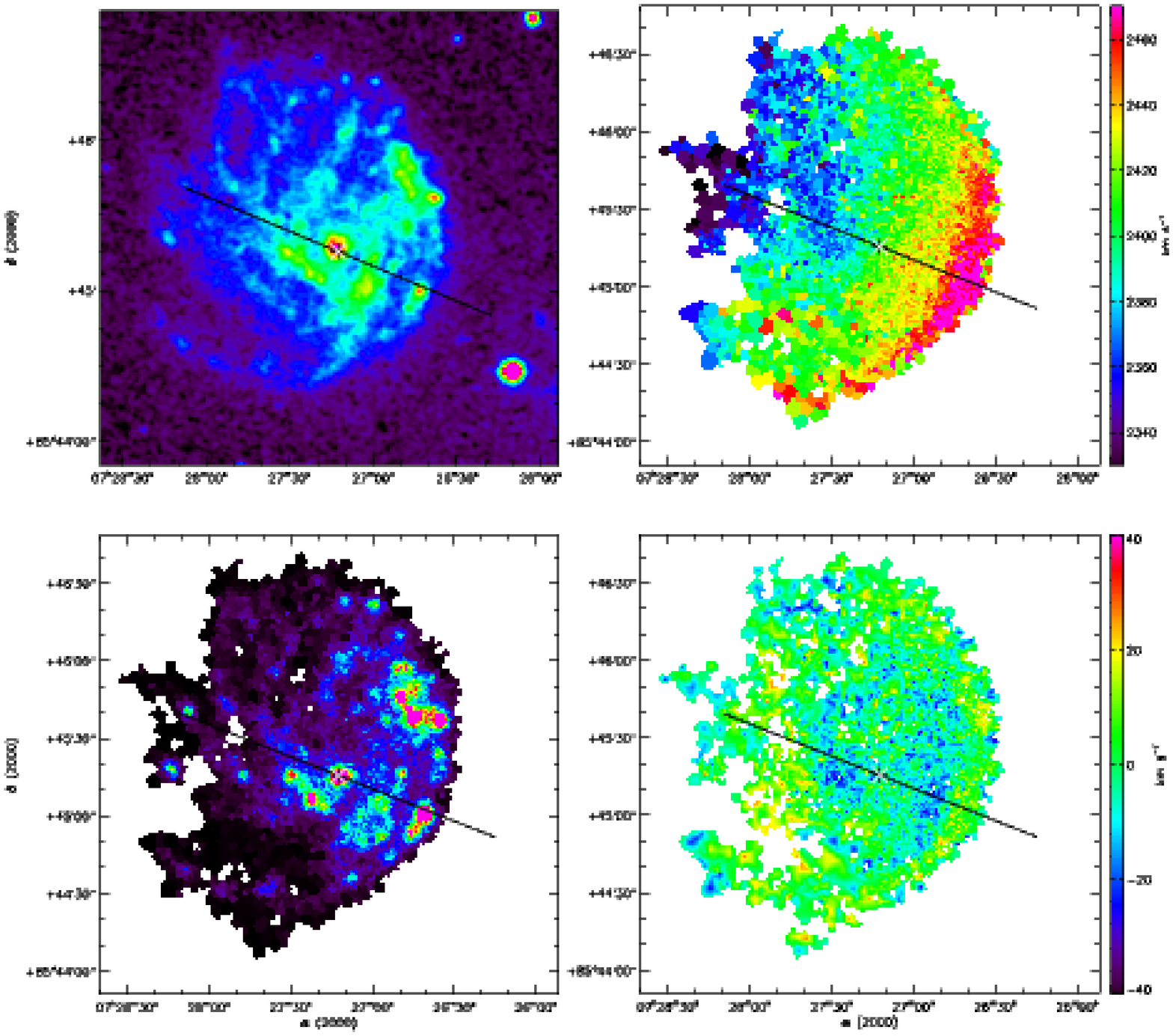}
   \includegraphics[width=19cm]{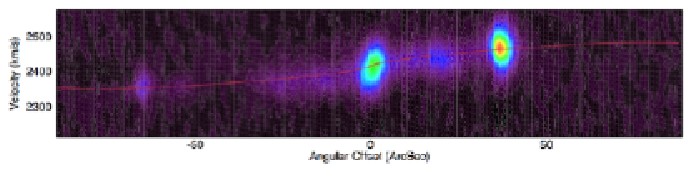}
\end{center}
\caption{UGC 3740. \textbf{Top left}: XDSS Blue Band image.
\textbf{Top right}: \ha~\VF. \textbf{Middle
left}: \ha~monochromatic image.
\textbf{Middle right}: \ha~residual \VF.
The white \& black cross is the kinematical center.
The black line is the major axis, its length represents the $D_{25}$.
\textbf{Bottom}: Position-velocity diagram along the major axis (full width of 7 pixels), arbitrary flux units.
The red line plots the \RC~computed from the model \VF~along the major axis (full width of 7 pixels).
} \label{ugc3740}
\end{minipage}
\end{figure*}
\clearpage
\setcounter{figure}{30}
\begin{figure*}
\begin{minipage}{180mm}
\begin{center}
   \includegraphics[width=17cm]{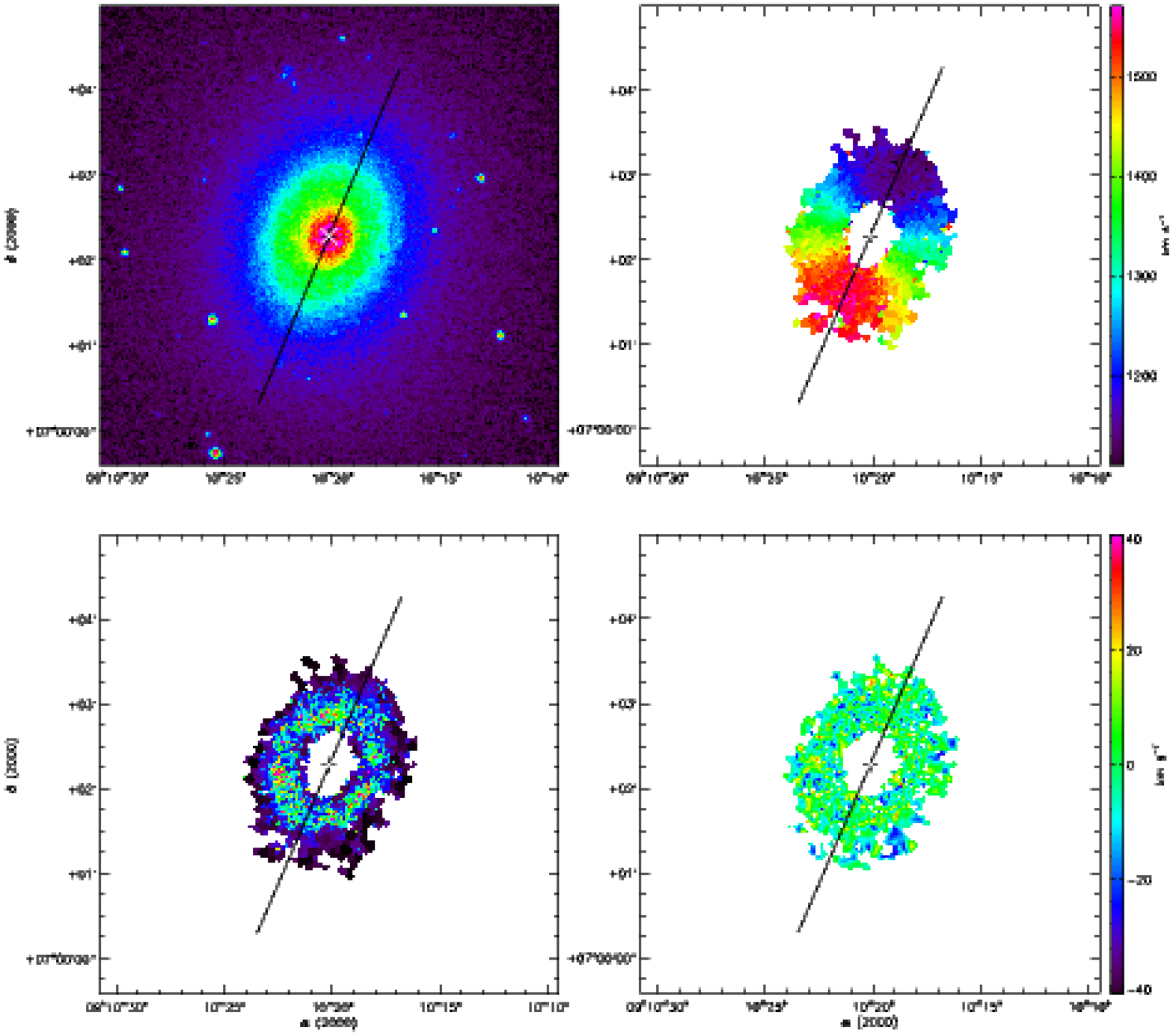}
   \includegraphics[width=19cm]{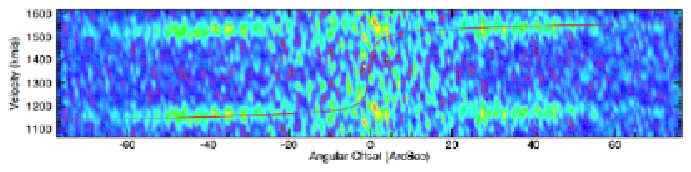}
\end{center}
\caption{UGC 4820. \textbf{Top left}: XDSS Blue Band image.
\textbf{Top right}: \ha~\VF. \textbf{Middle
left}: \ha~monochromatic image.
\textbf{Middle right}: \ha~residual \VF.
The white \& black cross is the kinematical center.
The black line is the major axis, its length represents the $D_{25}$.
\textbf{Bottom}: Position-velocity diagram along the major axis (full width of 7 pixels), arbitrary flux units.
The red line plots the \RC~computed from the model \VF~along the major axis (full width of 7 pixels).
} \label{ugc4820}
\end{minipage}
\end{figure*}
\clearpage
\setcounter{figure}{44}
\begin{figure*}
\begin{minipage}{180mm}
\begin{center}
   \includegraphics[width=17cm]{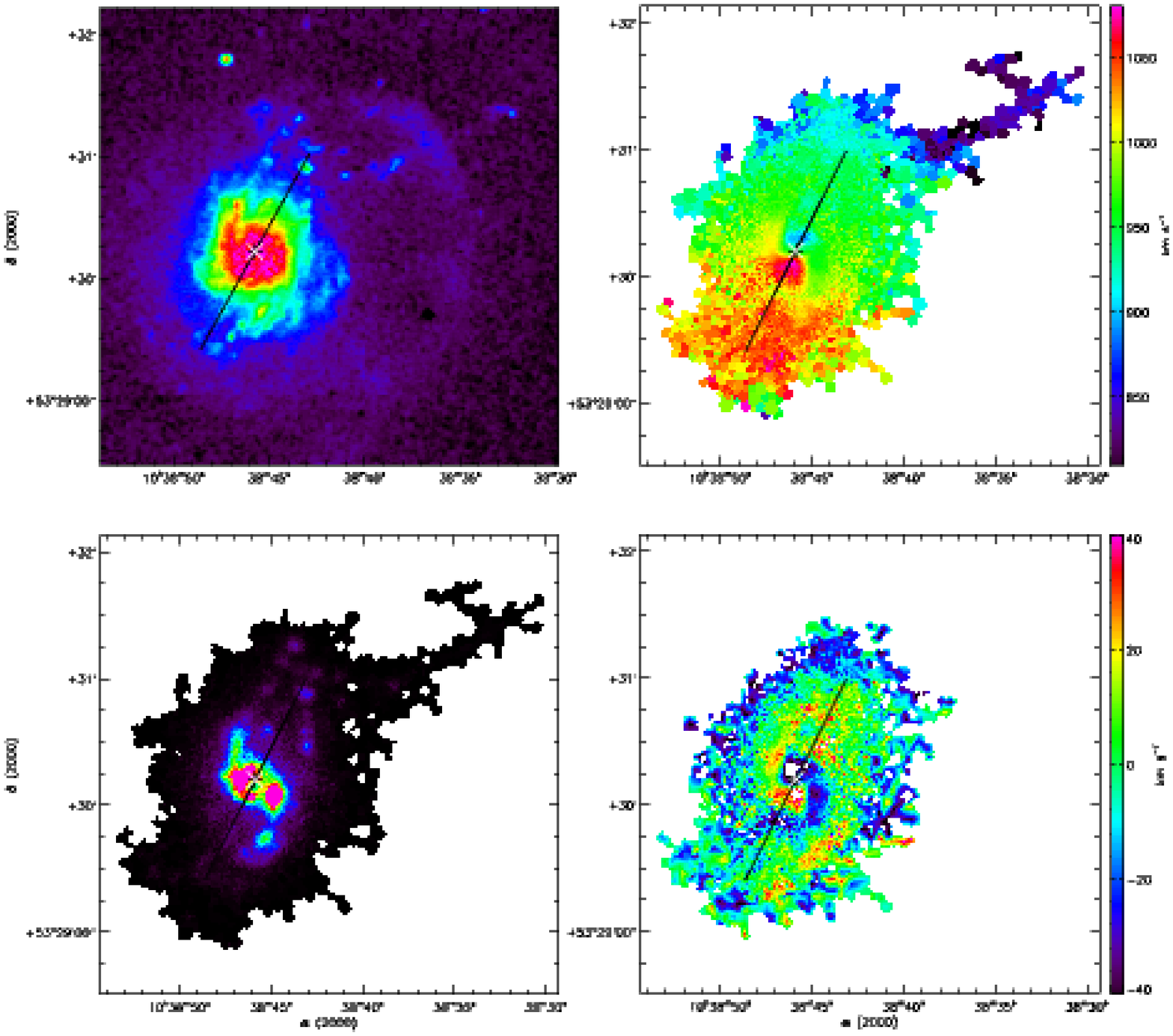}
   \includegraphics[width=19cm]{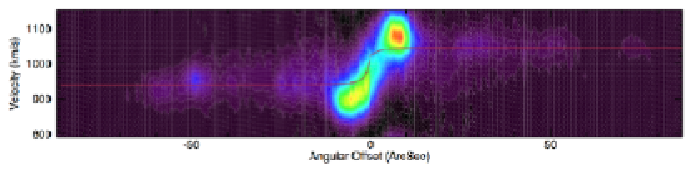}
\end{center}
\caption{UGC 5786. \textbf{Top left}: XDSS Blue Band image.
\textbf{Top right}: \ha~\VF. \textbf{Middle
left}: \ha~monochromatic image.
\textbf{Middle right}: \ha~residual \VF.
The white \& black cross is the kinematical center.
The black line is the major axis, its length represents the $D_{25}$.
\textbf{Bottom}: Position-velocity diagram along the major axis (full width of 7 pixels), arbitrary flux units.
The red line plots the \RC~computed from the model \VF~along the major axis (full width of 7 pixels).
} \label{ugc5786}
\end{minipage}
\end{figure*}
\clearpage
\setcounter{figure}{55}
\begin{figure*}
\begin{minipage}{180mm}
\begin{center}
   \includegraphics[width=17cm]{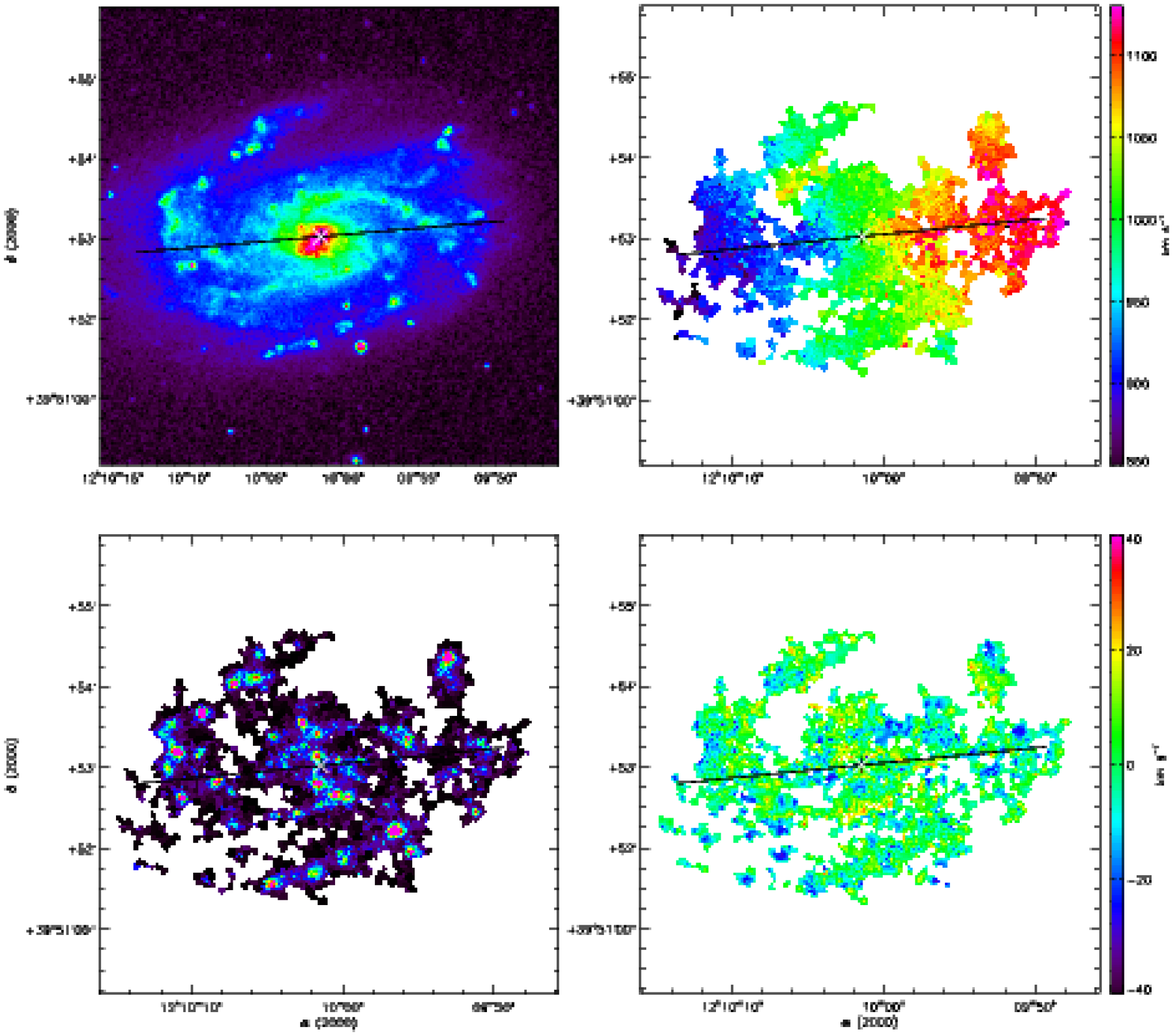}
   \includegraphics[width=19cm]{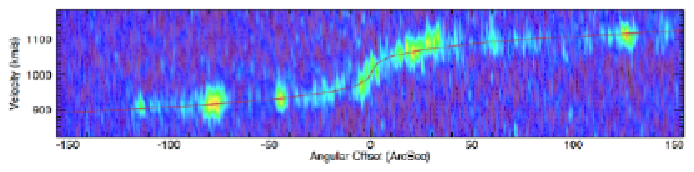}
\end{center}
\caption{UGC 7154. \textbf{Top left}: XDSS Blue Band image.
\textbf{Top right}: \ha~\VF. \textbf{Middle
left}: \ha~monochromatic image.
\textbf{Middle right}: \ha~residual \VF.
The white \& black cross is the kinematical center.
The black line is the major axis, its length represents the $D_{25}$.
\textbf{Bottom}: Position-velocity diagram along the major axis (full width of 7 pixels), arbitrary flux units.
The red line plots the \RC~computed from the model \VF~along the major axis (full width of 7 pixels).
} \label{ugc7154}
\end{minipage}
\end{figure*}
\clearpage

\section{Rotation curves}
\label{rc}
\begin{figure*}
\begin{minipage}{180mm}
\begin{center}
   \includegraphics[width=8cm]{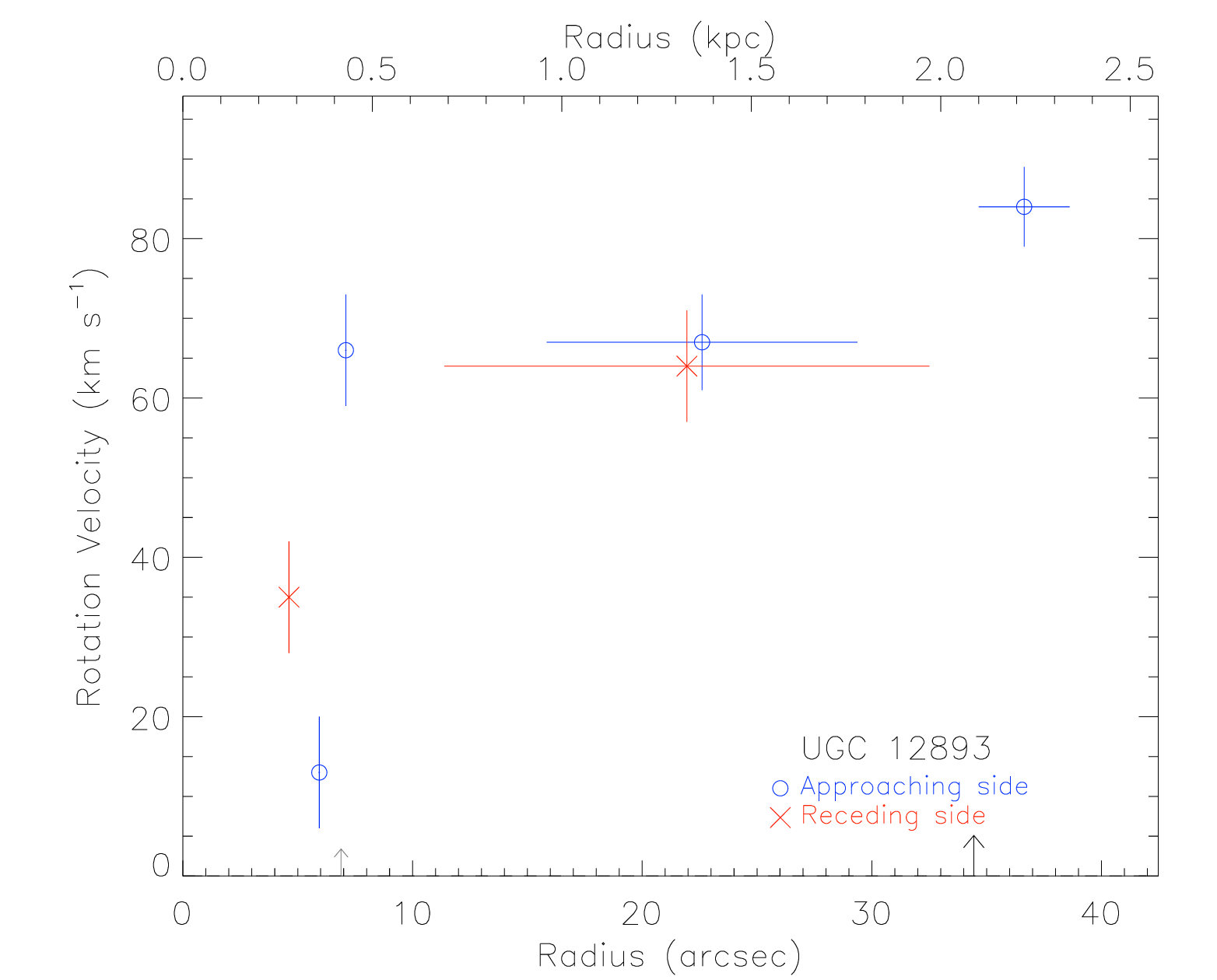}
   \includegraphics[width=8cm]{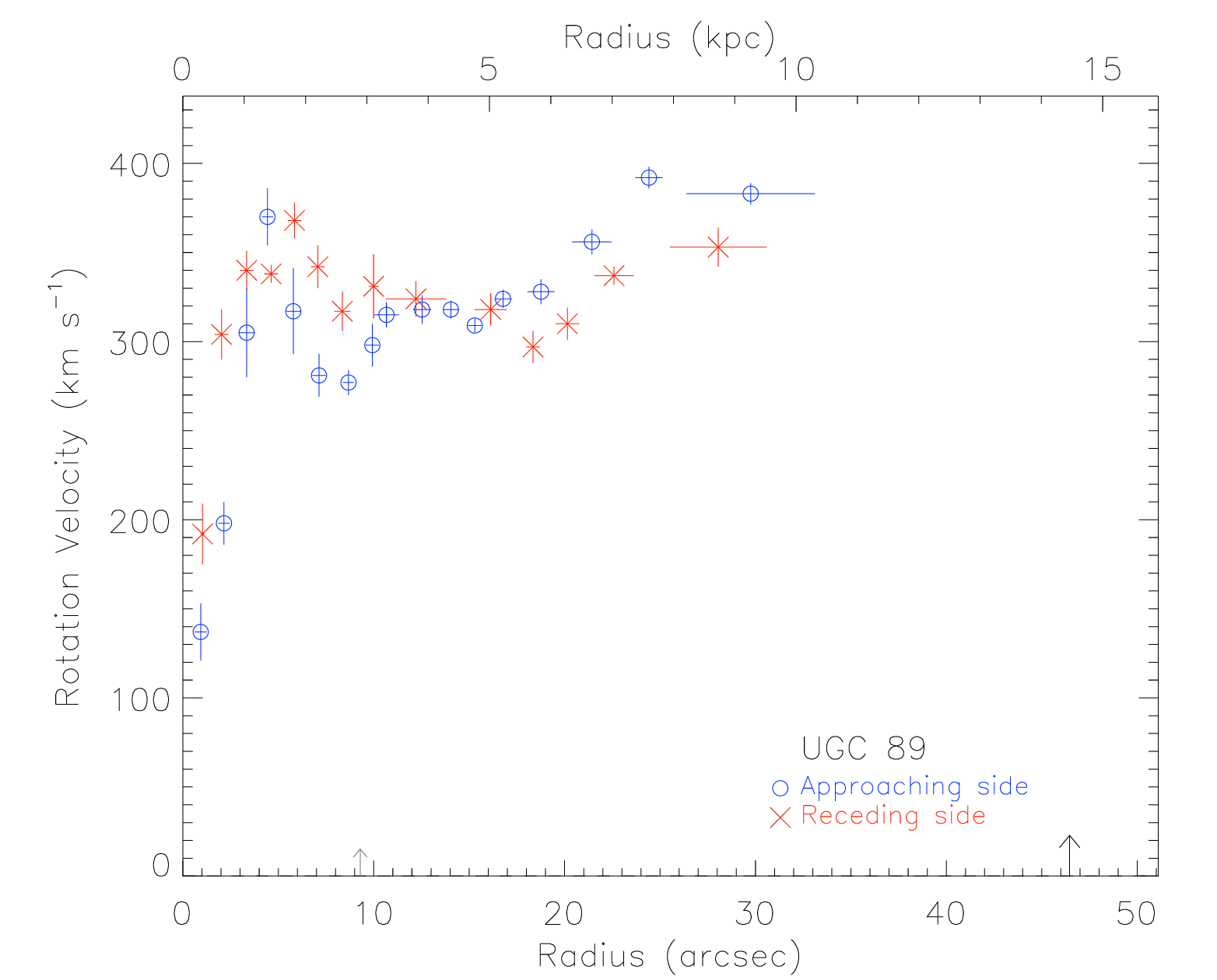}
   \includegraphics[width=8cm]{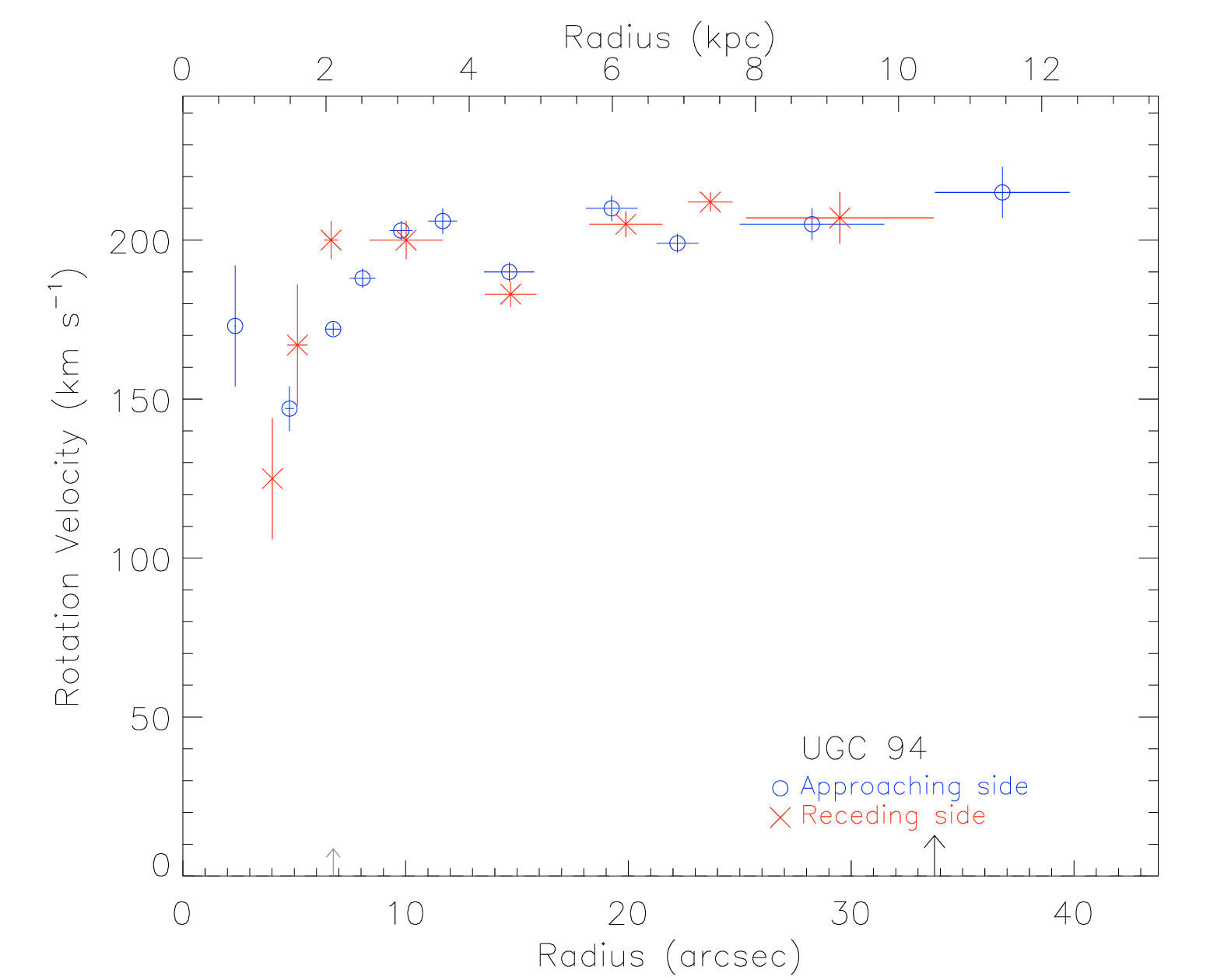}
   \includegraphics[width=8cm]{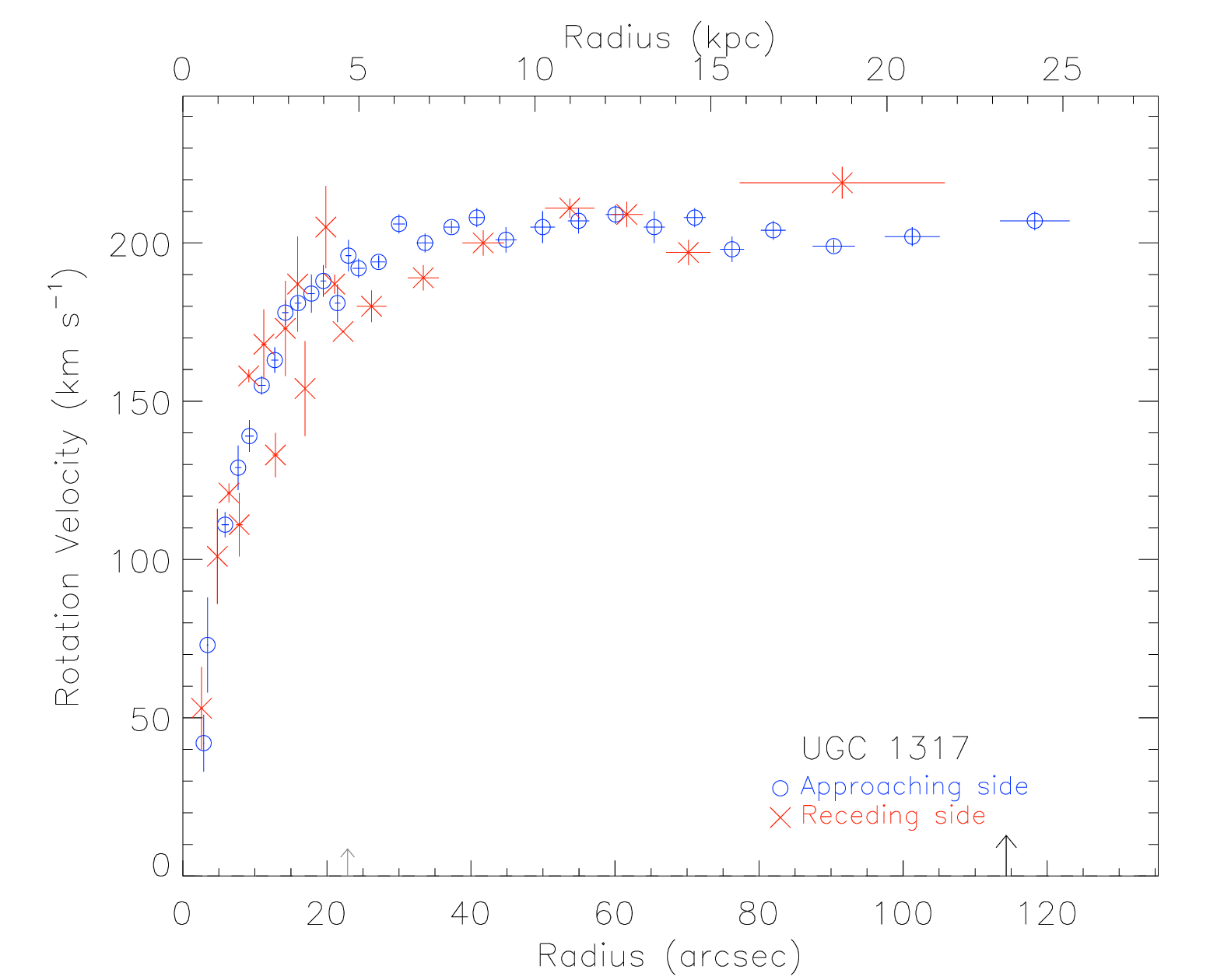}
   \includegraphics[width=8cm]{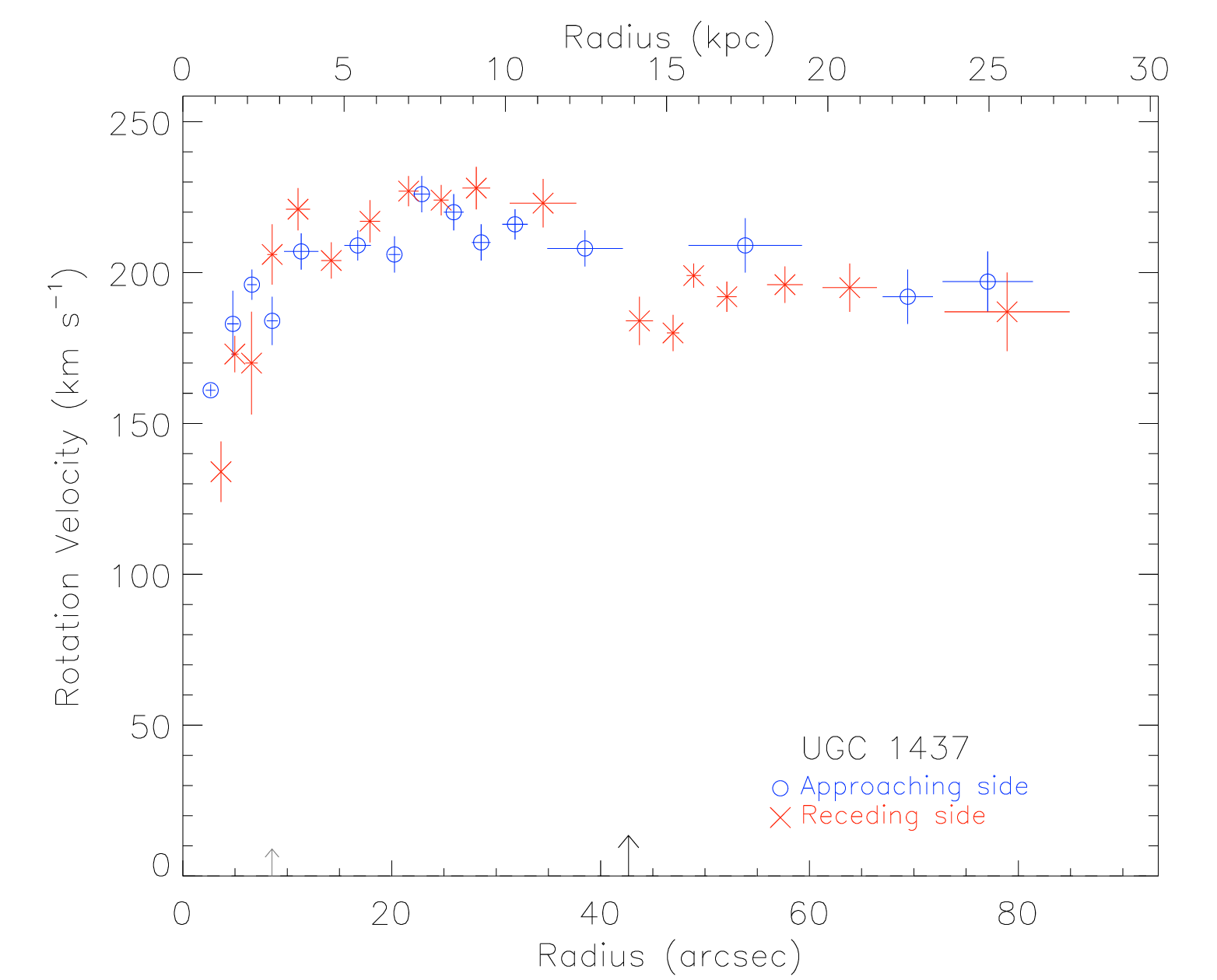}
   \includegraphics[width=8cm]{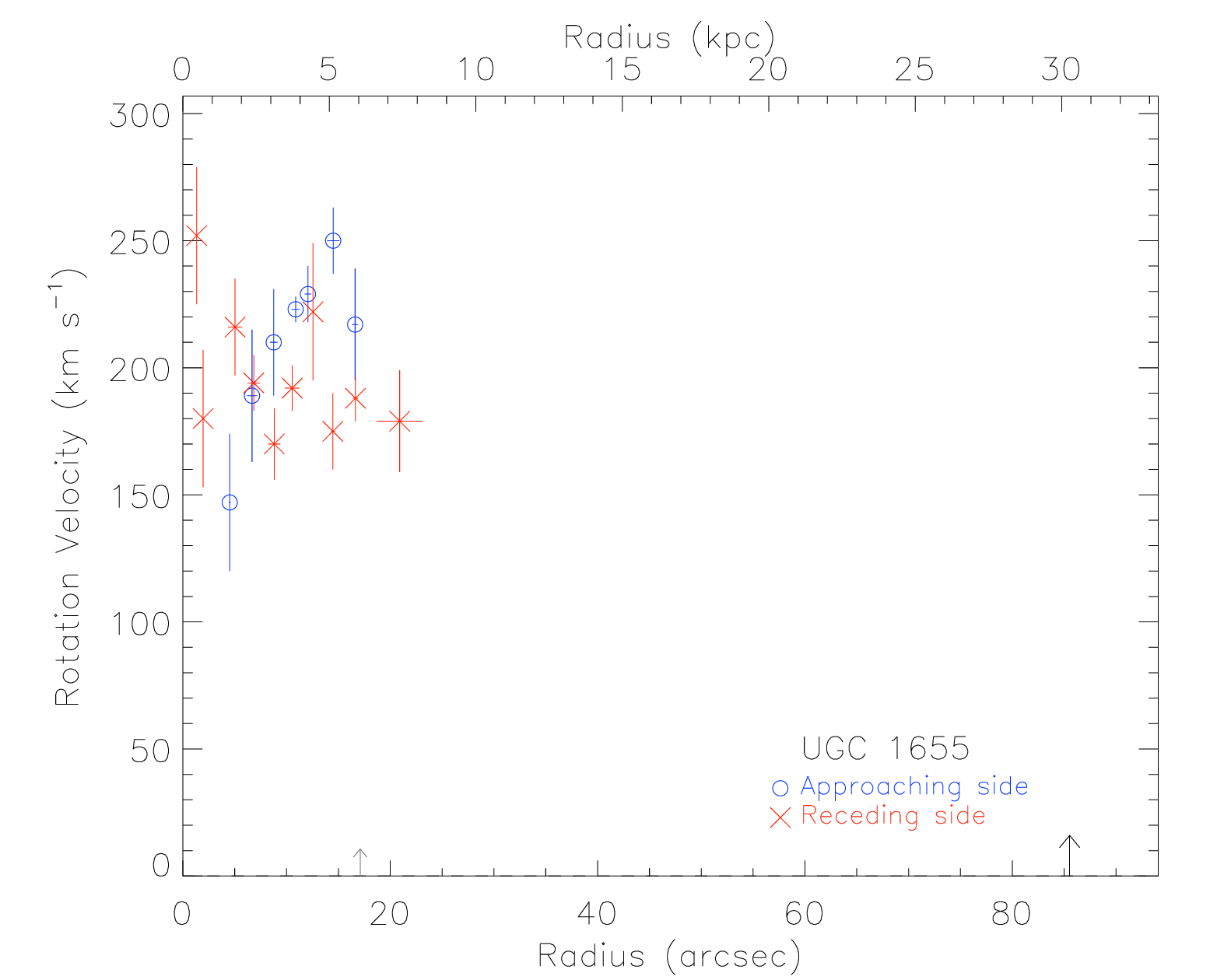}
\end{center}
\caption{From top left to bottom right: \ha~\RC~of UGC 12893, UGC 89, UGC 94, UGC 1317, UGC 1437, and UGC 1655.
}
\end{minipage}
\end{figure*}
\clearpage
\begin{figure*}
\begin{minipage}{180mm}
\begin{center}
   \includegraphics[width=8cm]{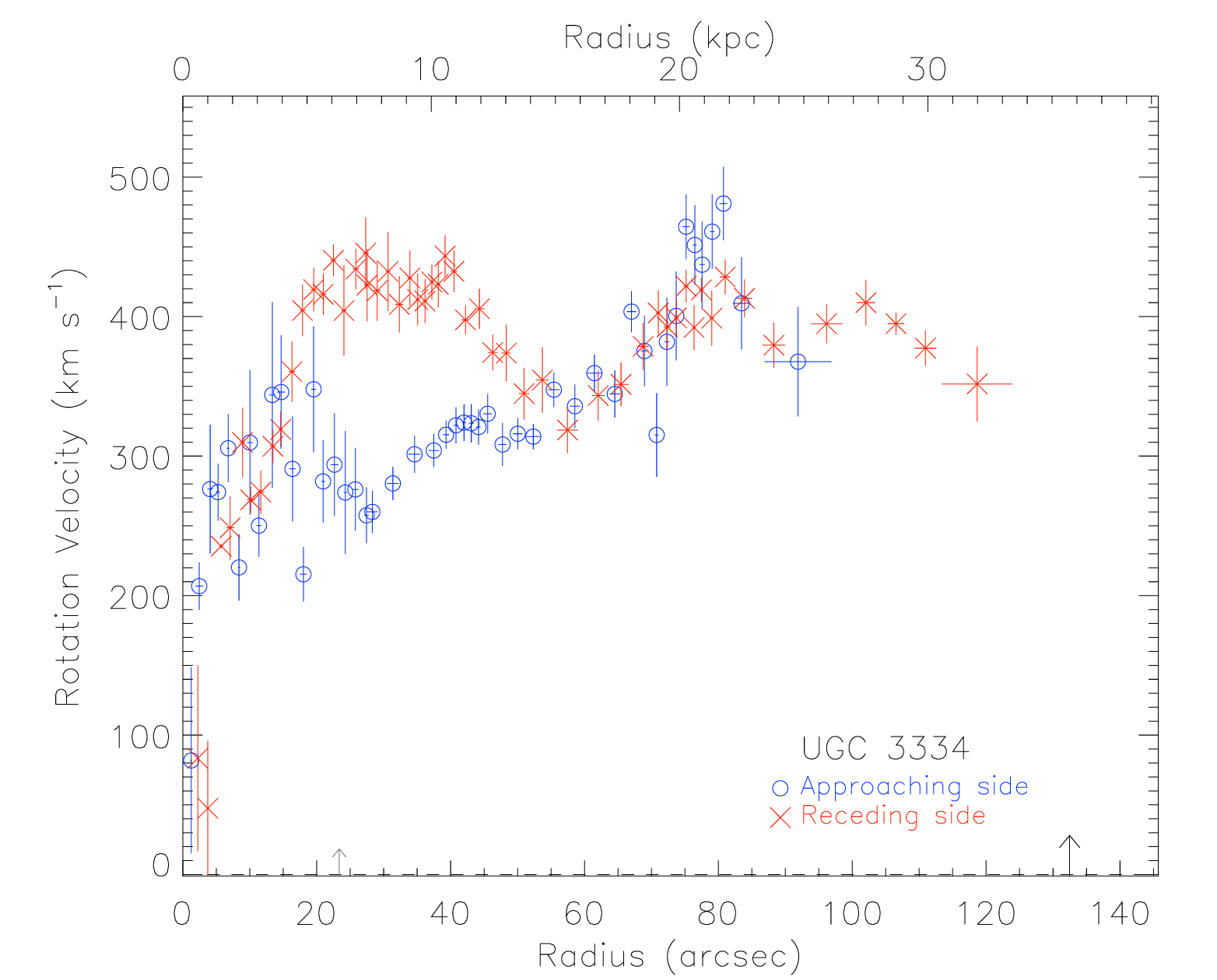}
   \includegraphics[width=8cm]{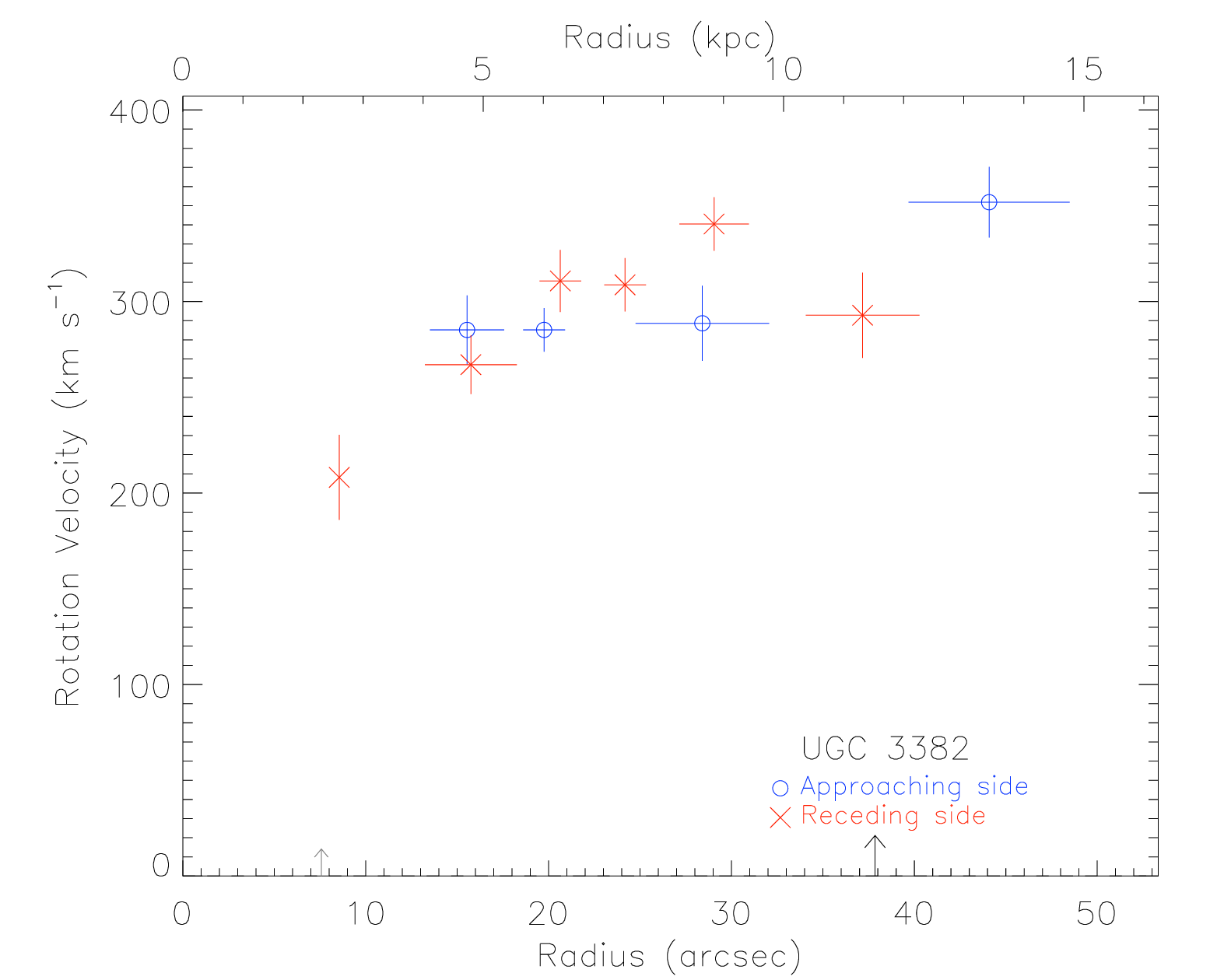}
   \includegraphics[width=8cm]{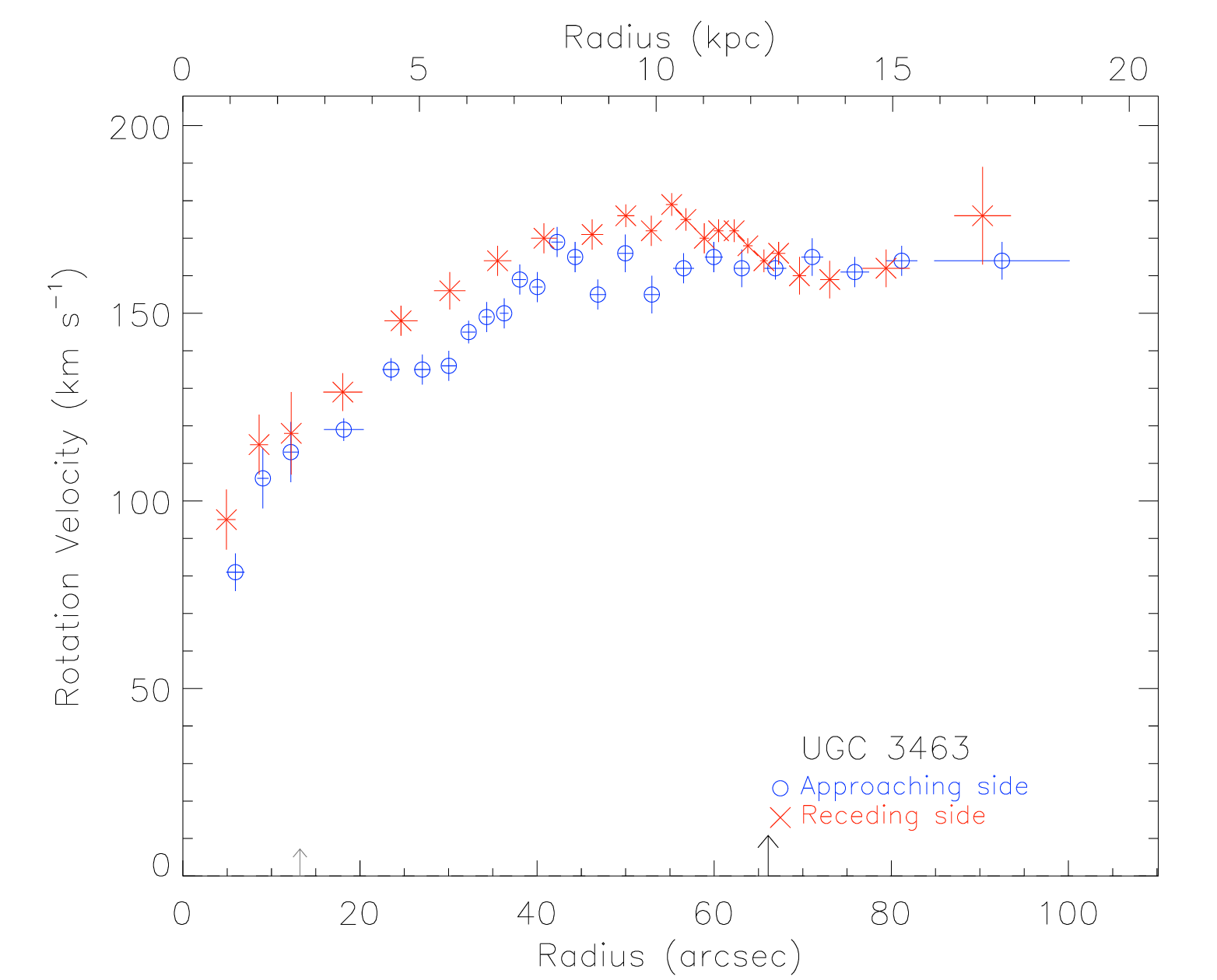}
   \includegraphics[width=8cm]{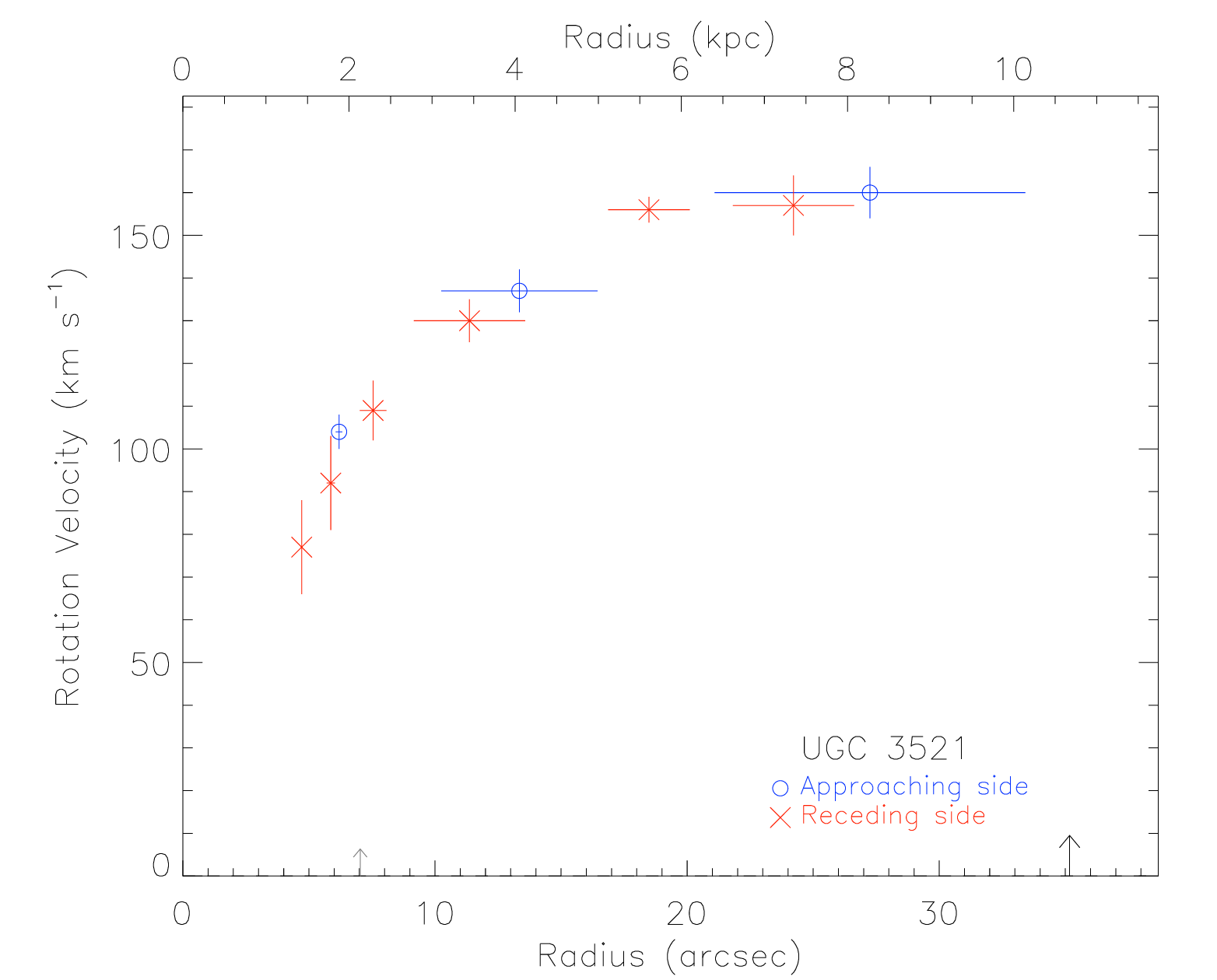}
   \includegraphics[width=8cm]{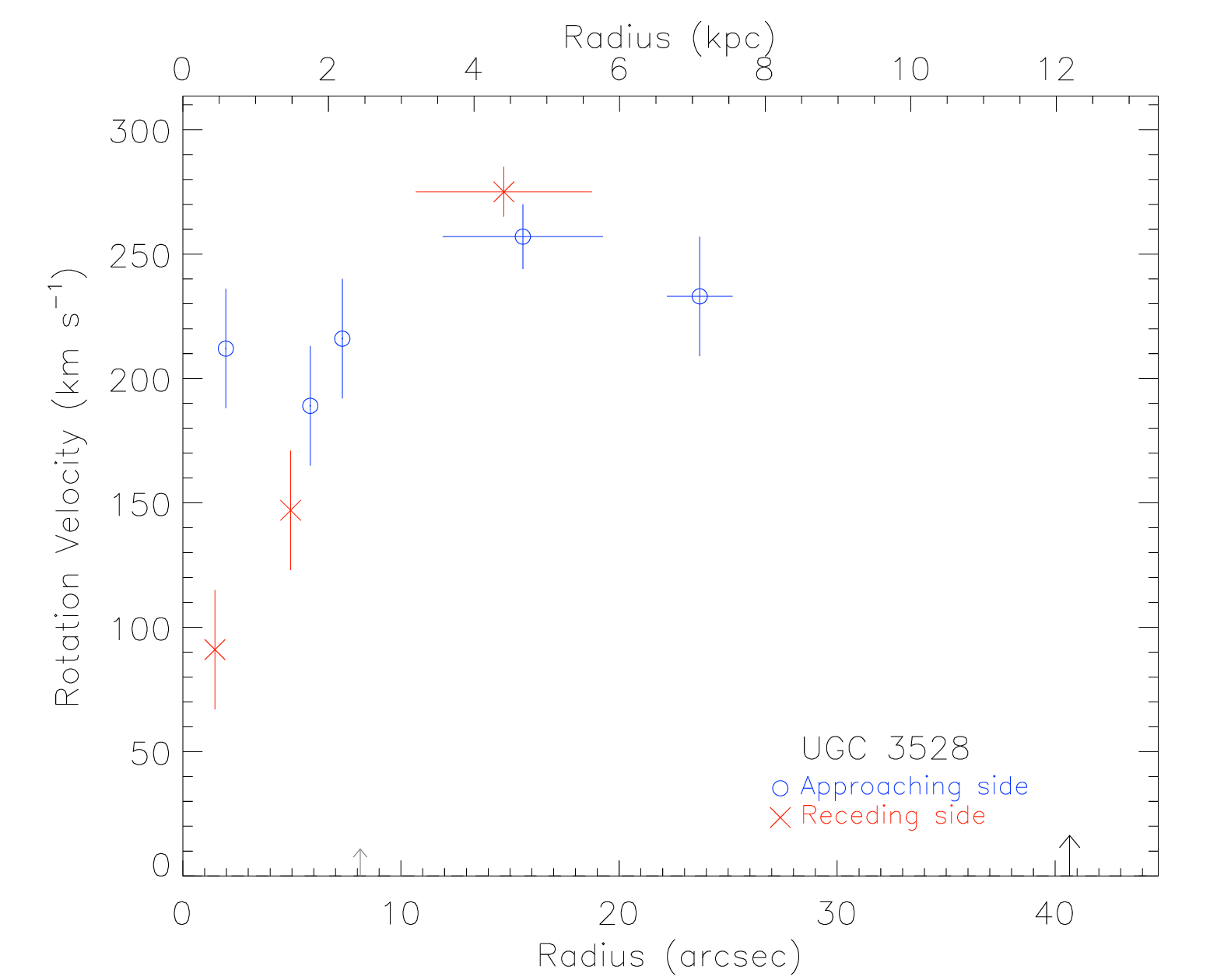}
   \includegraphics[width=8cm]{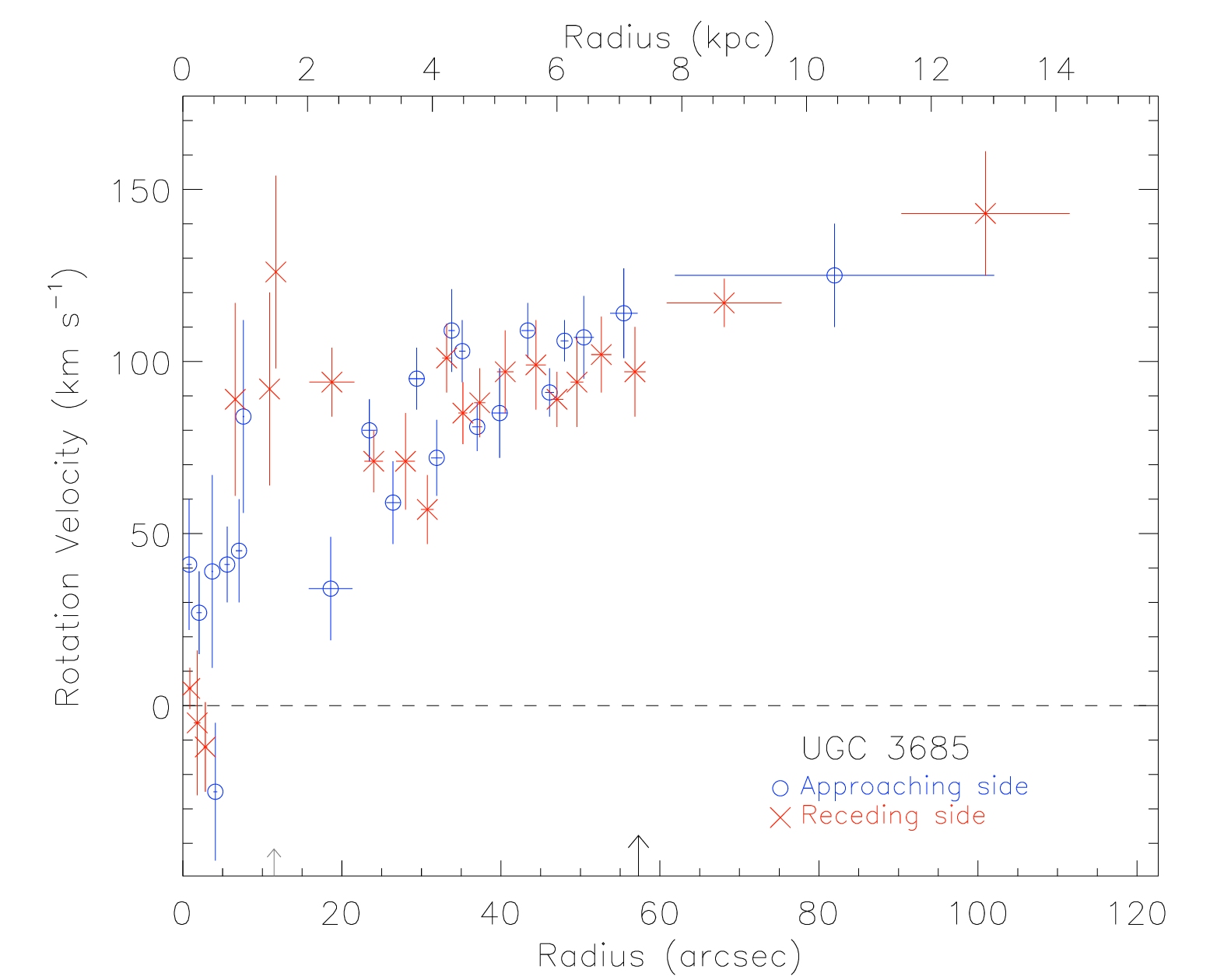}
\end{center}
\caption{From top left to bottom right: \ha~\RC~of UGC 3334, UGC 3382, UGC 3463, UGC 3521, UGC 3528, and UGC 3685.
}
\end{minipage}
\end{figure*}
\clearpage
\begin{figure*}
\begin{minipage}{180mm}
\begin{center}
   \includegraphics[width=8cm]{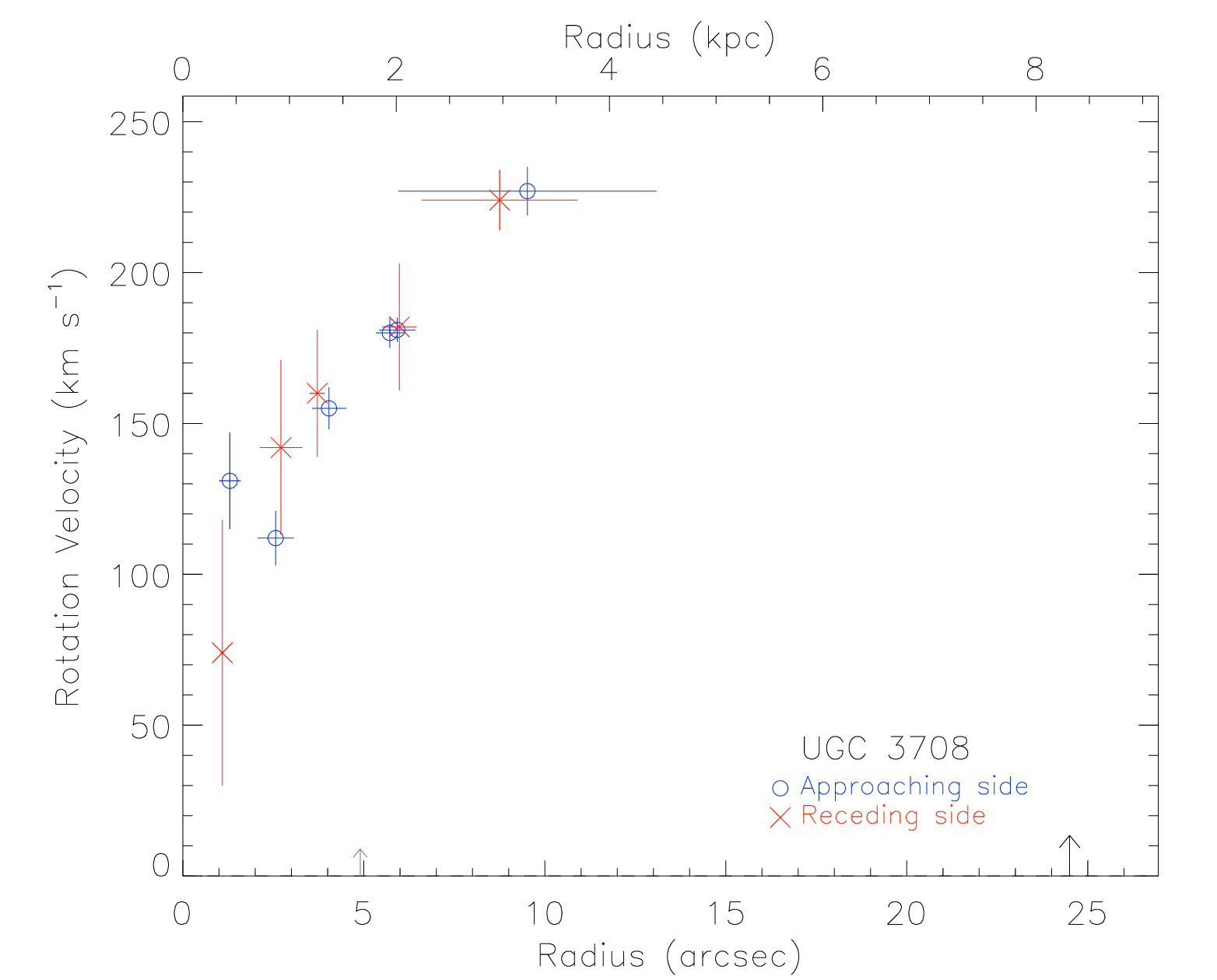}
   \includegraphics[width=8cm]{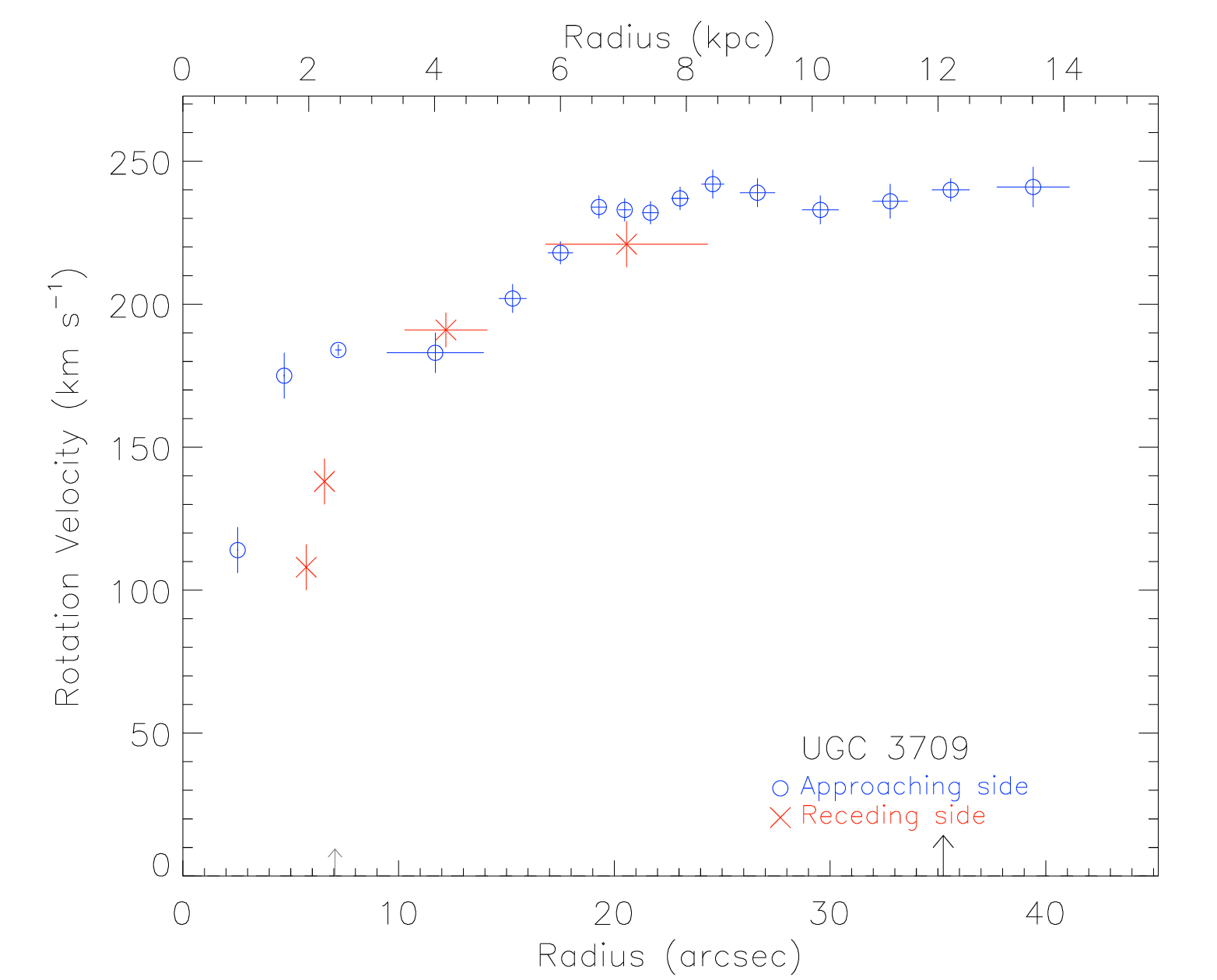}
   \includegraphics[width=8cm]{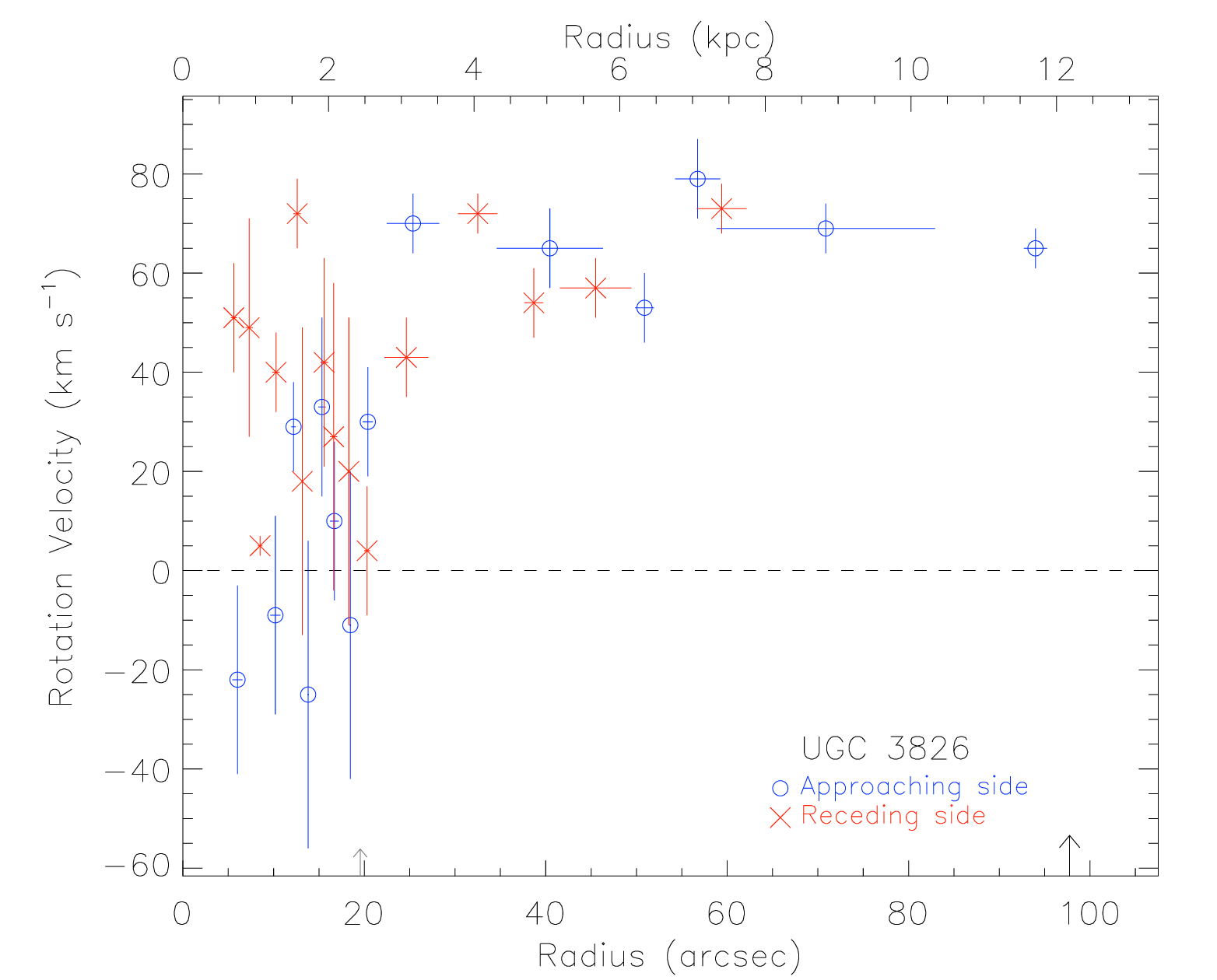}
   \includegraphics[width=8cm]{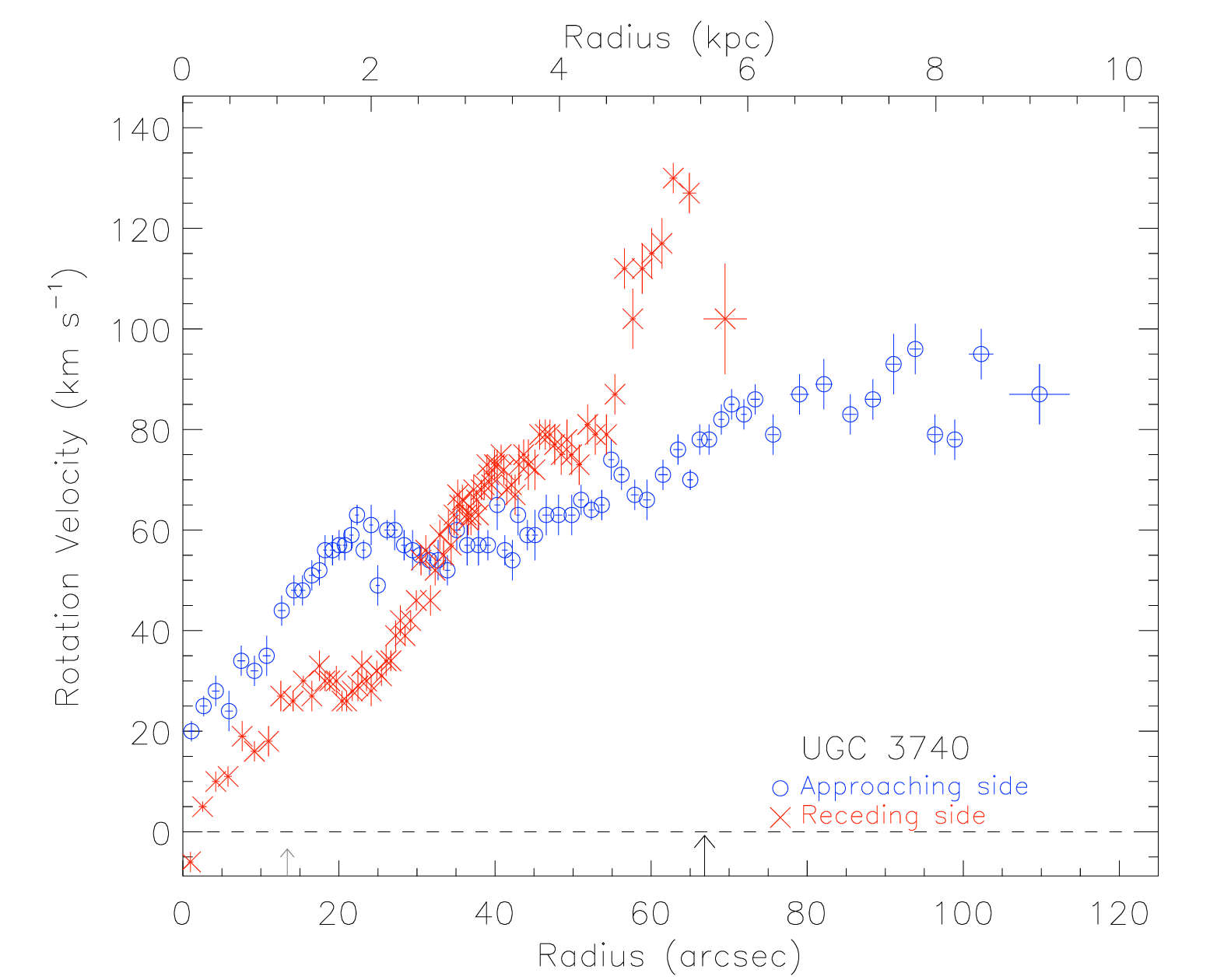}
   \includegraphics[width=8cm]{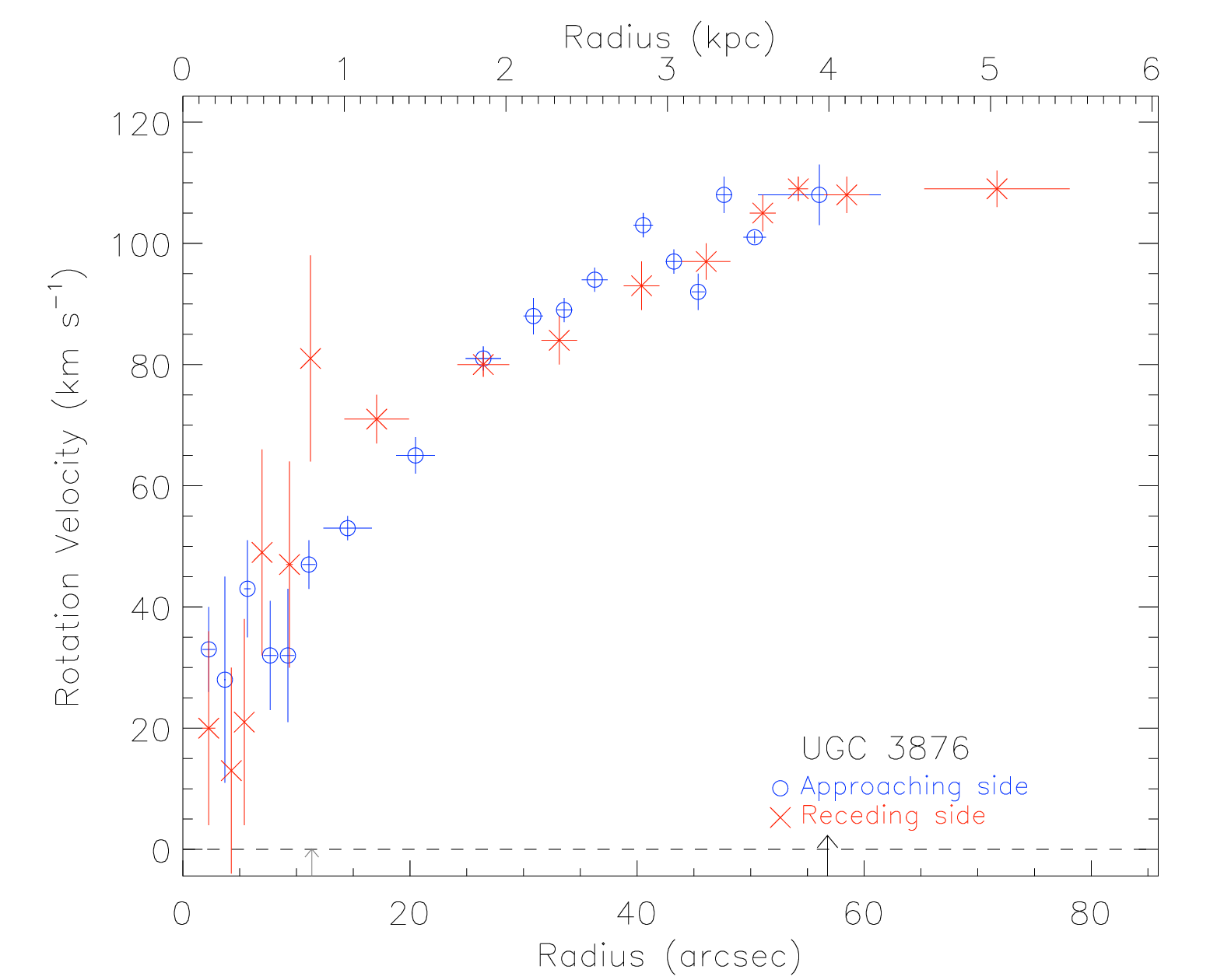}
   \includegraphics[width=8cm]{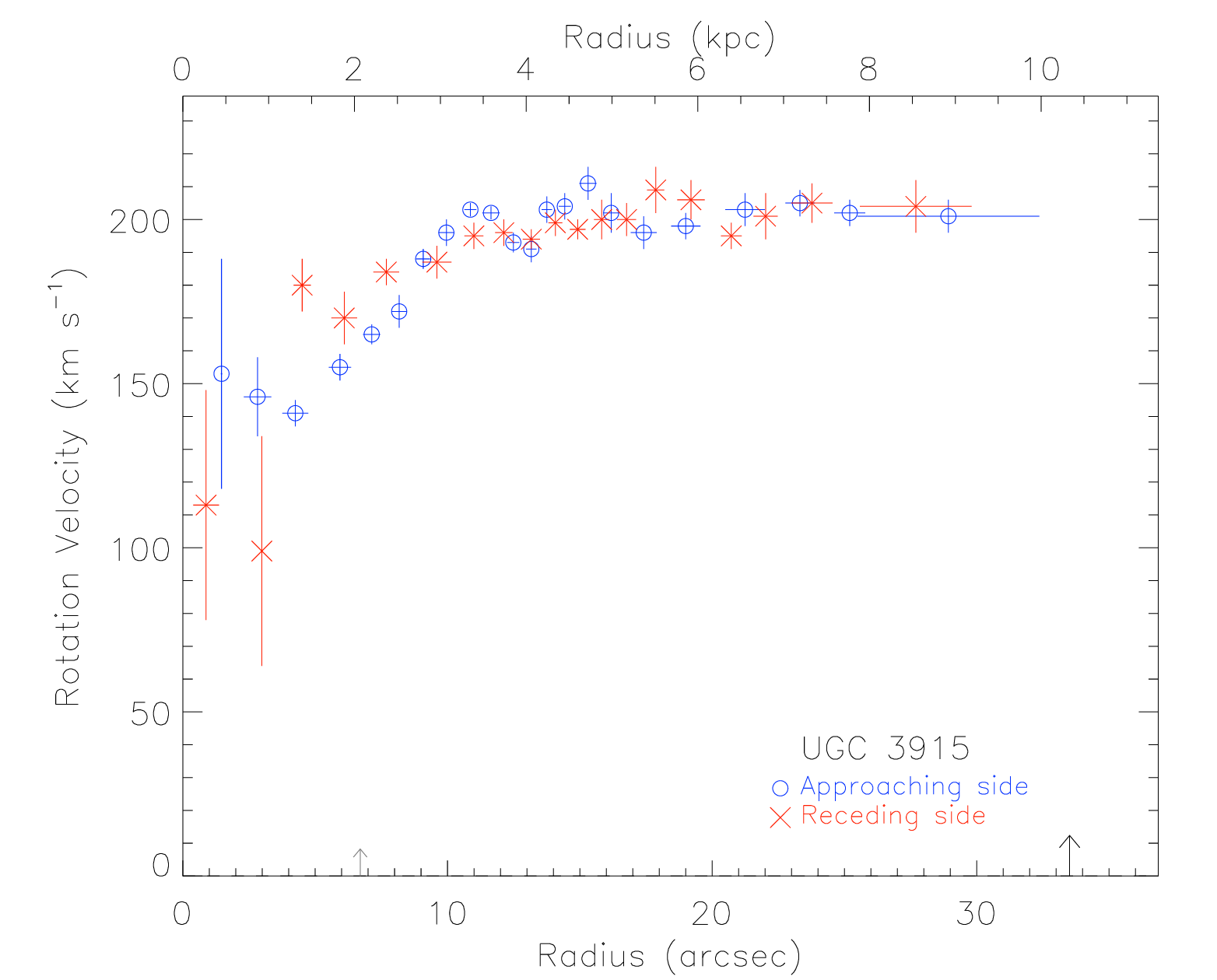}
\end{center}
\caption{From top left to bottom right: \ha~\RC~of UGC 3708, UGC 3709, UGC 3826, UGC 3740, UGC 3876, and UGC 3915.
}
\end{minipage}
\end{figure*}
\clearpage
\begin{figure*}
\begin{minipage}{180mm}
\begin{center}
   \includegraphics[width=8cm]{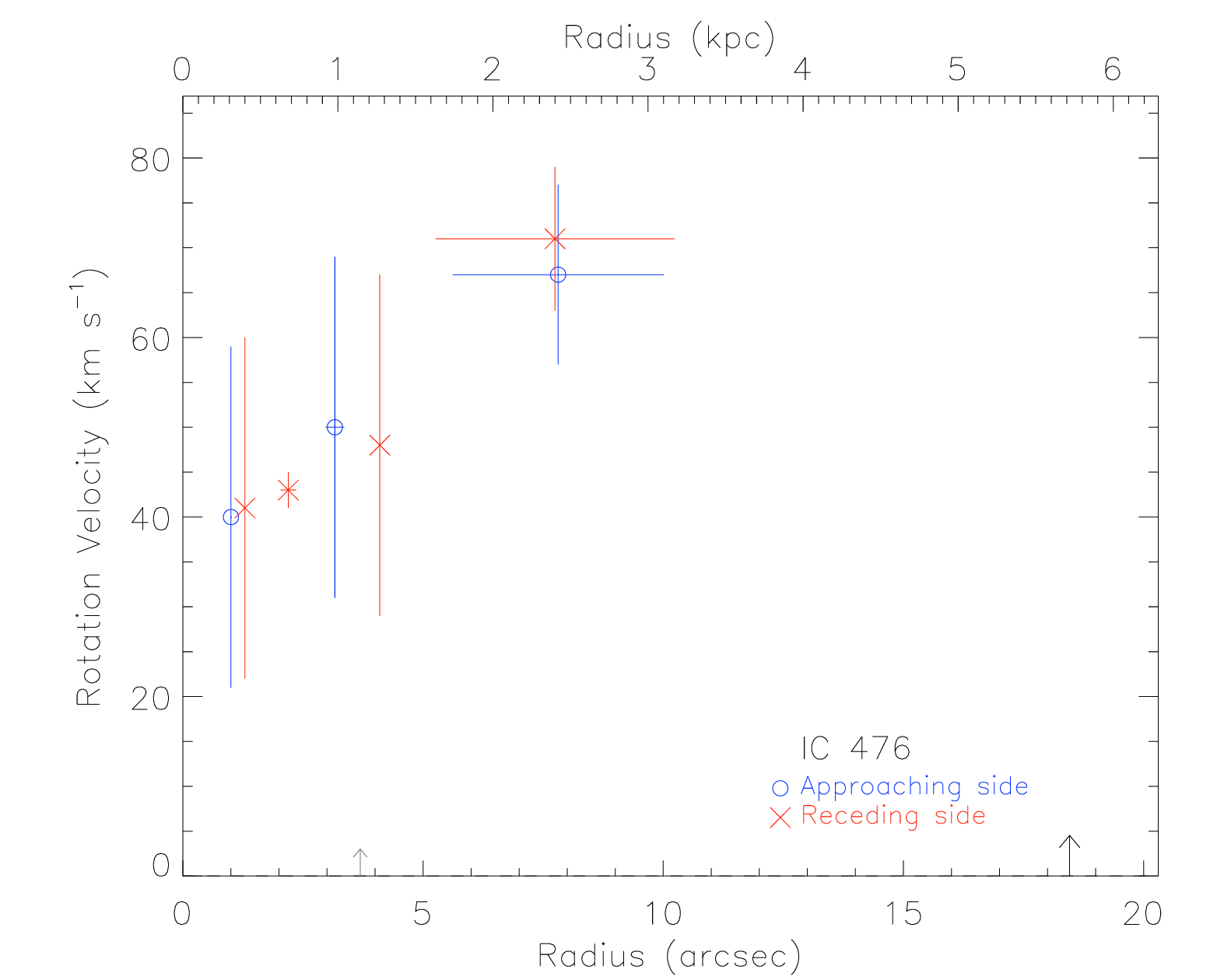}
   \includegraphics[width=8cm]{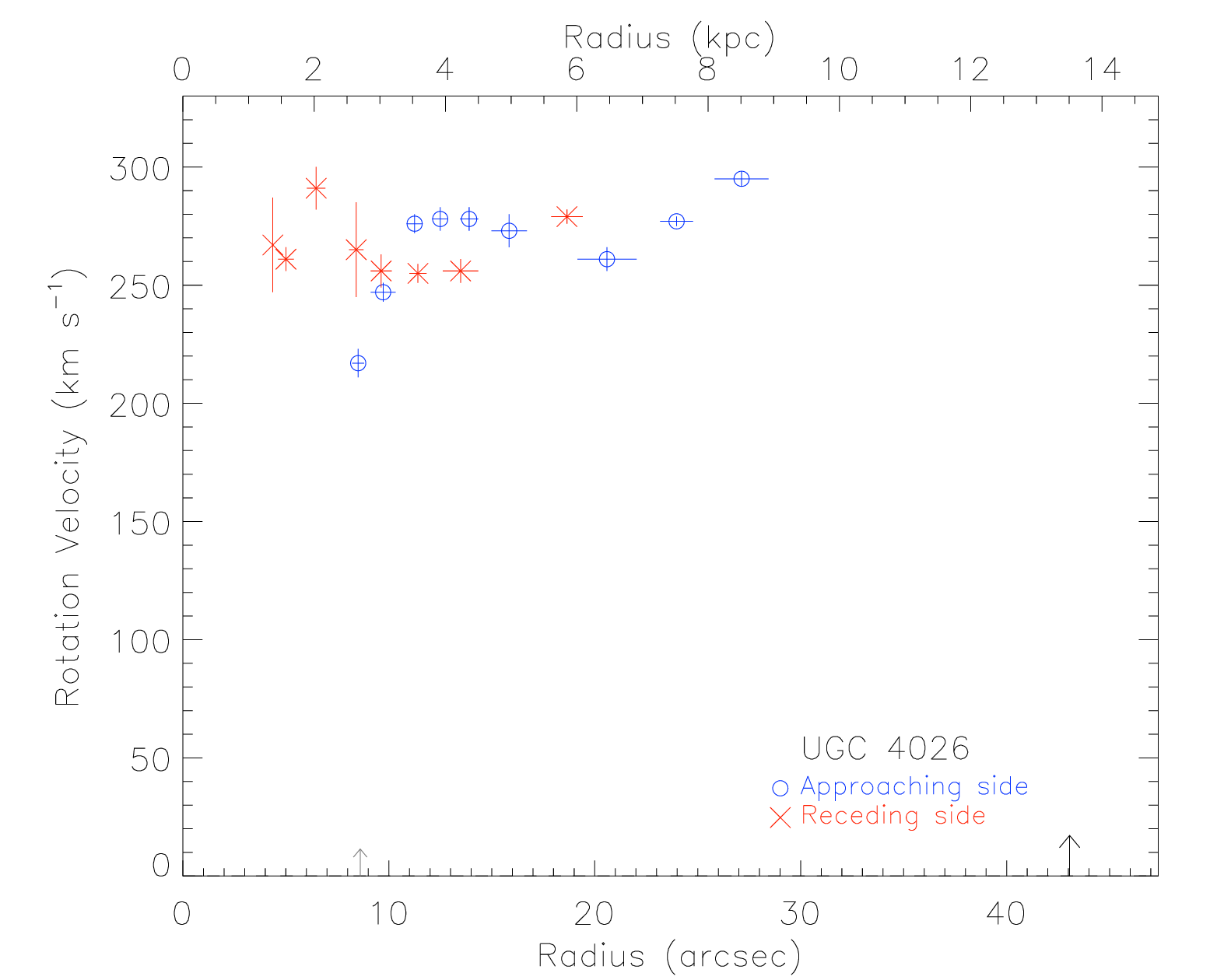}
   \includegraphics[width=8cm]{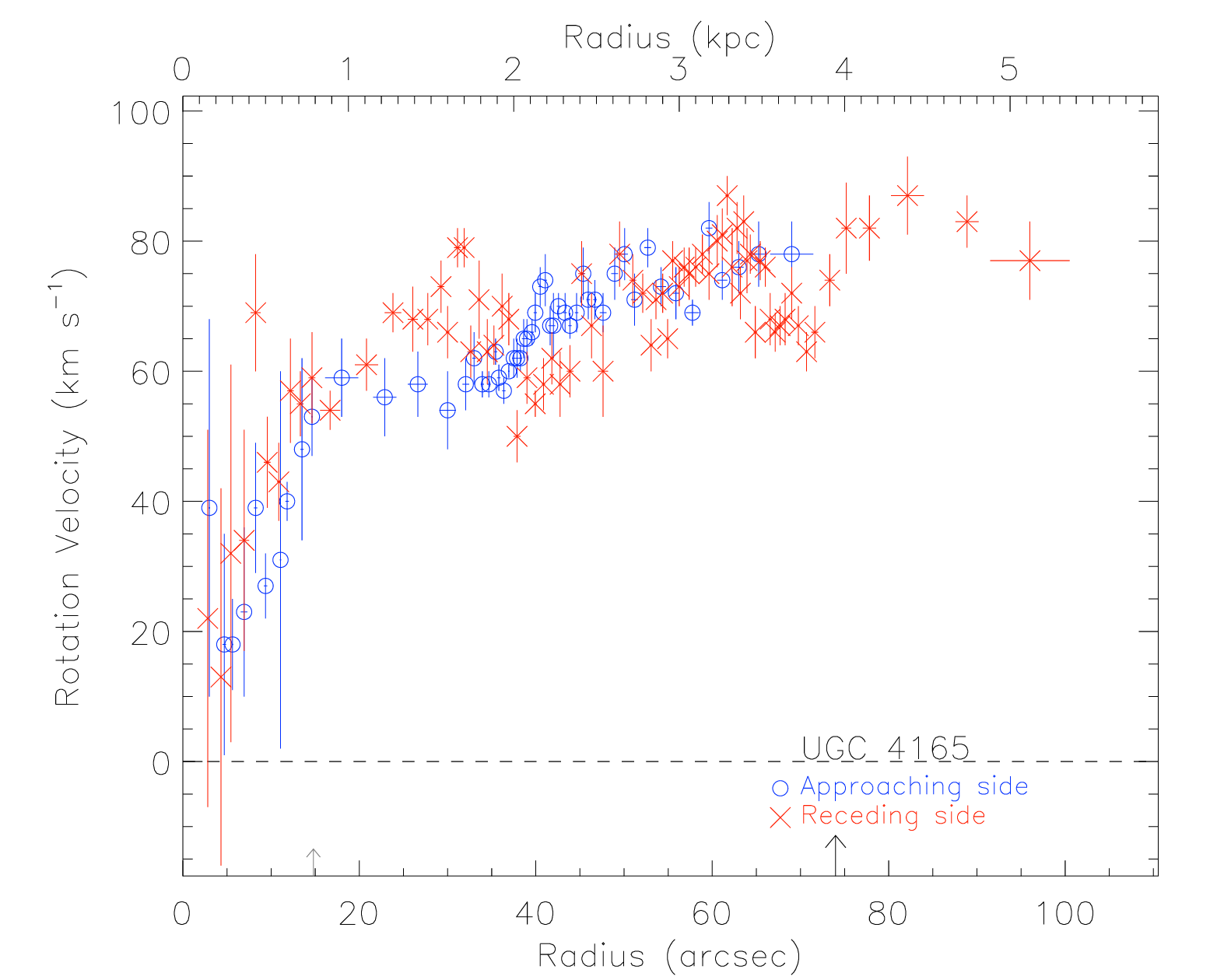}
   \includegraphics[width=8cm]{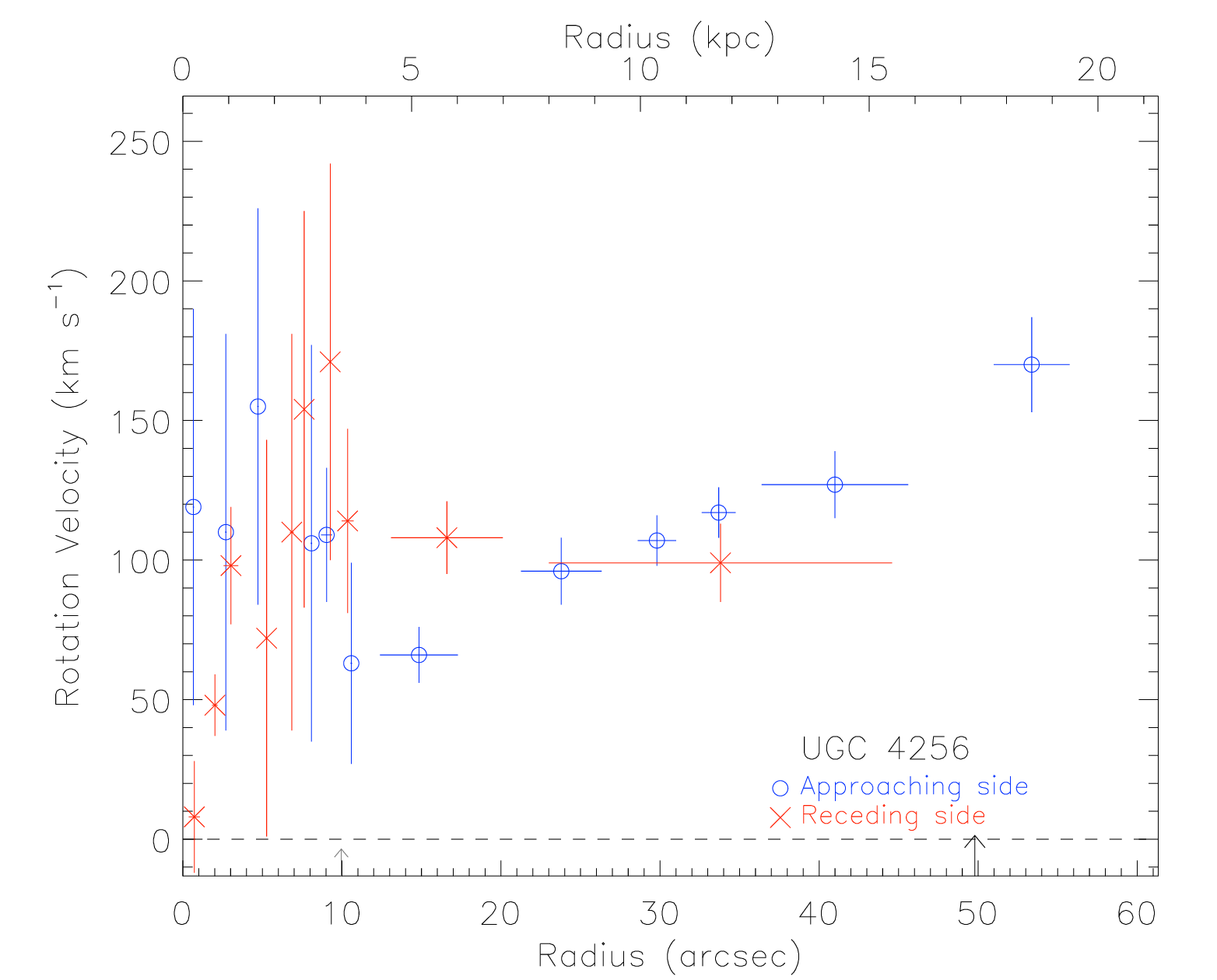}
   \includegraphics[width=8cm]{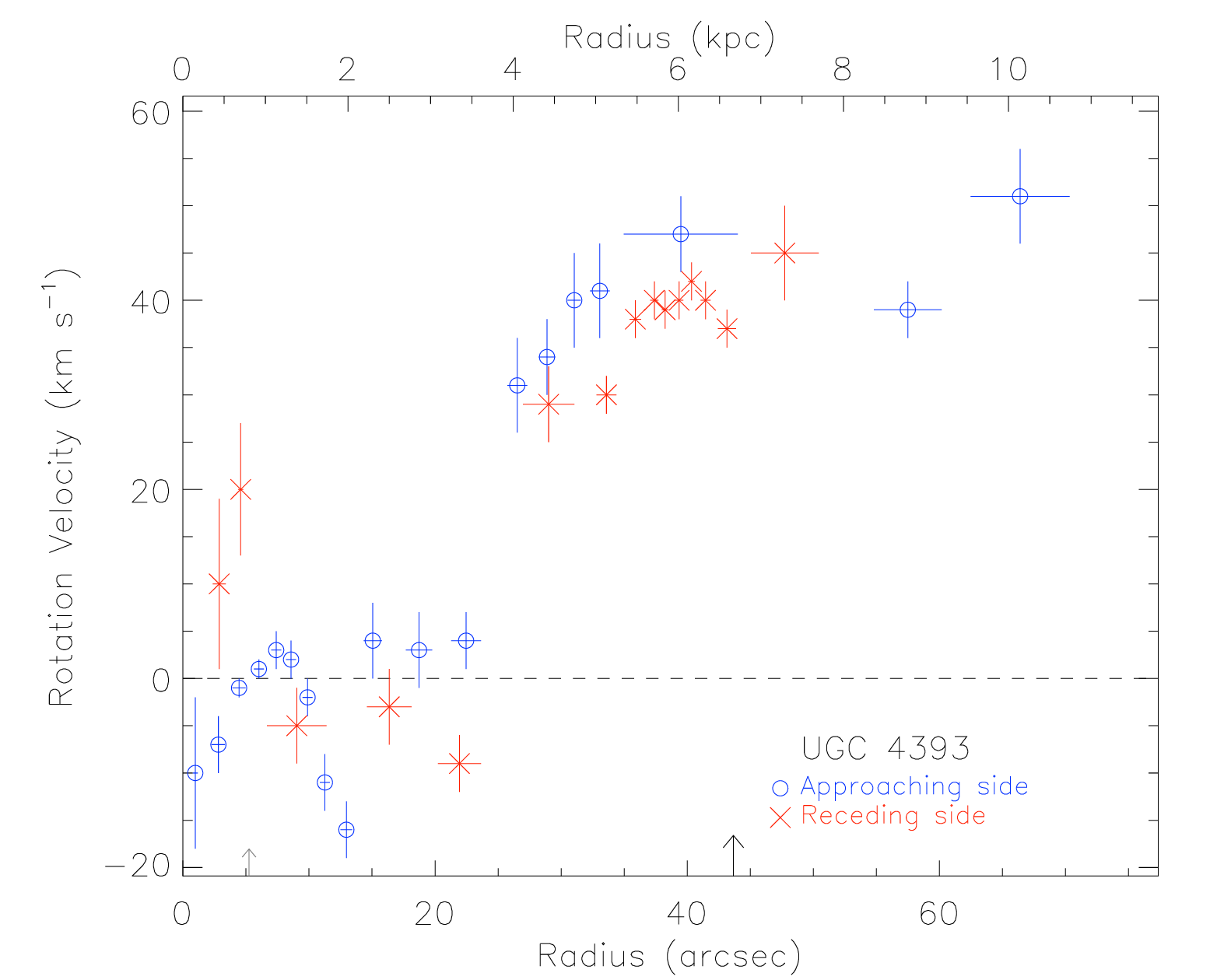}
   \includegraphics[width=8cm]{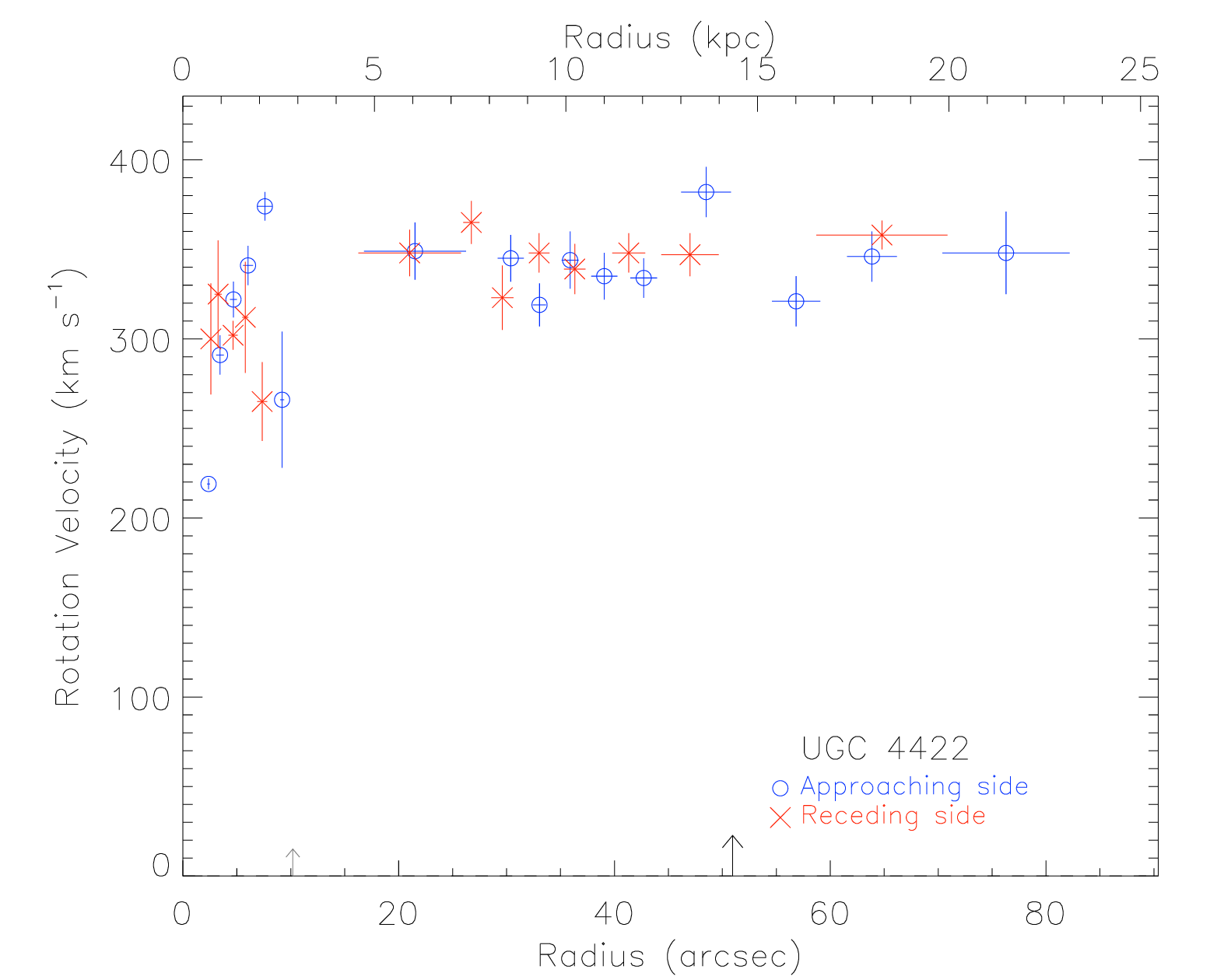}
\end{center}
\caption{From top left to bottom right: \ha~\RC~of IC 476, UGC 4026, UGC 4165, UGC 4256, UGC 4393, and UGC 4422.
}
\end{minipage}
\end{figure*}
\clearpage
\begin{figure*}
\begin{minipage}{180mm}
\begin{center}
   \includegraphics[width=8cm]{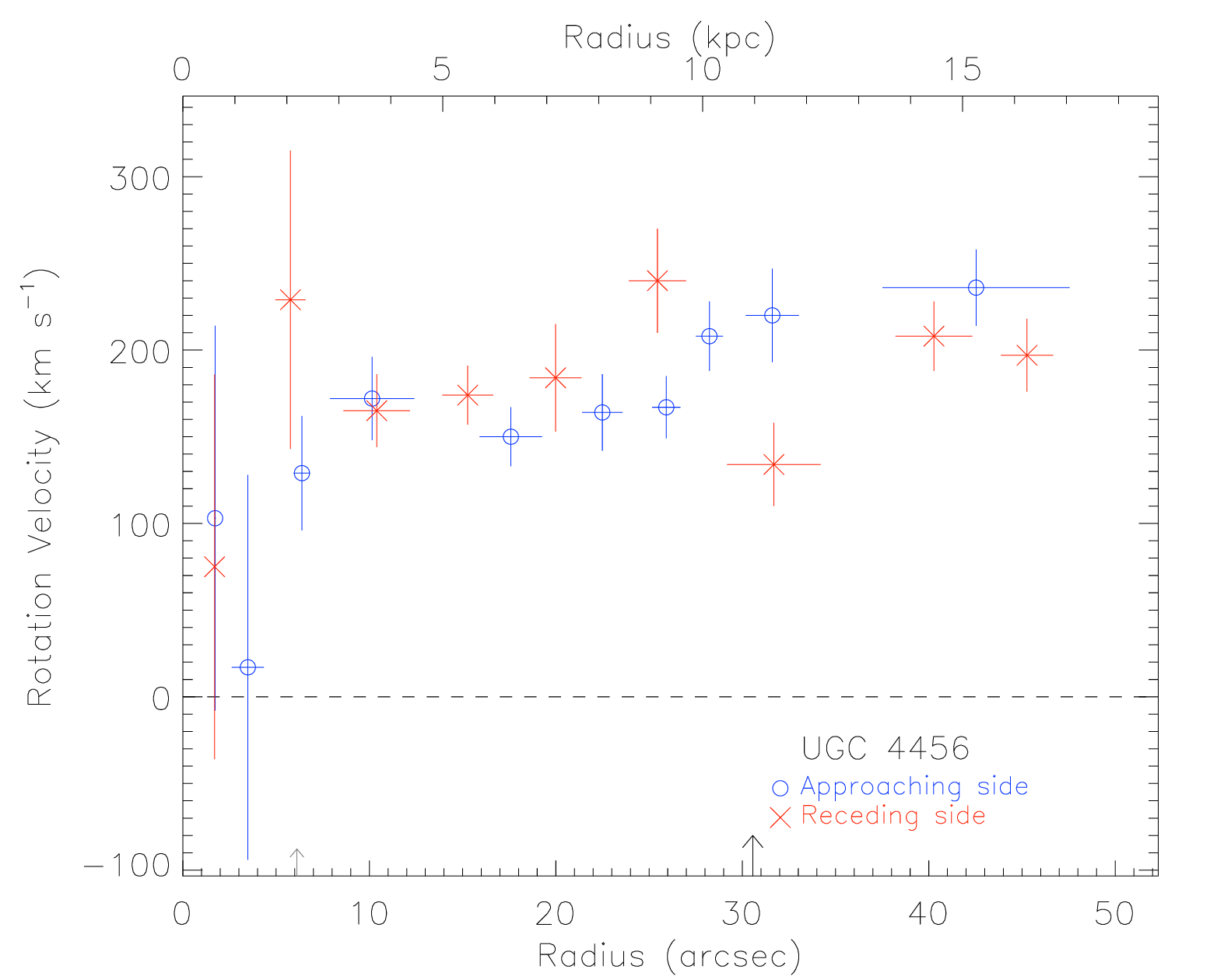}
   \includegraphics[width=8cm]{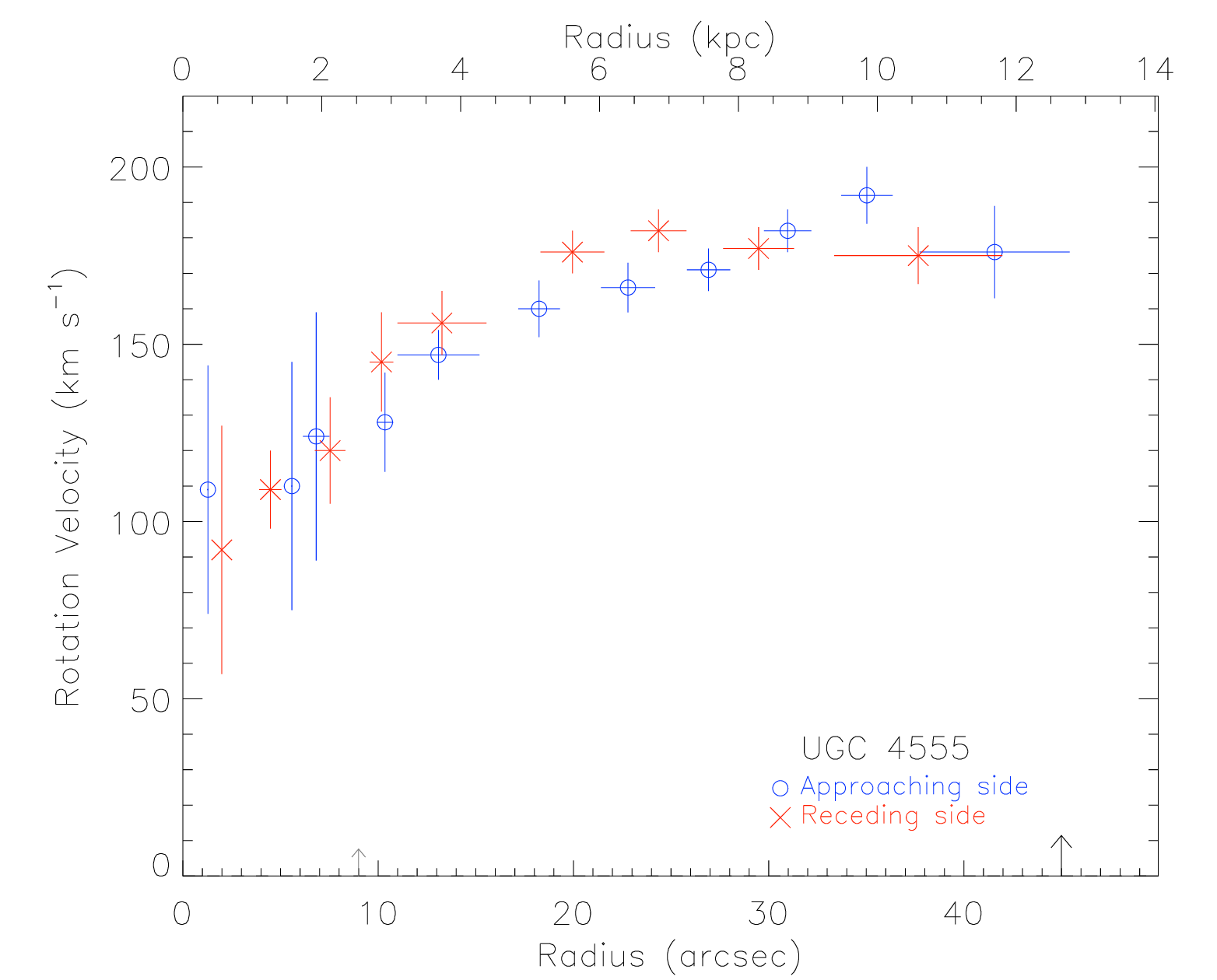}
   \includegraphics[width=8cm]{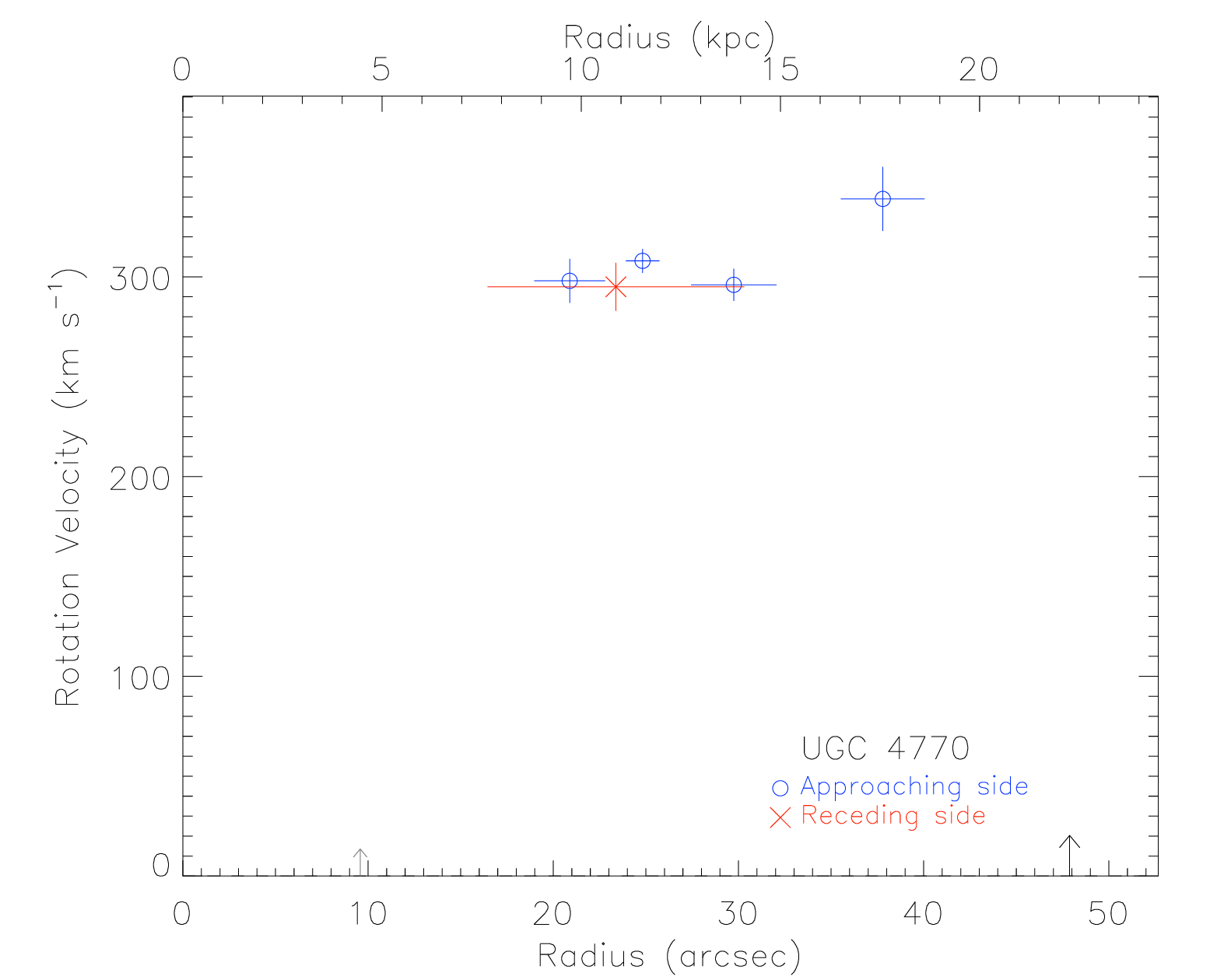}
   \includegraphics[width=8cm]{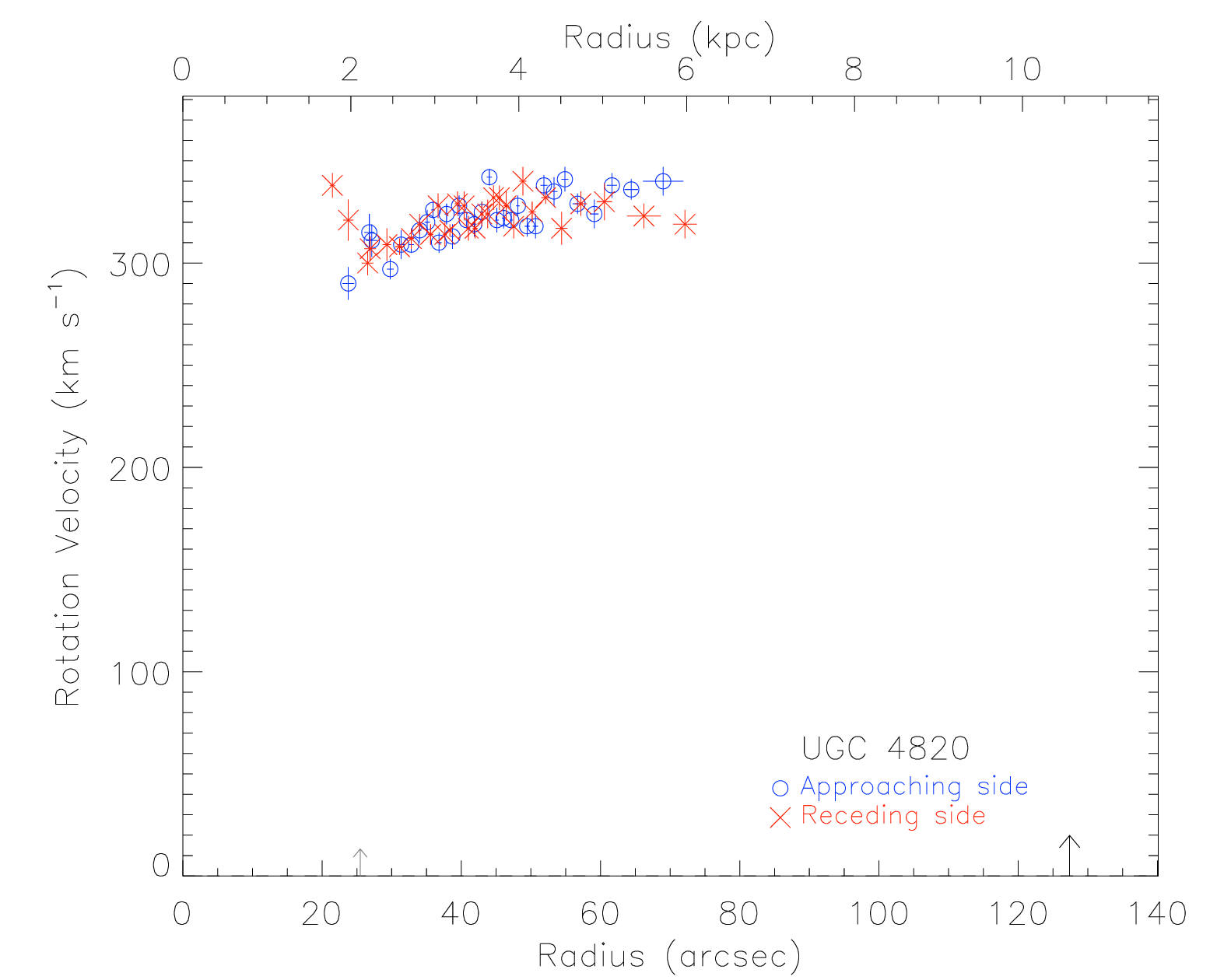}
   \includegraphics[width=8cm]{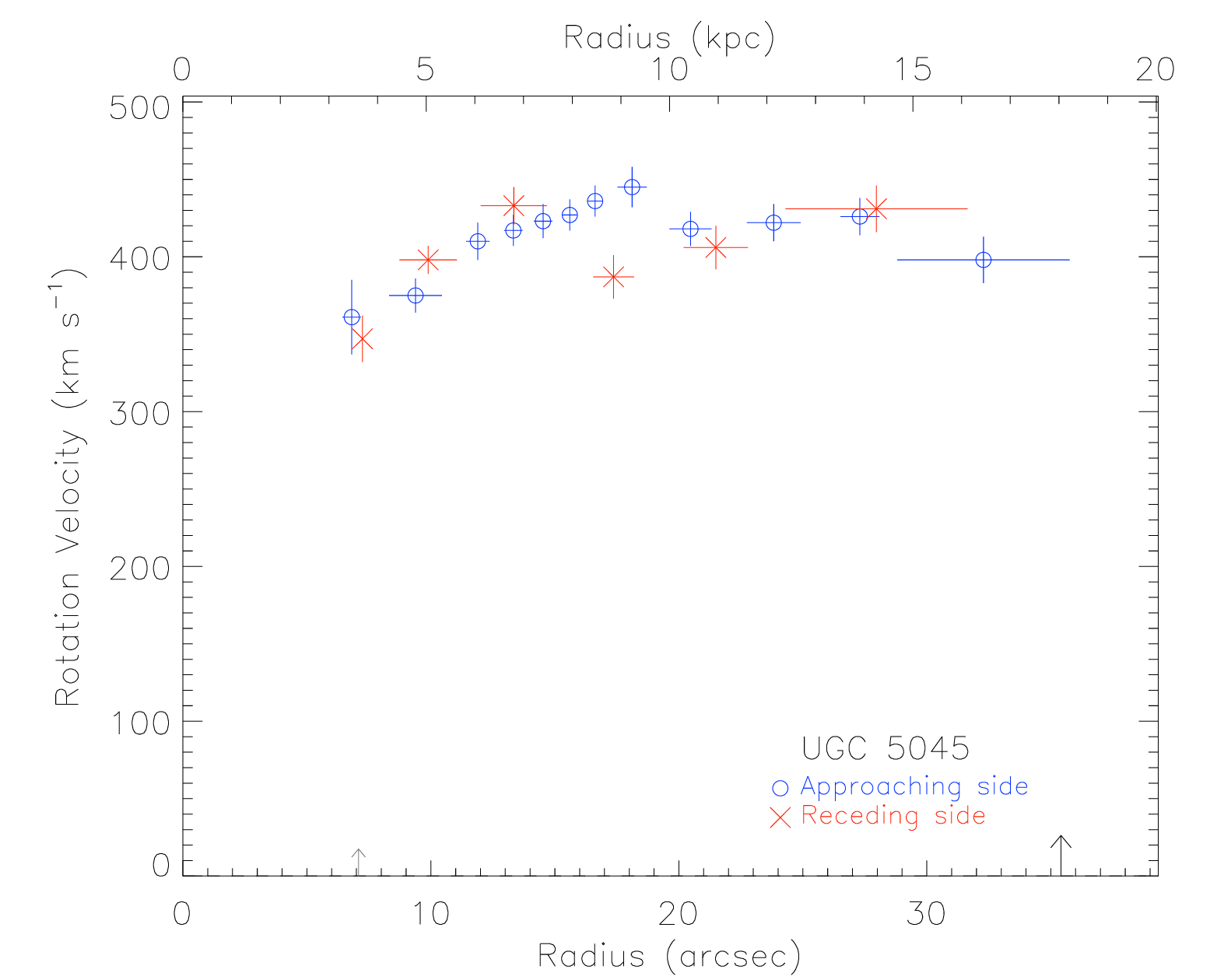}
   \includegraphics[width=8cm]{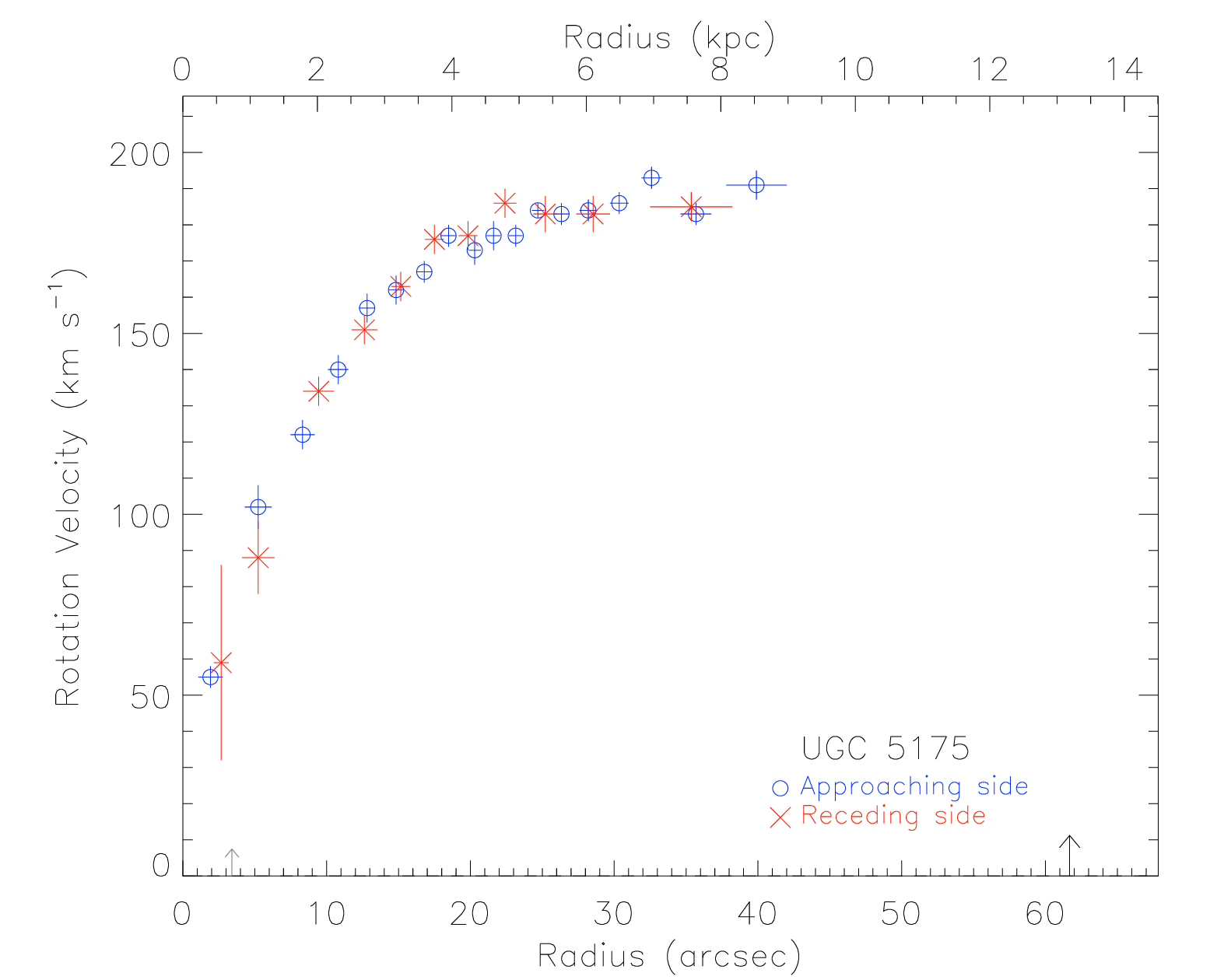}
\end{center}
\caption{From top left to bottom right: \ha~\RC~of UGC 4456, UGC 4555, UGC 4770, UGC 4820, UGC 5045, and UGC 5175.
}
\end{minipage}
\end{figure*}
\clearpage
\begin{figure*}
\begin{minipage}{180mm}
\begin{center}
   \includegraphics[width=8cm]{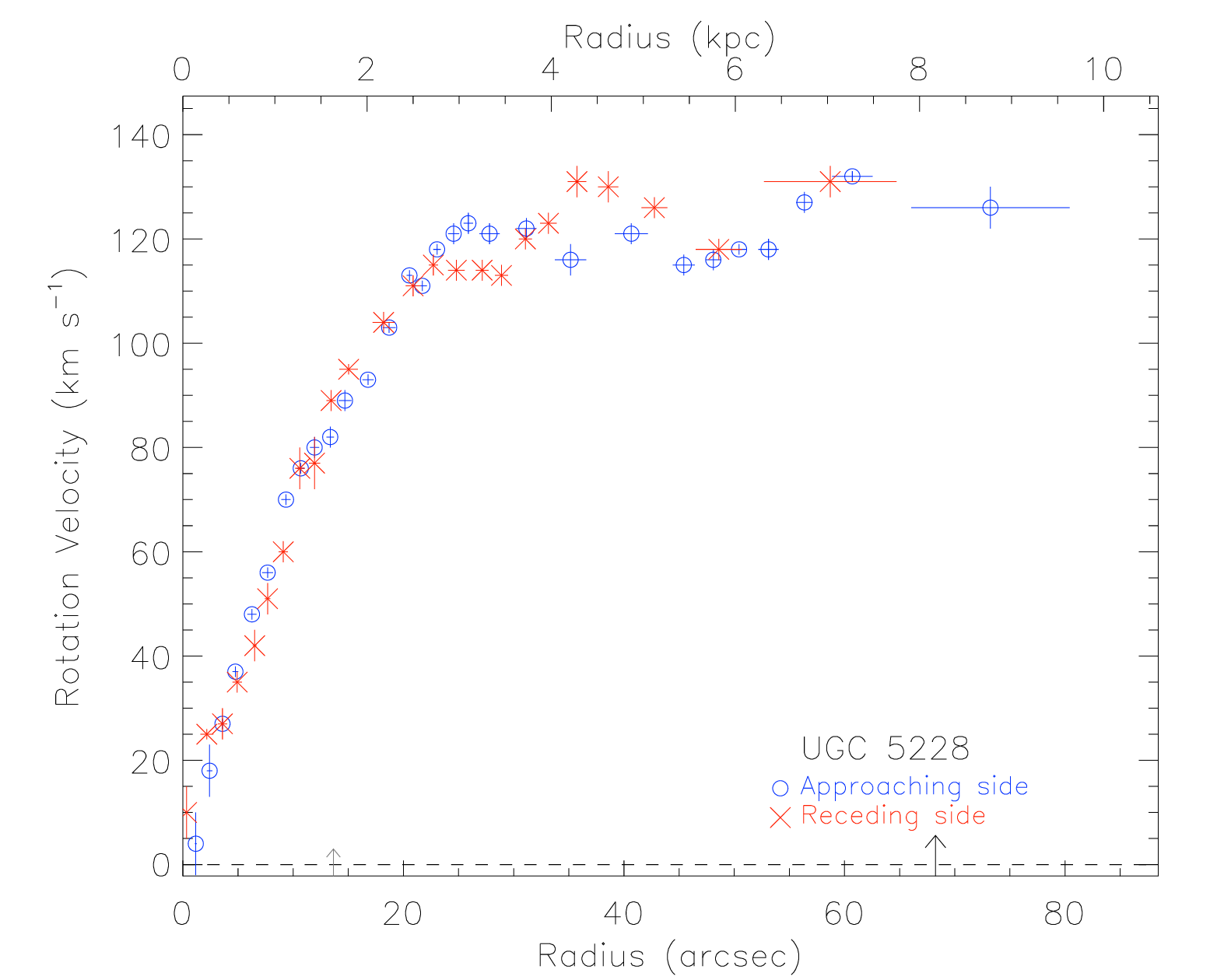}
   \includegraphics[width=8cm]{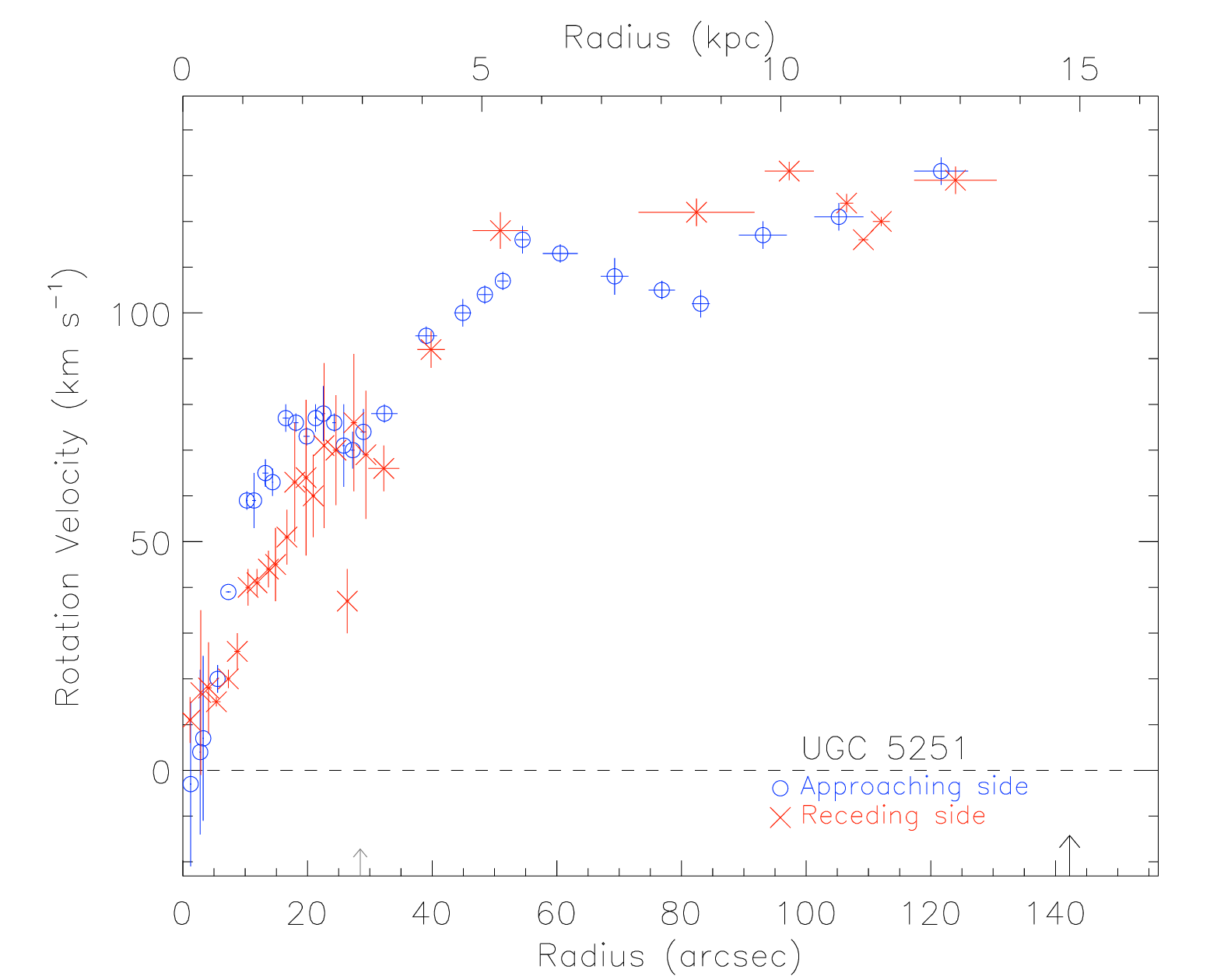}
   \includegraphics[width=8cm]{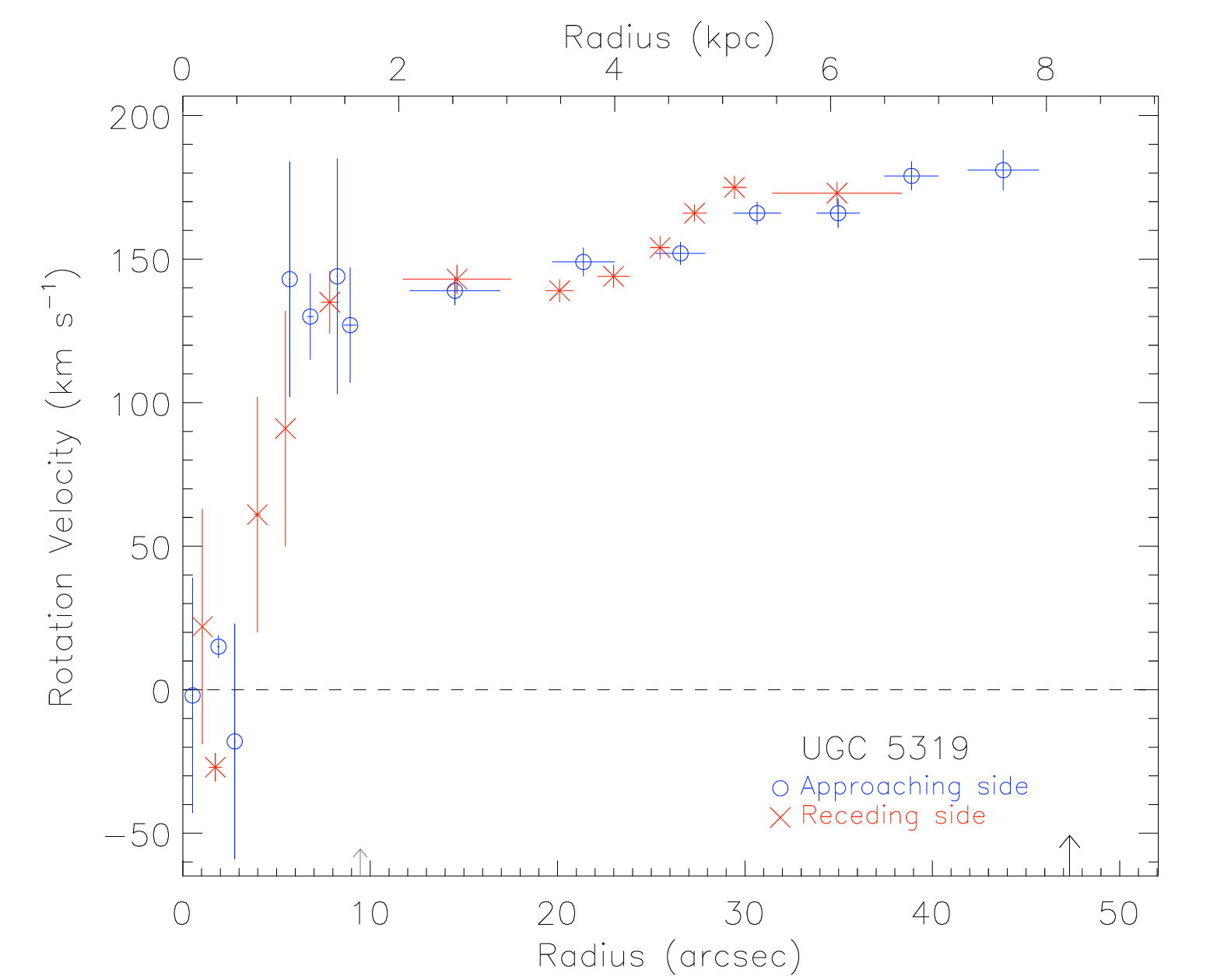}
   \includegraphics[width=8cm]{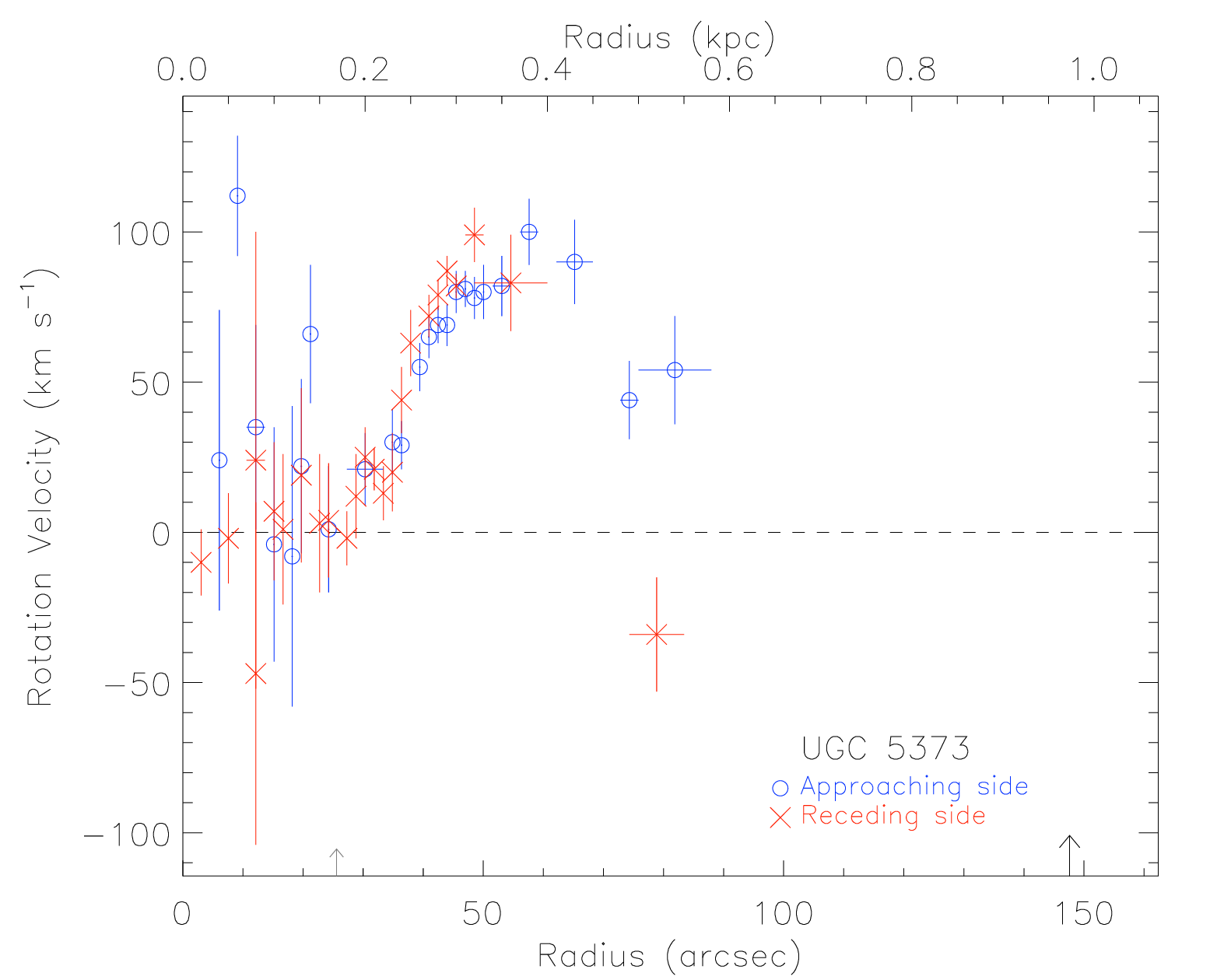}
   \includegraphics[width=8cm]{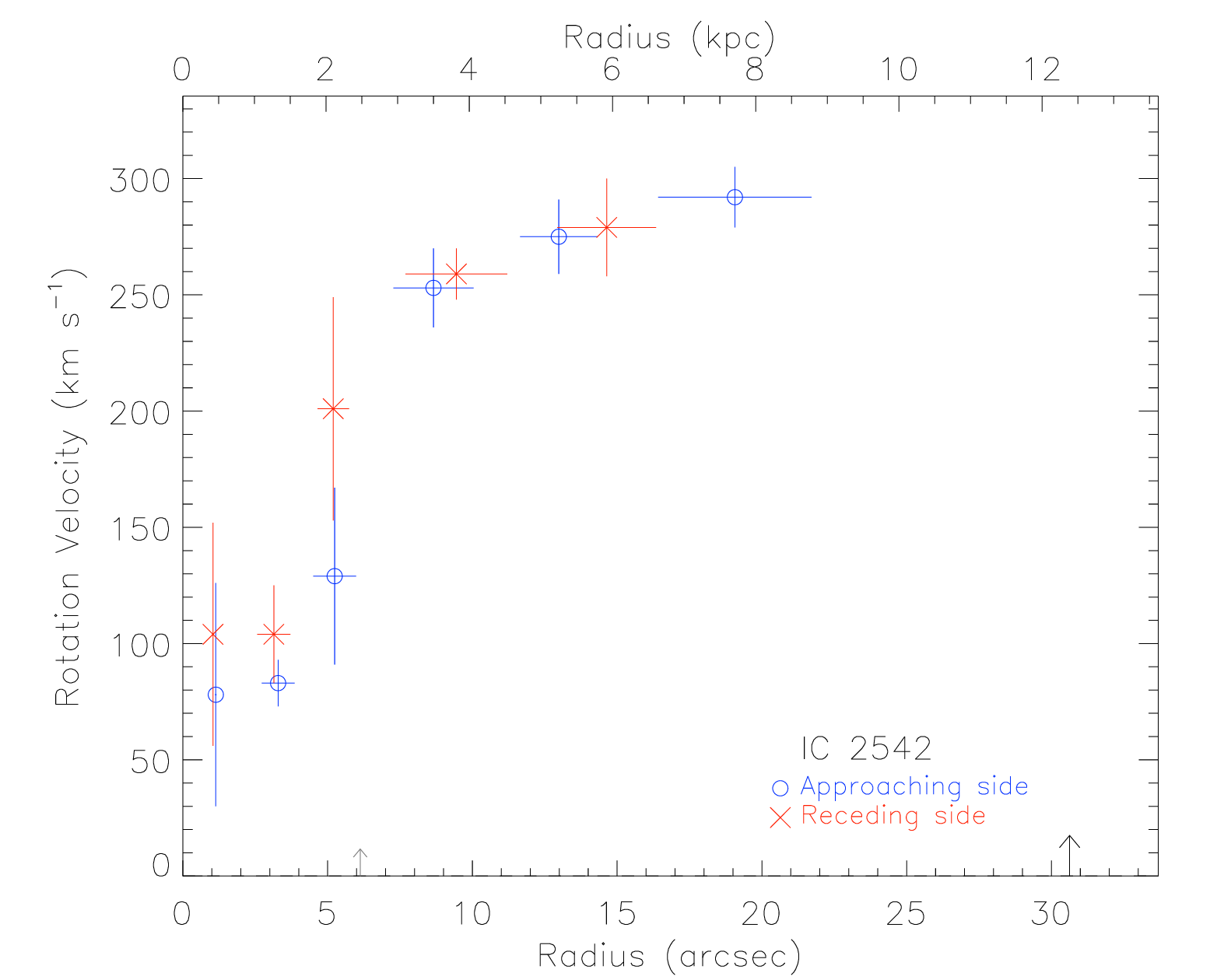}
   \includegraphics[width=8cm]{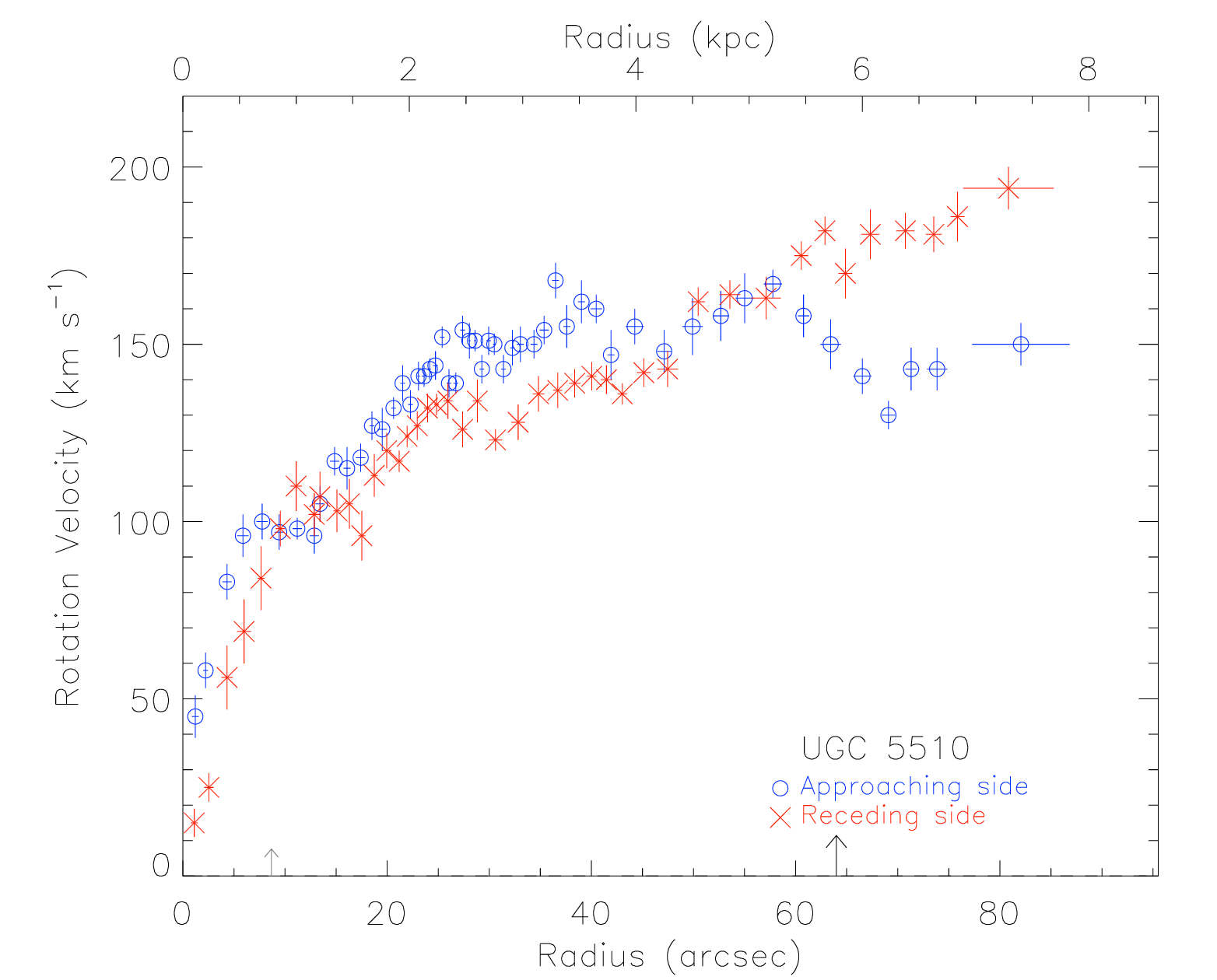}
\end{center}
\caption{From top left to bottom right: \ha~\RC~of UGC 5228, UGC 5251, UGC 5319, UGC 5373, IC 2542, and UGC 5510.
}
\end{minipage}
\end{figure*}
\clearpage
\begin{figure*}
\begin{minipage}{180mm}
\begin{center}
   \includegraphics[width=8cm]{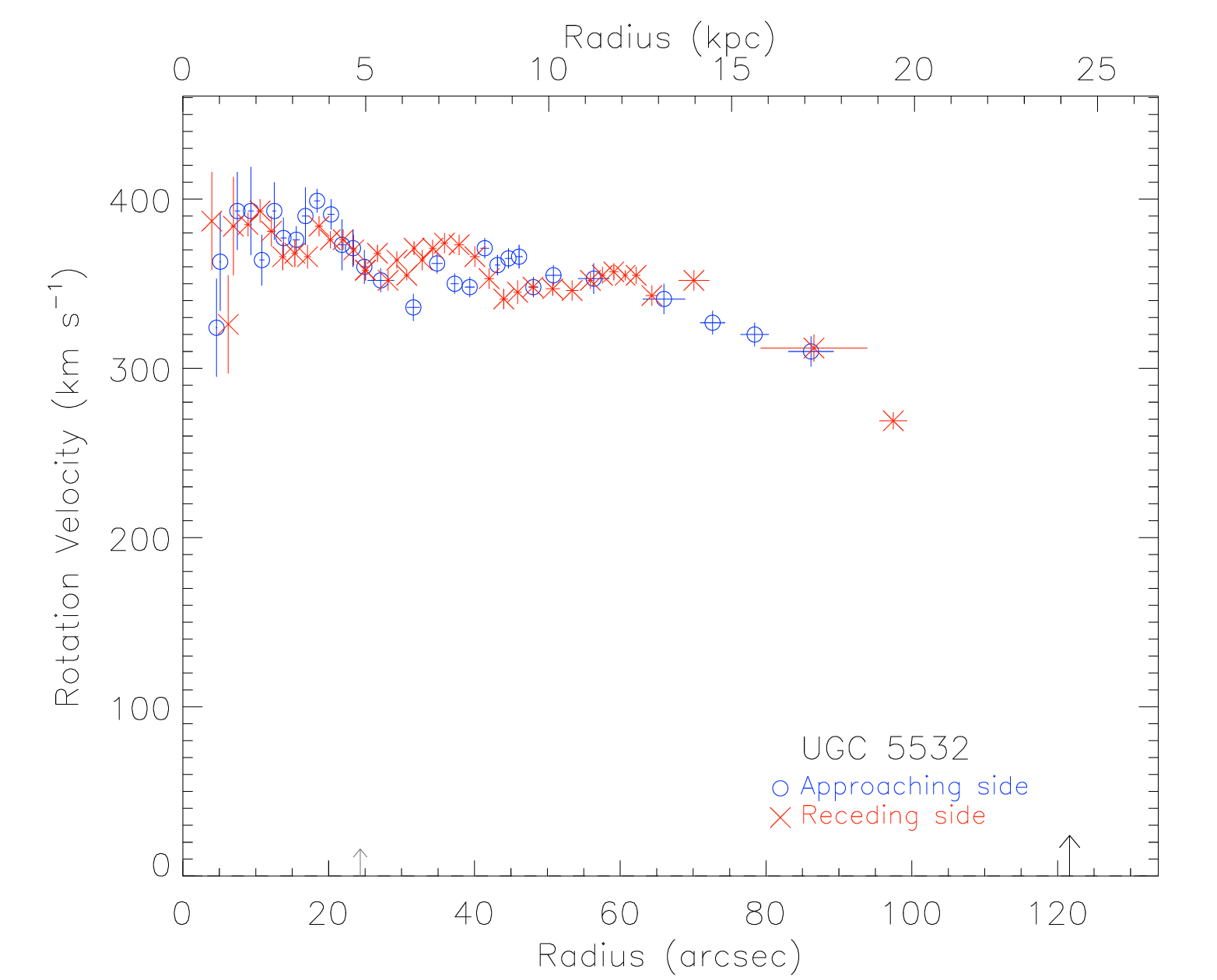}
   \includegraphics[width=8cm]{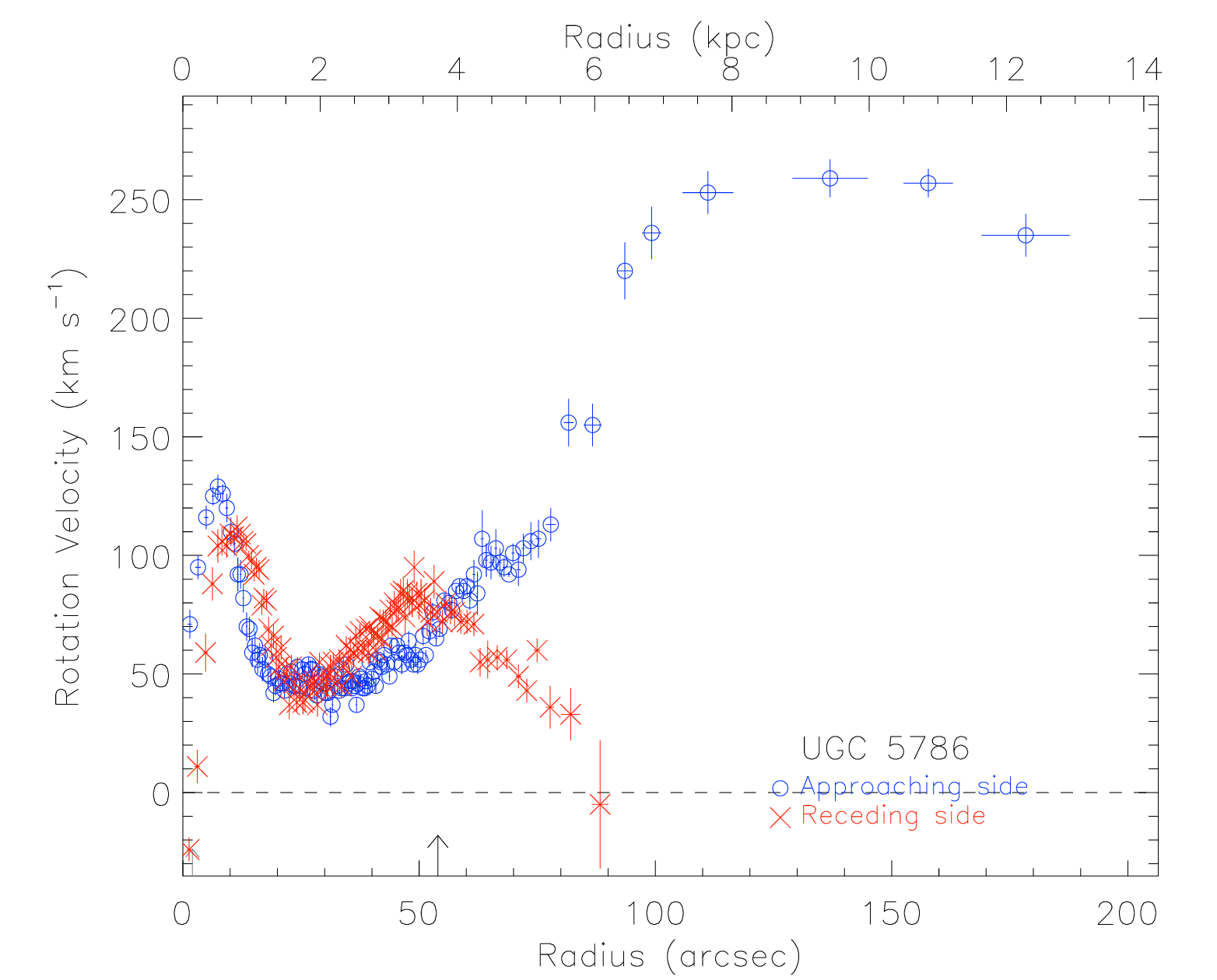}
   \includegraphics[width=8cm]{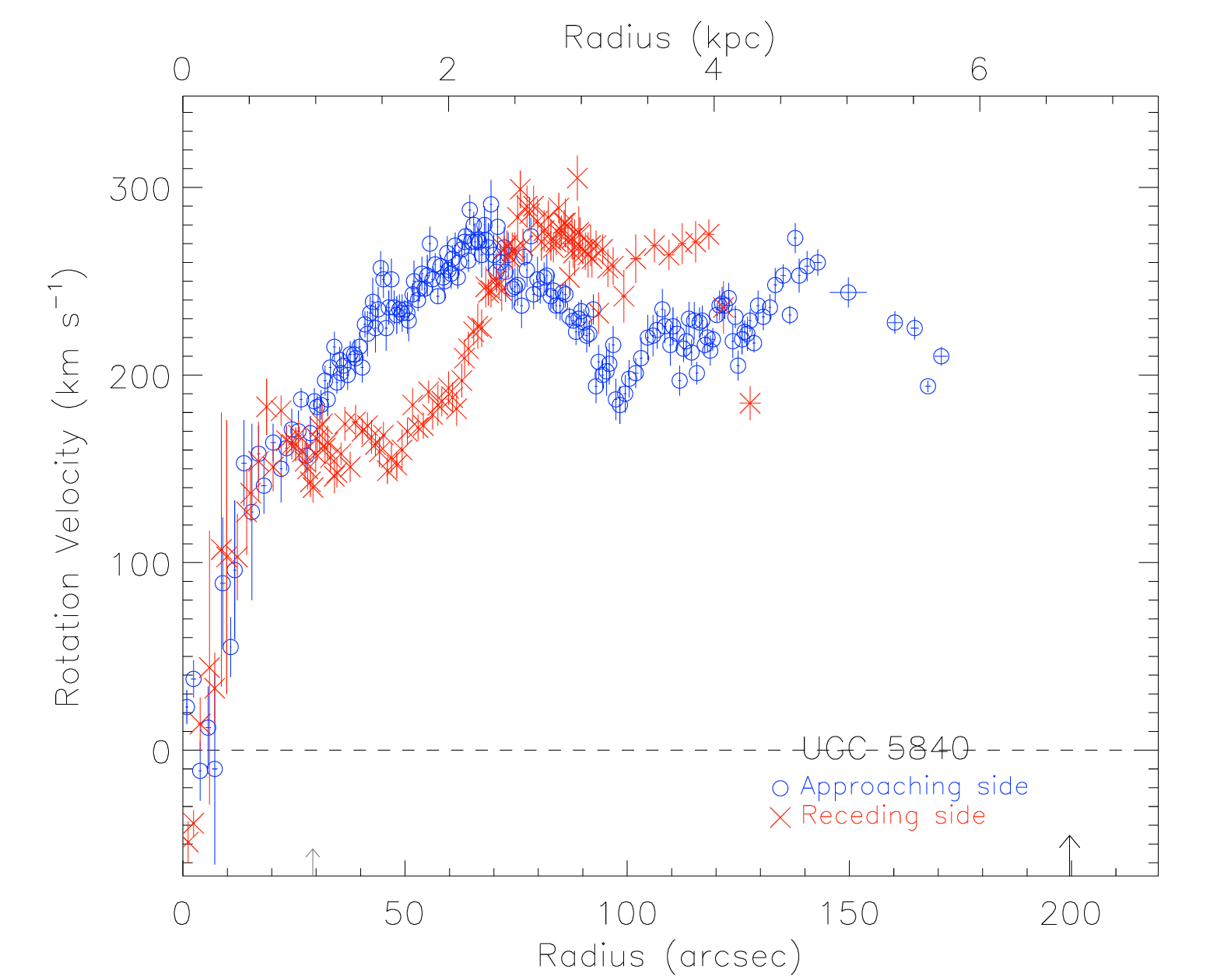}
   \includegraphics[width=8cm]{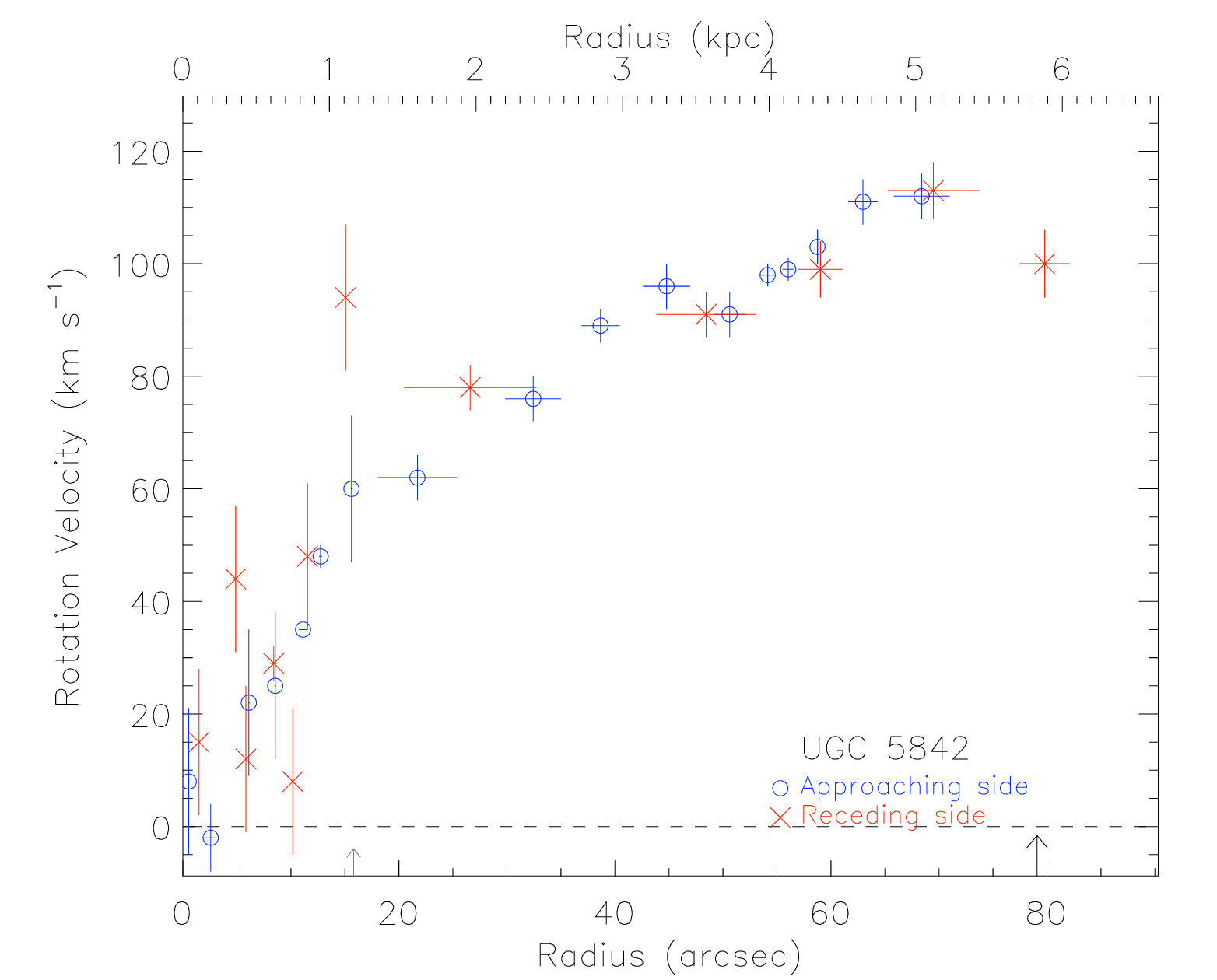}
   \includegraphics[width=8cm]{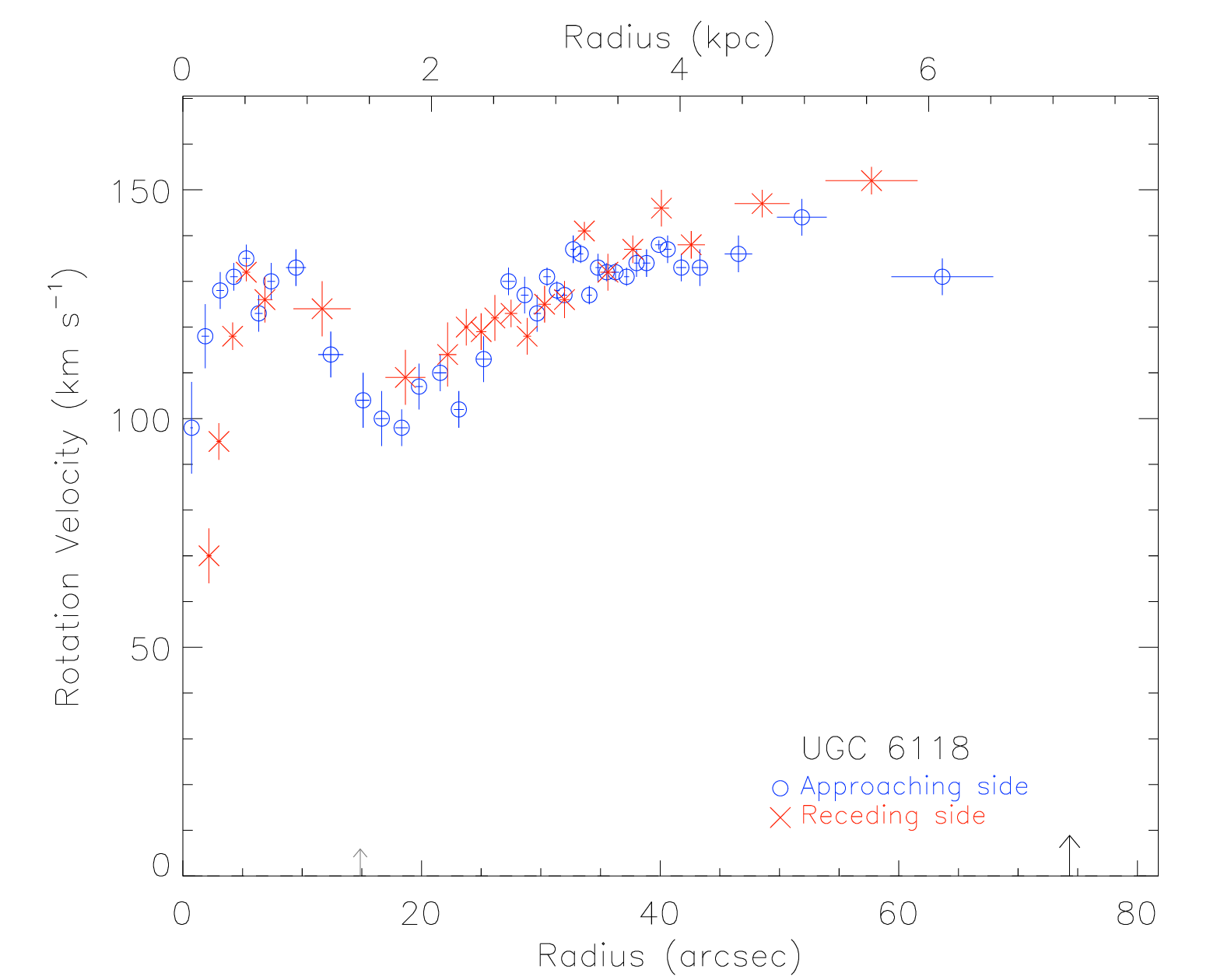}
   \includegraphics[width=8cm]{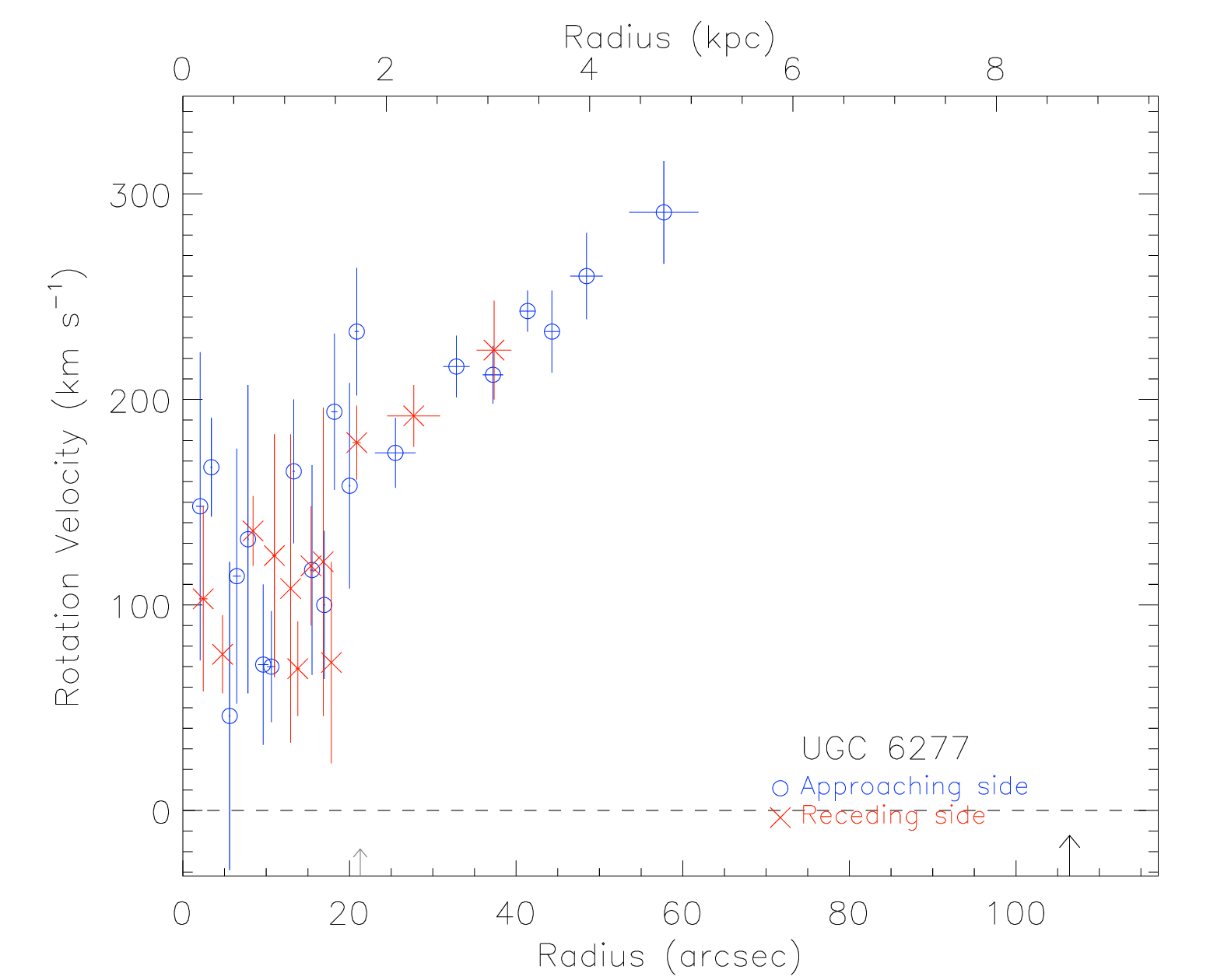}
\end{center}
\caption{From top left to bottom right: \ha~\RC~of UGC 5532, UGC 5786, UGC 5840, UGC 5842, UGC 6118, and UGC 6277.
}
\end{minipage}
\end{figure*}
\clearpage
\begin{figure*}
\begin{minipage}{180mm}
\begin{center}
   \includegraphics[width=8cm]{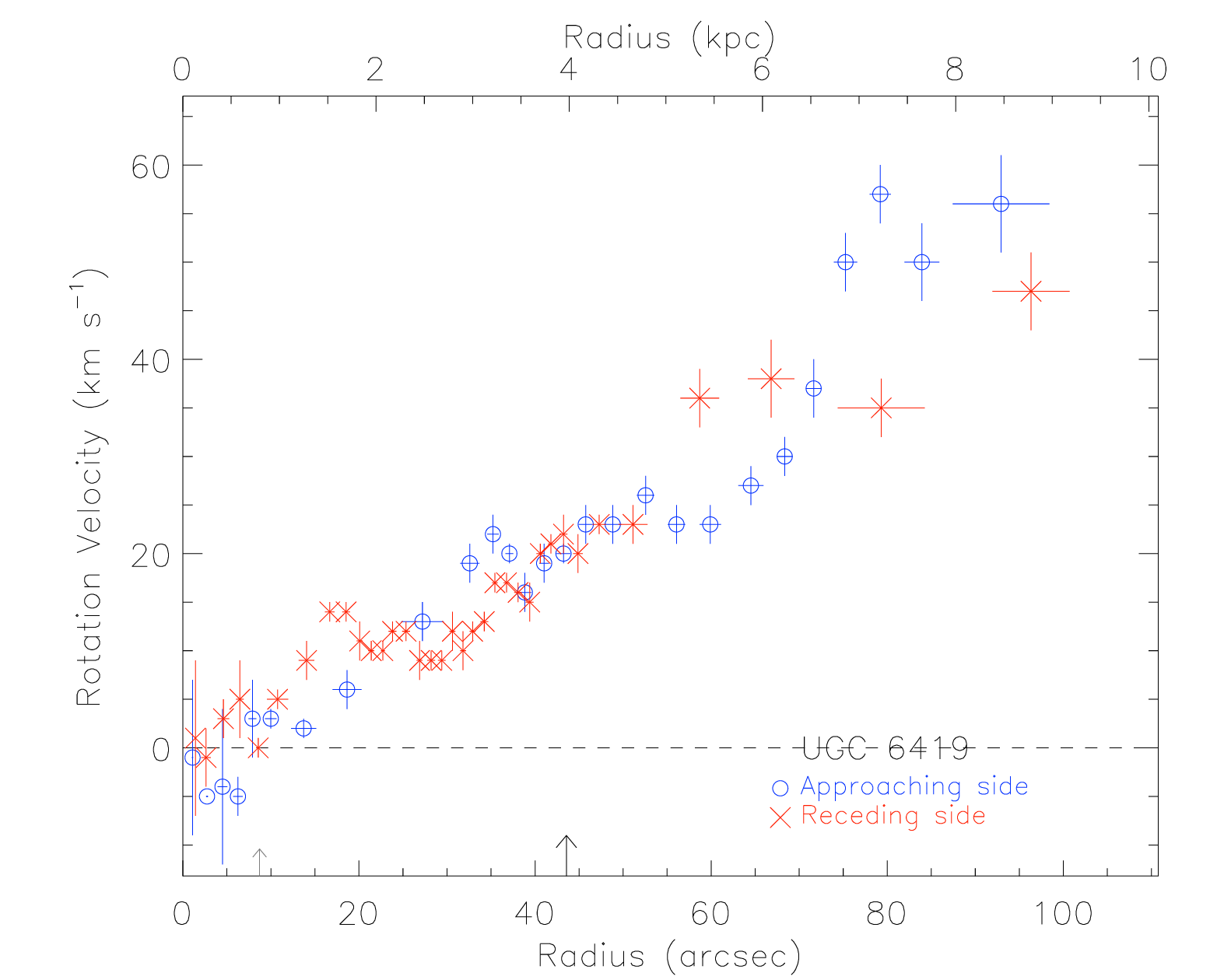}
   \includegraphics[width=8cm]{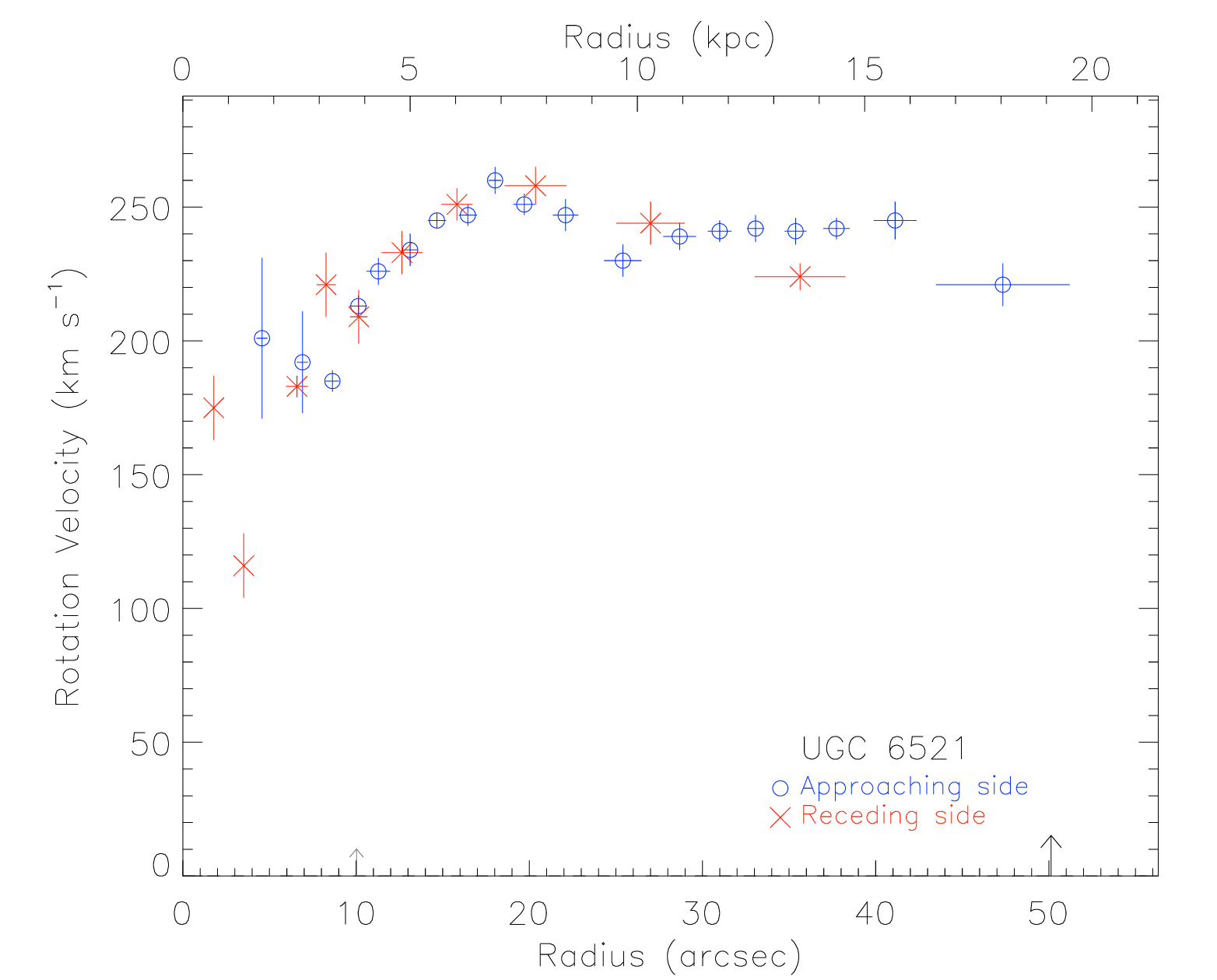}
   \includegraphics[width=8cm]{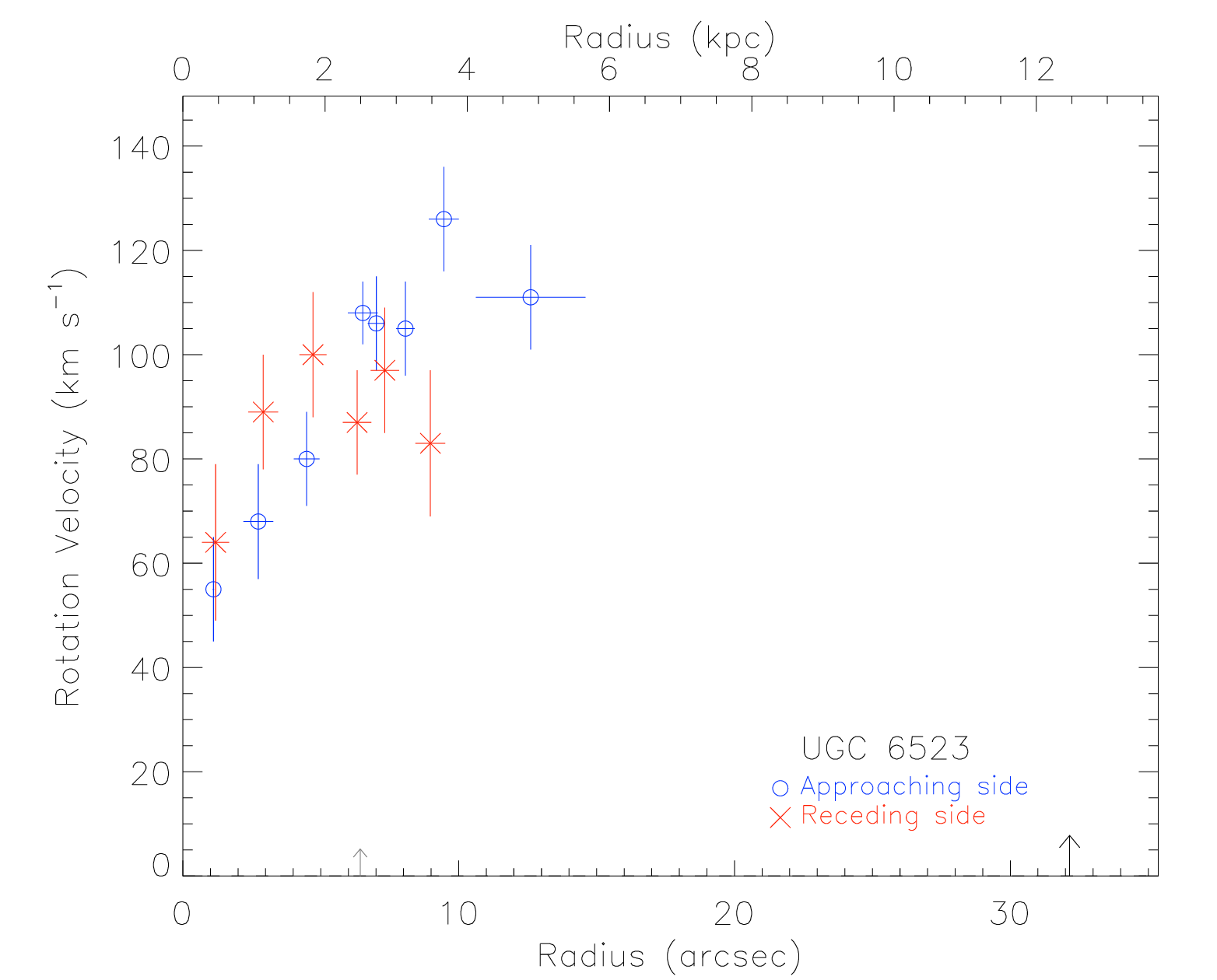}
   \includegraphics[width=8cm]{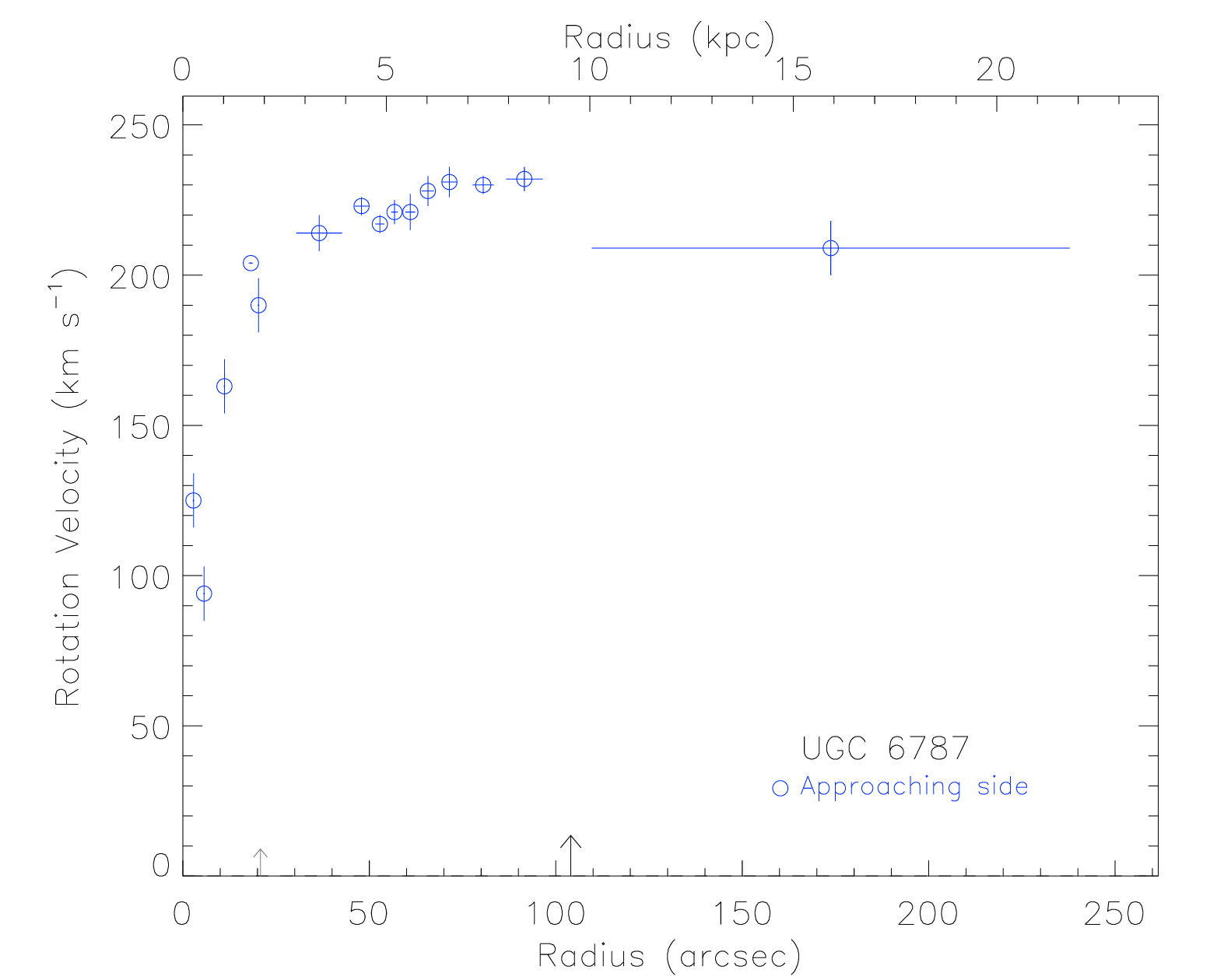}
   \includegraphics[width=8cm]{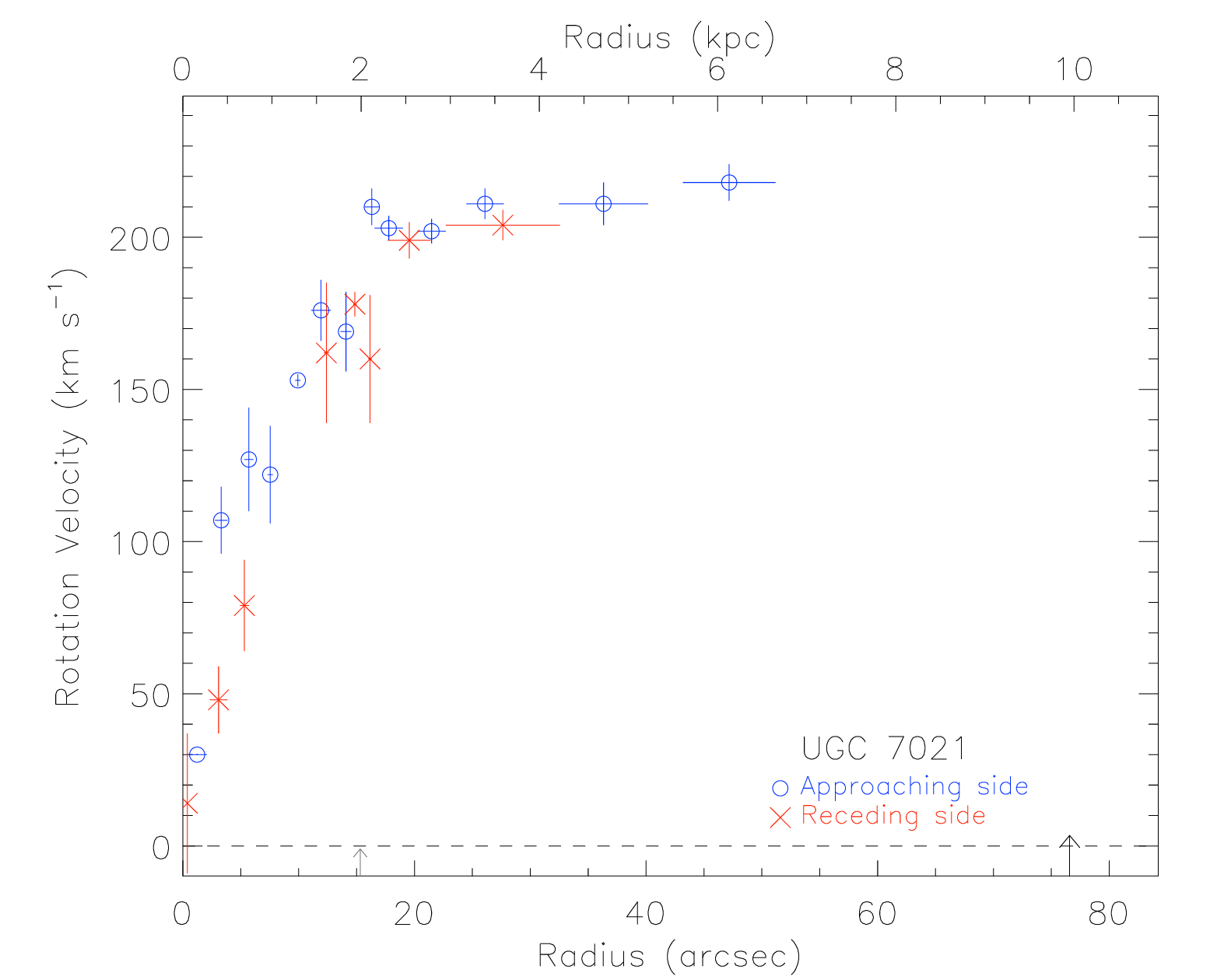}
   \includegraphics[width=8cm]{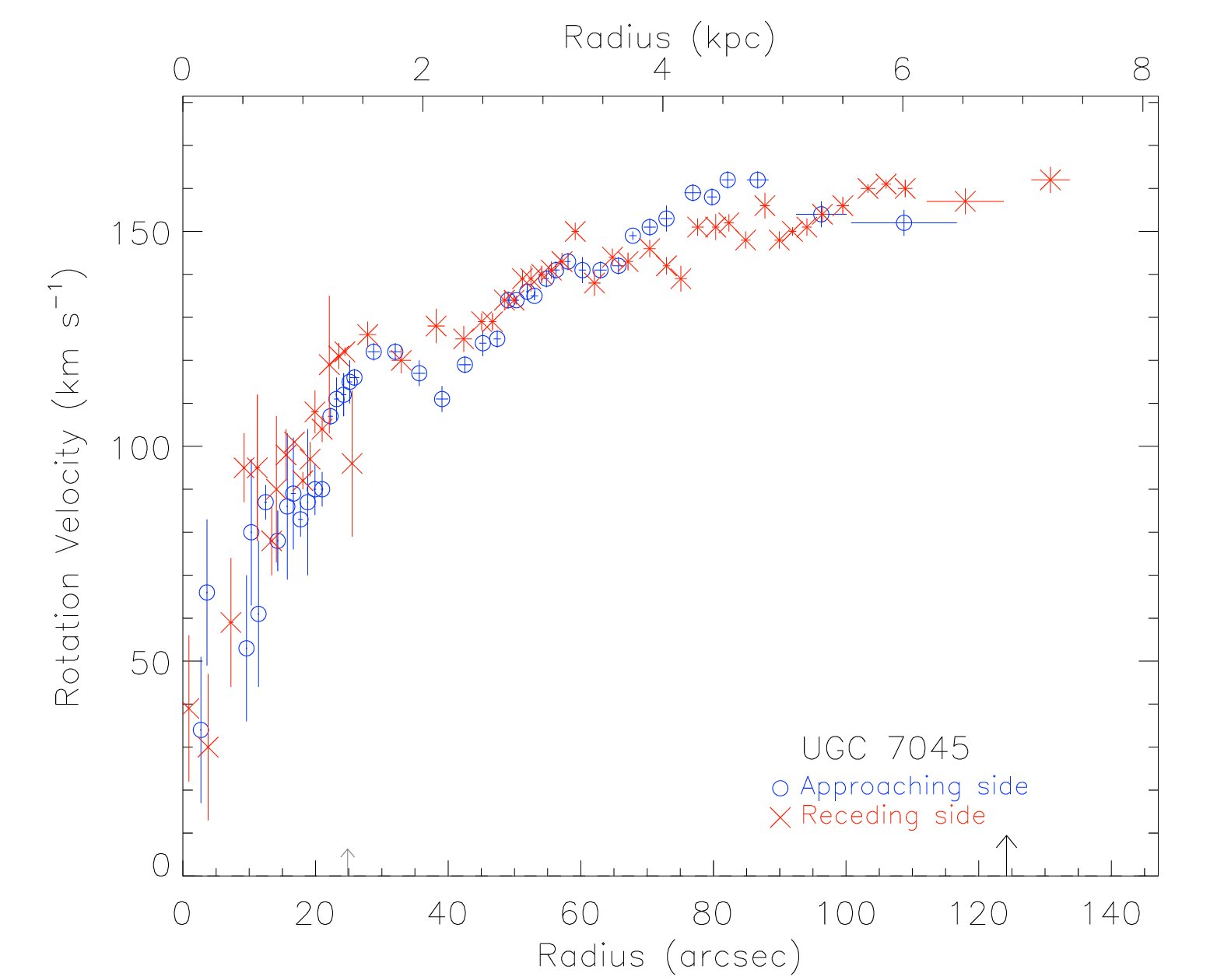}
\end{center}
\caption{From top left to bottom right: \ha~\RC~of UGC 6419, UGC 6521, UGC 6523, UGC 6787, UGC 7021, and UGC 7045.
}
\end{minipage}
\end{figure*}
\clearpage
\begin{figure*}
\begin{minipage}{180mm}
\begin{center}
   \includegraphics[width=8cm]{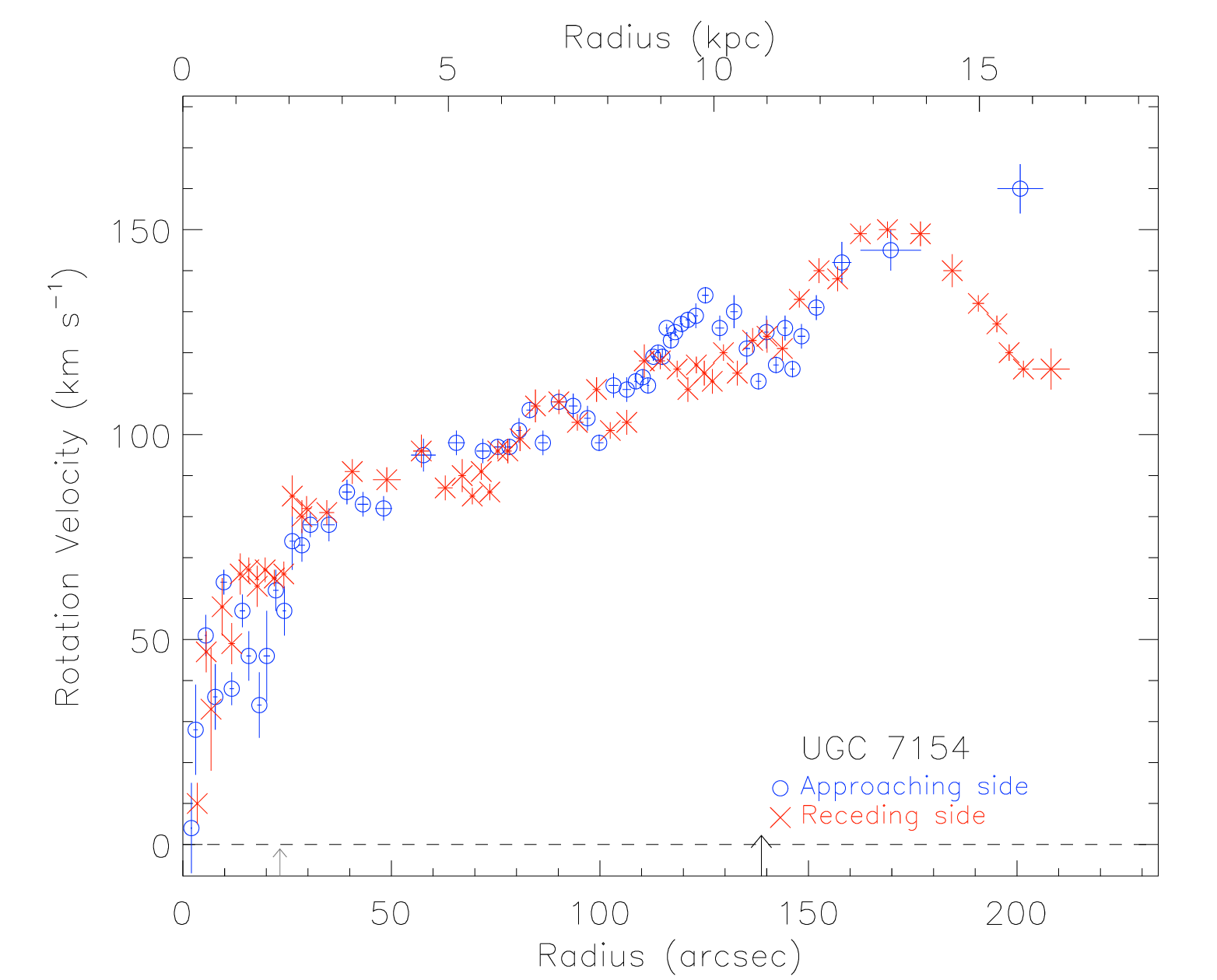}
   \includegraphics[width=8cm]{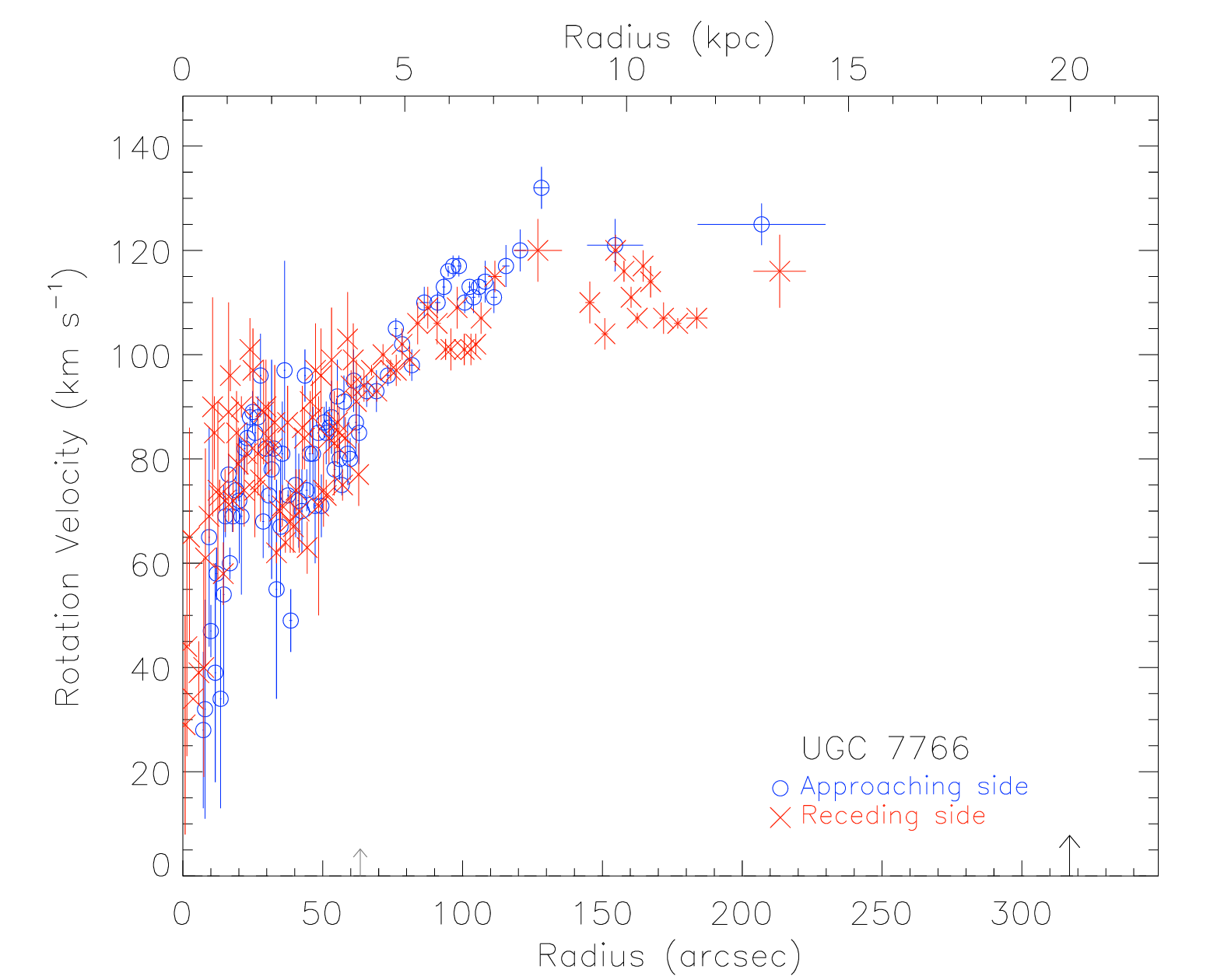}
   \includegraphics[width=8cm]{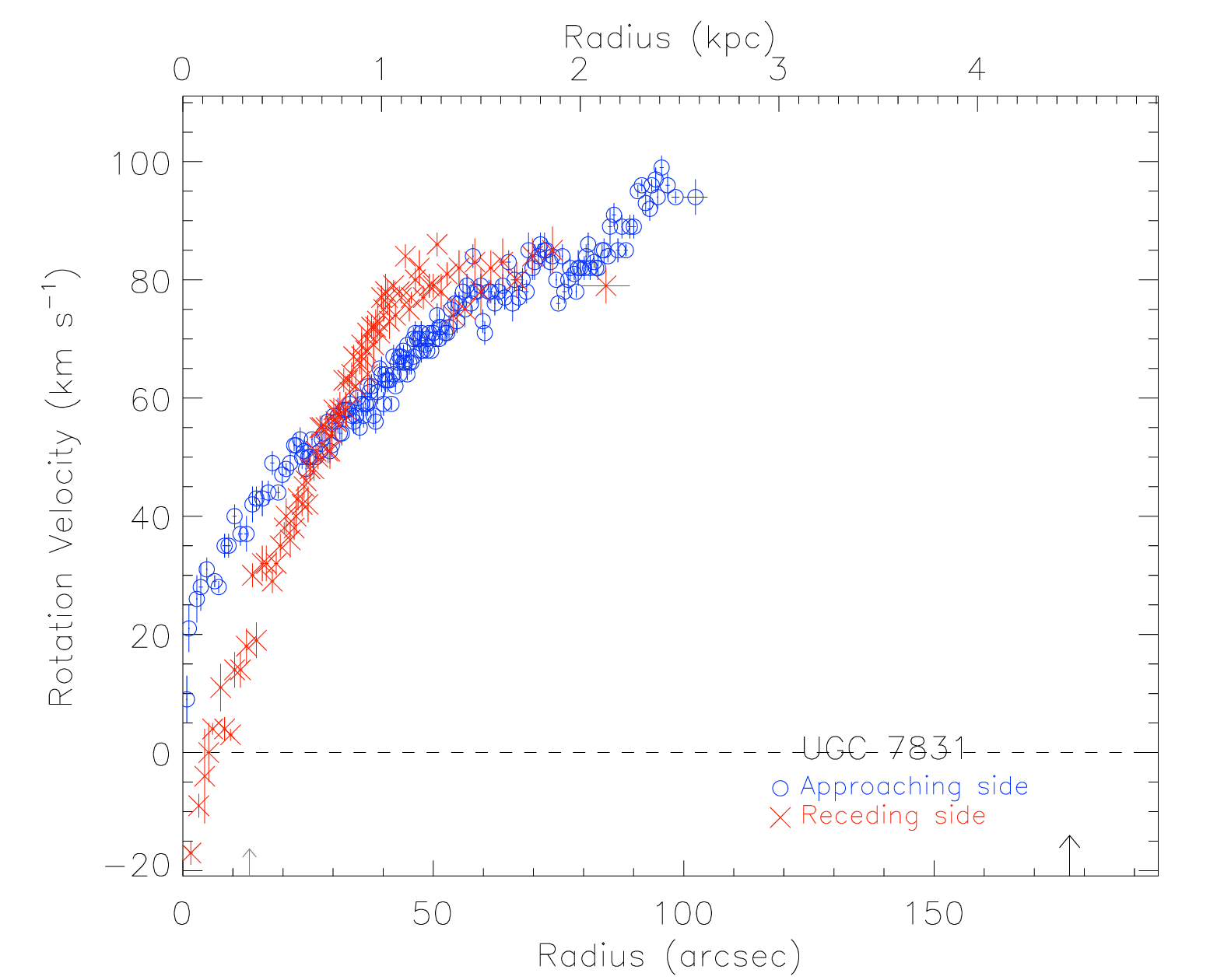}
   \includegraphics[width=8cm]{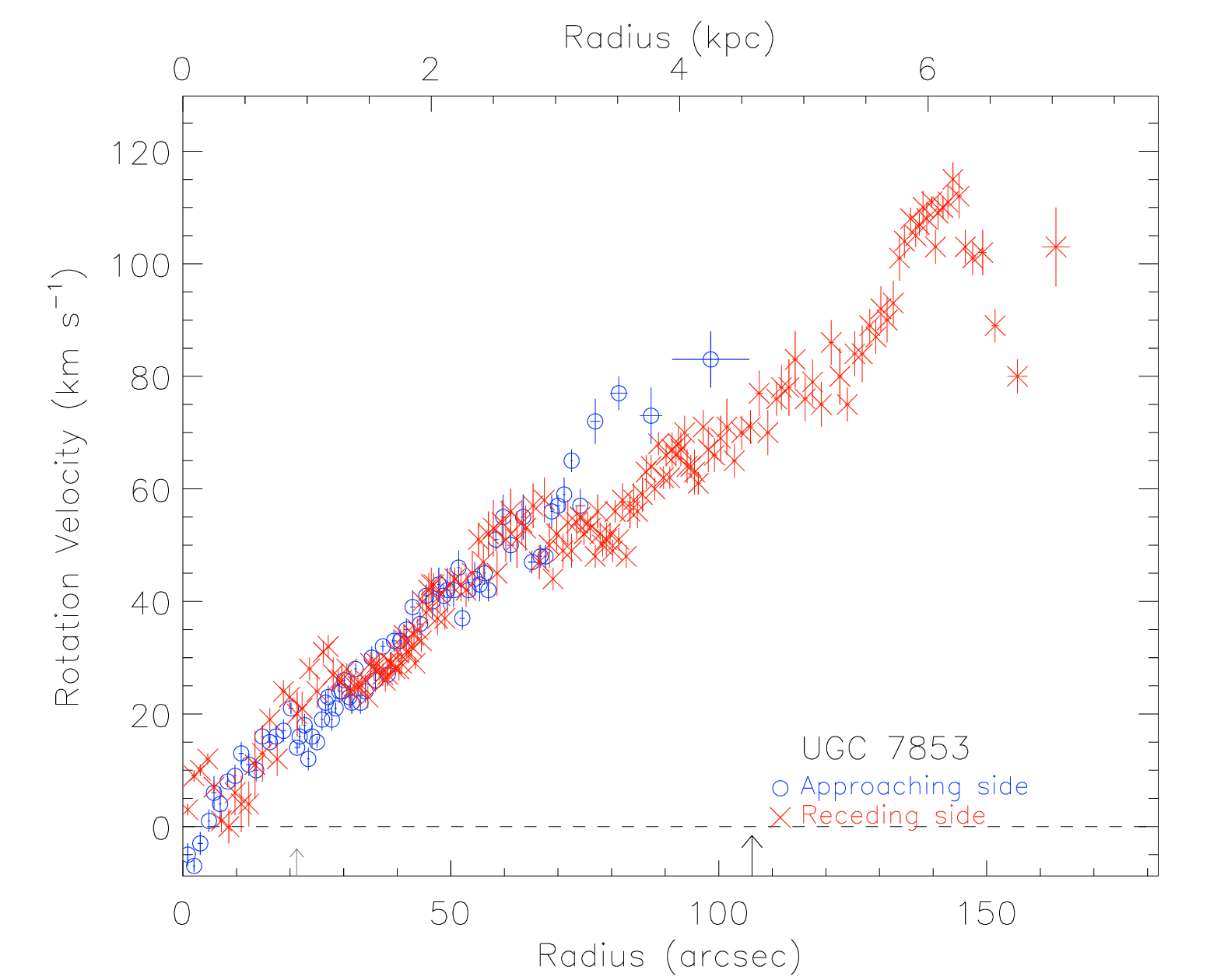}
   \includegraphics[width=8cm]{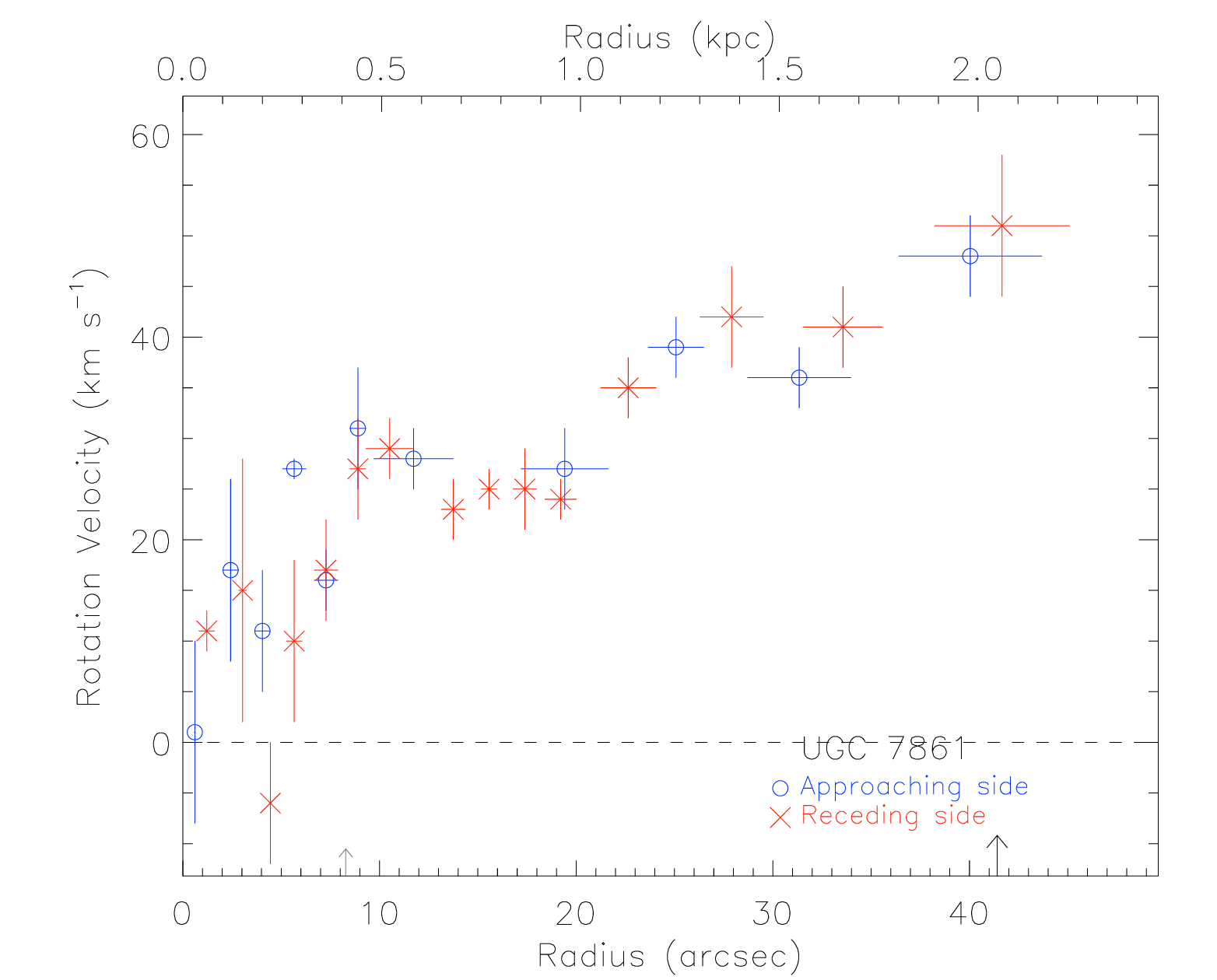}
   \includegraphics[width=8cm]{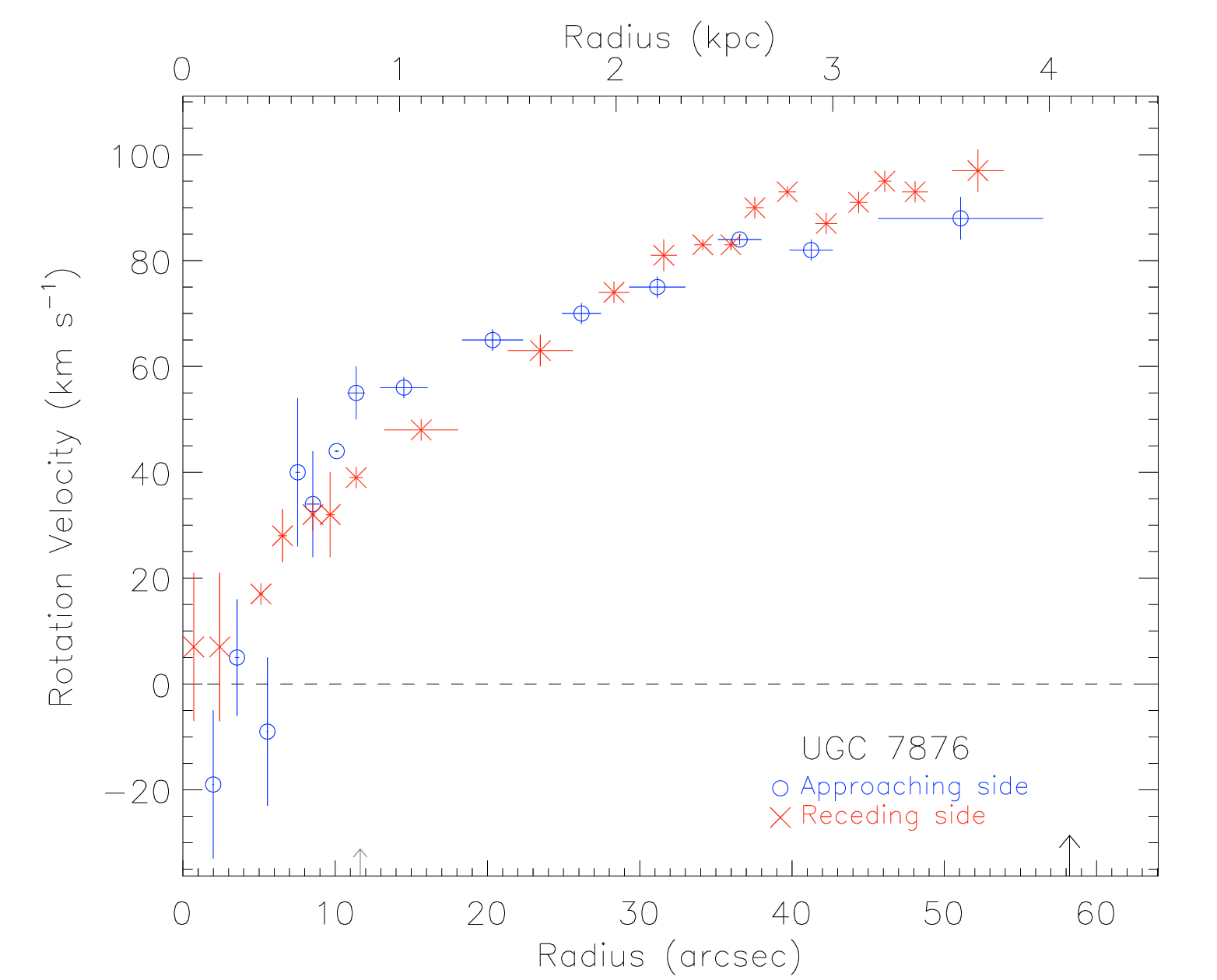}
\end{center}
\caption{From top left to bottom right: \ha~\RC~of UGC 7154, UGC 7766, UGC 7831, UGC 7853, UGC 7861, and UGC 7876.
}
\end{minipage}
\end{figure*}
\clearpage
\begin{figure*}
\begin{minipage}{180mm}
\begin{center}
   \includegraphics[width=8cm]{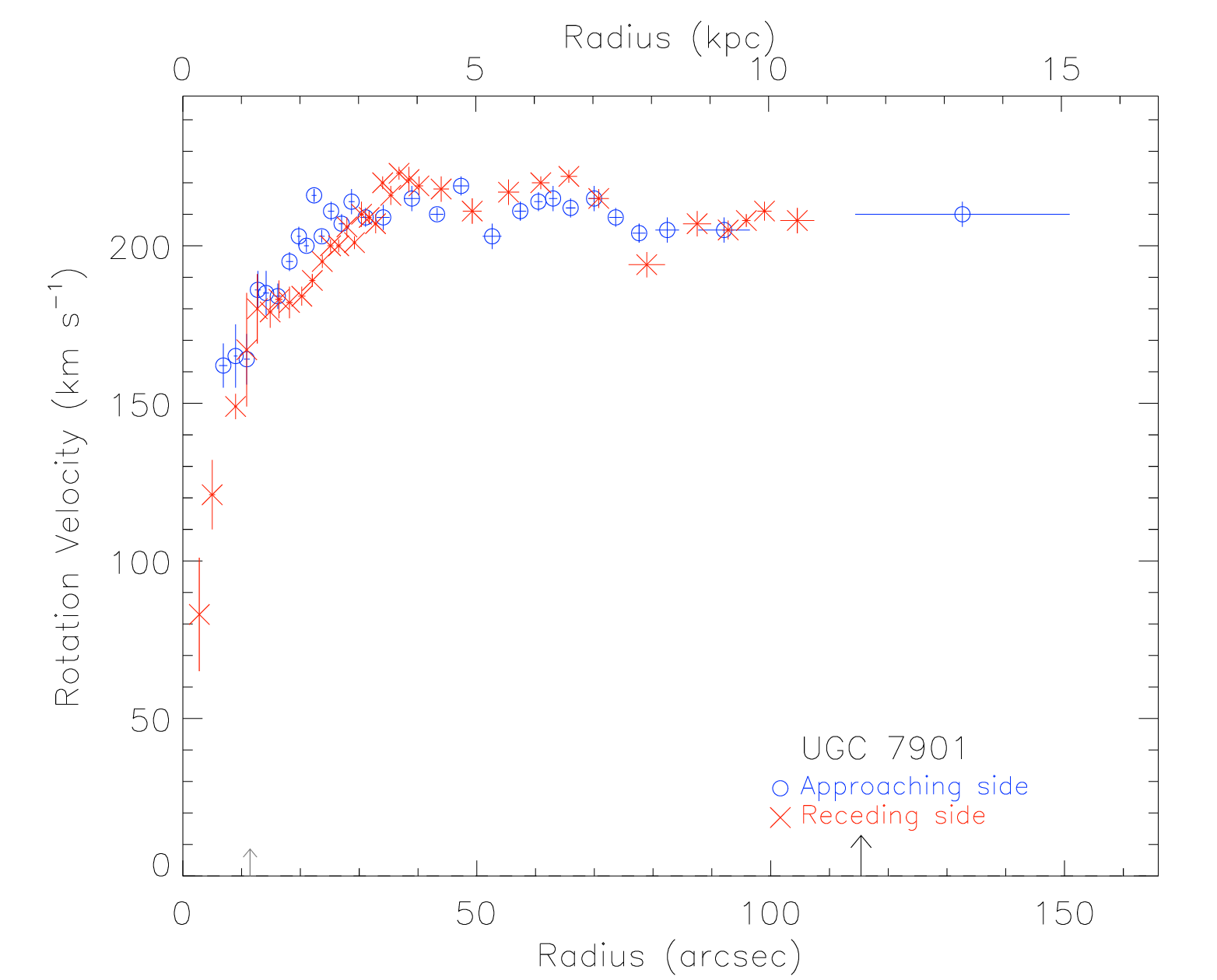}
   \includegraphics[width=8cm]{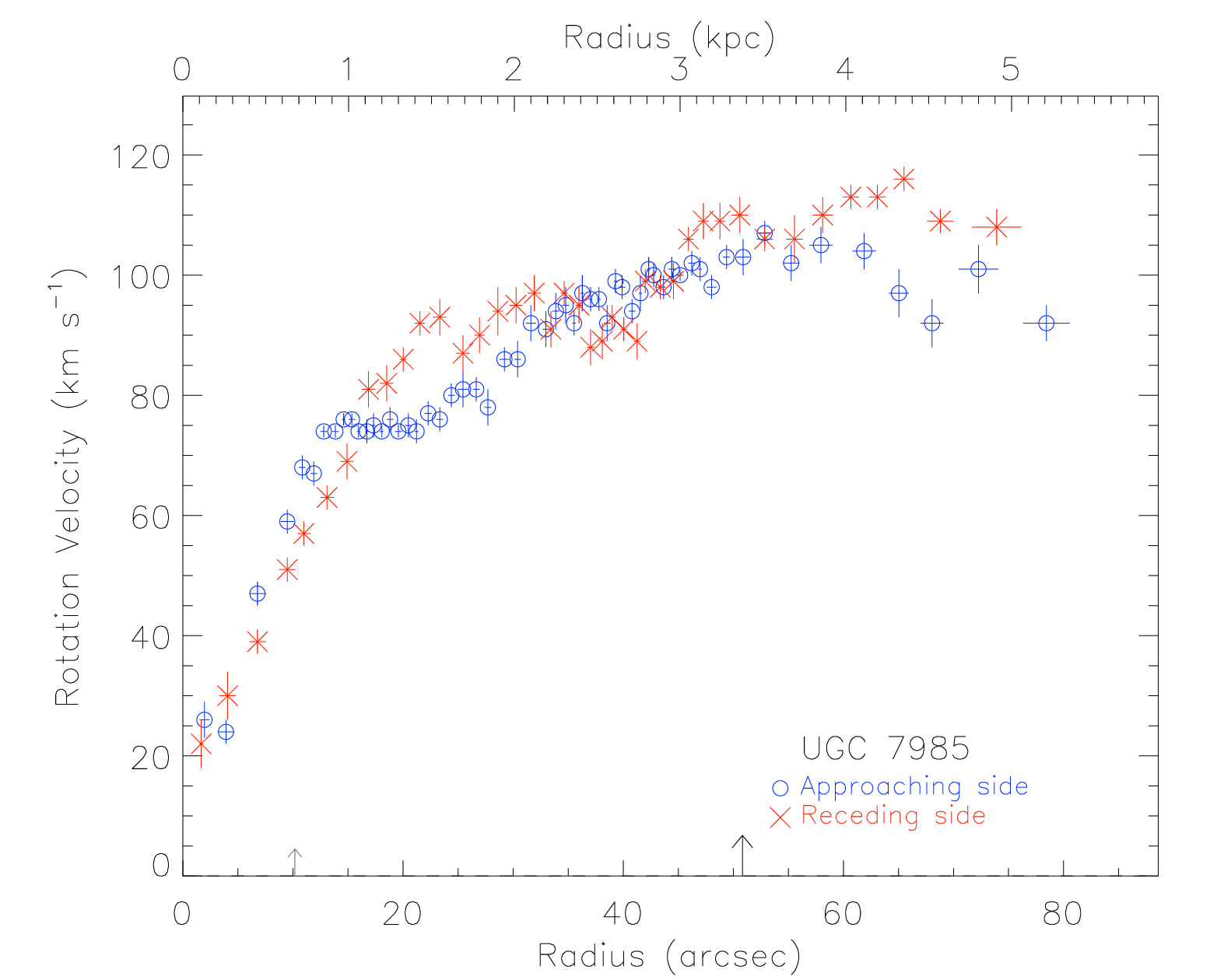}
   \includegraphics[width=8cm]{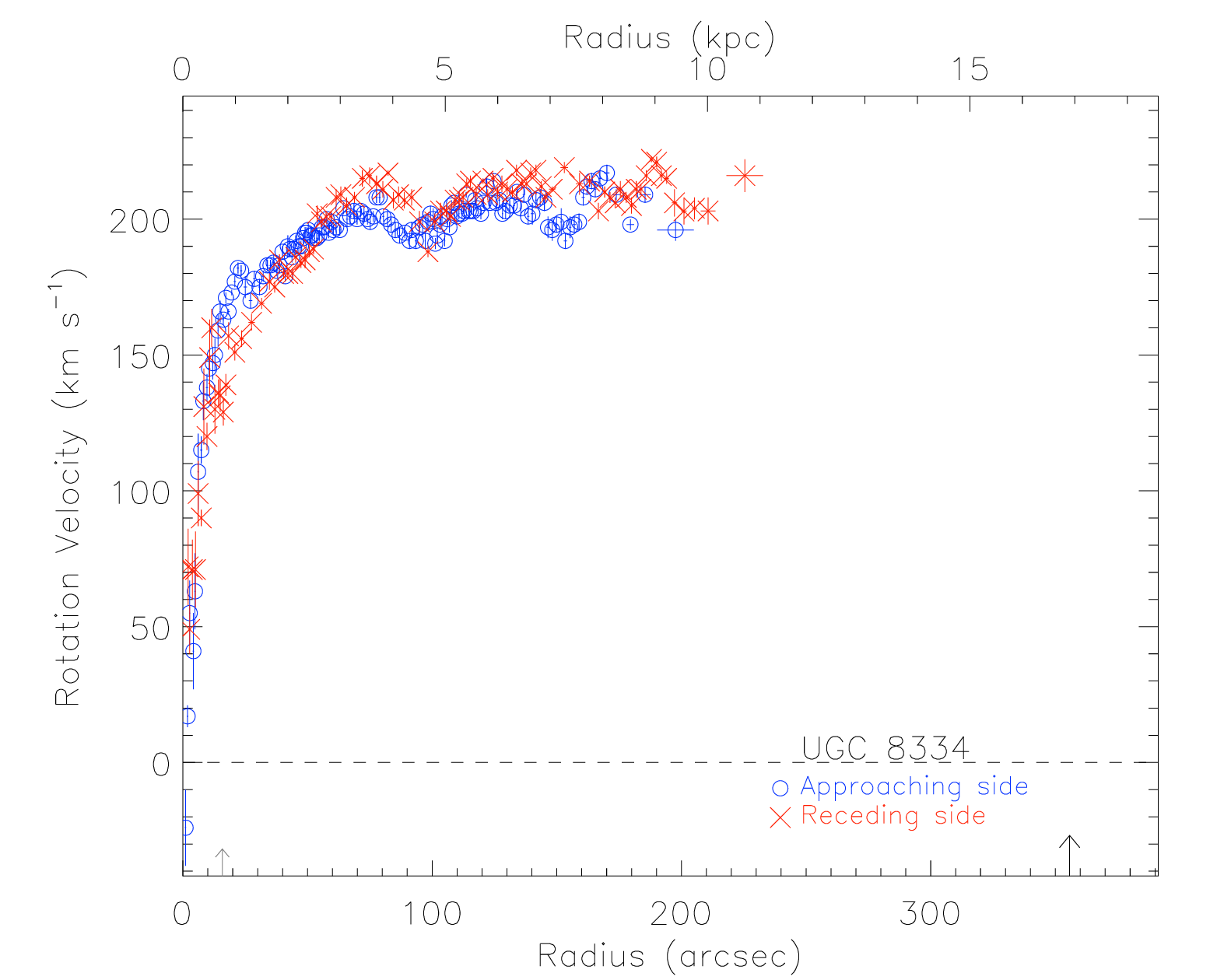}
   \includegraphics[width=8cm]{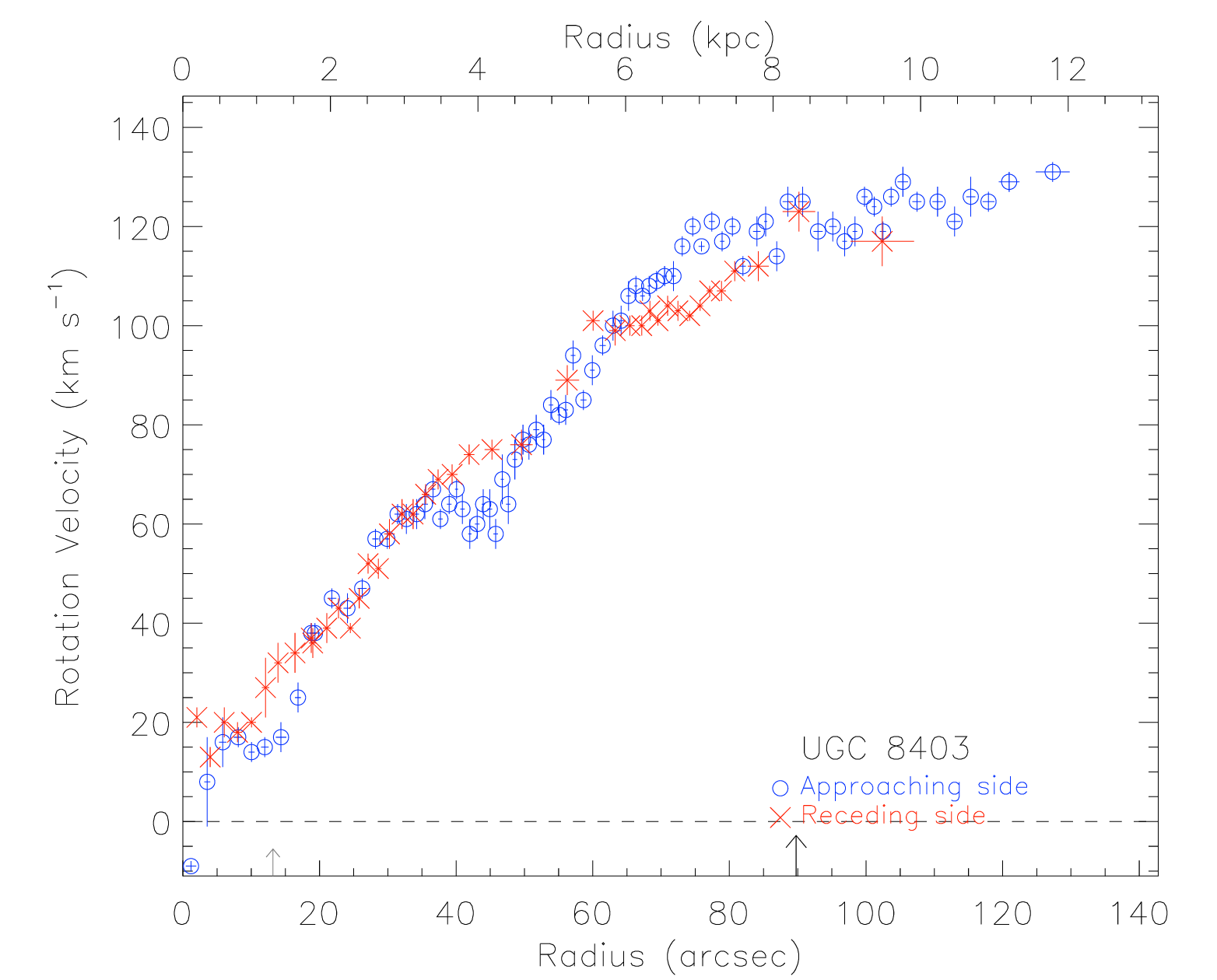}
   \includegraphics[width=8cm]{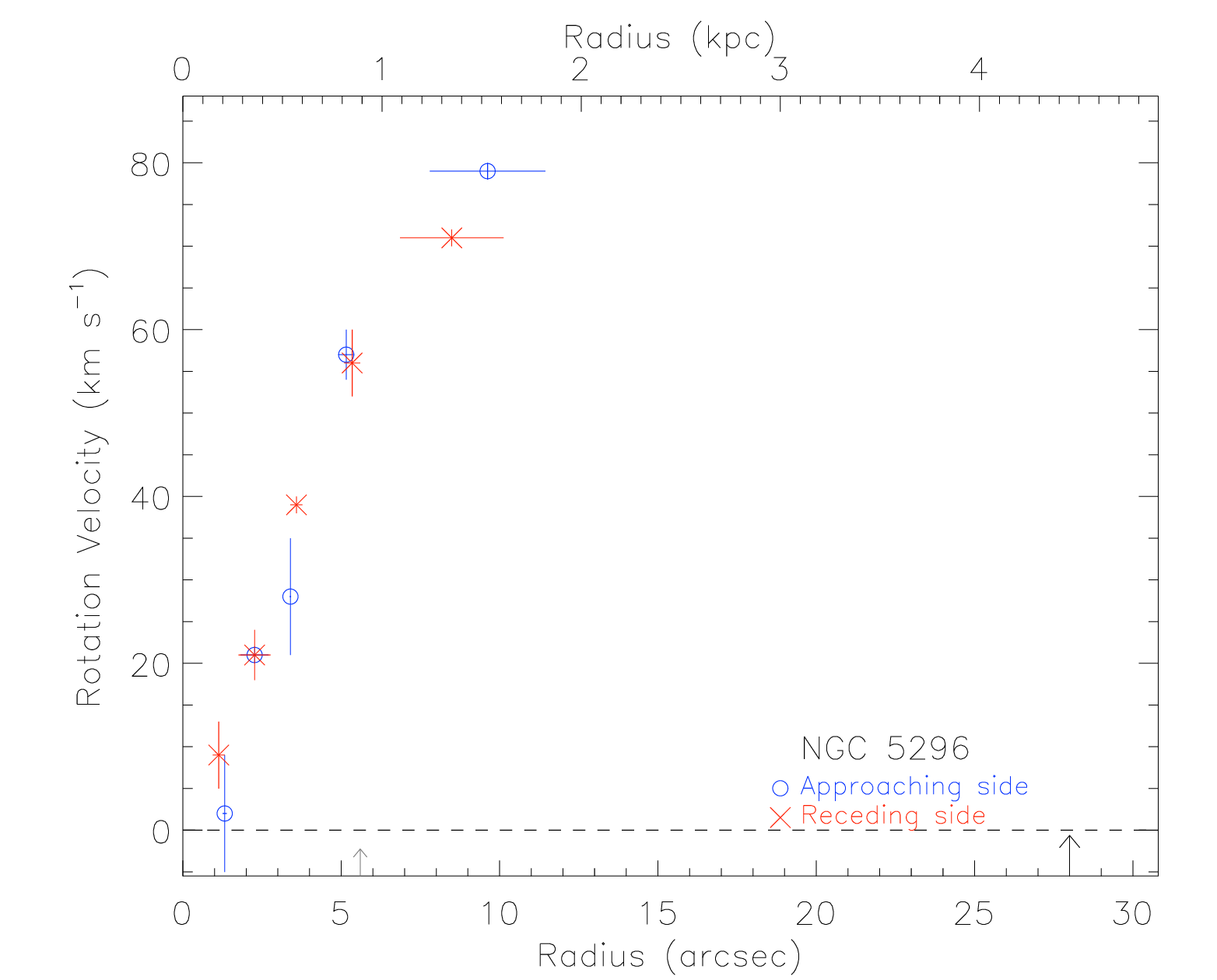}
   \includegraphics[width=8cm]{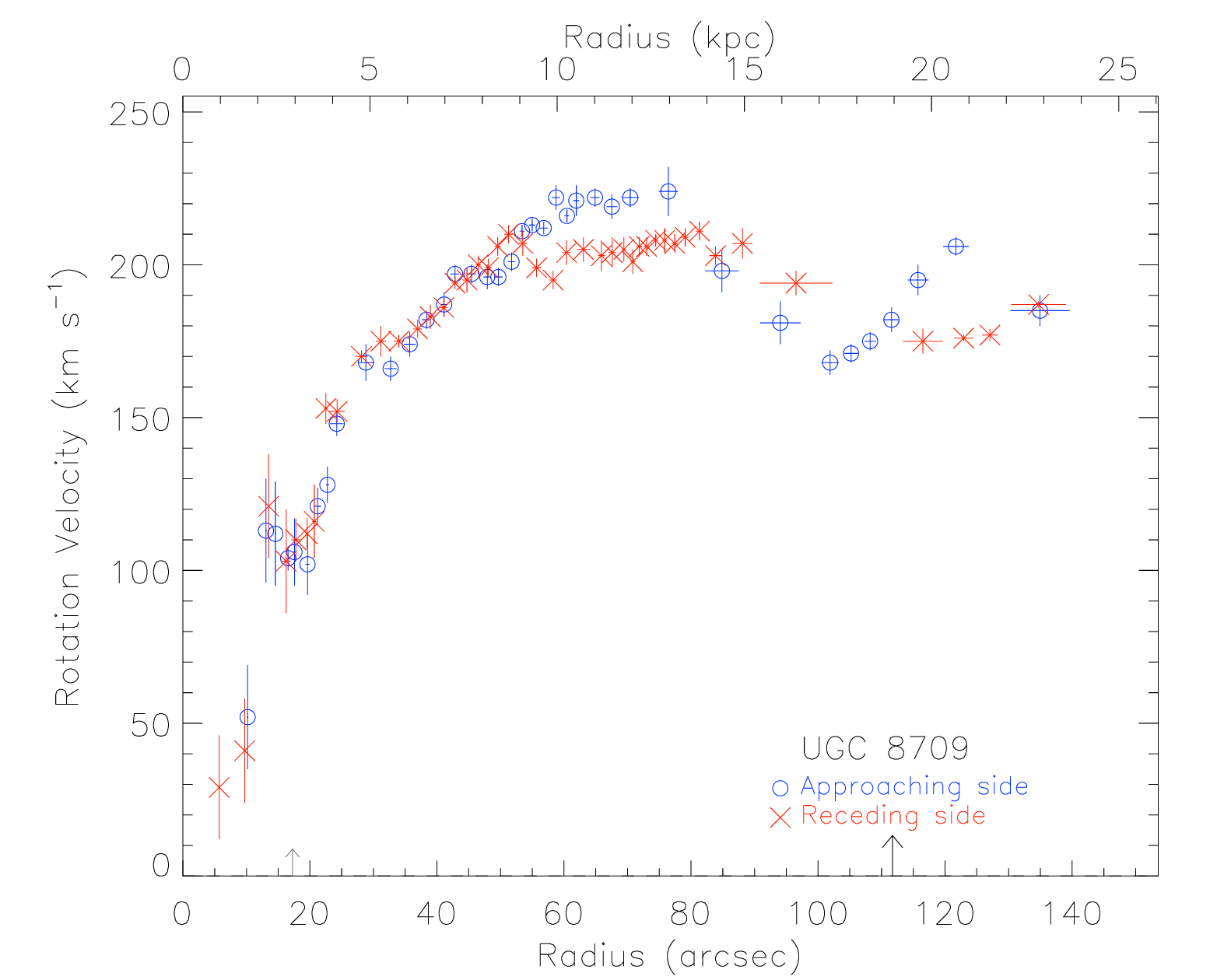}
\end{center}
\caption{From top left to bottom right: \ha~\RC~of UGC 7901, UGC 7985, UGC 8334, UGC 8403, NGC 5296, and UGC 8709.
}
\end{minipage}
\end{figure*}
\clearpage
\begin{figure*}
\begin{minipage}{180mm}
\begin{center}
   \includegraphics[width=8cm]{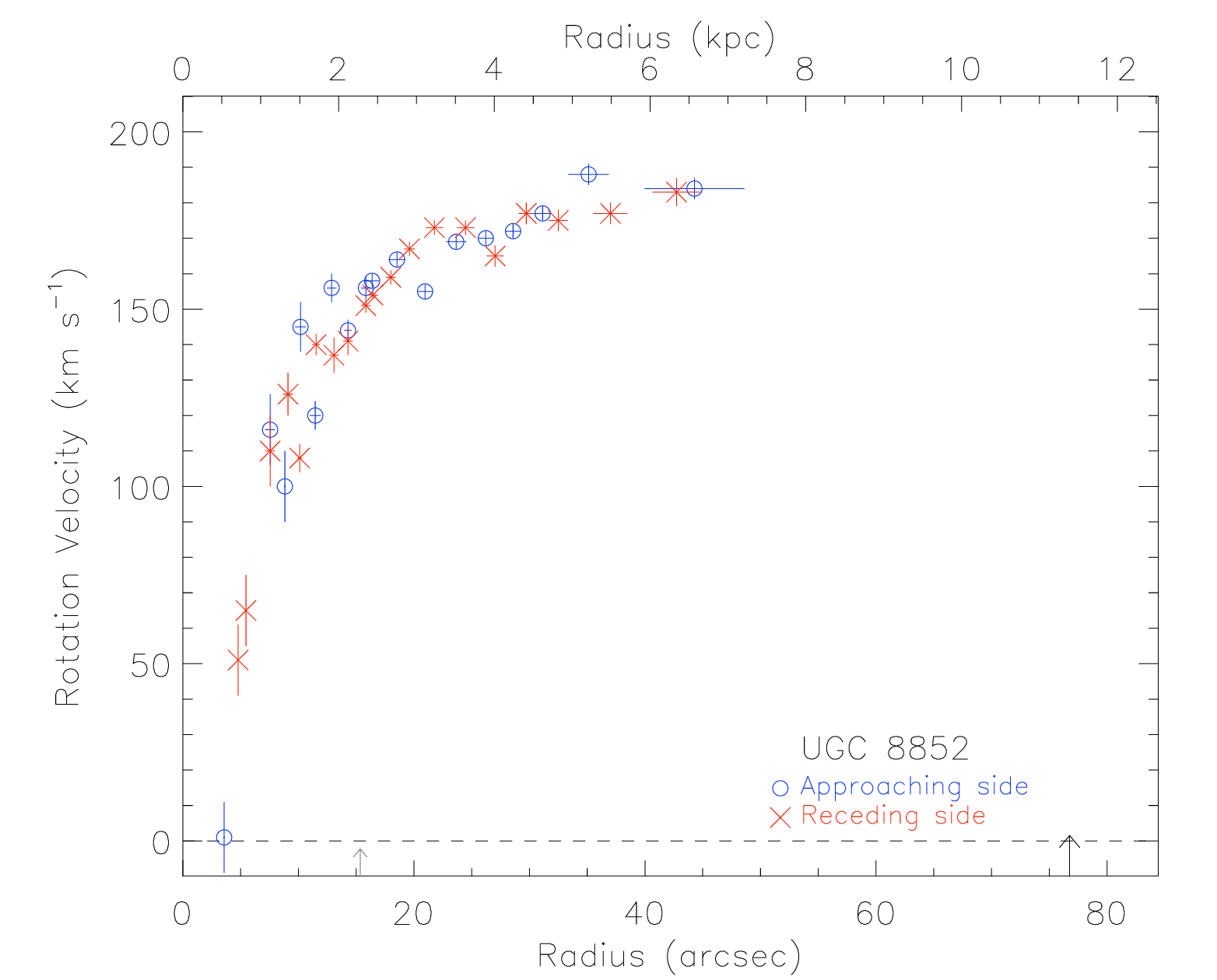}
   \includegraphics[width=8cm]{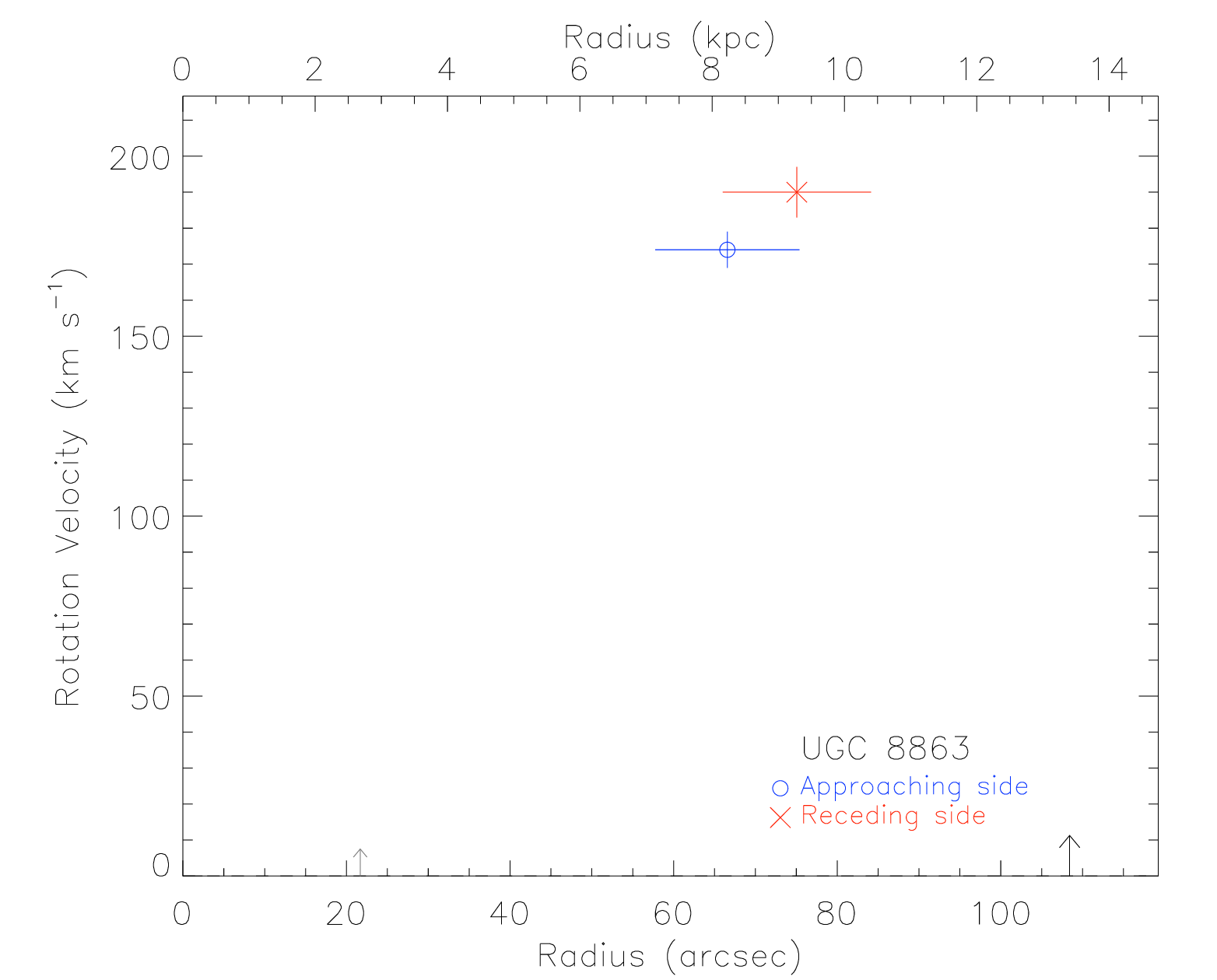}
   \includegraphics[width=8cm]{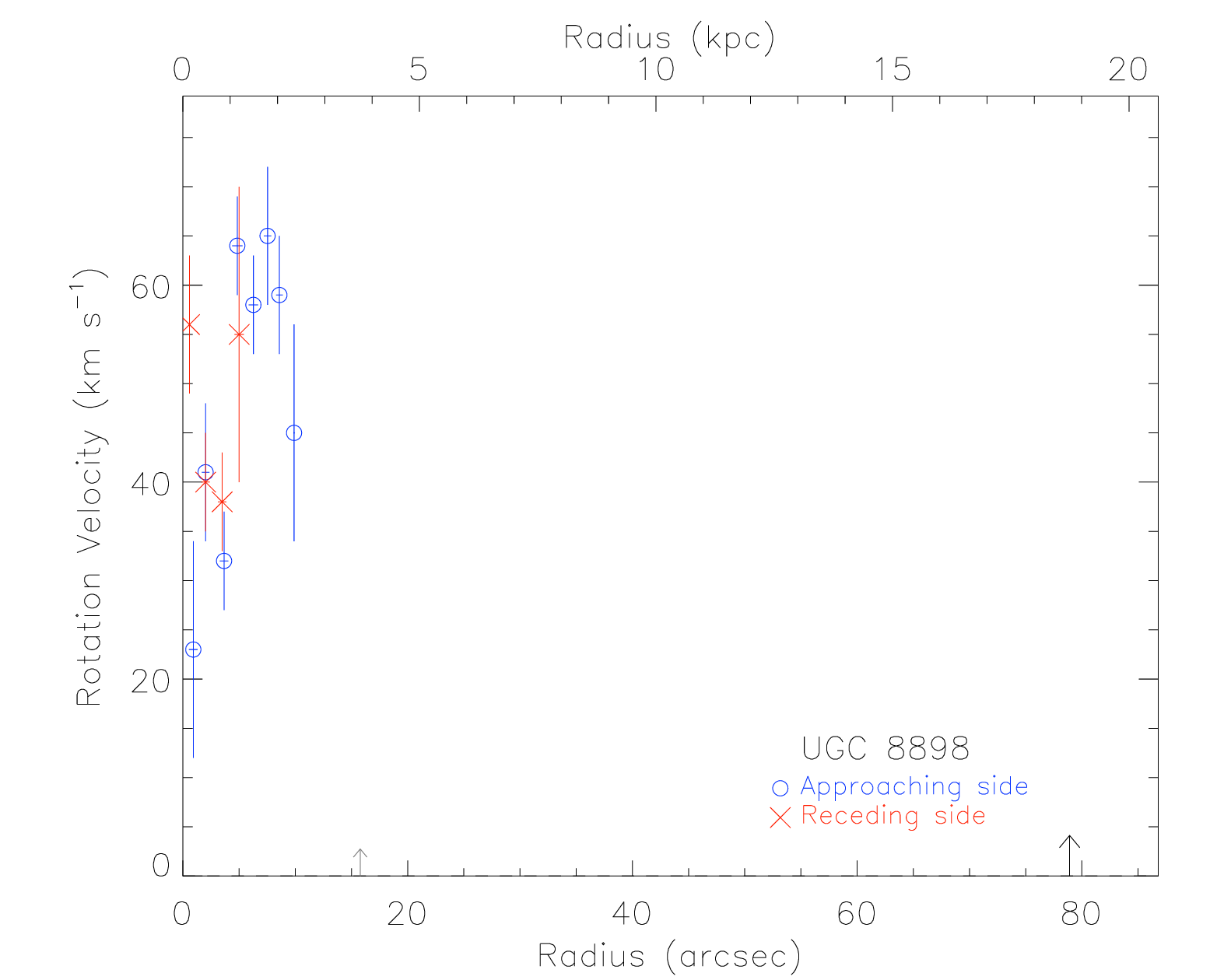}
   \includegraphics[width=8cm]{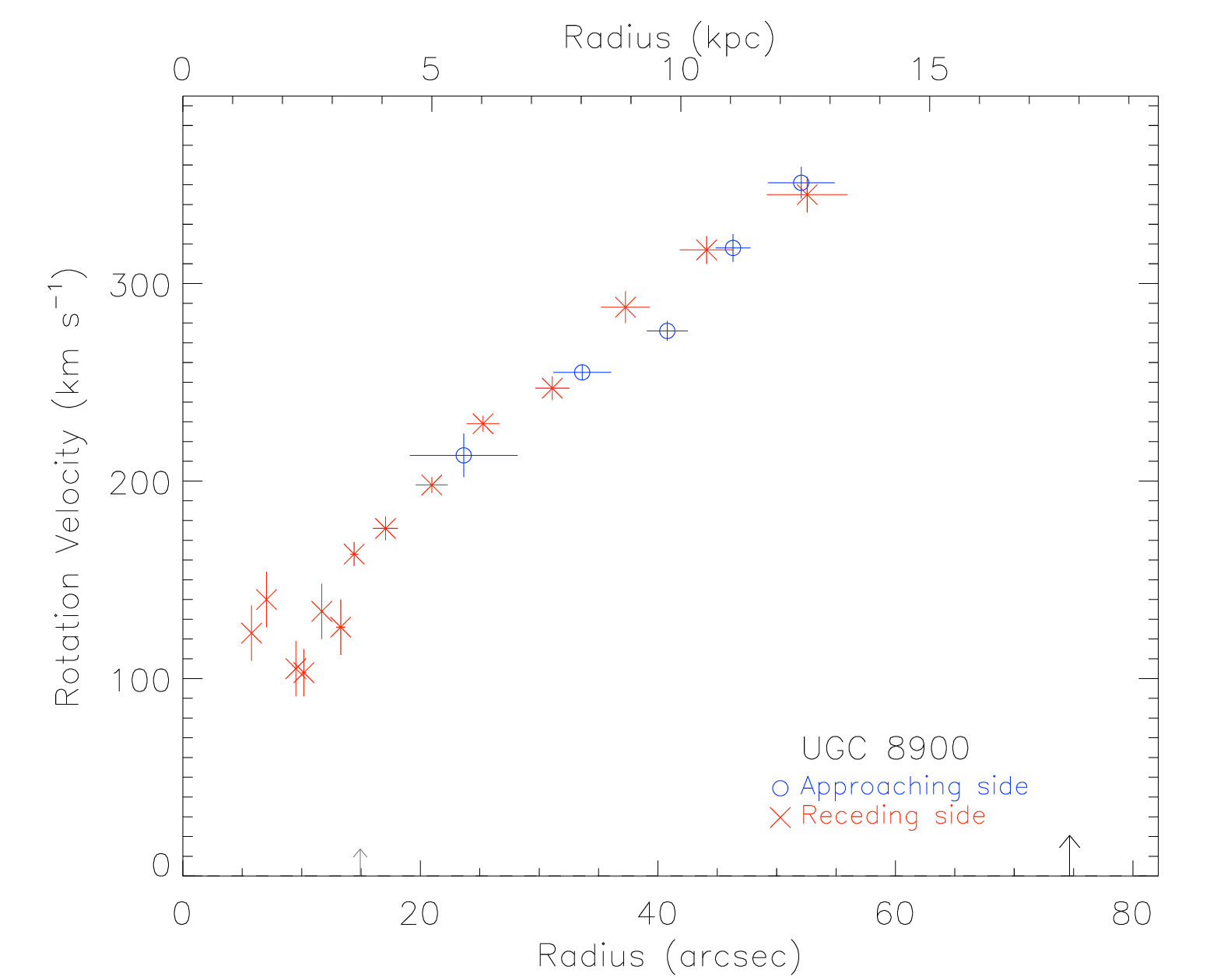}
   \includegraphics[width=8cm]{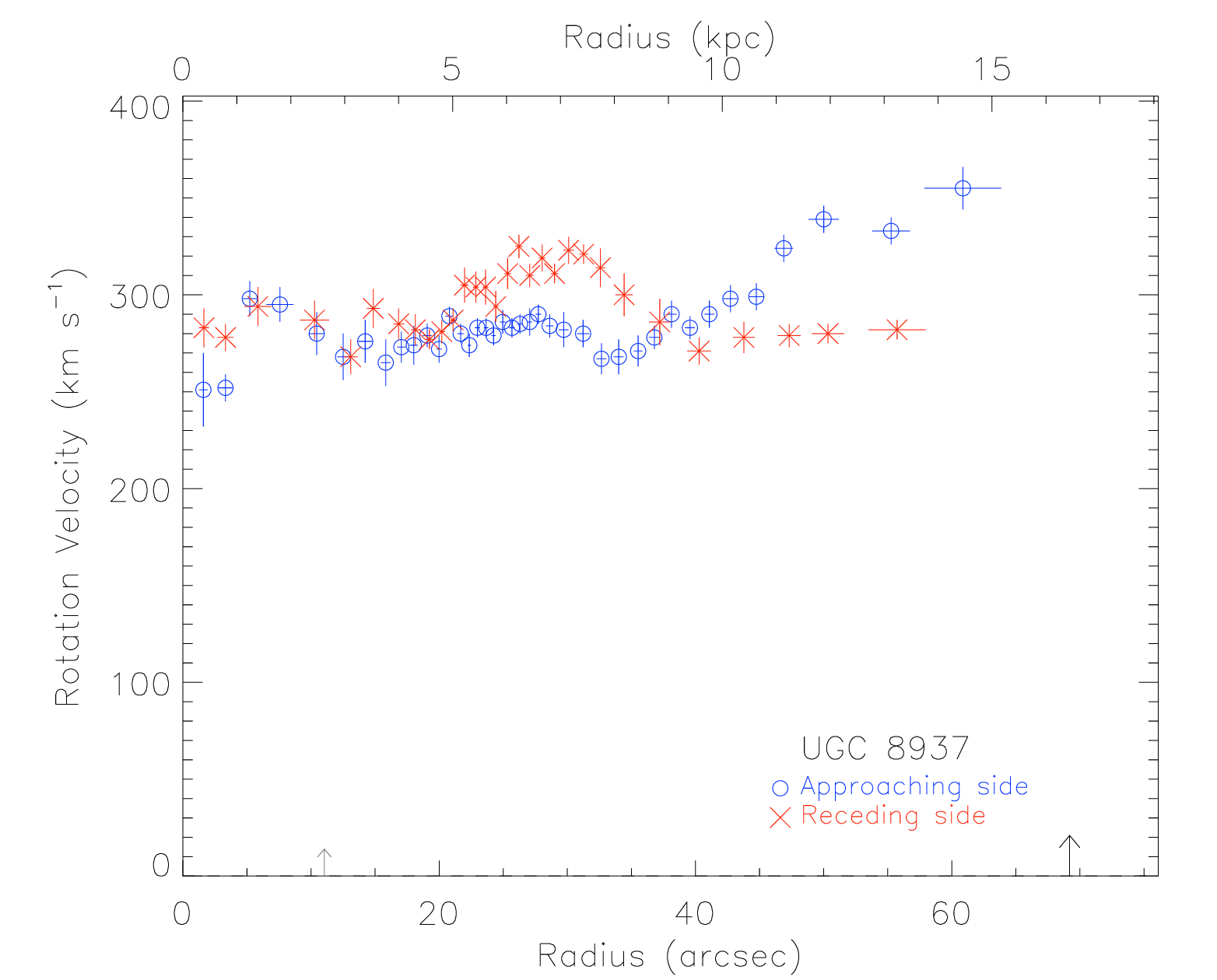}
   \includegraphics[width=8cm]{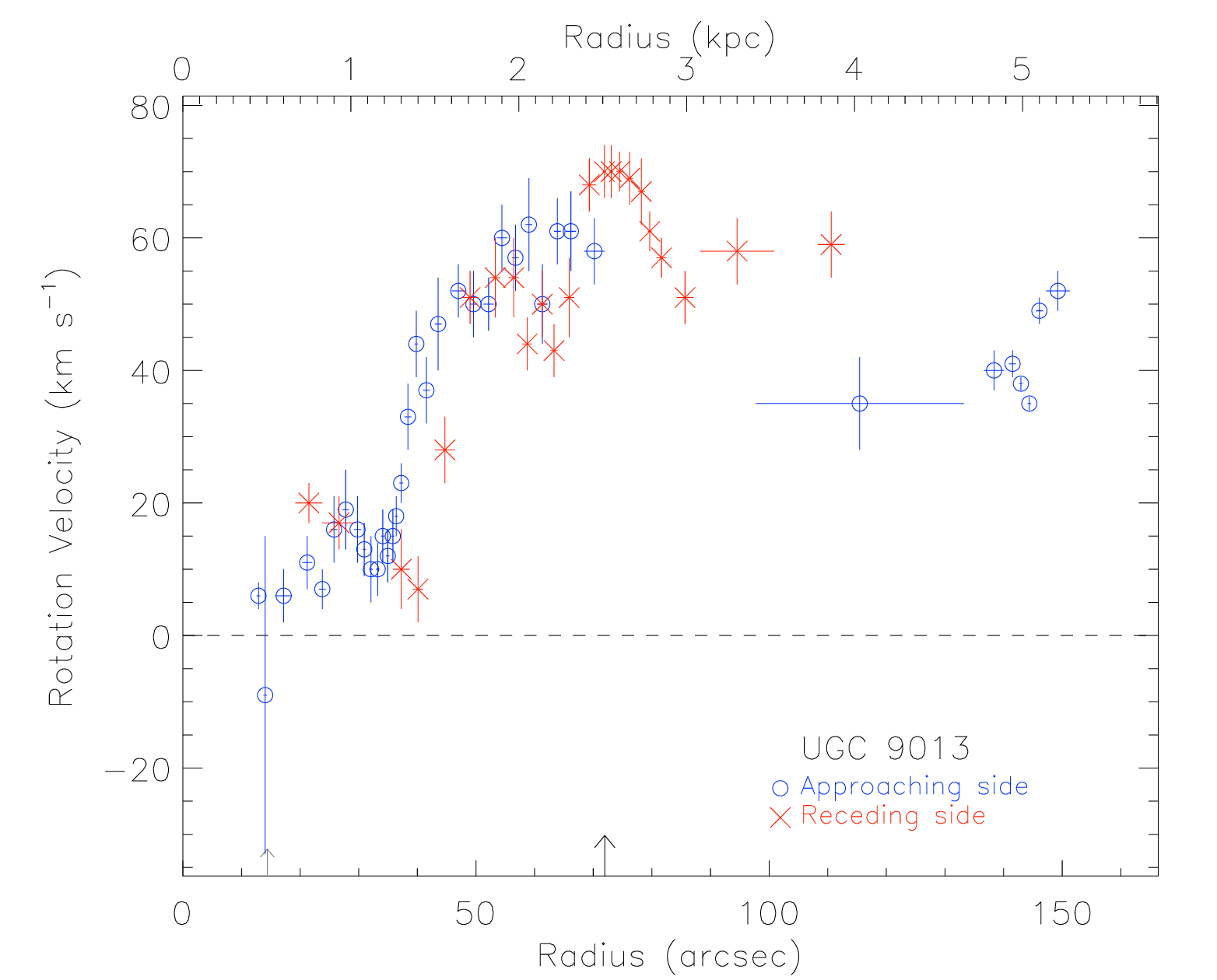}
\end{center}
\caption{From top left to bottom right: \ha~\RC~of UGC 8852, UGC 8863, UGC 8898, UGC 8900, UGC 8937, and UGC 9013.
}
\end{minipage}
\end{figure*}
\clearpage
\begin{figure*}
\begin{minipage}{180mm}
\begin{center}
   \includegraphics[width=8cm]{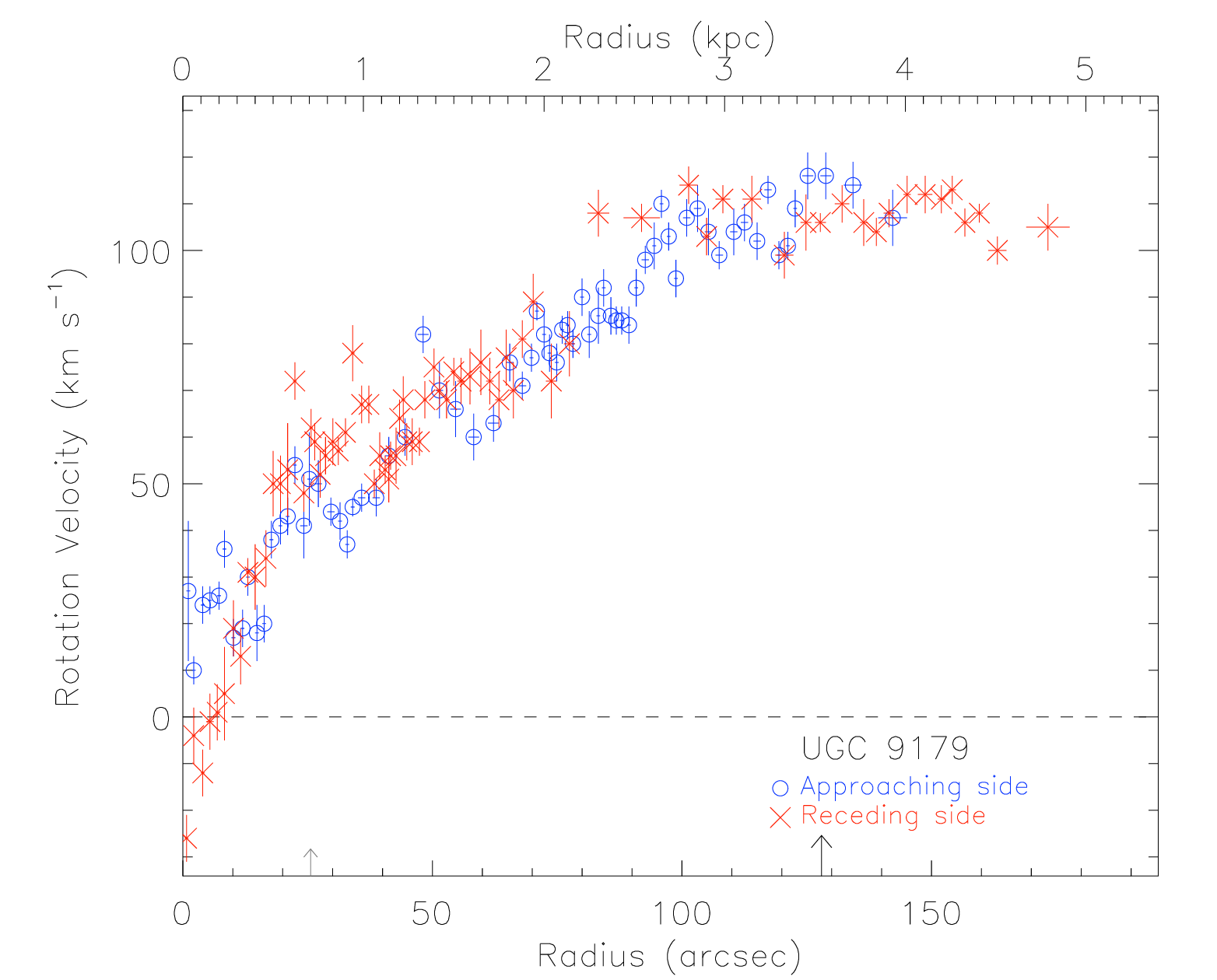}
   \includegraphics[width=8cm]{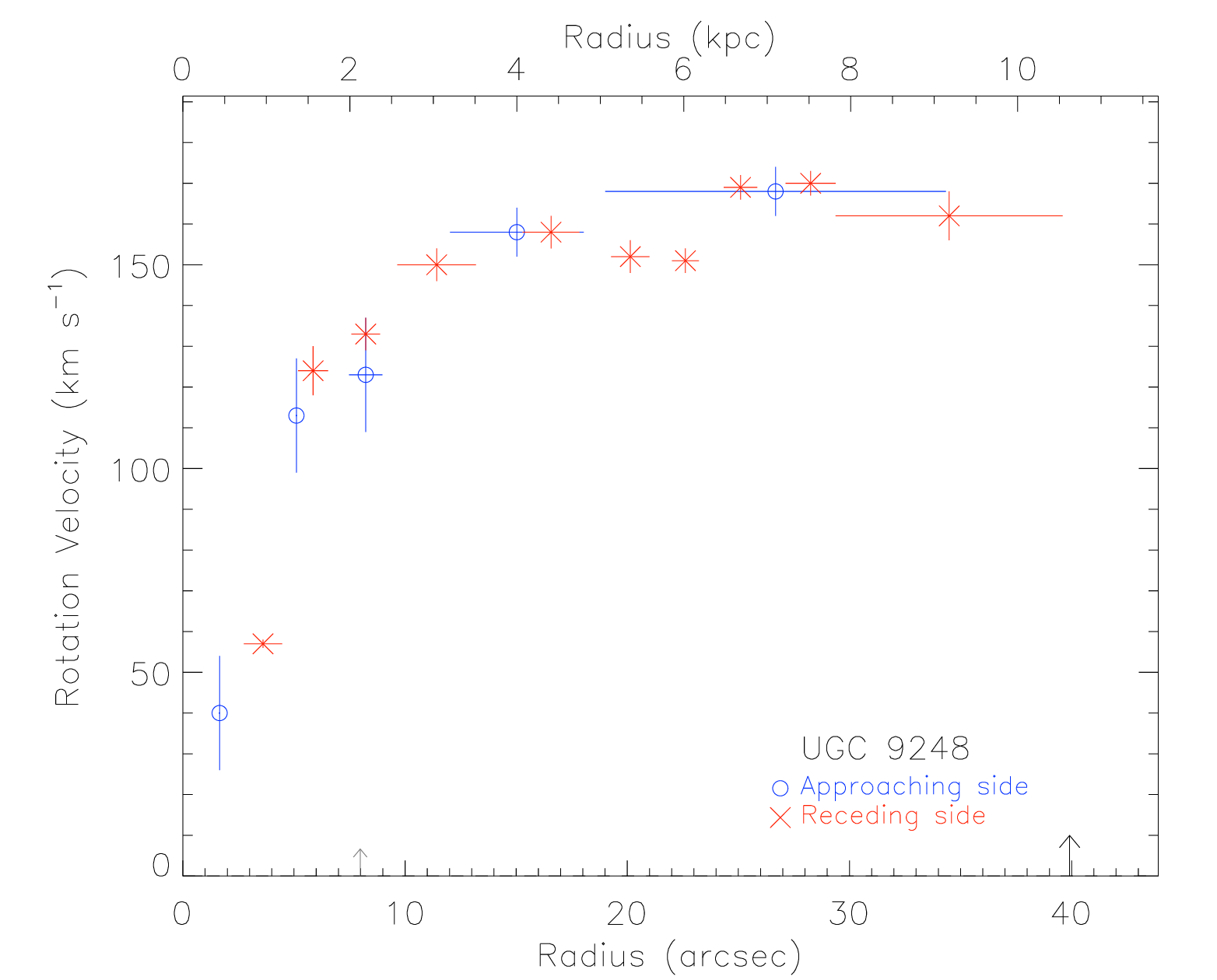}
   \includegraphics[width=8cm]{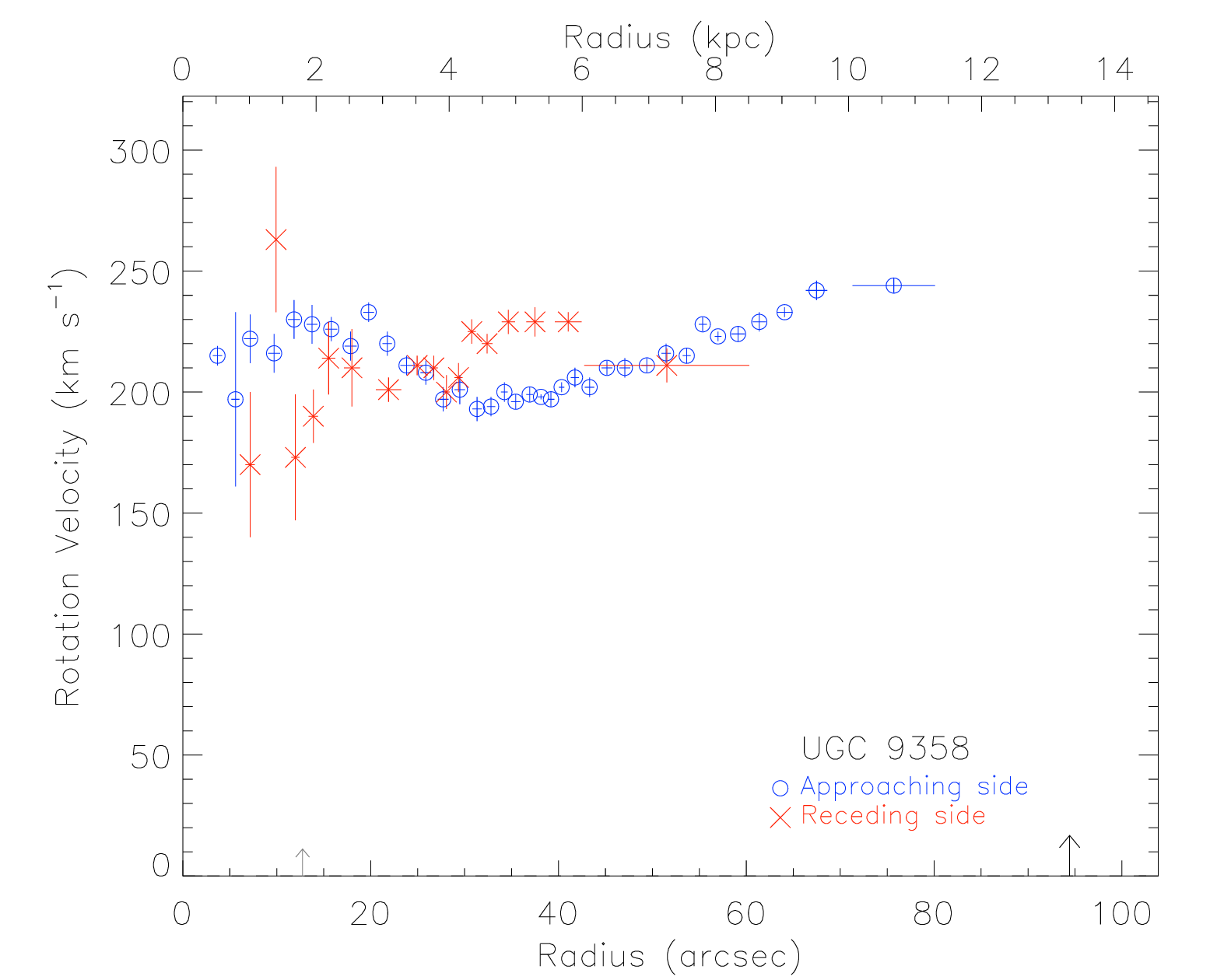}
   \includegraphics[width=8cm]{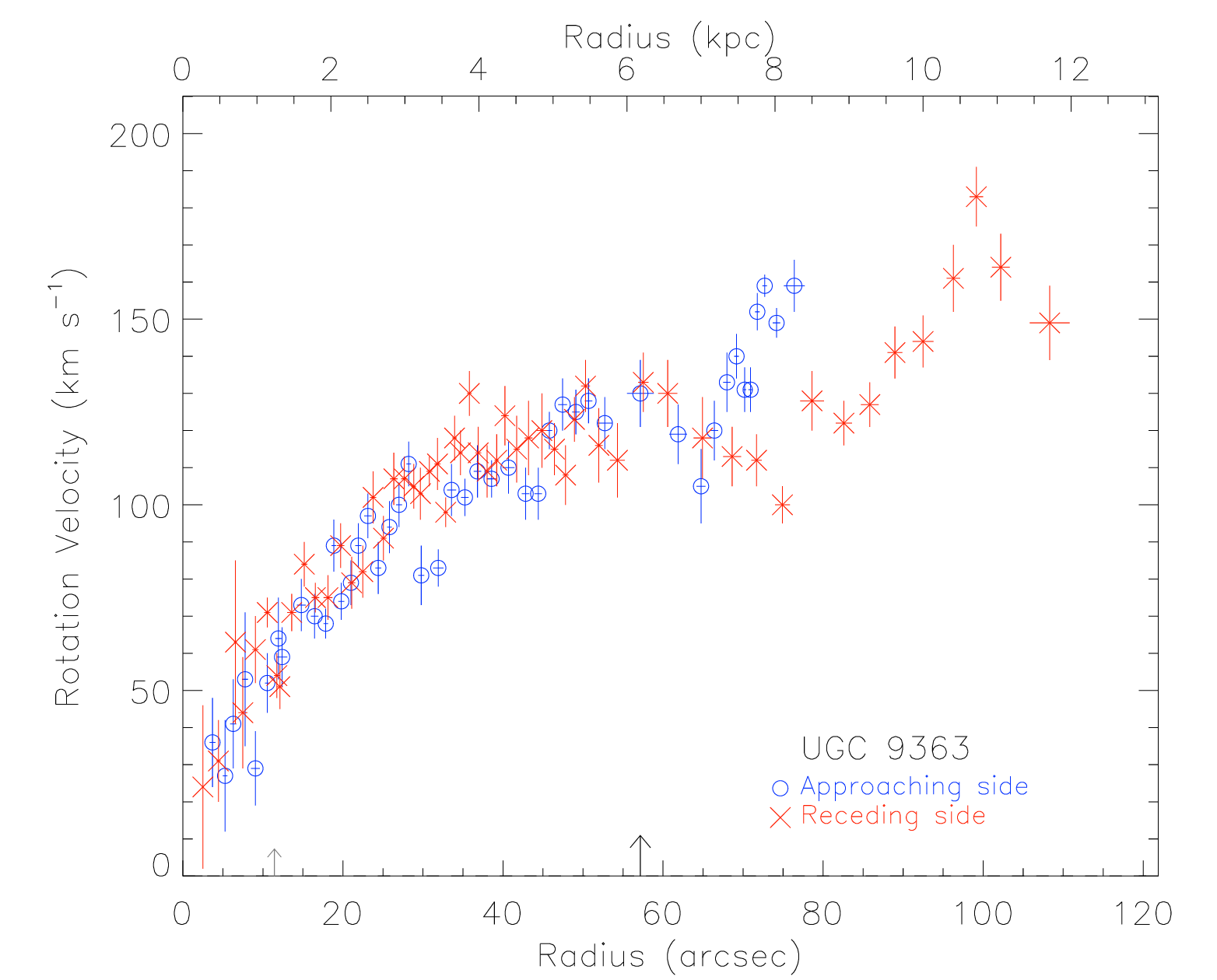}
   \includegraphics[width=8cm]{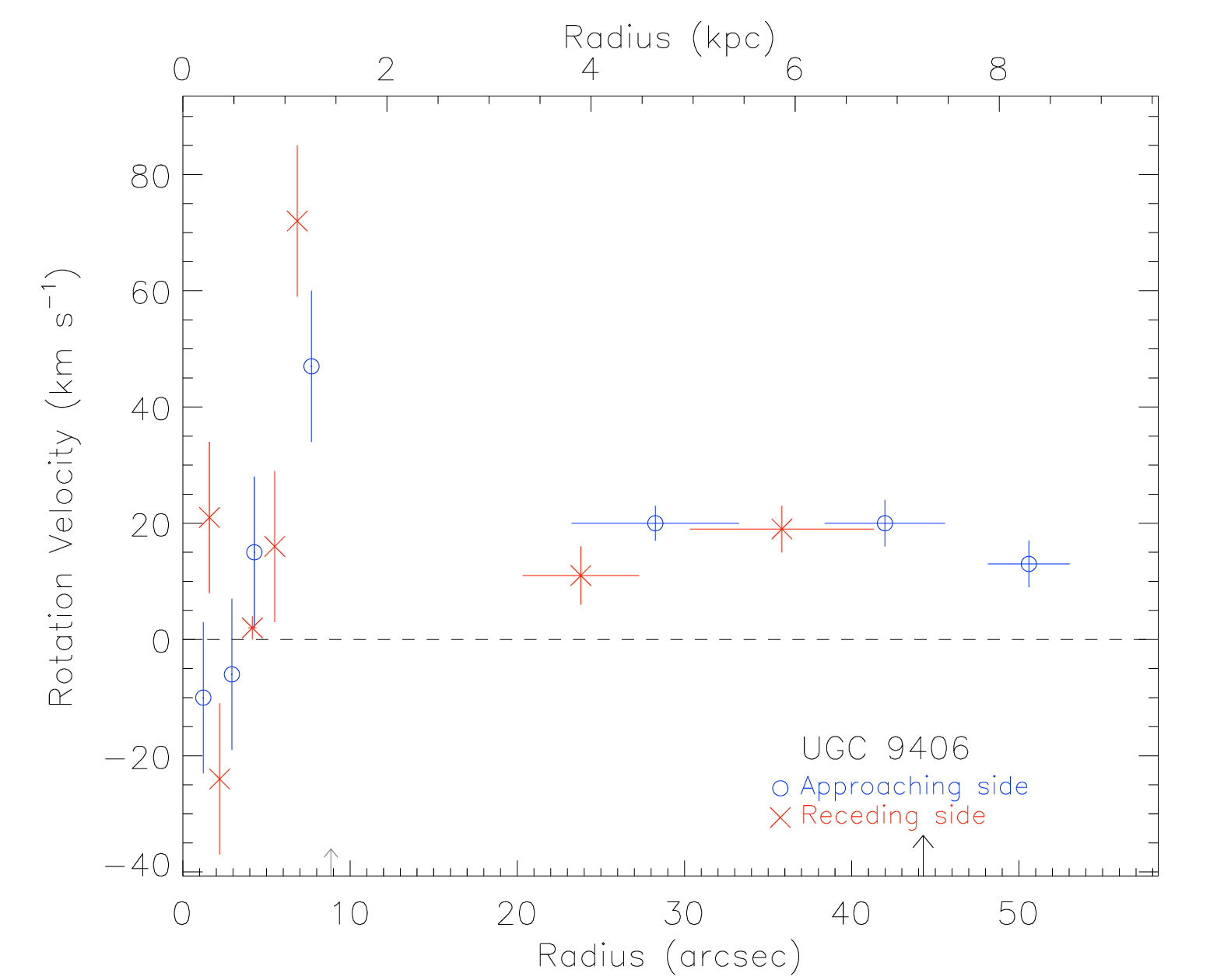}
   \includegraphics[width=8cm]{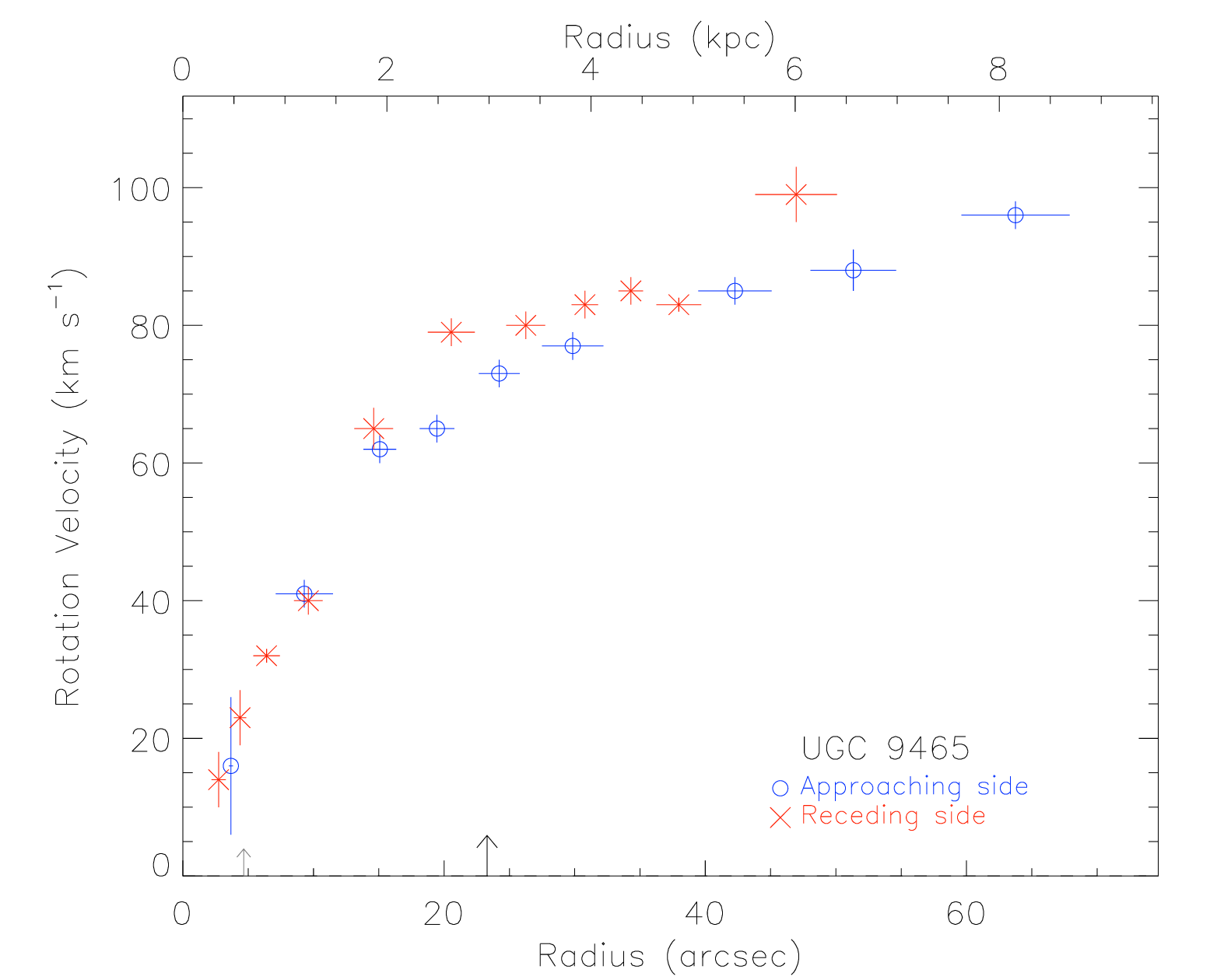}
\end{center}
\caption{From top left to bottom right: \ha~\RC~of UGC 9179, UGC 9248, UGC 9358, UGC 9363, UGC 9406, and UGC 9465.
}
\end{minipage}
\end{figure*}
\clearpage
\begin{figure*}
\begin{minipage}{180mm}
\begin{center}
   \includegraphics[width=8cm]{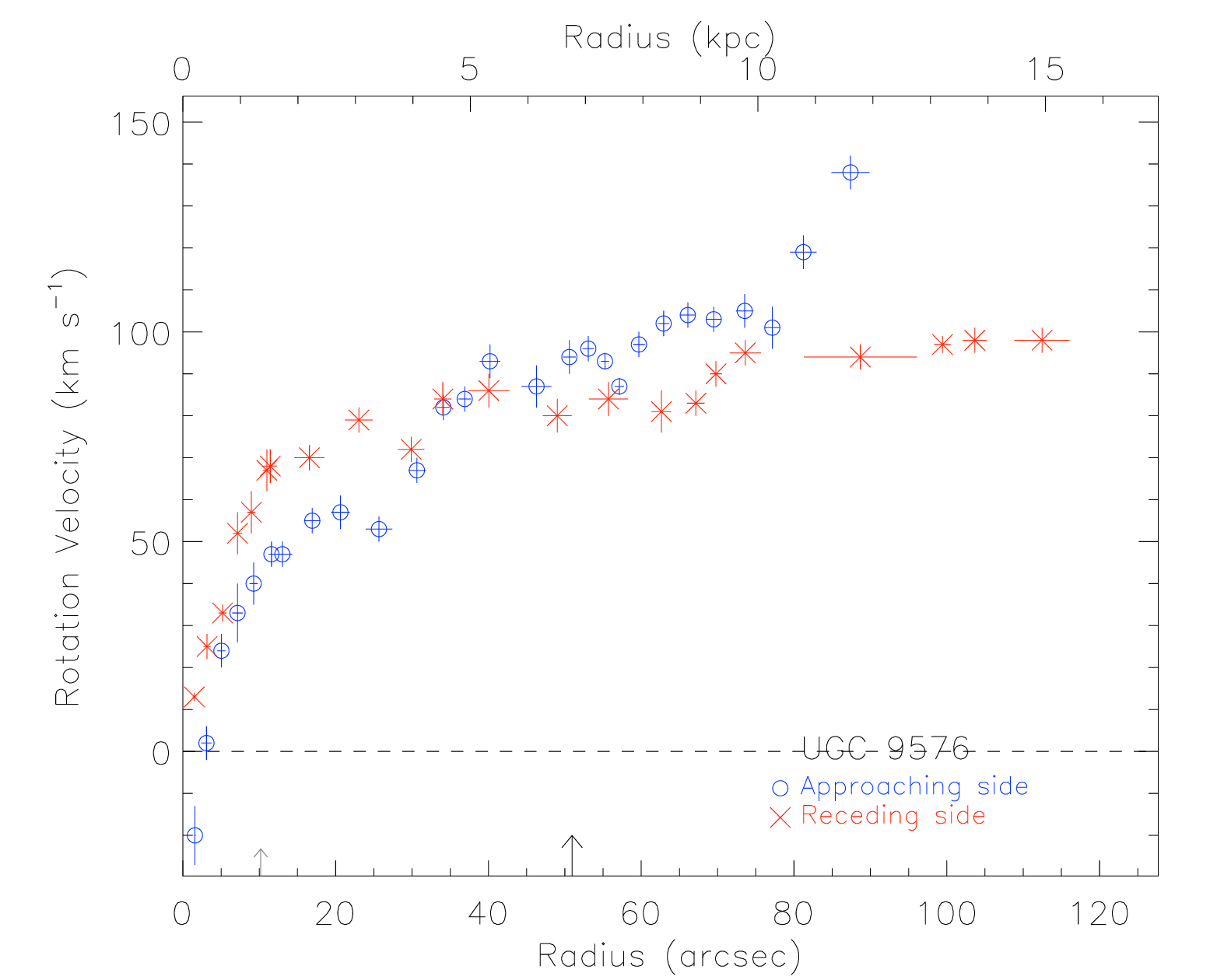}
   \includegraphics[width=8cm]{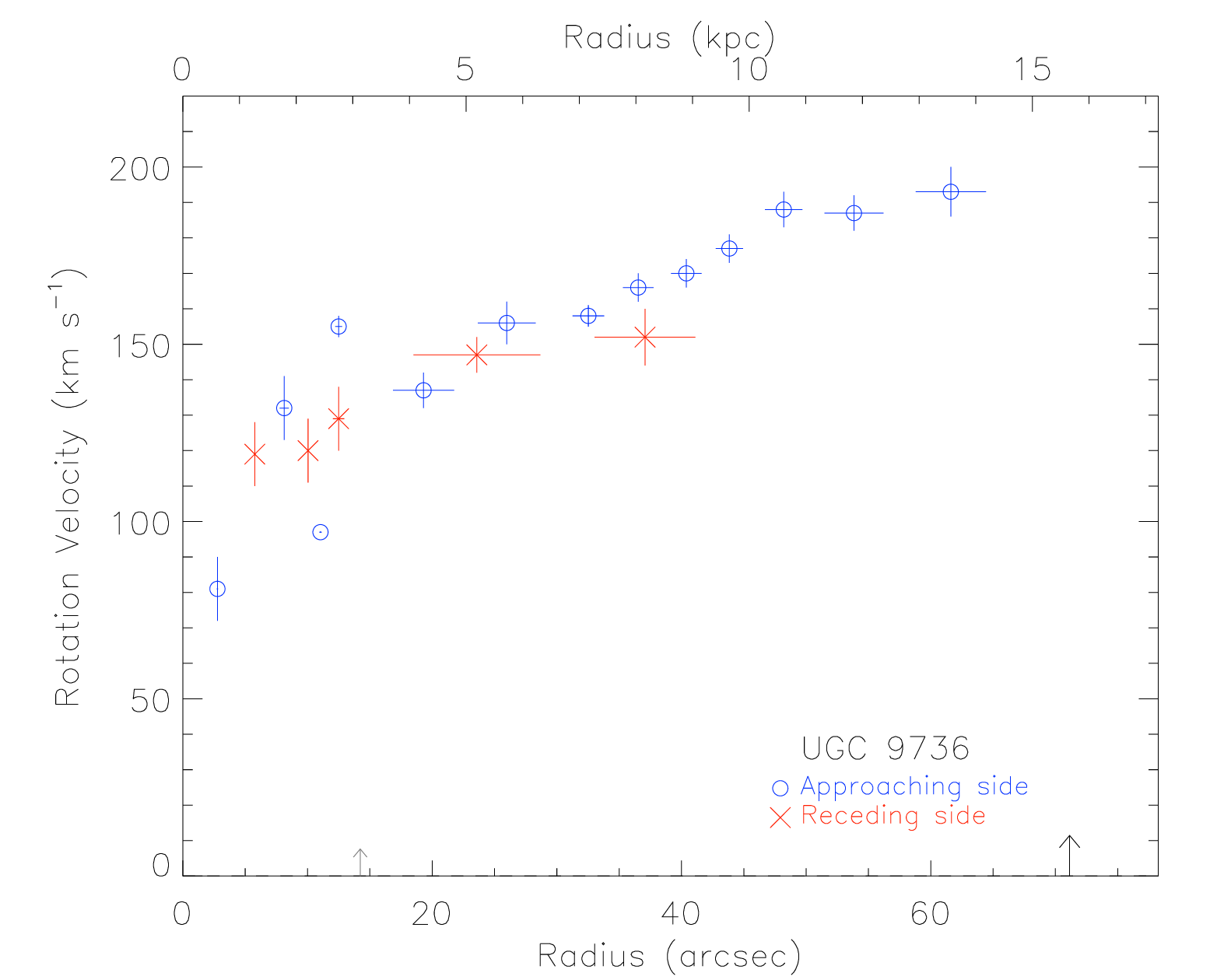}
   \includegraphics[width=8cm]{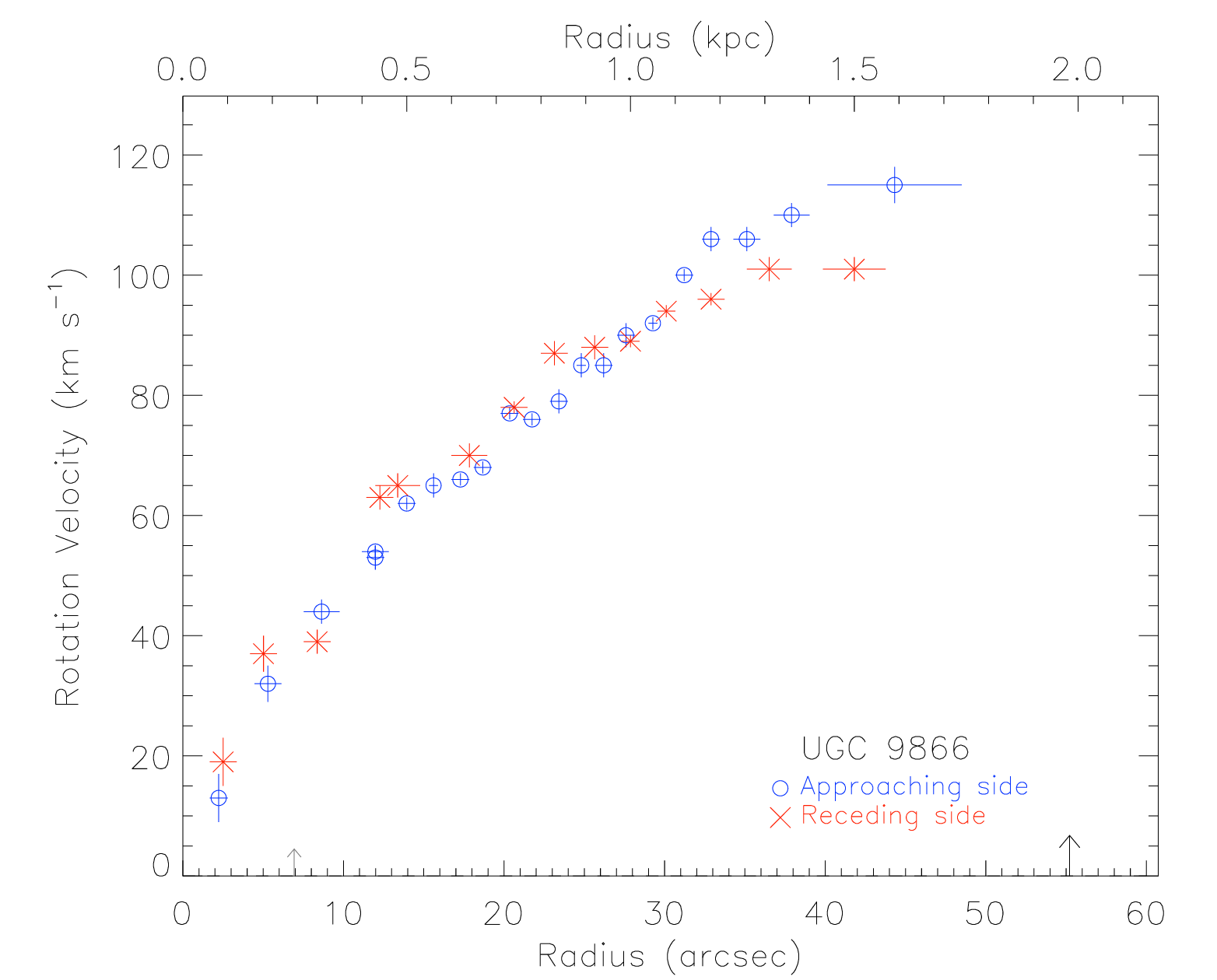}
   \includegraphics[width=8cm]{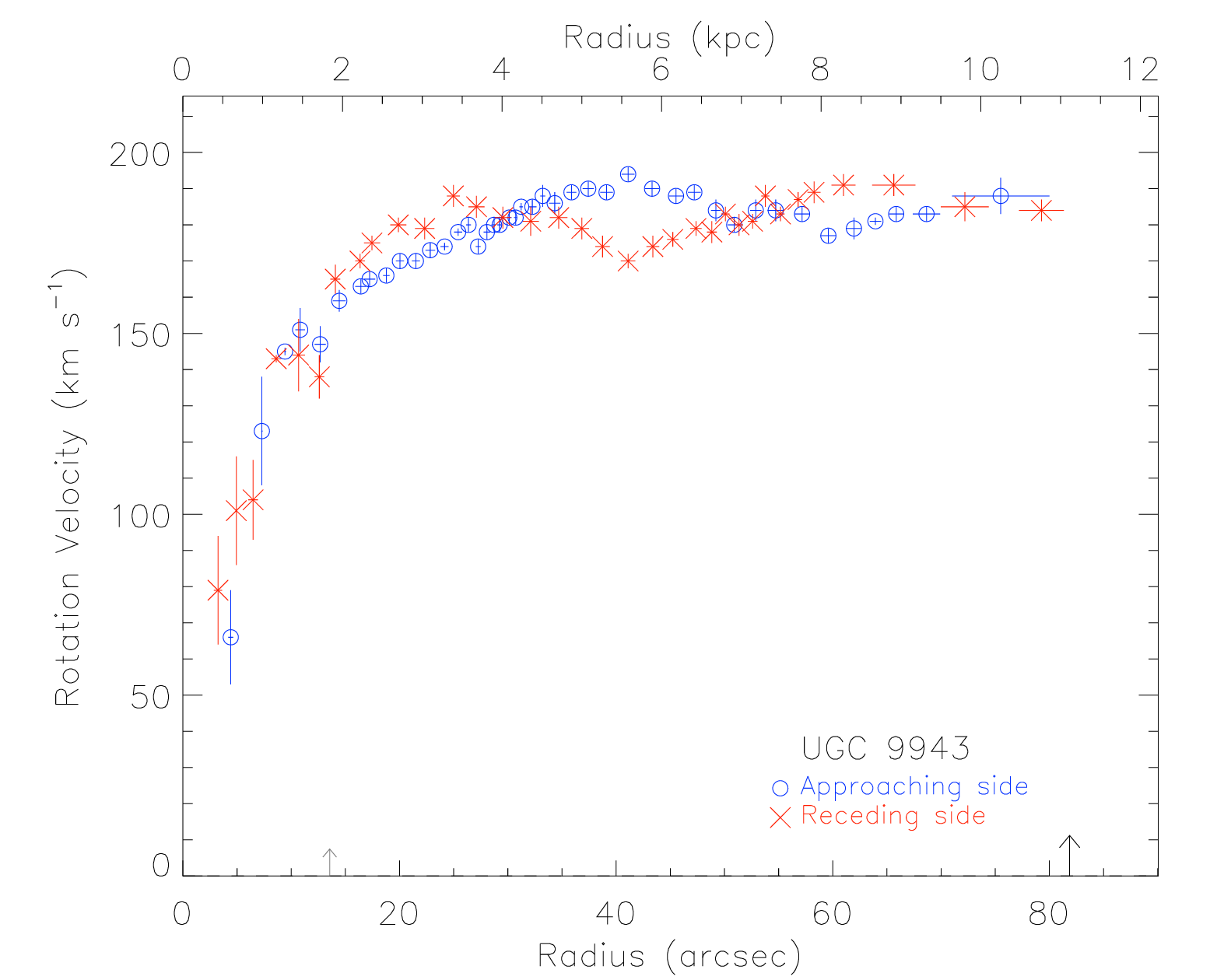}
   \includegraphics[width=8cm]{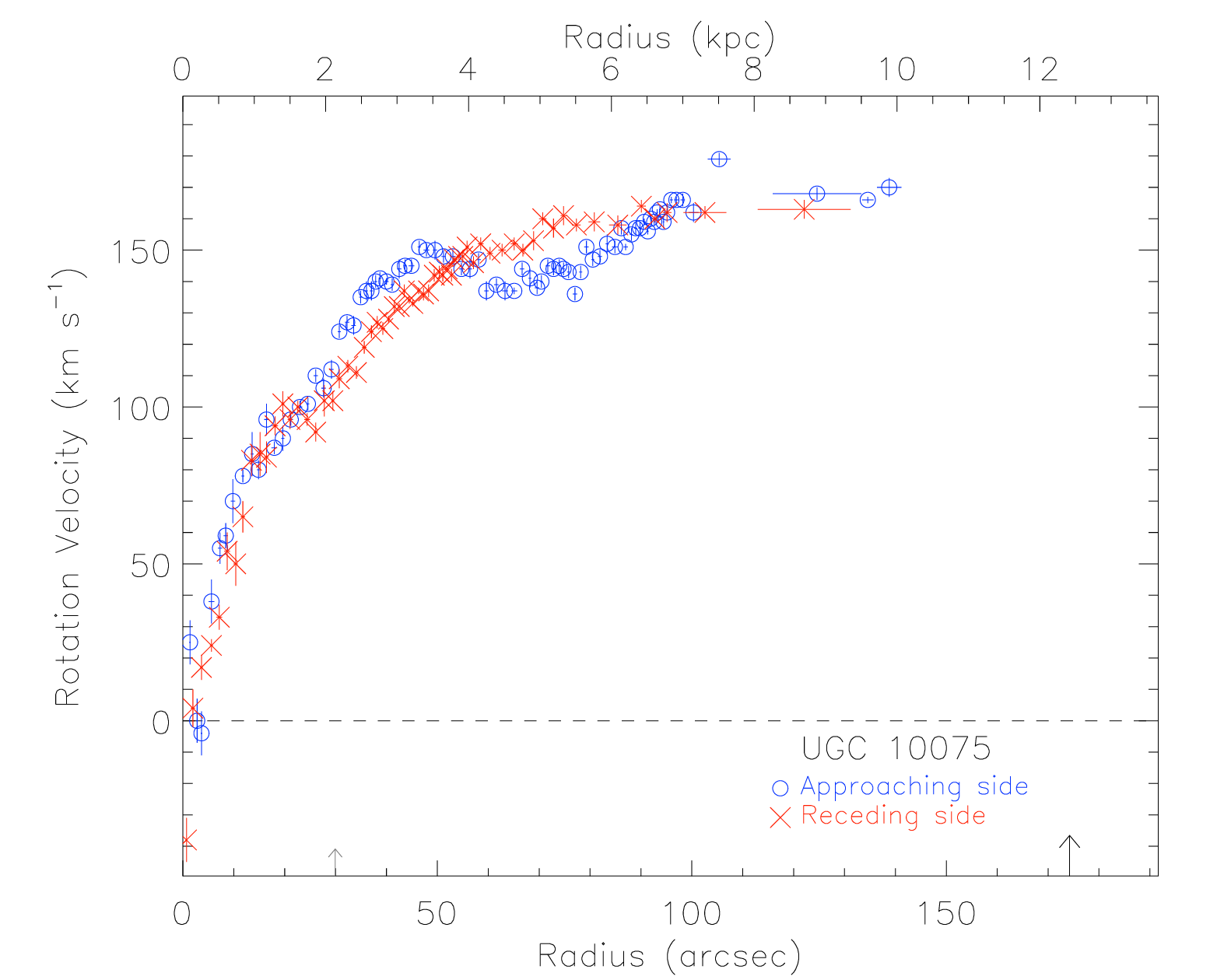}
   \includegraphics[width=8cm]{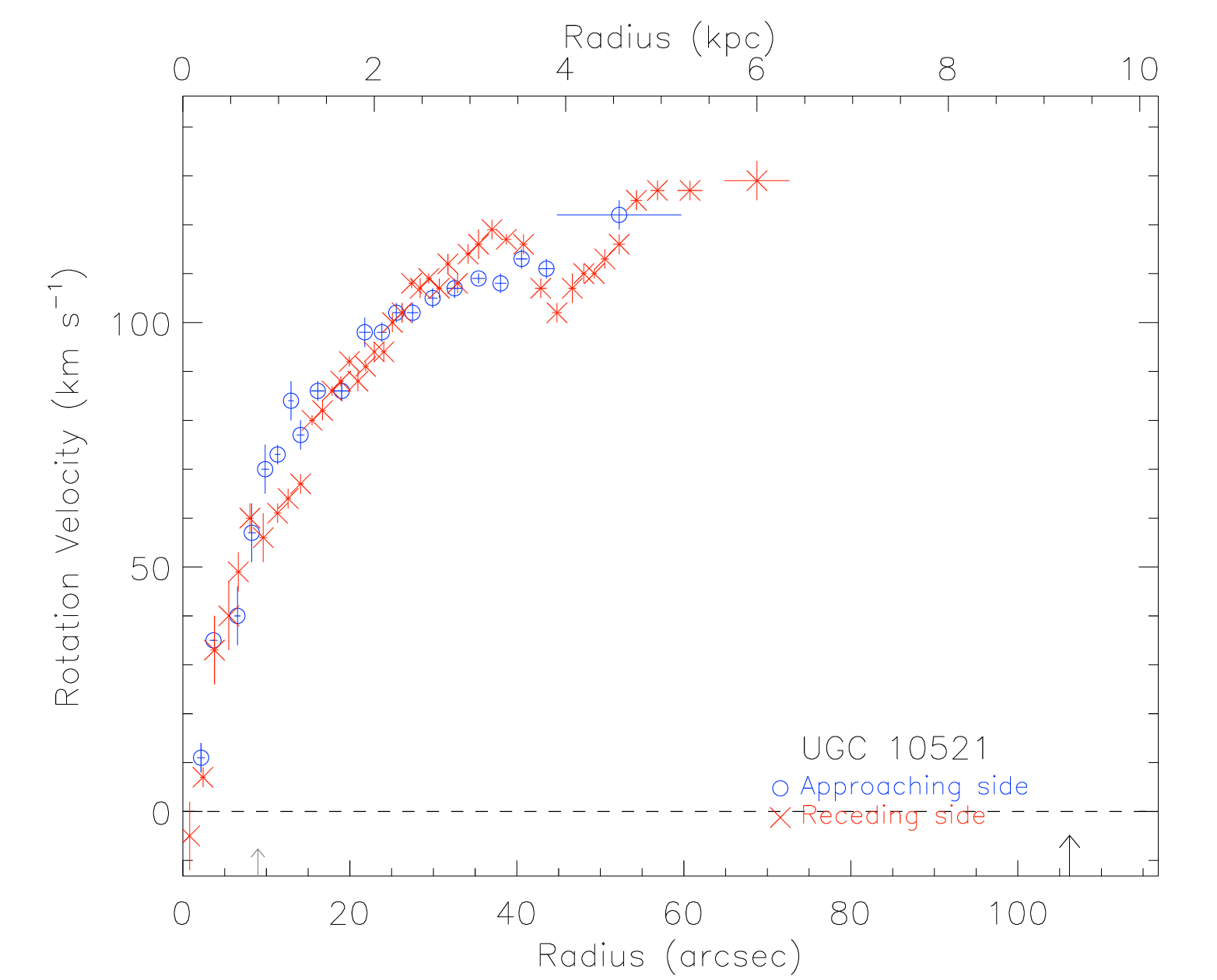}
\end{center}
\caption{From top left to bottom right: \ha~\RC~of UGC 9576, UGC 9736, UGC 9866, UGC 9943, UGC 10075, and UGC 10521.
}
\end{minipage}
\end{figure*}
\clearpage
\begin{figure*}
\begin{minipage}{180mm}
\begin{center}
   \includegraphics[width=8cm]{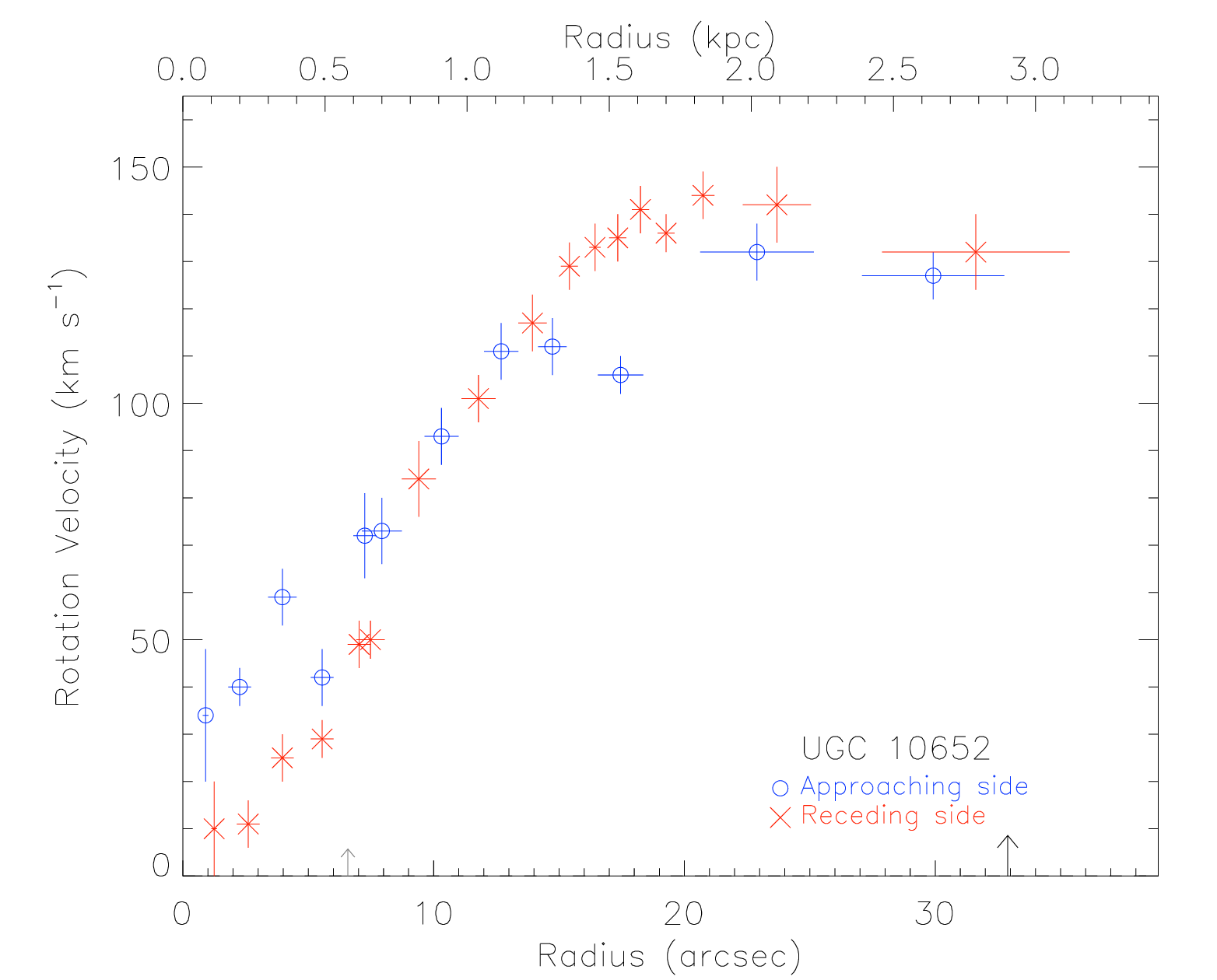}
   \includegraphics[width=8cm]{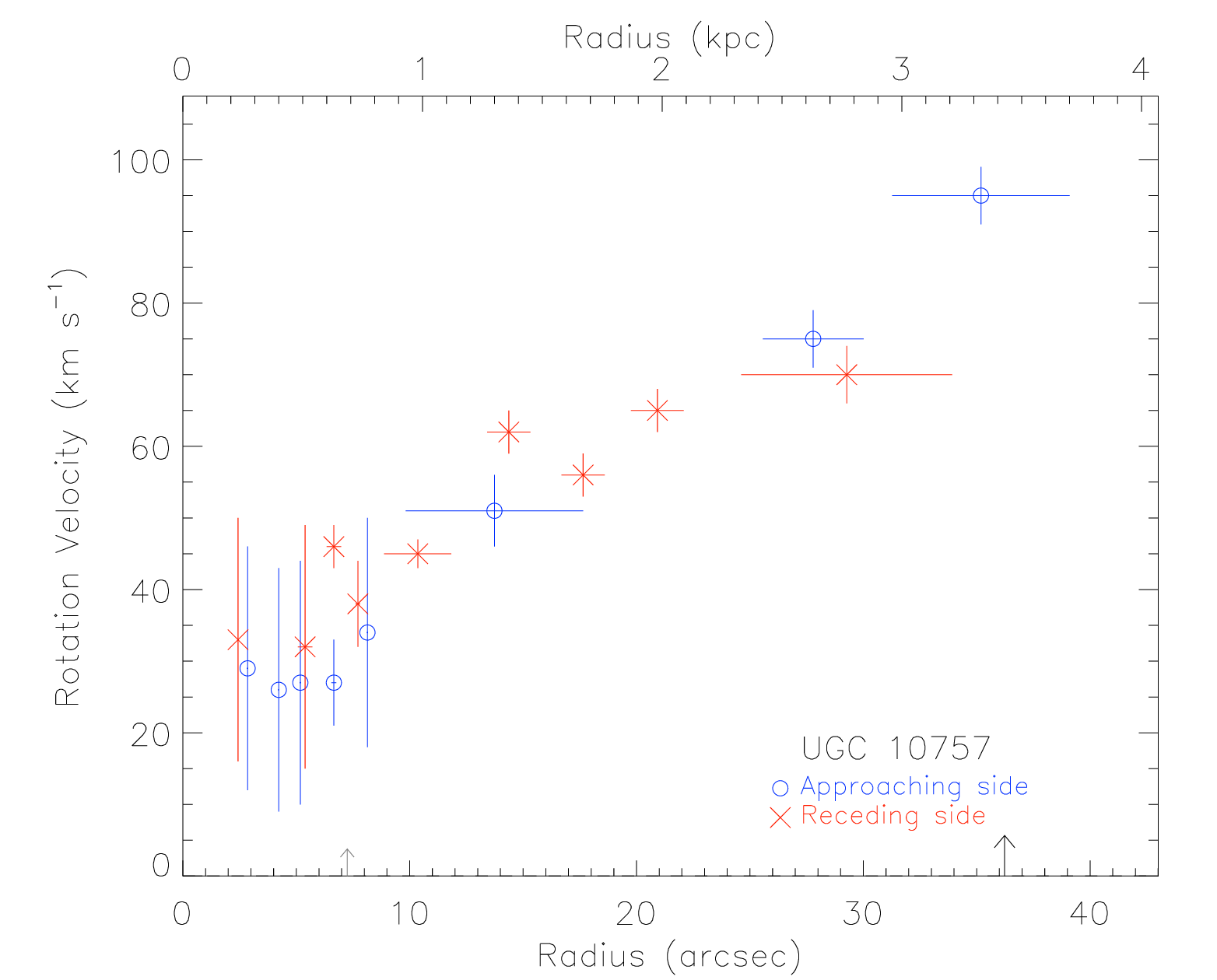}
   \includegraphics[width=8cm]{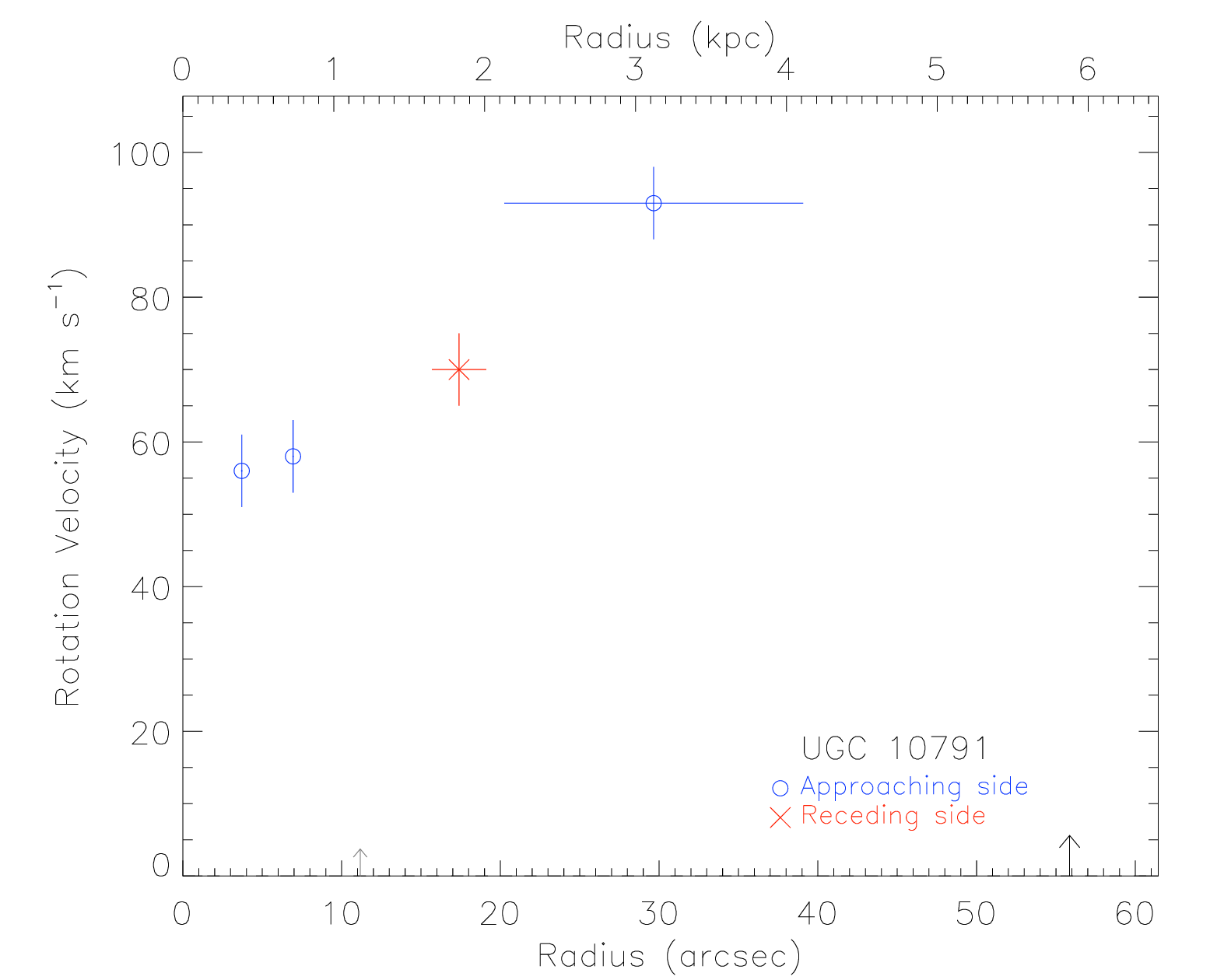}
   \includegraphics[width=8cm]{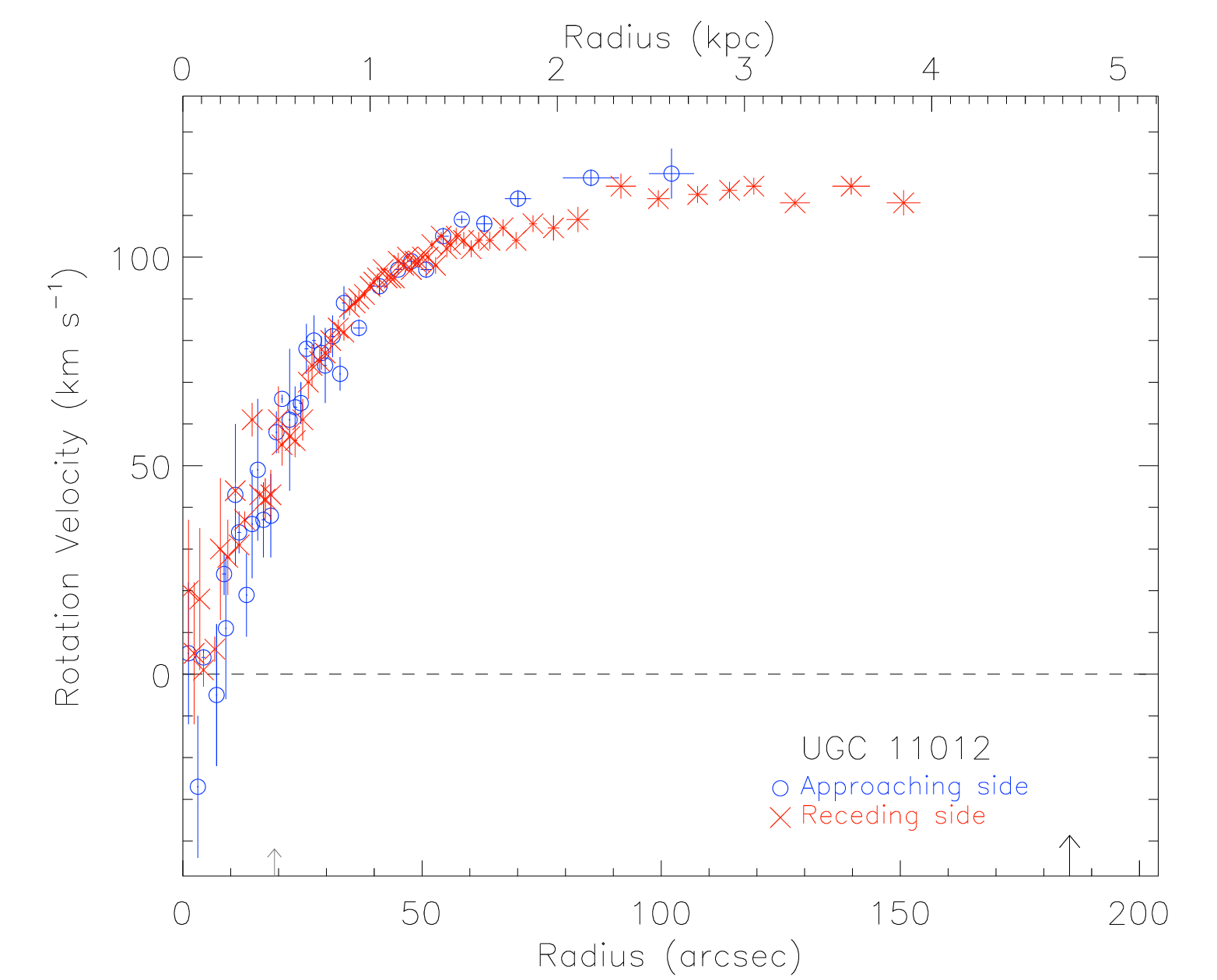}
   \includegraphics[width=8cm]{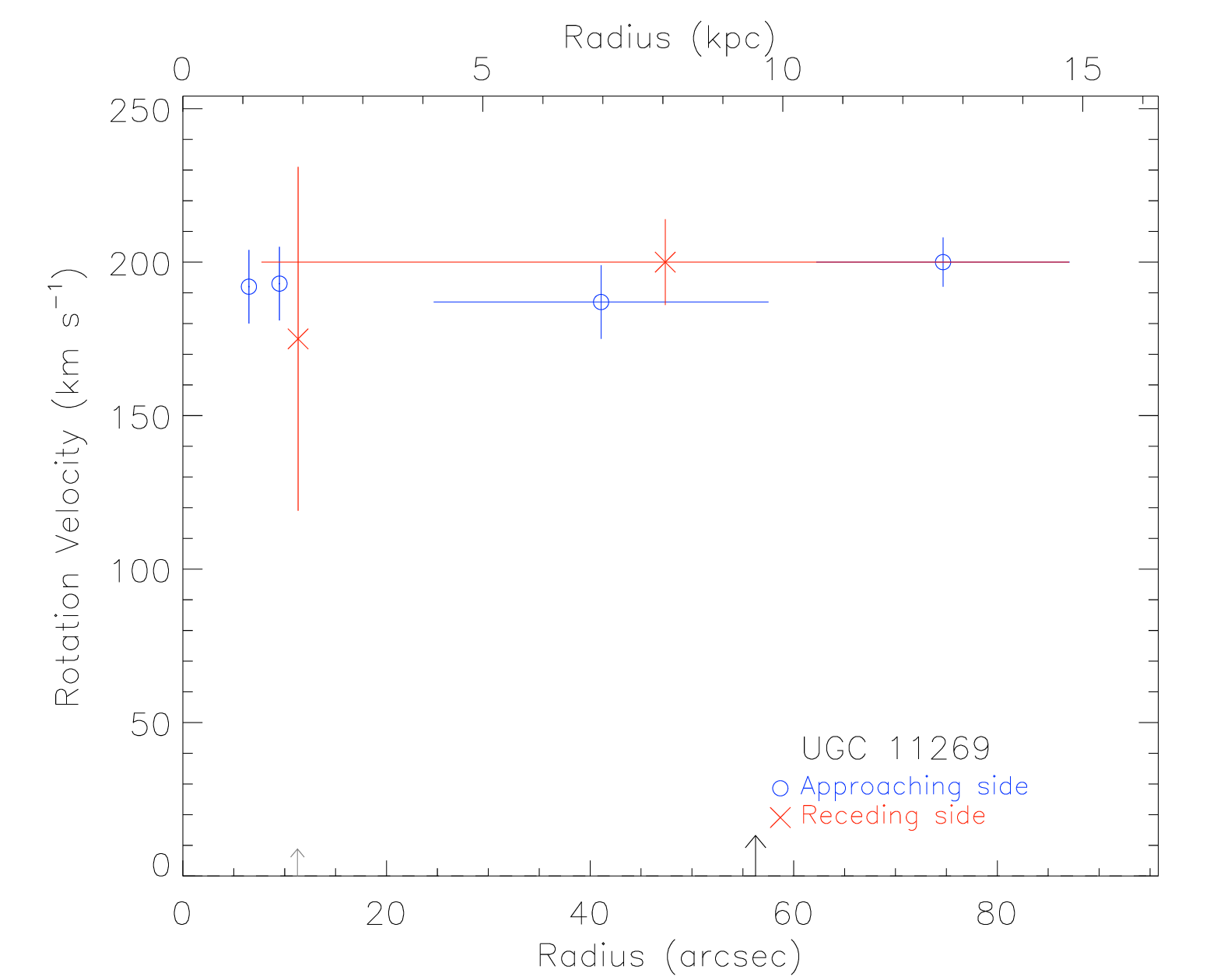}
   \includegraphics[width=8cm]{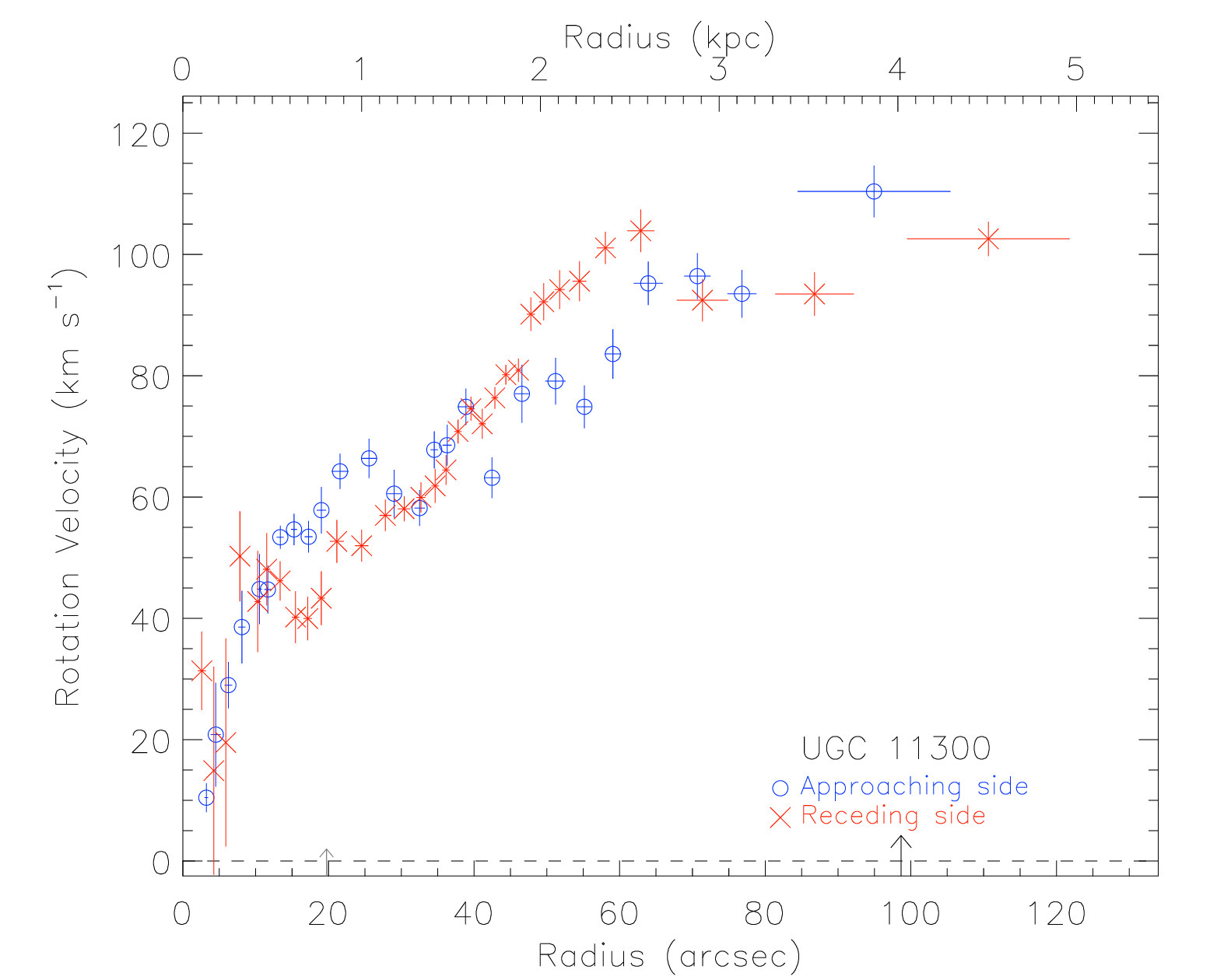}
\end{center}
\caption{From top left to bottom right: \ha~\RC~of UGC 10652, UGC 10757, UGC 10791, UGC 11012, UGC 11269, and UGC 11300.
}
\end{minipage}
\end{figure*}
\clearpage
\begin{figure*}
\begin{minipage}{180mm}
\begin{center}
   \includegraphics[width=8cm]{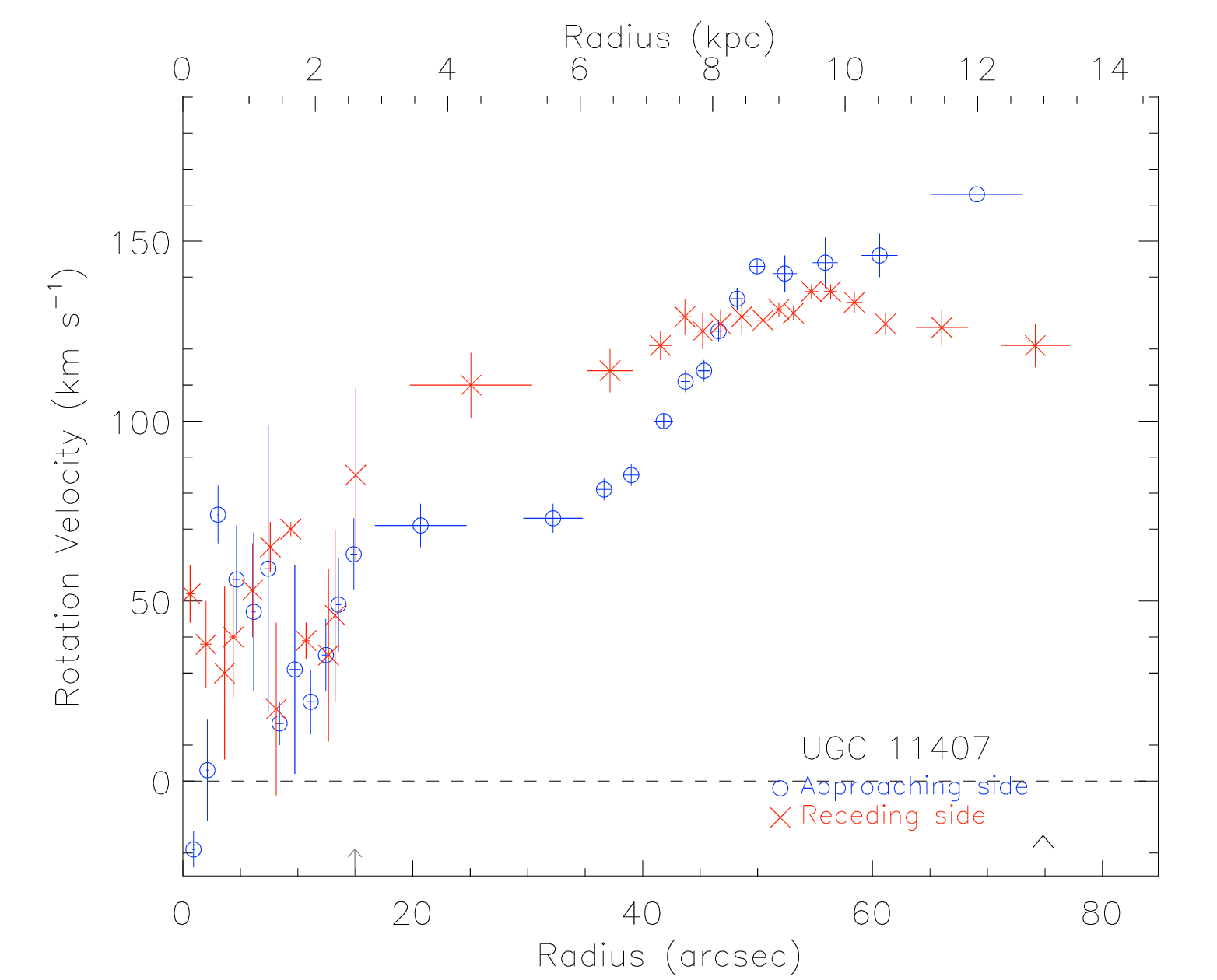}
   \includegraphics[width=8cm]{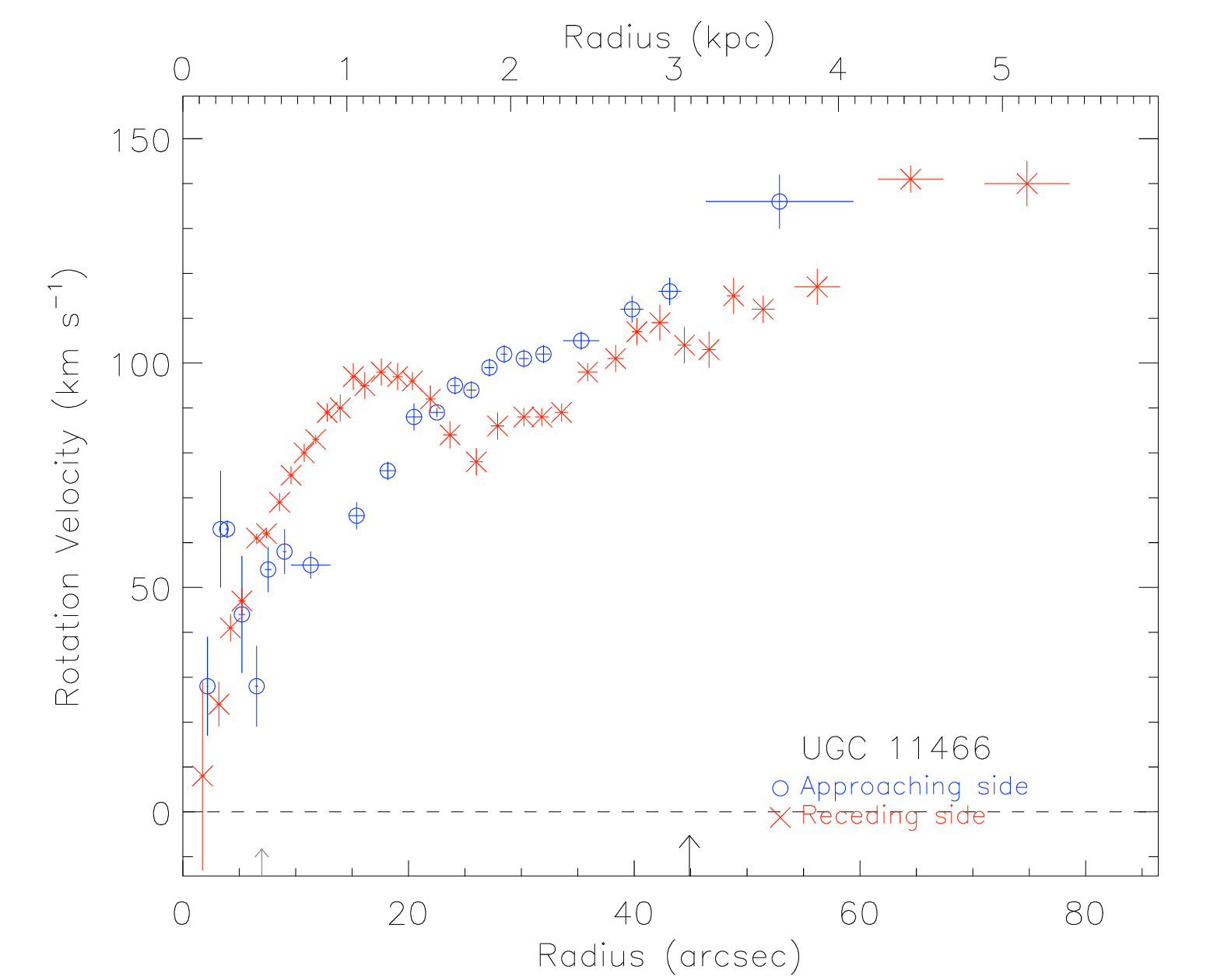}
   \includegraphics[width=8cm]{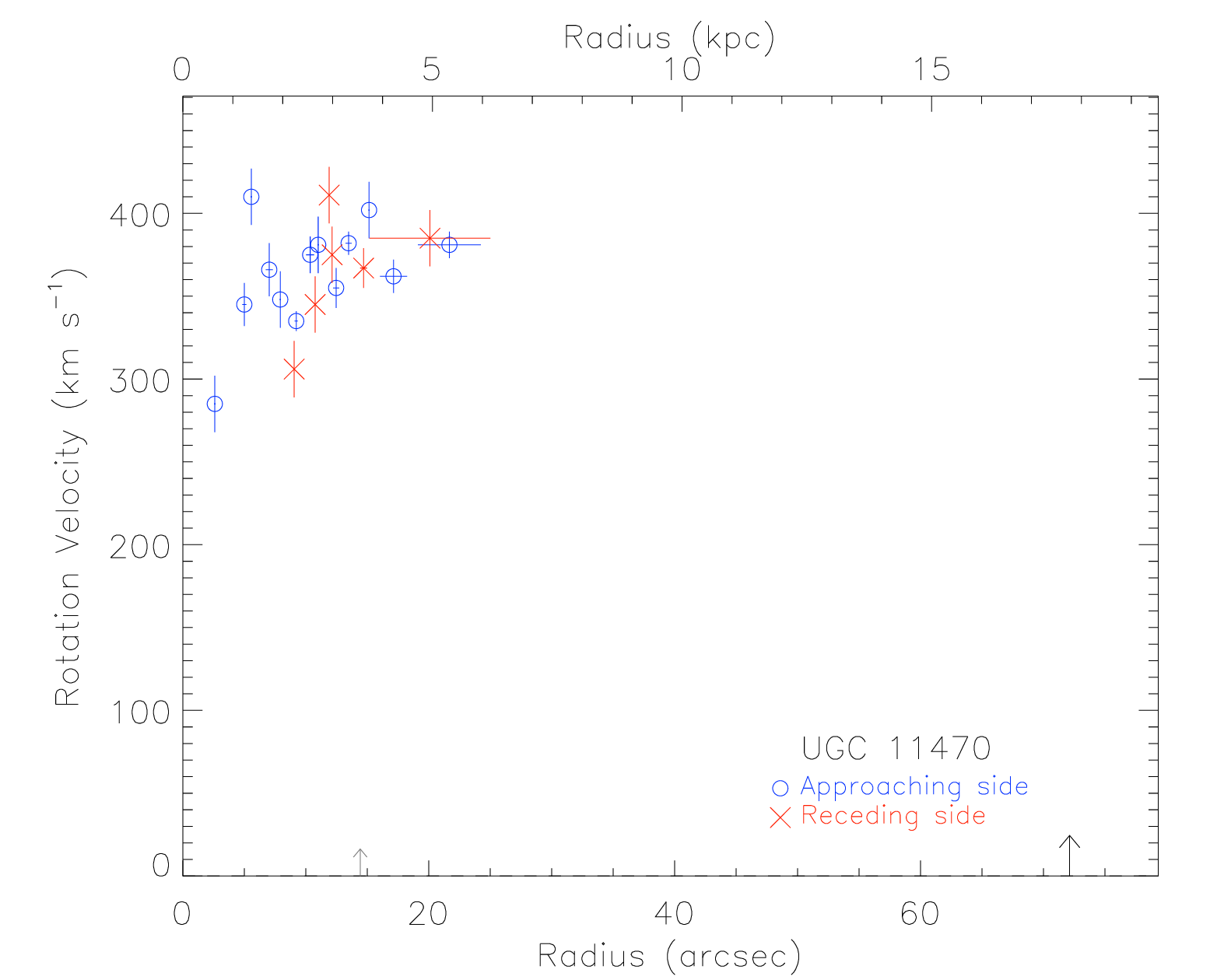}
   \includegraphics[width=8cm]{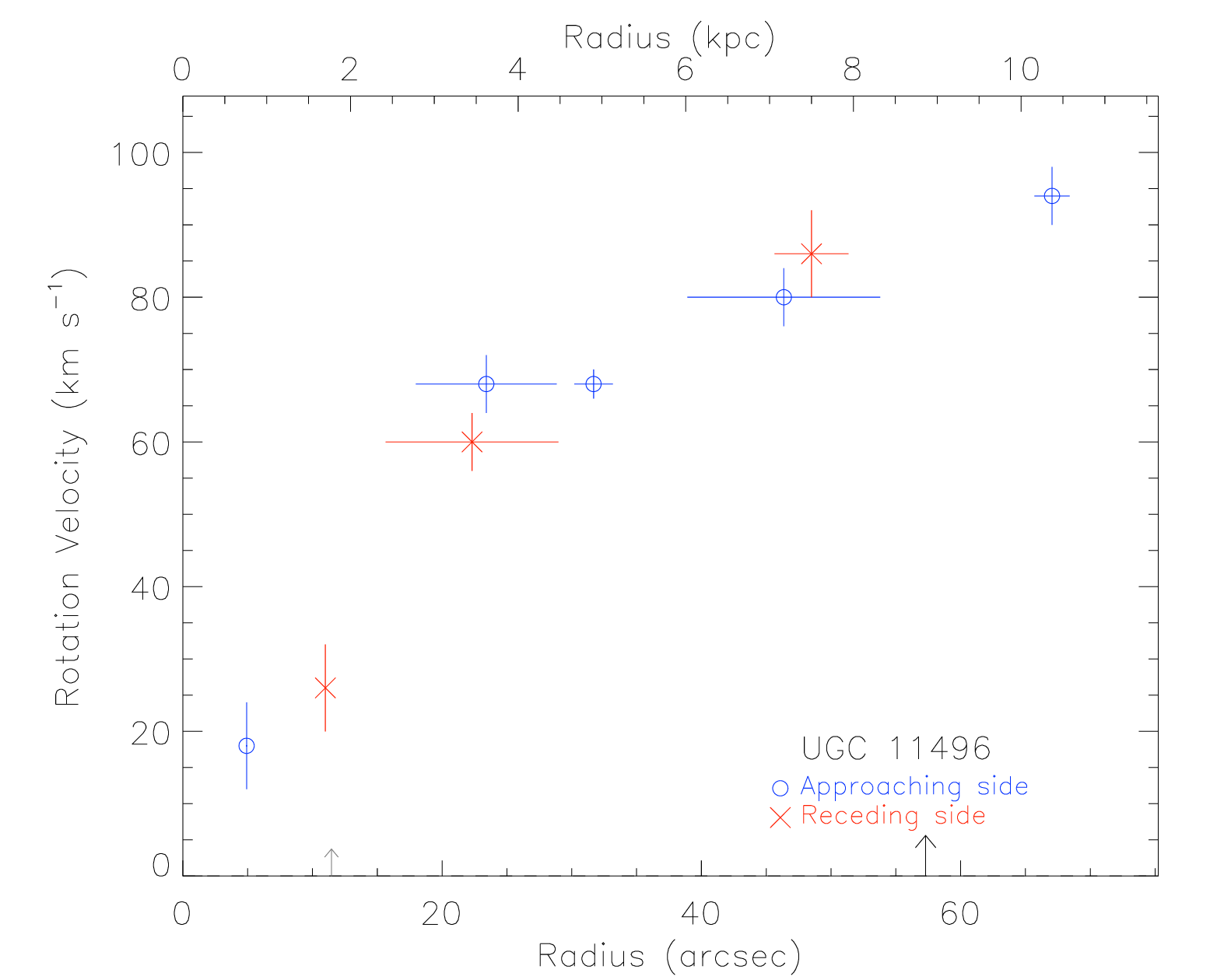}
   \includegraphics[width=8cm]{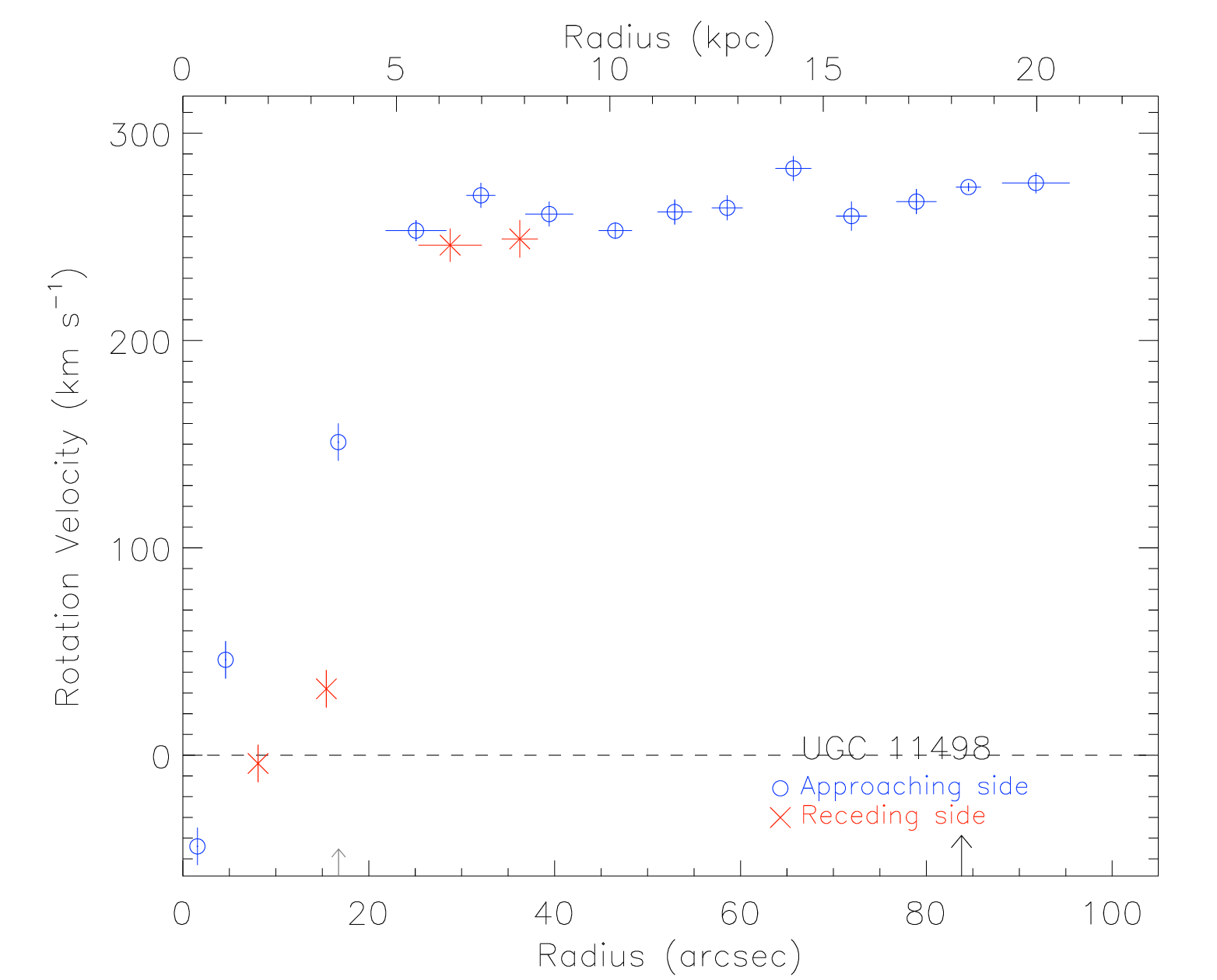}
   \includegraphics[width=8cm]{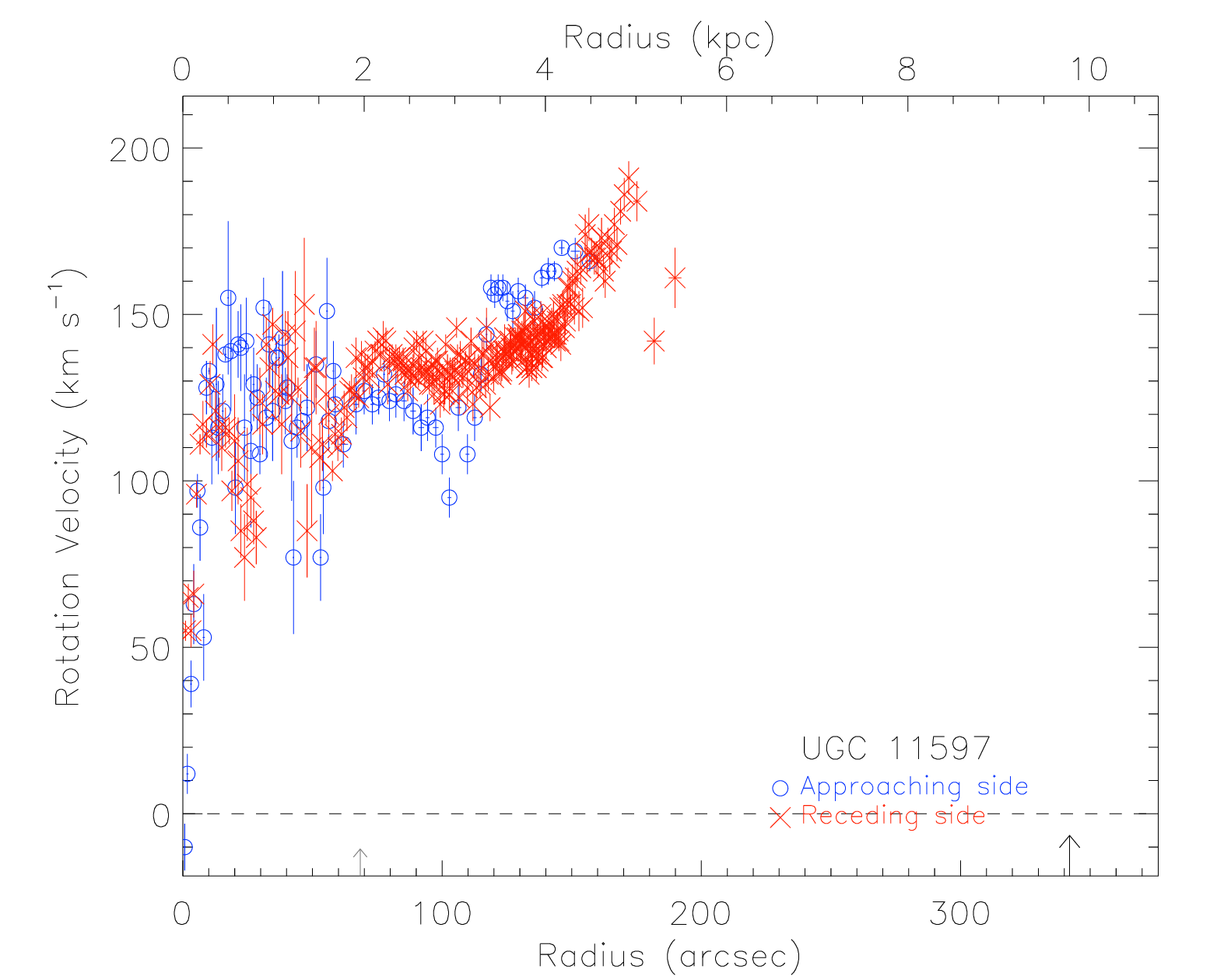}
\end{center}
\caption{From top left to bottom right: \ha~\RC~of UGC 11407, UGC 11466, UGC 11470, UGC 11496, UGC 11498, and UGC 11597.
}
\end{minipage}
\end{figure*}
\clearpage
\begin{figure*}
\begin{minipage}{180mm}
\begin{center}
   \includegraphics[width=8cm]{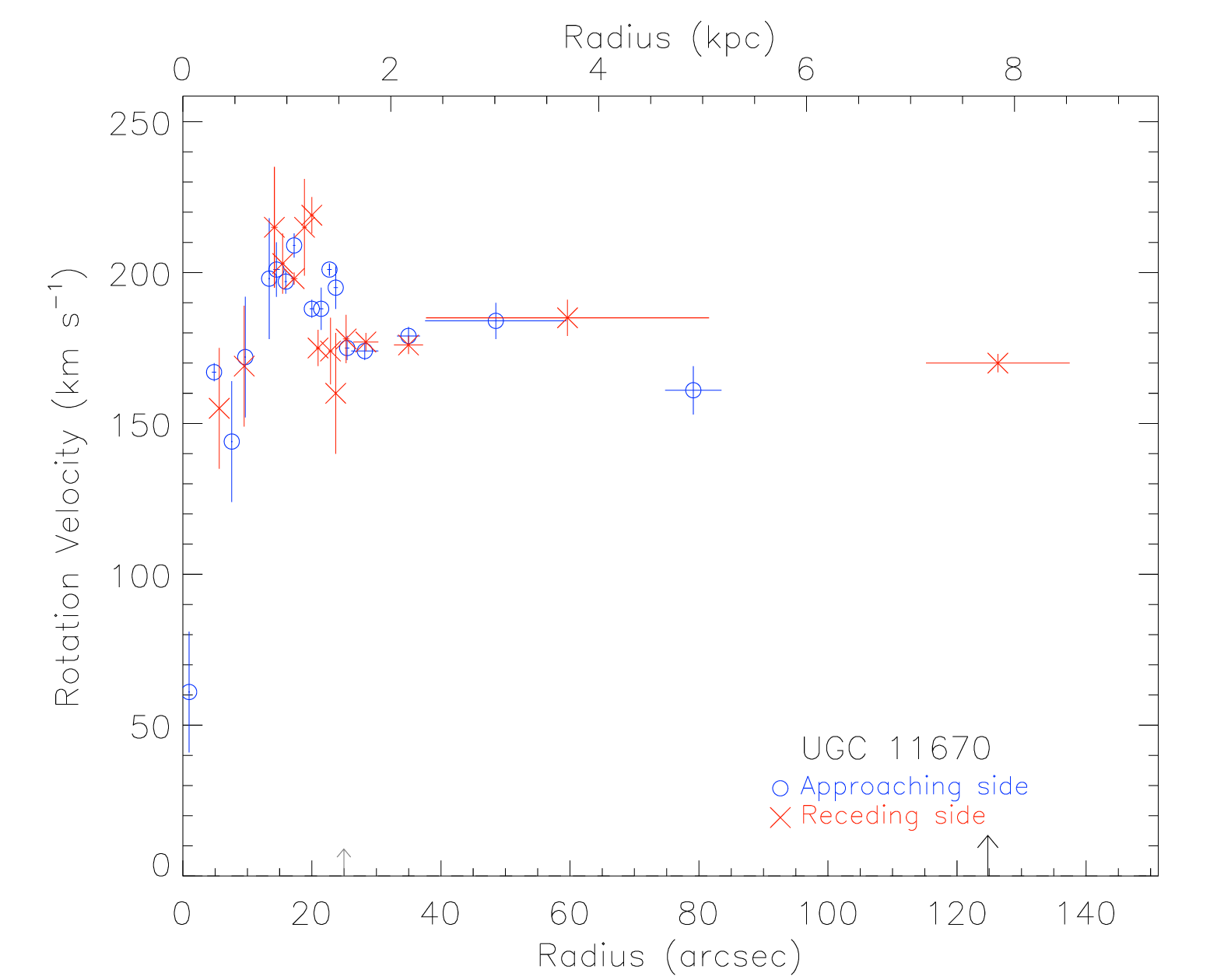}
   \includegraphics[width=8cm]{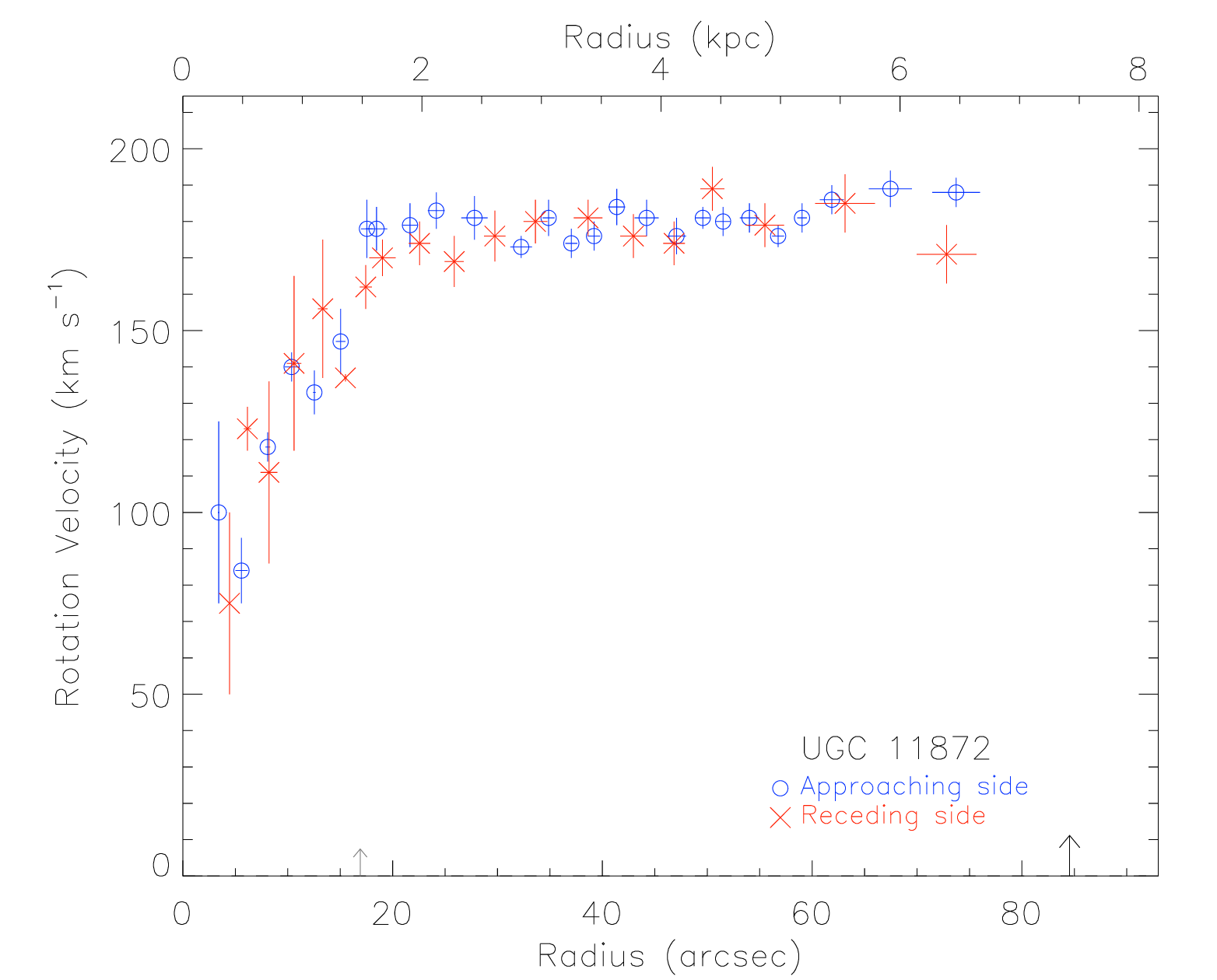}
   \includegraphics[width=8cm]{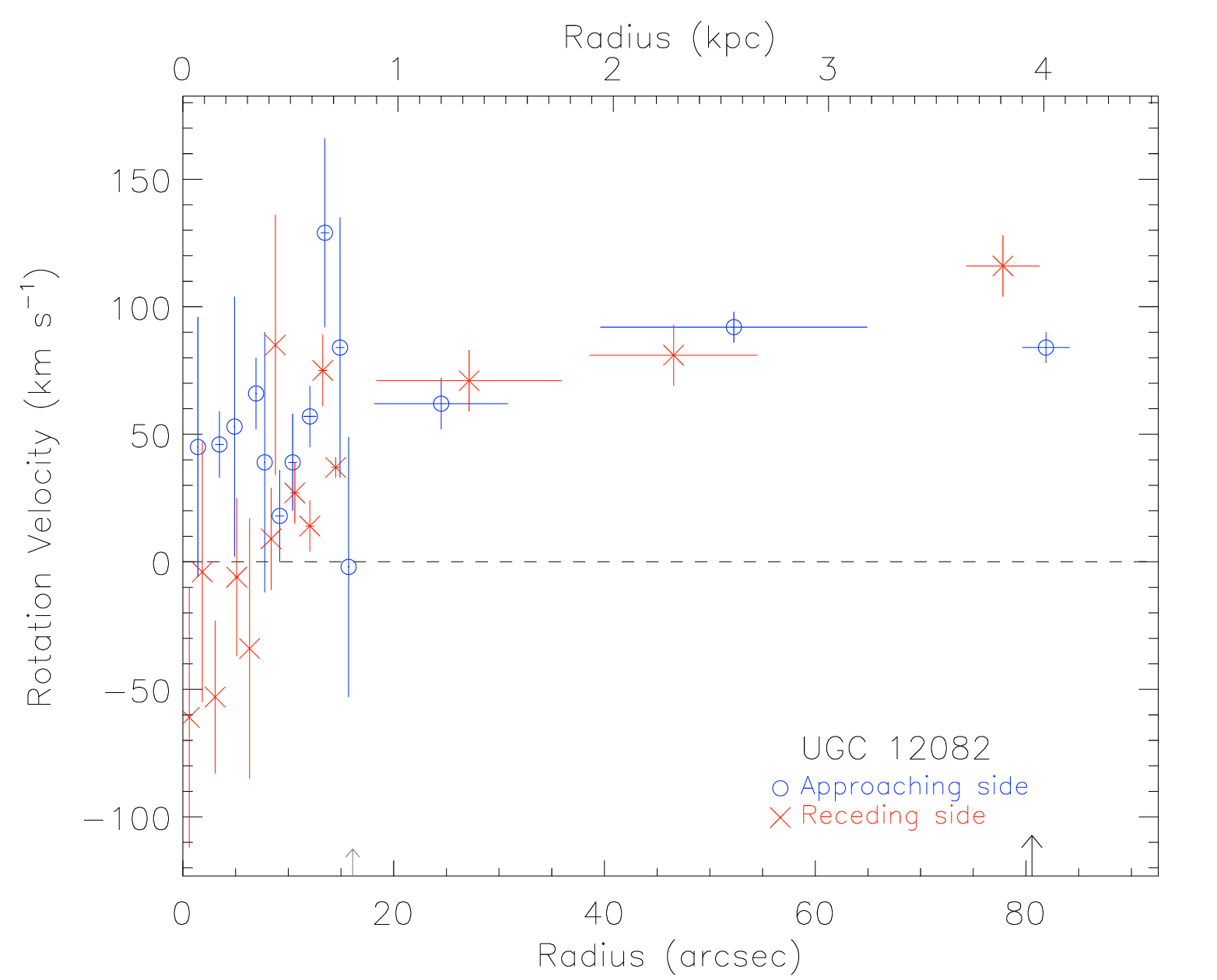}
\end{center}
\caption{From top left to bottom right: \ha~\RC~of UGC 11670, UGC 11872, UGC 12082.
}
\end{minipage}
\end{figure*}
\clearpage

\section{Rotation curves tables}
\label{rc_tables}
\begin{table*}
\caption{UGC 12893 \RC}
\begin{tabular}{cccccccc}
\noalign{\medskip} \hline
           r (kpc)  &  $\sigma_r$ (kpc)  &             r (")  &    $\sigma_r$ (")  &          v (\kms)  & $\sigma_v$ (\kms)  &            N bins  &              side \\
(1)&(2)&(3)&(4)&(5)&(6)&(7)&(8)\\
\hline
            0.28  &            0.00  &             4.6  &             0.0  &              35  &               7  &               1  &               r \\
            0.36  &            0.00  &             5.9  &             0.0  &              13  &               7  &               1  &               a \\
            0.43  &            0.00  &             7.1  &             0.0  &              66  &               7  &               1  &               a \\
            1.33  &            0.64  &            21.9  &            10.6  &              64  &               7  &              22  &               r \\
            1.37  &            0.41  &            22.6  &             6.8  &              67  &               6  &              22  &               a \\
            2.22  &            0.12  &            36.6  &             2.0  &              84  &               5  &               9  &               a \\
\hline
\label{rclabel_ugc12893}
\end{tabular}
\\(1), (3) Galactic radius. (2), (4) Dispersion around the galactic radius. (5) Rotation velocity. (6) Dispersion on the rotation velocity. (7) Number of velocity bins. (8) Receding -- r -- or approaching -- a -- side.
\end{table*}

\begin{table*}
\caption{UGC 89 \RC}
\begin{tabular}{cccccccc}
\noalign{\medskip} \hline
           r (kpc)  &  $\sigma_r$ (kpc)  &             r (")  &    $\sigma_r$ (")  &          v (\kms)  & $\sigma_v$ (\kms)  &            N bins  &              side \\
(1)&(2)&(3)&(4)&(5)&(6)&(7)&(8)\\
\hline
            0.29  &            0.10  &             0.9  &             0.3  &             137  &              16  &               4  &               a \\
            0.32  &            0.03  &             1.0  &             0.1  &             192  &              17  &               2  &               r \\
            0.63  &            0.11  &             2.0  &             0.4  &             304  &              14  &              11  &               r \\
            0.67  &            0.09  &             2.2  &             0.3  &             198  &              12  &               5  &               a \\
            1.04  &            0.13  &             3.3  &             0.4  &             305  &              25  &              11  &               a \\
            1.04  &            0.11  &             3.3  &             0.4  &             340  &              11  &              14  &               r \\
            1.38  &            0.09  &             4.4  &             0.3  &             370  &              16  &               7  &               a \\
            1.44  &            0.11  &             4.6  &             0.4  &             338  &               5  &              13  &               r \\
            1.80  &            0.12  &             5.8  &             0.4  &             317  &              24  &               8  &               a \\
            1.82  &            0.11  &             5.8  &             0.4  &             368  &              10  &               8  &               r \\
            2.20  &            0.11  &             7.1  &             0.4  &             342  &              12  &               7  &               r \\
            2.22  &            0.11  &             7.1  &             0.4  &             281  &              12  &               9  &               a \\
            2.60  &            0.13  &             8.4  &             0.4  &             317  &              11  &               3  &               r \\
            2.70  &            0.09  &             8.7  &             0.3  &             277  &               7  &              11  &               a \\
            3.09  &            0.11  &             9.9  &             0.4  &             298  &              12  &              10  &               a \\
            3.11  &            0.08  &            10.0  &             0.3  &             331  &              18  &               5  &               r \\
            3.32  &            0.20  &            10.7  &             0.6  &             315  &               7  &              23  &               a \\
            3.80  &            0.49  &            12.2  &             1.6  &             324  &              10  &              23  &               r \\
            3.90  &            0.15  &            12.5  &             0.5  &             318  &               8  &              23  &               a \\
            4.37  &            0.13  &            14.0  &             0.4  &             318  &               5  &              23  &               a \\
            4.76  &            0.11  &            15.3  &             0.4  &             309  &               4  &              23  &               a \\
            5.02  &            0.26  &            16.1  &             0.8  &             318  &               9  &              23  &               r \\
            5.22  &            0.14  &            16.8  &             0.4  &             324  &               5  &              23  &               a \\
            5.71  &            0.11  &            18.3  &             0.4  &             297  &               9  &              23  &               r \\
            5.84  &            0.22  &            18.8  &             0.7  &             328  &               7  &              23  &               a \\
            6.27  &            0.19  &            20.1  &             0.6  &             310  &               9  &              23  &               r \\
            6.67  &            0.32  &            21.4  &             1.0  &             356  &               7  &              23  &               a \\
            7.03  &            0.32  &            22.6  &             1.0  &             337  &               5  &              23  &               r \\
            7.60  &            0.22  &            24.4  &             0.7  &             392  &               6  &              23  &               a \\
            8.73  &            0.79  &            28.0  &             2.5  &             353  &              11  &              20  &               r \\
            9.26  &            1.05  &            29.8  &             3.4  &             383  &               6  &              23  &               a \\
\hline
\label{rclabel_ugc89}
\end{tabular}
\\(1), (3) Galactic radius. (2), (4) Dispersion around the galactic radius. (5) Rotation velocity. (6) Dispersion on the rotation velocity. (7) Number of velocity bins. (8) Receding -- r -- or approaching -- a -- side.
\end{table*}


\end{document}